\pdfoutput = 1

\documentclass[letterpaper, nopdfoutputerror,accepted=2025-03-12]{quantumarticle}
\usepackage{geometry}

\usepackage{tikz}
\usetikzlibrary{decorations,decorations.pathmorphing,decorations.markings,calc,snakes,positioning,fixedpointarithmetic,shapes,shapes.geometric,patterns}

\usepackage{etoolbox}
\AtBeginEnvironment{tikzpicture}{\catcode`$=3 } 

\tikzset{alignmid/.style={baseline={([yshift=-.5ex]current bounding box.center)}}} 
\tikzset{every picture/.append style=alignmid}

\tikzset{
bottomzigzag/.style={postaction={draw,decorate, decoration={zigzag,amplitude=1pt,segment length=3pt,raise=1pt}}},
zigzag/.style={draw,decorate, decoration={zigzag,amplitude=1pt,segment length=3pt}},
rc/.style=rounded corners,
}

\tikzset{
    -|/.style={to path={-| (\tikztotarget)}},
    |-/.style={to path={|- (\tikztotarget)}},
}

\usepackage{ifthen}
\usepackage{xparse,pgffor}

\tikzset{
mark/.code={
\tikzset{postaction={/network/mark/.cd,#1,/tikz/.cd,decorate,decoration={name=markings,mark=at position \netmarkpos with{
\begin{scope}[netmarktrafo]
\netmarkcode
\end{scope}
}}}}
\def\netmarkpos{0.5}
},
}

\def\netmarkpos{0.5}
\def\netmarkposoff{0cm}
\def\netmarkcode{}

\tikzset{
netmarktrafo/.style={},
netmarkstyle/.style={solid,semithick,sharp corners},
}

\pgfkeys{
/network/mark/.cd,
sty/.code={
\tikzset{netmarkstyle/.style={#1}}
},
asty/.code={
\tikzset{netmarkstyle/.append style={#1}}
},
f/.style={asty=fill},
p/.code={ 
\def\netmarkpos{#1}
},
poff/.code={ 
\def\netmarkposoff{#1cm}
},
e/.code={ 
\def\netmarkpos{\pgfdecoratedpathlength-0.005cm-\netmarkposoff}

\tikzset{netmarktrafo/.append style={shift={(-\netmarkwidth,0)}}}
},
s/.code={ 
\def\netmarkpos{0.005cm+\netmarkposoff}

\tikzset{netmarktrafo/.append style={shift={(\netmarkwidth,0)},xscale=-1,yscale=-1}}
},
a/.code={ 
\def\netmarkpos{\pgfdecoratedpathlength-0.005cm}

\tikzset{netmarktrafo/.append style={xscale=-1,shift={(-\netmarkwidth,0)}}}
},
b/.code={ 
\def\netmarkpos{0.005cm}

\tikzset{netmarktrafo/.append style={xscale=-1,shift={(\netmarkwidth,0),yscale=-1}}}
},
-/.code={ 
\tikzset{netmarktrafo/.append style={xscale=-1}}
},
r/.code={ 
\tikzset{netmarktrafo/.append style={yscale=-1}}
},
sideoff/.code={ 
\tikzset{netmarktrafo/.append style={shift={(0,#1)}}}
},
slab/.code={ 
\def\netmarkwidth{0}
\def\netmarkcode{
\node[inner sep=0.04cm,netmarkstyle,draw=none] (mylabelwidthtest) at (0,0){\phantom{#1}};
\path let \p1=(mylabelwidthtest.north east), \p2=(mylabelwidthtest.south east), \n1 = {max(abs(\y1),abs(\y2))} in node[inner sep=0.04cm,netmarkstyle] at (0,\n1) {#1};
}
},
lab/.code={ 
\def\netmarkwidth{0}
\def\netmarkcode{
\node[inner sep=0.04cm,anchor=\netmarkanchor] (mylabelwidthtest) at (0,0) {\phantom{#1}};
\draw[white] (mylabelwidthtest.\pgfdecoratedangle)--(mylabelwidthtest.\pgfdecoratedangle+180);
\node[inner sep=0.04cm,anchor=\netmarkanchor,netmarkstyle] at (0,0) {#1};
}
},
rotlab/.code={ 
\def\netmarkwidth{0}
\def\netmarkcode{
\node[inner sep=0.04cm,fill=white,transform shape,rotate=90,anchor=\netmarkrotanchor,netmarkstyle] (mydecorationnodename) at (0,0) {#1};
}
},
mrotlab/.style={ 
rotlab={$\scriptstyle{#1}$}
},
arr/.code={ 
\def\netmarkwidth{0.04}
\def\netmarkcode{\draw[netmarkstyle] (-0.04,0.08)--(0.04,0)--(-0.04,-0.08);}
},
arrbar/.code={ 
\def\netmarkwidth{0.08}
\def\netmarkcode{\draw[netmarkstyle] (-0.08,0.08)--(0,0)--(-0.08,-0.08) (0.04,0.08)--(0.04,-0.08);}
},
rarr/.code={ 
\def\netmarkwidth{0.04}
\def\netmarkcode{\draw[netmarkstyle] (-0.04,-0.08)arc(90-180:90:0.08);}
},
dot/.code={ 
\def\netmarkwidth{0.08}
\def\netmarkcode{\draw[netmarkstyle] (0,0)circle(0.08);}
},
flag/.code={ 
\def\netmarkwidth{0.06}
\def\netmarkcode{\draw[netmarkstyle] (-0.06,0)--(0,0.09)--(0.06,0)--cycle;}
},
darr/.code={ 
\def\netmarkwidth{0.08}
\def\netmarkcode{\draw[netmarkstyle] (-0.04,0)--(0.04,0)--(-0.04,0.08)--cycle;}
},
circ/.code={
\def\netmarkwidth{0.1}
\def\netmarkcode{\draw[netmarkstyle] (0,0) circle (0.1);}
},
hcirc/.code={ 
\def\netmarkwidth{0.1}
\def\netmarkcode{\draw[netmarkstyle] (-0.1,0) arc (180:0:0.1);}
},
diamond/.code={
\def\netmarkwidth{0.1}
\def\netmarkcode{\draw[netmarkstyle] (-0.1,0)--(0,-0.1)--(0.1,0)--(0,0.1)--cycle;}
},
bar/.code={ 
\def\netmarkwidth{0.05}
\def\netmarkcode{
\draw[netmarkstyle] (0,-0.08cm-0.5*\pgflinewidth)--(0,0.08cm+0.5*\pgflinewidth);
}
},
dbar/.code={ 
\def\netmarkwidth{0.13}
\def\netmarkcode{
\draw[netmarkstyle] (-0.04cm,-0.08cm-0.5*\pgflinewidth)--(-0.04cm,0.08cm+0.5*\pgflinewidth) (0.04cm,-0.08cm-0.5*\pgflinewidth)--(0.04cm,0.08cm+0.5*\pgflinewidth);
}
},
hbar/.code={ 
\def\netmarkwidth{0.05}
\def\netmarkcode{
\draw[netmarkstyle] (0, 0.5*\pgflinewidth)--++(0,0.12);
}
},
mill/.code={
\def\netmarkwidth{0.16}
\def\netmarkcode{
\draw[netmarkstyle] (0,-0.5*\pgflinewidth)--++(-0.08,-0.08)--++(0,0.08);
\draw[netmarkstyle] (0,0.5*\pgflinewidth)--++(0.08,0.08)--++(0,-0.08);
}
},
three dots/.code={
\def\netmarkwidth{0.2}
\def\netmarkcode{
\fill (-0.12,0) circle (0.5*0.05) (0,0) circle (0.5*0.05) (0.12,0) circle (0.5*0.05);
}
},
}

\tikzset{wid/.style={minimum width=#1cm}}
\tikzset{hei/.style={minimum height=#1cm}}

\tikzset{sx/.style={xshift=#1cm}}
\tikzset{sy/.style={yshift=#1cm}}

\tikzset{box/.style={draw,rectangle}}
\tikzset{fbox/.style={draw,rectangle, line width=1.1}}
\tikzset{roundbox/.style={draw,rectangle,rounded corners}}
\tikzset{froundbox/.style={draw,rectangle, rounded corners, line width=1.1}}
\tikzset{rounddiamond/.style={draw,diamond,rounded corners}}
\tikzset{dot/.style={draw, shape=circle, fill=black, scale=0.5}}

\tikzset{
netbox/.code={
\node[draw,netbdstyle] (\atomname) at (0,0) {#1};
\coordinate (\atomname-r) at (\atomname.east);
\coordinate (\atomname-l) at (\atomname.west);
\coordinate (\atomname-t) at (\atomname.north);
\coordinate (\atomname-b) at (\atomname.south);
\coordinate (\atomname-tr) at (\atomname.north east);
\coordinate (\atomname-br) at (\atomname.south east);
\coordinate (\atomname-tl) at (\atomname.north west);
\coordinate (\atomname-bl) at (\atomname.south west);
},
}

\tikzset{bdlw/.code={\tikzset{mybdstyle/.style={draw, line width=#1}}}}
\tikzset{bdcol/.code={\tikzset{mybdstyle/.append style={#1}}}}

\newcommand\setelements[1]{
\pgfkeys{/network/atom/.cd,#1}
}

\newcommand\setmarks[1]{
\pgfkeys{/network/mark/.cd,#1}
}

\newcommand\atoms[2]{
\foreach \name/\keys in {#2}{
\expandafter\atom\expandafter{\keys,#1}{\name}
}
}

\newcommand\atom[2]{
\def\atomname{#2}
\tikzset{
nettrafo/.style={},
netatompos/.style={},
netdeco/.style={},
netpostdeco/.style={},
}

\pgfkeys{/network/atom/.cd,#1}

\begin{scope}[netatompos] 
\begin{scope}[nettrafo] 
\netshapecoords 
\fill[netbackstyle] \netshapepath;
\clip \netshapepath;
\tikzset{netdeco}
\draw[netbdstyle] \netshapepath;
\end{scope}
\tikzset{netpostdeco} 
\end{scope}

}

\pgfkeys{
/network/atom/.cd,
square/.code={

\def\netshapepath{(-\tempsize,-\tempsize)rectangle (\tempsize,\tempsize)}
\def\netshapecoords{
\node[rectangle,wid=2*\tempsize,hei=2*\tempsize,inner sep=0,transform shape](\atomname)at(0,0){};
\coordinate(\atomname-c) at (0,0);
\coordinate(\atomname-r) at (\tempsize,0);
\coordinate(\atomname-l) at (-\tempsize,0);
\coordinate(\atomname-t) at (0,\tempsize);
\coordinate(\atomname-b) at (0,-\tempsize);
\coordinate(\atomname-br) at (\tempsize,-\tempsize);
\coordinate(\atomname-tr) at (\tempsize,\tempsize);
\coordinate(\atomname-bl) at (-\tempsize,-\tempsize);
\coordinate(\atomname-tl) at (-\tempsize,\tempsize);
}},
circ/.code={

\def\netshapepath{(0,0)circle(\tempsize)}
\def\netshapecoords{
\node[circle,wid=2*\tempsize,hei=2*\tempsize,inner sep=0,transform shape](\atomname)at(0,0){};
\coordinate(\atomname-c) at (0,0);
\coordinate(\atomname-r) at (\tempsize,0);
\coordinate(\atomname-l) at (-\tempsize,0);
\coordinate(\atomname-t) at (0,\tempsize);
\coordinate(\atomname-b) at (0,-\tempsize);
}},
triang/.code={

\def\netshapepath{(-30:\tempsize)--(90:\tempsize)--(-150:\tempsize)--cycle}
\def\netshapecoords{
\node[regular polygon,regular polygon sides=3,wid=2*\tempsize,inner sep=0,transform shape](\atomname)at(0,0){};
\coordinate(\atomname-c) at (0,0);
\coordinate(\atomname-cr) at (-30:\tempsize);
\coordinate(\atomname-cl) at (-150:\tempsize);
\coordinate(\atomname-ct) at (90:\tempsize);
\coordinate(\atomname-mb) at (-90:0.5*\tempsize);
\coordinate(\atomname-mr) at (30:0.5*\tempsize);
\coordinate(\atomname-ml) at (150:0.5*\tempsize);
}},
diamond/.code={

\def\netshapepath{(0,-\tempsize)--(\tempsize,0)--(0,\tempsize)--(-\tempsize,0)--cycle}
\def\netshapecoords{
\node[rotate=45,rectangle,wid=sqrt(2)*\tempsize,hei=sqrt(2)*\tempsize,inner sep=0,transform shape](\atomname)at(0,0){};
\coordinate(\atomname-c) at (0,0);
\coordinate(\atomname-r) at (\tempsize,0);
\coordinate(\atomname-l) at (-\tempsize,0);
\coordinate(\atomname-t) at (0,\tempsize);
\coordinate(\atomname-b) at (0,-\tempsize);
}},
pentagon/.code={

\def\netshapepath{(-126:\tempsize)--(-54:\tempsize)--(18:\tempsize)--(90:\tempsize)--(162:\tempsize)--cycle}
\def\netshapecoords{
\node[regular polygon,regular polygon sides=5,wid=2*\tempsize,inner sep=0,transform shape](\atomname)at(0,0){};
\coordinate(\atomname-c) at (0,0);
\coordinate (\atomname-mb)at(-90:{\tempsize*cos(36)});
\coordinate (\atomname-mbr)at(-18:{\tempsize*cos(36)});
\coordinate (\atomname-mtr)at(54:{\tempsize*cos(36)});
\coordinate (\atomname-mtl)at(126:{\tempsize*cos(36)});
\coordinate (\atomname-mbl)at(-162:{\tempsize*cos(36)});
\coordinate (\atomname-cbr)at(-54:\tempsize);
\coordinate (\atomname-cr)at(18:\tempsize);
\coordinate (\atomname-ct)at(90:\tempsize);
\coordinate (\atomname-cl)at(162:\tempsize);
\coordinate (\atomname-cbl)at(-126:\tempsize);
}},
halfcirc/.code={

\def\netshapepath{(\tempsize,0)arc(0:180:\tempsize)--++(0,-0.04)-|cycle}
\def\netshapecoords{
\node[circle,wid=2*\tempsize,hei=2*\tempsize,inner sep=0,transform shape](\atomname)at(0,0){};
\coordinate(\atomname-c) at (0,0);
\coordinate(\atomname-r) at (\tempsize,0);
\coordinate(\atomname-l) at (-\tempsize,0);
\coordinate(\atomname-t) at (0,\tempsize);
\coordinate(\atomname-b) at (0,0);
}},
void/.code={
\def\netshapepath{}
\def\netshapecoords{
\coordinate(\atomname) at (0,0);
\coordinate(\atomname-c) at (0,0);
}},
labbox/.code={
\def\netshapepath{(0,0)}
\def\netshapecoords{}
\tikzset{netpostdeco/.append style={netbox=#1}}
},
}

\pgfkeys{
/network/atom/.cd,
p/.code={\tikzset{netatompos/.append style={shift={(#1)}}}},
xscale/.code={\tikzset{nettrafo/.append style={xscale=#1}}},
yscale/.code={\tikzset{nettrafo/.append style={yscale=#1}}},
scale/.code={\tikzset{nettrafo/.append style={scale=#1}}},
rot/.code={\tikzset{nettrafo/.append style={rotate=#1}}},
hflip/.style={xscale=-1},
vflip/.style={yscale=-1},
big/.style={scale=3/2},
small/.style={scale=2/3},
huge/.style={scale=9/4},
tiny/.style={scale=4/9},
flat/.style={yscale=2/3},
fflat/.style={yscale=4/9},
}

\tikzset{
netbdstyle/.style={line width=0.15em}, 
netdecstyle/.style={},
netpostdecstyle/.style={},
netbackstyle/.style={white},
}
\pgfkeys{
/network/atom/.cd,
bdstyle/.code={\tikzset{netbdstyle/.style={#1}}},
bdastyle/.code={\tikzset{netbdstyle/.append style={#1}}},
decstyle/.code={\tikzset{netdecstyle/.style={#1}}},
decastyle/.code={\tikzset{netdecstyle/.append style={#1}}},
postdecstyle/.code={\tikzset{netpostdecstyle/.style={#1}}},
postdecastyle/.code={\tikzset{netpostdecstyle/.append style={#1}}},
backstyle/.code={\tikzset{netbackstyle/.style={#1}}},
backastyle/.code={\tikzset{netbackstyle/.append style={#1}}},
style/.style={bdstyle=#1, decstyle=#1, postdecstyle=#1},
astyle/.style={bdastyle=#1, decastyle=#1, postdecastyle=#1},
whiteblack/.style={decstyle=white,backstyle=black},
nobd/.style={bdstyle={line width=0}},
}

\tikzset{
netbscope/.code={\begin{scope}[#1]},
netescope/.code={\end{scope}},
}

\pgfkeys{
/network/atom/.cd,
bscope/.code={\tikzset{netdeco/.append style={netbscope={#1}}}},
escope/.code={\tikzset{netdeco/.append style={netescope}}},
dec/.style 2 args={bscope={#1},#2,escope}, 
vbar/.style={dec={rotate=90}{hbar}},
dcross/.style={dec={rotate=45}{hbar},dec={rotate=-45}{hbar}},
cross/.style={hbar,vbar},
sect6/.style={sect=60},
sect8/.style={sect=45},
sect10/.style={sect=36},
}

\def\regdec#1{\pgfkeys{/network/atom/.cd,#1/.code={\tikzset{netdeco/.append style={net#1}}}}}
\regdec{all}\regdec{rhalf}\regdec{rquart}\regdec{brquart}\regdec{dot}\regdec{spiral}\regdec{swirl}\regdec{hstripe}\regdec{hbar}\regdec{rrey}
\pgfkeys{/network/atom/.cd,sect/.code={\tikzset{netdeco/.append style={netsect=#1}}}}

\tikzset{
netall/.code={\fill[netdecstyle] (-0.3,-0.3)rectangle (0.3,0.3);}, 
netrhalf/.code={\fill[netdecstyle] (0,-0.3)rectangle (0.3,0.3);}, 
netrquart/.code={\fill[netdecstyle] (0.075,-0.3)rectangle (0.3,0.3);}, 
netbrquart/.code={\fill[netdecstyle] (0,0)rectangle (0.3,-0.3);}, 
netsect/.code={\fill[netdecstyle] (0,0)--(0,-0.3)arc(-90:-90+#1:0.3)--cycle;}, 
netdot/.code={\fill[netdecstyle] (0,0)circle(0.07);}, 
netspiral/.code={\draw[netdecstyle] plot [variable=\t,domain=0:4] ({0.075*\t*cos(pi*(\t-0.5) r)},{0.075*\t*sin(pi*(\t-0.5) r)});}, 
netswirl/.code={\fill[netdecstyle] plot [variable=\t,domain=0:2] ({0.15*\t*cos(pi*(\t-0.5) r)},{0.15*\t*sin(pi*(\t-0.5) r)}) arc(-90:-450:0.3)--cycle;}, 
nethstripe/.code={\fill[netdecstyle] (-0.3,-0.05)rectangle(0.3,0.05);}, 
nethbar/.code={\draw[netdecstyle] (-0.3,0)--(0.3,0);}, 
netrrey/.code={\draw[netdecstyle] (0,0)--(0.3,0);} 
}

\pgfkeys{
/network/atom/.cd,
postbscope/.code={\tikzset{netpostdeco/.append style={netbscope={#1}}}},
postescope/.code={\tikzset{netpostdeco/.append style={netescope}}},
postdec/.style 2 args={postbscope={#1},#2,postescope}, 
arc/.code={\tikzset{netpostdeco/.append style={netarc=#1}}},
lab/.code={\tikzset{netpostdeco/.append style={netlab={#1}}}},
shadecirc/.code={\tikzset{netpostdeco/.append style=netshadecirc}},
shaderect/.code={\tikzset{netpostdeco/.append style={netshaderect={#1}}}},
debug/.code={\tikzset{netpostdeco/.append style=netdebug}},
markline/.code 2 args={\tikzset{netpostdeco/.append style={netmarkline={#1}{#2}}}},
postcirc/.code={\tikzset{netpostdeco/.append style=netpostcirc}},
}

\tikzset{
netlab/.code={
\pgfkeys{/network/atom/lab/.cd,#1}
\node[netpostdecstyle] at (\ifdefined\netlabpos\netlabpos\else\netlabang:\netlabdist\fi) {\netlabwrap{\netlabtext}};
},
netarc/.code args={#1:#2:#3}{
\draw[netpostdecstyle] (#1:#3) arc (#1:#2:#3);
},
netshadecirc/.code= {
\fill[opacity=0.4,netpostdecstyle] (0,0)circle(0.4);
},
netpostcirc/.code= {
\draw[netpostdecstyle] (0,0)circle(0.15);
},
netshaderect/.code= {
\fill[rc,opacity=0.4,netpostdecstyle] ($-1*(#1)$) rectangle (#1);
},
netdebug/.code= {
\node[red] at (0,0){\atomname};
},
netmarkline/.code 2 args= {
\draw (\atomname)edge[mark={#2}]++(#1);
},
}

\def\netlabwrap#1{#1}
\pgfkeys{
/network/atom/lab/.cd,
t/.code={\def\netlabtext{#1}}, 
p/.code={\def\netlabpos{#1}}, 
ang/.code={\def\netlabang{#1}},
dist/.code={\def\netlabdist{#1}},
wrap/.code={\def\netlabwrap##1{#1}}, 
}

\usetikzlibrary{patterns.meta}

\usepackage{xcolor}
\usepackage{amsmath}
\usepackage{amssymb}
\usepackage{amsthm}
\theoremstyle{definition}
\newtheorem{mydef}{Definition}
\newtheorem{myprop}{Proposition}
\newtheorem{mythm}{Theorem}

\usepackage{bbold}
\usepackage{mathtools}
\usepackage{hyperref}
\usepackage{makecell}
\usepackage{braket}
\usepackage{diagbox}

\newcommand\tcr[1]{\textcolor{gray}{#1}} 

\def\zz{\mathbb{Z}}

\def\rr{\mathbb{R}}
\def\idop{\mathbb{1}}

\def\Abar{\overline A}
\def\bbar{\overline b}
\def\cbar{\overline c}
\def\Sbar{\overline S}
\def\lagr{\mathcal L}
\def\ovl{\overline}
\def\mmod{\operatorname{mod}}
\def\sup{\operatorname{Sup}}

\def\mwpm{\operatorname{Fix}_{\text{min}}}
\def\fix{\operatorname{Fix}}
\def\onelog{\mathbb1_{\text{log}}}

\def\cupsymb{\scalebox{1.5}{$\cup$}}
\def\bichaincola{red}
\def\bichaincolb{cyan}
\def\bichaincolhelp{green}

\def\ditto{\multicolumn{1}{c|}{''}}

\def\gditto{\multicolumn{1}{c|}{\textcolor{gray}{''}}}

\tikzset{
varlab/.style 2 args={mark={slab=$\scriptstyle{#1}$,#2}},
varlabo/.style 2 args={mark={lab=$\scriptstyle{#1}$,asty={fill=white,inner sep=0},#2}},
ind/.style={mark={lab=#1,a}}, 
startind/.style={mark={lab=#1,b}}, 
classical/.style={line width=1.5},
quantum/.style={double,double distance=0.5mm},
auxiliary/.style={red},
actualedge/.style={line width=2},
anyon1gon/.style={fill=red,fill opacity=0.2,text opacity=1},
e or not/.style={preaction={draw,very thick,red!50}},
m or not/.style={pattern={Hatch[distance=1.2mm,line width=0.3mm,angle=45]},pattern color=red,opacity=0.3},
front/.style={preaction={draw,white,line width=2}},
bichain1/.style={\bichaincola,dash pattern=on 1pt off 2pt,line width=3},
bichain2/.style={\bichaincolb, line width=1.5},
bichainhelp/.style={\bichaincolhelp,dash pattern=on 1pt off 2pt,line width=3},
bichain1f/.style={pattern=north west lines, pattern color=\bichaincola},
bichain2f/.style={cyan, opacity=0.3},
bichainhelpf/.style={pattern=north west lines, pattern color=\bichaincolhelp},
bichain1ff/.style={pattern={Dots[radius=0.025cm]},pattern color=\bichaincola},
bichain2ff/.style={cyan, opacity=0.6},
channelpoint/.pic={\draw (-0.1,-0.1)--(0.1,0.1) (0.1,-0.1)--(-0.1,0.1);},
}

\tikzset{
bichain1p/.pic={\draw[\bichaincola, line width=1] (-0.15,-0.15)--(0.15,0.15) (-0.15,0.15)--(0.15,-0.15);},
bichain2p/.pic={\fill[\bichaincolb] (0,0)circle(0.14);},
bichainhelpp/.pic={\draw[\bichaincolhelp, line width=1] (-0.15,-0.15)--(0.15,0.15) (-0.15,0.15)--(0.15,-0.15);},
}

\setmarks{
ar/.style={arr,asty=fill},
}

\definecolor{darkgreen}{rgb}{0,0.6,0.2}
\setelements{
z2/.style={small,circ},
delta/.style={small,circ,all}, 
z2odd/.style={small,circ,bdastyle=red},
deltaodd/.style={small,circ,all,bdastyle=red}, 
deltatens/.style={tiny,circ,all}, 
hadamard/.style={small,square,dcross},
vertex/.style={tiny,circ,all},
vca/.style={astyle=red},
vcb/.style={style=darkgreen},
vcc/.style={style=blue},
}

\newcommand\drawedge{
\atoms{vertex}{0/, 1/p={0.8,0}}
\draw (0)edge[mark=arr] (1);
}

\newcommand\drawtriangle{
\atoms{vertex}{0/, 1/p=60:0.8, 2/p=0:0.8}
\draw (0)edge[mark=arr](1) (1)edge[mark=arr](2) (0)edge[mark=arr](2);
}

\newcommand\drawtetrahedron{
\atoms{vertex}{0/p={-150:1}, 1/p=90:1, 3/p=-30:1, 2/}
\draw (0)edge[mark=arr](1) (1)edge[mark=arr](2) (0)edge[mark=arr](2) (0)edge[mark=arr](3) (1)edge[mark=arr](3) (2)edge[mark=arr](3);
}

\newcommand\drawsquare{
\atoms{vertex}{0/, 1/p={0.8,0}, {2/p={0,0.8}}, 3/p={0.8,0.8}}
\draw (0)edge[](1) (1)edge[](3) (0)edge[](2) (2)edge[](3);
}

\newcommand\drawcube{
\atoms{vertex}{{100/p={1,0}}, {010/p={0,1}}, {001/p={0.6,0.4}}, 000/, {110/p={$(100)+(010)$}}, {101/p={$(100)+(001)$}}, {011/p={$(010)+(001)$}}, 111/p={$(100)+(010)+(001)$}}
\draw (000)edge[](100) (100)edge[](110) (000)edge[](010) (010)edge[](110);
\draw (001)edge[dashed](101) (101)edge[](111) (001)edge[dashed](011) (011)edge[](111);
\draw (000)edge[dashed](001) (100)edge[](101) (010)edge[](011) (110)edge[](111);
}

\newcommand\drawshearedcube{
\atoms{void}{001/p={1.1/2,0.7/2}}
\atoms{vertex}{100/p={1.5/2,0}, 010/p={0,1.5/2}, 110/p={$(100)+(010)$}, 011/p={$(010)+(001)$}, 101/p={$(100)+(001)$}, 111/p={$(100)+(010)+(001)$}, 020/p={$2*(010)$}, 021/p={$2*(010)+(001)$}}
\draw (100)edge[mark=arr](010) (110)edge[mark=arr](020) (100)edge[mark=arr](101) (110)edge[mark=arr](111) (110)edge[mark=arr](100) (020)edge[mark=arr](010) (111)edge[mark=arr](101) (020)edge[mark=arr](021);
\draw[dashed] (010)edge[mark=arr](011) (101)edge[mark=arr](011) (111)edge[mark=arr](021) (021)edge[mark=arr](011);}

\newcommand\drawmodcube{
\atoms{vertex}{{0/}, {1/p={1.5,0}}, {2/p={0.75,0.45}}, {3/p={2.25,0.45}}}
\atoms{vertex}{{0x/p={0,1.5}}, {1x/p={1.5,1.5}}, {2x/p={0.75,1.95}}, {3x/p={2.25,1.95}}}
\draw[dashed] (0)to[bend left=20](2) (0)to[bend right=20](2) (2x)--(2)--(3) (3)to[bend left=30](3x) (2x)to[bend right=30](3x);
\draw (0)to[bend left=5](1) (0)to[bend left=30](1) (0)to[bend right=5](0x) (0)to[bend right=30](0x) (2x)--(0x)--(1x)--(1)--(3) (3)to[bend left=5](3x) (2x)to[bend right=5](3x) (1x)to[bend left=20](3x) (1x)to[bend right=20](3x);
}
\begin{document}

\title{Low-overhead non-Clifford fault-tolerant circuits for all non-chiral abelian topological phases}
\author{Andreas Bauer}
\email{andib@mit.edu}
\affiliation{Freie Universit{\"a}t Berlin, Arnimallee 14, 14195 Berlin, Germany}
\affiliation{MIT Department of Mechanical Engineering, 77 Massachusetts Avenue, Cambridge, MA 02139, USA}

\begin{abstract}
We propose a family of explicit geometrically local circuits on a 2-dimensional planar grid of qudits, realizing any abelian non-chiral topological phase as an actively error-corrected fault-tolerant memory.
These circuits are constructed from measuring 1-form symmetries in discrete fixed-point path integrals, which we express through cellular cohomology and higher-order cup products.
The specific path integral we use is the abelian Dijkgraaf-Witten state sum on a 3-dimensional cellulation, which is a spacetime representation of the twisted quantum double model.
The resulting circuits are based on a syndrome extraction circuit of the (qudit) stabilizer toric code, into which we insert non-Clifford phase gates that implement the ``twist''.
The overhead compared to the toric code is moderate, in contrast to known constructions for twisted abelian phases.
The simplest non-trivial example is a fault-tolerant circuit for the double-semion phase, defined on the same set of qubits as the stabilizer toric code, with $12$ controlled-$S$ gates in addition to the $8$ controlled-$X$ gates and $2$ single-qubit measurements of the toric code per spacetime unit cell.
We also show that other architectures for the (qudit) toric code phase, like measurement-based topological quantum computation or Floquet codes, can be enriched with phase gates to implement twisted quantum doubles instead of their untwisted versions.
As a further result, we prove fault tolerance under arbitrary local (including non-Pauli) noise for a very general class of topological circuits that we call 1-form symmetric fixed-point circuits.
This notion unifies the circuits in this paper as well as the stabilizer toric code, subsystem toric code, measurement-based topological quantum computation, or the (CSS) honeycomb Floquet code.
We also demonstrate how our method can be adapted to construct fault-tolerant circuits for specific non-Abelian phases.
In the appendix we present an explicit combinatorial procedure to define formulas for higher cup products on arbitrary cellulations, which might be interesting in its own right to the TQFT and topological-phases community.
\end{abstract}

\maketitle
\tableofcontents

\section{Introduction}
Topological quantum computation is one of the most promising routes towards scalable universal fault-tolerant quantum computation.
In topological quantum computation, quantum information is robustly encoded in the ground state of a topologically ordered medium on a topologically non-trivial spatial configuration \cite{Kitaev1997,Dennis2001,Nayak2007}.
This configuration may also include defects, such as boundaries \cite{Bravyi1998}, anyons, domain walls, or twist defects \cite{Bombin2010}.
Computation is performed by adiabatically deforming the spatial configuration of the medium, including the defects.
One can distinguish between two levels on which topological quantum computation might be implemented in experiment, which we will refer to as \emph{active} and \emph{passive} topological fault tolerance.
Passive fault tolerance assumes the existence of a material in a topological phase that can be cooled to effectively remain in its ground state.
One of the most important candidate materials are fractional quantum Hall systems \cite{Tsui1982,Laughlin1983}.
Unfortunately, the creation and controlled manipulation of anyons in such models remains challenging \cite{Willett2013,Camino2005}.
Furthermore, non-Abelian fractional quantum Hall states \cite{Moore1991,Wen1991}, which would be required for universal quantum computation, have not yet been unambiguously confirmed in experiment.
For these reasons, much of the current research is focusing on active fault tolerance instead.
Active fault tolerance combines a large number of highly-controllable gadgets, such as superconducting qubits \cite{Google2022}, in order to merely simulate the cooling process of a topological material.
In this context, cooling takes place through measurements that lower the entropy of the system by extracting classical information.

Even though active and passive topological quantum computation are very different in experiment, one may use the same microscopic model to describe both the actual material and its simulation in theory.
However, an important difference is that for passive fault tolerance we do not actually need to know this microscopic model.
Instead, we simply trust that cooling keeps the material in its topologically ordered ground state, which changes adiabatically when braiding anyons, moving twist defects, interfacing with other blocks of material, etc.
The logical operation that is fault-tolerantly performed by some braiding or deformation protocol can be inferred without knowing the microscopic model for the cooling process, using one of the following two possibilities.
The first possibility is to use microscopic toy models that describe the ground state space, but not the the dynamic cooling process of a topological phase.
These models are usually commuting-projector Hamiltonians, such as \emph{string-net models} \cite{Levin2004} or \emph{quantum double models} \cite{Kitaev1997}.
Some papers that study topological quantum computation on this level include Refs.~\cite{Mochon2003,Koenig2010,Cong2016,Laubscher2018,Ren2023}.
The other possibility is to use higher-level invariant data describing the topological phase, such as the \emph{unitary modular tensor category} describing its anyons.
Examples for papers that study topological quantum computation on this level include Refs.~\cite{Freedman2001,Nayak2007,Naidu2009,Cui2014}.
Since the ground-state physics of topological phases as such is well understood, these works usually focus on achieving a universal logical gate set for quantum computation.
To this end, exotic topological phase are used, in particular all universal protocols seem to involve non-Abelian phases in some way.

On the other hand, to implement active topological fault-tolerance we need to actually know the microscopic model for the full dynamic cooling process, such that we can engineer it using our controllable gadgets.
This microscopic model is usually given as a geometrically local circuit consisting of unitaries, projective measurements, and classically controlled operations.
\footnote{Note that state-of-the-art fault-tolerant topological circuits require a decoder that chooses the controls based on the measurement outcomes, which is a global classical algorithm that is non-local and not affected by noise.
For a faithful simulation of cooling of a topological material, we would expect the complete process including the decoder to be a geometrically local circuit of quantum channels.
For this, the decoder would have to be local, that is, a classical cellular automaton, which so far only exists in $4+1$ dimensions \cite{Kubica2018}.
This might not be a practical issue since current classical information technology is much faster and reliable than its quantum counterpart.
However, it is an important fundamental question that we will not address in this paper.}
The most popular way to construct such fault-tolerant circuits has been to repeatedly measure the stabilizer operators of \emph{Pauli stabilizer codes}.
The most well-known topological stabilizer codes are the \emph{toric or surface code} \cite{Kitaev1997,Dennis2001,Bravyi1998}, and the \emph{color code} \cite{Bombin2006}.
A generalization of stabilizer codes are \emph{subsystem codes} \cite{Kribs2004}, which can give rise to fault-tolerant circuits as well \cite{Bravyi2012}.
Another route towards fault-tolerant circuits is \emph{measurement-based topological quantum computation} \cite{Raussendorf2007} or \emph{fusion-based topological quantum computation} \cite{Bartolucci2021}.
Recently, a new type of dynamic Pauli-measurement code known as \emph{Floquet codes} has been discovered \cite{Hastings2021,Kesselring2022,Davydova2022,Aasen2022,Dua2023}.
Apart from a few exceptions that we will discuss below, the topological phases
\footnote{Here we exclude fracton phases in $3+1$ dimensions, some of which can be represented by stabilizer codes, but which are not directly topological in a TQFT sense.}
implemented by known explicit fault-tolerant circuits are restricted to the toric-code phase or multiple copies thereof.
This is in particular true for all examples given above \cite{path_integral_qec,Bombin2023}.
For example, the color code is equivalent to two copies of the toric code \cite{Kubica2015}.
Of course, the generalization of the toric code from qubits to qudits, and from $2+1$ to $3+1$ dimensions is straight-forward, but not much is known about implementing more exotic topological phases.
The restriction to the toric code phase severely limits the logical gates that can be performed purely topologically.
As a consequence, the most common way to achieve a universal gate set is to supplement topologically protected gates with \emph{magic state distillation} \cite{Bravyi2004,Bravyi2012a,Litinski2019}, which is costly and becomes the bottleneck of the architecture.

We see that there is a large discrepancy between the theory of active and passive fault tolerance:
Whereas arbitrary topological phases are studied in passive fault tolerance, only the toric-code phase is routinely implemented as an explicit microscopic circuit.
Finding fault-tolerant circuits for more exotic topological phases is an important goal for two reasons:
First, it allows us to implement a larger set of logical gates purely topologically, which is more natural and perhaps more efficient than magic state distillation.
Second, it answers an important fundamental physics question, namely providing a microscopic dynamical model for the process of cooling exotic topological materials.
\footnote{Again, to fully answer this fundamental question we would need to find a local classical decoder, which we do not address in this paper.}
This paper aims to help resolve the discrepancy between the phases considered for universal passive topological quantum computation, and those implemented in active fault-tolerant circuits.
To this end, we provide a general and versatile method to construct fault-tolerant circuits for exotic topological phases.
Specifically, we find explicit circuits for arbitrary non-chiral abelian topological phases, at a moderate overhead compared to the qudit toric code.
For example, the simplest non-trivial case is the double-semion phase, which yields a circuit where in addition to the $8$ controlled-$X$ gates and $2$ single-qubit measurements per unit cell in the toric code, we need to apply $12$ controlled-$S$ gates and a few single-qubit $S$ gates.
Since the implemented phases are still abelian, we believe that they cannot yet be used to purely-topologically implement a gate set that is universal.
However, going from the qudit toric code to arbitrary abelian non-chiral topological phases certainly enriches the implementable gate set, and is an important step towards purely-topological universality.
Further, the versatility of our method makes it a powerful tool to further explore this direction.
As a first step, we show how our method can be used to construct fault-tolerant circuits for a simple non-Abelian phase in Section~\ref{sec:z23_twisted}.
Or rather, we point out that such protocols \cite{Bombin2018,Brown2019} have already been considered in the literature, with the goal of constructing a purely-topological universal gate set but without acknowledging the connection to non-Abelian phases.

We will now summarize what is already known in the literature on active error correction for topological phases beyond the (qudit) toric code.
First of all, there have been two approaches for constructing quantum codes for abelian topological phases.
The first is to extend the local ground-state projectors of abelian twisted quantum doubles to commuting non-Pauli stabilizers.
Such an extension was first found in the case of the double-semion model in Ref.~\cite{Dauphinais2018}, and then generalized to arbitrary twisted quantum doubles in Ref.~\cite{Magdalena2020}.
The second approach, presented in Ref.~\cite{Ellison2021}, is to start from the $\zz_n$ Pauli stabilizer code of a larger, untwisted ``parent'' quantum double model.
One then condenses a subgroup of bosonic anyons by adding their short string operators to the stabilizer group, resulting in a stabilizer code for a ``smaller'' but potentially ``twisted'' anyon theory.
Subsystem codes related to chiral anyon models have also been considered \cite{Bombin2009,Roberts2020,Ellison2022}, but which phases they represent depends a measurement schedule that needs to be chosen.
For non-abelian phases, there are some works that consider decoding \cite{Wootton2013,Dauphinais2016,Schotte2022} using a phenomenological noise model, but do not construct a microscopic circuit for syndrome extraction.
To the best of our knowledge, Ref.~\cite{Schotte2020} is the only work in the literature that proposes an explicit microscopic circuit for syndrome extraction in a non-abelian phase, namely the double-Fibonacci phase.
Finally, it is known that non-Clifford gates can be performed transversally in three layers of the $3+1$-dimensional toric code \cite{Vasmer2018}, or equivalently \cite{Kubica2015} in the color code \cite{Bombin2006a,Bombin2018a}.
Applying this gate corresponds to a temporal $3+0$-dimensional domain wall inside the $3+1$-dimensional bulk model.
While the surrounding toric code is abelian, the domain wall itself has a non-abelian twist.
Refs.~\cite{Bombin2018,Brown2019} use just-in-time decoding to turn the $3+0$-dimensional application of the transversal gate into $2+1$-dimensional protocols.
It thus seems conceivable that these $2+1$-dimensional protocols are related to a non-abelian phase, and we will indeed find that this is the case as discussed below.

Let us now describe our methods and results.
We will construct fault-tolerant topological circuits by relating them to fixed-point path integrals, and will accordingly call such circuits \emph{fixed-point circuits}.
\footnote{We use the name \emph{fixed-point path integral code} for the same concept in Ref.~\cite{path_integral_qec}.}
A fixed-point circuit is a geometrically local uniform circuit of unitaries, projective measurements, and classically controlled unitaries.
When fixing a configuration of measurement outcomes and controls we get a circuit of unitaries and projectors.
This non-unitary circuit then equals a discrete path integral representing a topological phase at zero correlation length, together with a pattern of topological defects.
These path integrals are state-sum TQFTs \cite{Fukuma1992,Dijkgraaf1990,Turaev1992,Barrett1993,Crane1993}, or equivalently tensor-network path integrals \cite{liquid_intro}.
For the specific circuits constructed in this paper, the path integral is the Dijkgraaf-Witten state sum \cite{Dijkgraaf1990} with an abelian gauge group, twisted by a type-I or type-II group 3-cocycle.
It is the path-integral formulation of the (abelian) \emph{twisted quantum double model} \cite{Hu2012}.
These path integrals are known to represent arbitrary non-chiral
\footnote{
By \emph{non-chiral}, we mean topological phases with gapped boundary, whose anyon theory is a Drinfeld center.
The vanishing of the chiral central charge is a necessary condition for this, but is not in general sufficient.
}
abelian topological orders.
The topological defects are projective 1-form symmetries in this path integral, which physically correspond to the worldlines of the abelian anyons.
The Dijkgraaf-Witten path integral equals that of the qudit toric code, apart from phase factors given by a group 3-cocycle.
When transforming the path integral into a fault-tolerant circuit, these phase factors become controlled-phase gates.
So our circuits are given by known syndrome-extraction circuits for the qudit toric code together with some additional phase gates.

While it has been known for a long time that active topological fault tolerance is best thought of in spacetime \cite{Dennis2001}, this point of view has been more appreciated since the discovery of Floquet codes \cite{Hastings2021,Kesselring2022,Davydova2022,Aasen2022}.
Floquet codes are periodic circuits of Pauli measurements that manage to fault-tolerantly preserve logical qubits, despite being trivial when the measurements are interpreted as gauge checks of a subsystem code \cite{Kribs2004,Bombin2009,Bravyi2012}.
In Ref.~\cite{path_integral_qec}, we have shown how path integrals can be used to understand conventional stabilizer codes, topological subsystem codes, Floquet codes, and measurement-based topological quantum computation, in a unified way:
All these codes correspond to the same toric-code path integral, which is put onto different lattices and traversed in different time directions.
A similar point of view is taken in Refs.~\cite{Bombin2023,Teague2023}, using the ZX-calculus.
In this paper, we apply the same mechanisms to our new fault-tolerant circuits:
While the circuits presented in Sections~\ref{sec:double_semion} and \ref{sec:general_doubles} are based on a standard syndrome-extraction protocol for the stabilizer toric code, Section~\ref{sec:measurement_floquet} discusses how to obtain circuits based on the Floquet and measurement-based architectures.

While already Ref.~\cite{Dennis2001} contained a proof for fault-tolerance of the toric code under Pauli and measurement errors, not so much has been explicitly spelled out in the literature about how this proof carries over to other settings.
For example, we are not aware of an explicit proof that the toric code is robust also to non-Pauli errors.
In Section~\ref{sec:fault_tolerance}, we provide a fault-tolerance proof for arbitrary 1-form symmetric fixed-point circuits, under arbitrary local noise.
This contains, for example, the stabilizer toric code, color code, subsystem toric code, measurement-based topological quantum computation, as well as Floquet codes \cite{path_integral_qec}.
Most importantly, it also contains the fault-tolerant circuits constructed in this paper.
In addition to showing fault tolerance of our circuits and providing explicit proofs for folklore theorems about fault tolerance, our proof might be insightful to readers since it follows a new paradigm:
Instead of arguing that the circuit will detect a Pauli error, our proof is based on the fact that any local perturbation in a fixed-point path integral disappears when surrounded with unperturbed path integral.

We would like to stress that, unlike the vast majority of the QEC literature, our path integral framework is completely agnostic towards the Pauli basis.
Thus, our methods overcome the necessity to express everything in terms of Pauli measurements and Clifford operations.
This generates a lot of flexibility when designing new topologically fault-tolerant circuits.
After all, in a real device, performing a single-qubit or two-qubit Clifford operation is as hard as performing a non-Clifford operation.

After having summarized our own results, let us compare them to the existing literature mentioned earlier.
Refs.~\cite{Dauphinais2018,Magdalena2020} construct complex 12-qubit stabilizer operators.
A decomposition of these measurements into elementary gates is not provided, and might be very complicated and difficult to find.
In contrast, our circuit only involves basic 2-qudit unitaries and single-qudit measurements.
The stabilizer codes in Ref.~\cite{Ellison2021} are constructed by condensing anyons in an untwisted model with a larger gauge group.
Therefore, the qudit dimension in these codes is larger than in our circuit.
For example, their double-semion code consists of 4-dimensional qudits, whereas our double-semion circuit is defined on the same array of 2-dimensional qubits as the toric code.
Our circuits thus have considerably lower overhead, and our double-semion circuit may be the simplest candidate so far for fault-tolerantly implementing a non-toric-code phase.
None of Refs.~\cite{Dauphinais2018,Magdalena2020,Ellison2021} allow to freely switch between stabilizer, measurement-based, or Floquet-style architectures, though certain Floquet variants of Ref.~\cite{Ellison2021} have been proposed recently \cite{Ellison2023}.
Such Floquet variants could also be constructed using our method, by performing anyon condensation on the level of path integrals.
Furthermore, none of Refs.~\cite{Dauphinais2018,Magdalena2020,Ellison2021} explicitly discuss fault tolerance, decoding and corrections.
The non-Abelian circuit proposed in Ref.~\cite{Schotte2020} can be understood in a path integral picture as well, even though we do not explicitly discuss this in this paper:
The fault-tolerant circuit that we constructed in Ref.~\cite{path_integral_qec} using path integrals for the (abelian) double-semion phase generalizes straight-forwardly to arbitrary Turaev-Viro path integrals \cite{Turaev1992}, and this generalization includes the circuit in Ref.~\cite{Schotte2020}.
However, these circuits have a very large overhead, and in the non-abelian case it is not fully clear yet how to best perform decoding and corrections.
Also the just-in-time decoded protocols in Refs.~\cite{Bombin2018,Brown2019} can be understood in the path integral formalism:
As we will show explicitly in Section~\ref{sec:z23_twisted}, analyzing these protocols with our formalism reveals that they secretly implement fault-tolerant circuits for a non-Abelian topological phase.
Overall, we find that path integrals provide a versatile, unifying, efficient, and particularly direct method for constructing and analyzing fault-tolerant topological circuits.
While in this paper we focus on constructing low-overhead abelian fault-tolerant circuits, our method has potential for also constructing non-Abelian fault-tolerant circuits.

Finally, we want to highlight some results that might be of separate interest to the topological-phases and TQFT community.
First, in Appendix~\ref{sec:cup_product}, we give a fully combinatorial procedure for defining Steenrod's \emph{higher-order cup products} \cite{Steenrod1947} on arbitrary cellulations.
These cup products have recently appeared in the study of topological phases, such as spin TQFTs \cite{Gaiotto2015}.
In Ref.~\cite{Steenrod1947}, Steenrod defines higher-order cup products only for branching-structure triangulations.
In Ref.~\cite{Chen2021}, higher-cup product formulas on a cubic lattice were defined, using a continuous-geometric interpretation of the cup product given in Ref.~\cite{Tata2020}.
Here, we give a purely combinatorial procedure to obtain any consistent cochain-level formula for higher-order cup products on arbitrary cellulations, using the basic recursive relation between higher-order cup products given in Ref.~\cite{Steenrod1947}.
Also the explicit form of 1-form symmetries in abelian Dijkgraaf-Witten path integrals, which is central to our construction, might be of separate interest.
Furthermore, we provide a new way of writing down the action of these path integrals, which we use in Appendix~\ref{sec:cyww} to elegantly express the Drinfel'd center of these models in terms of quotients of $\zz^k$.

The remainder of the paper is organized as follows.
We start out by reviewing the general formalism of fixed point path integrals and circuits in Section~\ref{sec:path_integral_codes}.
Then we discuss the simplest non-trivial representative of our family of fault-tolerant circuits in Section~\ref{sec:double_semion}, based on the double-semion model.
In Section~\ref{sec:general_doubles}, we generalize the same procedure to general twisted abelian quantum doubles.
In Section~\ref{sec:measurement_floquet}, we construct measurement-based architectures for all twisted abelian quantum doubles, and Floquet-like circuits for the type-II twisted $\zz_2\times \zz_2$ quantum double.
In Section~\ref{sec:fault_tolerance}, we prove fault tolerance for all 1-form symmetric fixed-point circuits.
In Section~\ref{sec:z23_twisted}, we discuss how to adapt our methods to implement a specific non-abelian phase.
We conclude in Section~\ref{sec:outlook}.
In Appendix~\ref{sec:cup_product}, we give an explicit procedure to define higher order cup products purely combinatorially on arbitrary cellulations.
In Appendix~\ref{sec:cyww}, we discuss the relation of the 1-form symmetries we construct in abelian Dijkgraaf-Witten theories and the corresponding anyon theories.

\section{Fixed-point path integrals and circuits}
\label{sec:path_integral_codes}
In this section, we will formally define fixed-point path integrals with projective 1-form symmetries, and fault-tolerant circuits related to these path integrals.
These concepts were previously introduced in Ref.~\cite{path_integral_qec}.

\subsection{Preliminaries on cellular (co-)homology}
\label{sec:cohomology}
Before we get to the formalism, let us introduce some basic vocabulary of cellular (co-)homology that we will need throughout the document.
Consider a $d$-dimensional cellulation $M$, whose set of $i$-cells we denote by $S_i[M]$.
For an abelian group $G$ that we will denote additively, a $G$-valued \emph{simplicial $a$-chain} $A$ is a map
\begin{equation}
A: S_a[M]\rightarrow G\;.
\end{equation}
Sometimes, we will instead view an $a$-chain as a formal sum over $a$-cells with $G$-valued coefficients, and identify a $a$-cell with a chain whose value is $1$ (a fixed generator of $G$) at this $a$-cell and $0$ everywhere else.
The central operation in (co-)homology is a (co-)boundary map on the set of chains.
In order to define these we need some extra structure.
First, we assume there is some set of $a-1$-cellulations of the $a-1$-sphere that we call \emph{$a$-cell representatives}, and every $a$-cell $\alpha$ is identified with one $a$-cell representative $\widetilde\alpha$.
Further, we assume that every $a$-cell representative $\widetilde\alpha$ (as well as $M$ itself) carries an orientation $\sigma[\widetilde\alpha]$ (and $\sigma[M]$) which is a $\zz_2$-valued $a-1$-chain ($d$-chain).
$\sigma[\widetilde\alpha](\beta)$ is $0$ or $1$ depending on whether the $a-1$ cell $\beta$ inside $\widetilde\alpha$ is oriented positively ($0$) or negatively ($1$).
With this, the \emph{coboundary} $dA$ is the $a+1$-chain whose value on an $a+1$-cell $\beta$ is given by the sum of $G$-elements on all $a$-cells $\alpha$ contained in $\beta$,
\begin{equation}
dA(\beta) = \sum_{\alpha\in S_a[\beta]} (-1)^{\sigma[\widetilde\beta](\alpha)} A(\alpha)\;.
\end{equation}
Here, for a $b$-cell $\beta$, $S_a[\beta]$ denotes the $a$-cells contained in $\beta$ if $a\leq b$, and the $a$-cells containing $\beta$ if $a\geq b$.
\footnote{Note that if multiple $a$-cells of $\widetilde\beta$ correspond to the same $a$-cell of $\beta$ in $M$, then $S_a[\beta]$ contains this $a$-cell of $M$ multiple times.}
$A$ is called an \emph{$a$-cocycle} if $dA=0$, and an \emph{$a$-coboundary} if $A=dB$ for some $a-1$-chain $B$.
In the context of taking the coboundary of chains, the latter are usually referred to as \emph{cochains}, but we will not always make this distinction.
The \emph{boundary} $\delta A$ is the $a-1$-chain whose value on an $a-1$-cell $\beta$ is given by the sum of $G$-elements on all $a$-cells $\alpha$ containing $\beta$,
\begin{equation}
\delta A(\beta) = \sum_{\alpha\in S_a[\beta]} (-1)^{\sigma[\widetilde\alpha](\beta)} A(\alpha)\;.
\end{equation}
$A$ is called an \emph{$a$-cycle} if $\delta A=0$, and an \emph{$a$-boundary} if $A=\delta B$ for some $a+1$-chain $B$.
A $\zz_2$-valued 1-cycle $A$ can be visualized as a superposition of closed loops by coloring all edges with $A=1$, a 2-cycle can be visualized as a superposition of closed membranes, and so on.
For general $G$, the lines or membranes can fuse or branch according to the group multiplication of $G$.
$i$-cocycles are equivalent to $n-i$-cycles on the Poincar\'e dual cellulation.
\footnote{Note, however, that cohomology is different from homology on the Poincar\'e dual cellulation on non-orientable manifolds for groups $G$ with elements of order greater than $2$.}
The central identity of (co-)homology is
\begin{equation}
\label{eq:double_boundary}
\delta\circ\delta=0\;,\qquad (d\circ d=0)\;,
\end{equation}
where $0$ denotes the function that maps every (co-)cycle to the trivial (co-)cycle with all $G$-elements equal to $0$.
The sets of (co-)chains, (co-)cycles, and (co-)boundaries form groups under cell-wise group multiplication.
Due to Eq.~\eqref{eq:double_boundary}, the group of $a$-(co-)boundaries is a subgroup of the group of $a$-(co-)cycles.
The quotient group is known as the \emph{$a$th (co-)homology group}, and its elements are known as \emph{cohomology classes}.
The cohomology groups do not depend on the cellulation but only on the topology of the manifold.

In addition to $\delta$ and $d$, we will need one other basic operation on chains known as \emph{higher-order cup product} \cite{Steenrod1947}.
More precisely, the \emph{$x$th cup product} is a 2-valent operation mapping an $a$-chain $A$ and a $b$-chain $B$ to an $a+b-x$-chain $A\cup_x B$.
All chains are valued in $G=\zz$.
The $\zz$-value of $A\cup_x B$ on a $a+b-x$-cell $\gamma$ is given by an expression of the form
\begin{equation}
\label{eq:cup_product_def}
\begin{multlined}
(A\cup_x B)(\gamma)\\
= \sum_{\alpha\in S_a[\gamma], \beta\in S_b[\gamma]} \cupsymb_x^{a,b}[\widetilde\gamma](\alpha, \beta) \cdot A(\alpha) \cdot B(\beta)\;,
\end{multlined}
\end{equation}
that is, the higher cup product is $\zz$-bilinear.
\footnote{Note that in slight abuse of notation, we let $\alpha$ and $\beta$ denote both cells of $\widetilde\gamma$ as well as cells of $M$.}
The coefficients of this bilinear form,
\begin{equation}
\cupsymb_x^{a,b}[\widetilde\gamma] :  S_a[\widetilde\gamma]\times S_b[\widetilde\gamma] \rightarrow \zz\;,
\end{equation}
will be constructed for general cellulations in Appendix~\ref{sec:cup_product}.
The higher-order cup products are defined through the relation \cite{Steenrod1947},
\begin{equation}
\label{eq:cup_product_defining_equation}
\begin{multlined}
d(A\cup_x B)= dA\cup_x B + (-1)^a A\cup_x dB \\
+ (-1)^{a+b+x} A\cup_{x-1} B + (-1)^{a+b+ab} B\cup_{x-1} A\;.
\end{multlined}
\end{equation}
In Appendix~\ref{sec:cup_product}, we show how this relation can be used to define $\cupsymb_x^{a,b}[\widetilde\gamma]$.
The ``ordinary'' cup product $\cup\eqqcolon\cup_0$ defines a (2-valent) \emph{cohomology operation}, that is, the cup product of two cocycles is again a cocycle, since Eq.~\eqref{eq:cup_product_defining_equation} reduces to
\begin{equation}
\label{eq:cup_product_boundary}
d(A\cup B)= dA\cup B + (-1)^a A\cup dB\;.
\end{equation}
For $x>0$, Eq.~\eqref{eq:cup_product_defining_equation} intuitively states that the $x$th order cup product fails to map cocycle pairs to cocycles, in the same way as the $x-1$th order cup product fails to be commutative.
So the higher-order cup products are not cohomology operations themselves, but can be combined in various ways to yield cohomology operations.

\subsection{Fixed-point path integrals}
\label{sec:fixed_points}
In this section, we will define fixed-point path integrals and their projective 1-form symmetries.
The path integrals under consideration are discrete path integrals, defined on some discrete $d$-dimensional \emph{spacetime cellulation} $M$.
There are two equivalent formulations for these discrete path integrals, namely as \emph{state sums} or as \emph{tensor-network path integrals}.
A state sum is a sum over configurations of discrete variables located on $M$.
Each summand is the product of weights located on $M$, which are complex numbers.
Each weight depends on the configuration of the variables, but only on these within a constant-size neighborhood.
Usually, the path integrals we consider will be translation invariant, and fully specified by a fixed set of variables and weights per unit cell on a hypercubic lattice $M$.
For example, there could be one variable taking values in $\{0,1\}$ at every edge of a 2-dimensional square lattice, and one weight $\omega_{a,b,c,d}$ at every plaquette depending on the configuration $a,b,c,d$ of the four surrounding edges.

Alternatively, a tensor-network path integral is a translation-invariant tensor network associated to $M$.
Unlike tensor networks like MPS or PEPS which parameterize states, tensor-network path integrals do not have any open indices.
When evaluating the tensor network we sum over all configurations of values for all contracted index pairs.
So we see that a tensor-network path integral is a state sum with one variable at each contracted index pair, and the tensor entries as weights.
Vice versa, every state sum where each variable has exactly two weights depending on it directly gives rise to a tensor network.
If there are more weights depending on the variable, we can implement the summation over this variable using a $\delta$-tensor, which is $1$ if all indices take the same value and $0$ otherwise.
In this paper, we will mostly use the state-sum formulation, but refer to the tensor-network formulation occasionally for illustrative purposes.

When we contract the tensor network or perform the state sum on a closed spacetime lattice $M$ with periodic boundary conditions, we get a number.
This number is what is usually known as the \emph{partition function} in physics, but is not of direct physical relevance for us.
What is more interesting is the evaluation on a lattice with a \emph{spatial boundary}.
At such a boundary, we keep the state-sum variables fixed, and only sum over the variables in the interior.
This way, we get a number for every boundary configuration, which we interpret as the amplitudes of a state with one qudit per boundary variable.
In the tensor-network formulation, we simply cut the bonds of the diagram along the spatial boundary, resulting in open indices.
The evaluation thus yields a tensor with indices distributed along the boundary, which we again interpret as the coefficient vector of a state.

Discrete path integrals can be used in different physical contexts, for example, the partition function of the classical Ising model is a discrete path integral in $2$ dimensions.
In this paper, the discrete path integrals represent the imaginary-time evolution of a local quantum Hamiltonian.
By Trotterization, the imaginary-time evolution can be approximated by a circuit of local operators, forming a tensor network in spacetime.
The tensor-network path integrals we consider share the same qualitative properties as these spacetime tensor networks, and are used in the same physical context.
The spatial-boundary state we get from evaluating the path integral is then analogous to a ground state of the model on the spatial boundary.
Furthermore, the path integrals we consider in this paper are defined on $2+1$-dimensional lattices, representing topological phases in $2$ spatial dimensions.

We are not interested in just any discrete path integral, but in ones that represent non-trivial topological phases in a particularly pure manner:
By a \emph{fixed-point path integral}, we mean one with zero correlation length.
That is, the correlations between any two points in spacetime vanish if the points are separated by more than some constant distance.
In a state sum, two-point correlations are computed by fixing the variables around two points and performing the summation over all other variables.
In a tensor network, we cut out two constant-size holes, or more generally manipulate the path integral around two points resulting in open indices.
The evaluation of the state sum depending on the fixed variables, or the contraction of the tensor network with additional open indices, can then be interpreted as a state or a tensor.
The qudits or indices of the resulting state or tensor at each point can be grouped together, such that we obtain a state on a bipartite system, or a matrix.
Zero correlation then means that the bipartite state is a product state, or that the matrix is rank-1.
In fact, we will use a slightly stronger condition, namely that the path integral on an ``annulus'' of topology $S_{d-1}\times [0,1]$ yields a product state, when grouping all degrees of freedom at $0$ and at $1$, respectively.
This ensures that correlations also vanish between any two extended regions, as long as they are separated from each other in a topologically trivial way.

After these explanations, let us make some more formal definitions.
\begin{mydef}
A \emph{discrete path integral} is a state sum (or tensor network) locally associated to a family of $d$-dimensional cellulations, which may or may not be restricted to $d$-dimensional hypercubic lattices.
Every weight (tensor) $r$ is associated to one $d$-cell $\widetilde r$ of $M$.
Further, there is some constant positive integer $\kappa$ such that the following holds:
For any cell $x$ of a cellulation $M$, let $x^{+\kappa}$ denote the set of $d$-cells that are connected to $x$ through $\kappa$ or less edges.
Define $X^{+\kappa}=\bigcup_{x\in X} x^{+\kappa}$ for any subset $X$ of cells analogously.
The weights (tensors) associated with one $d$-cell $m$, together with their variables (bonds), and the combinatorial structure they form, must only depend on the combinatorics of $m^{+\kappa}$.
For every sub-cellulation $M'$ (of one of the cellulations in the family), the evaluation of the path integral on $M'$ depends on the combinatorics of ${M'}^{+\kappa}$.
It consists as consisting of all weights (tensors) associated to the $d$-cells of $M'$, and sums over all variables with all dependent weights in $M'$ (bonds with both connected tensors in $M'$).
The evaluation depends on the configuration of the variables with some dependent weights inside of $M'$ and some dependent weights outside (bonds with one tensor inside and the other outside).
\end{mydef}

\begin{mydef}
\label{def:fixed_point_path_integral}
A \emph{fixed-point path integral} is a discrete path integral, for which there exists a positive integer $\chi$ called the \emph{correlation offset}, such that the following holds.
Consider the evaluation on a sub-cellulation $M'$ of topology $S_{d-1}\times [0,1]$, such that we need at least $\chi$ edges to connect the \emph{inside boundary} $S_{d-1}\times 0$ and the \emph{outside boundary} $S_{d-1}\times 1$.
Note that there is one such evaluation for any choice of ${M'}^{+\kappa}$.
The evaluation yields a state (tensor) with degrees of freedom (open indices) $i_0,i_1,\ldots$ at the inside boundary and $o_0,o_1,\ldots$ at the outside boundary.
After grouping the degrees of freedom (indices) at the inside boundary and the outside boundary, respectively, we obtain a matrix $T_{\vec i,\vec o}$.
This matrix $T$ has to be rank-1, that is,
\begin{equation}
T_{\vec i, \vec o} = v_{\vec i} \cdot w_{\vec o}\;,
\end{equation}
for some vectors $v$ and $w$.
\end{mydef}

As an example, consider a fixed-point path integral with $\chi=2$ defined on $1+1$ dimensional square lattices.
Let us formulate it as a tensor network that associates the same 4-index tensor to every plaquette.
Then the fixed-point condition would require that, for example, the following matrix $T$ is rank-1,
\begin{equation}
\begin{tikzpicture}
\foreach \x in {0,...,5}{
\foreach \y in {0,...,4}{
\atoms{circ,small}{\x\y/p={0.8*\x,0.8*\y}}
}}
\foreach \y in {0,...,4}{
\draw (0\y-l)--++(180:0.3) (5\y-r)--++(0:0.3);
\foreach[count=\xx from 1] \x in {0,...,4}{
\draw (\x\y)--(\xx\y);
}}
\foreach \x in {0,...,5}{
\draw (\x0-b)--++(-90:0.3) (\x4-t)--++(90:0.3);
\foreach[count=\yy from 1] \y in {0,...,3}{
\draw (\x\y)--(\x\yy);
}}
\fill[white] (1.1,1.1)rectangle(2.9,2.1);
\draw (21)edge[ind=$i_1$]++(90:0.3) (31)edge[ind=$i_2$]++(90:0.3) (42)edge[ind=$i_3$]++(180:0.3) (33)edge[ind=$\ldots$]++(-90:0.3);
\draw (00)edge[ind=$o_1$]++(-90:0.3) (10)edge[ind=$o_2$]++(-90:0.3) (20)edge[ind=$o_3$]++(-90:0.3) (30)edge[ind=$\ldots$]++(-90:0.3);
\end{tikzpicture}
=
T_{\vec i, \vec o}\;.
\end{equation}

An example for a concrete tensor-network fixed-point path integral describing the simplest non-trivial topological phase, the toric code, is described in Ref.~\cite{path_integral_qec}.
In this paper, the examples will be twisted discrete gauge theories in $2+1$ dimensions, also known as \emph{Dijkgraaf-Witten theories} \cite{Dijkgraaf1990}.
In the physics literature, the corresponding Hamiltonian models are known as \emph{twisted quantum double models} \cite{Hu2012}.
Dijkgraaf-Witten theories are best formulated as state sums.
Since these models are ``topological'' fixed-point path integrals, they are defined not just on regular spacetime lattices, but on arbitrary 3-dimensional cellulations $M$.
There is one $G$-valued variable at each edge of $M$ for some \emph{gauge group} $G$, such that a configuration of variables is given by a $G$-valued 1-cochain $A$.
In fact, the summation does not run over all 1-cochains $A$, but only the 1-cocycles.
Formally, this could be implemented by adding a weight at each face that is $1$ if $dA=0$ on this face, and $0$ otherwise.
The ``twist'' corresponds to an additional weight at each 3-cell, which is a phase factor.
That is, the configuration of weights forms a $U(1)$-valued $3$-cochain $\omega[A]$.
The value of $\omega[A]$ at a 3-cell depends only on the values of $A$ at the edges of this 3-cell.
The dependence of the weights on the local variables defines a local map from $G$-valued 1-cocycles to $U(1)$-valued 3-chains.
This local map needs to be a cohomology operation to ensure gauge invariance, which makes the path integral a fixed-point path integral.
Often, this cohomology operation can be decomposed into basic operations, like the coboundary operator and higher-order cup products.
In the most general case, $\omega$ is given by a \emph{group 3-cocycle} in $Z^3(BG,U(1))$.
In this context, it is common to specify the value of $\omega[A]$ on a tetrahedron as a function $\omega: G\times G\times G\rightarrow U(1)$, depending on the $G$-variables on 3 edges that determine these at all 6 edges of the tetrahedron through the constraint $dA=0$.

In explicit calculations, we prefer to replace $\omega[A]$ by the $\rr/\zz$-valued 3-cocycle $\lagr[A]$ such that $\omega=e^{2\pi i\lagr}$ element-wise.
Then the product of weights $\omega$ becomes a summation,
\begin{equation}
\label{eq:semion_action}
S[A] = \sum_{\gamma\in S_3[M]} \lagr[A](\gamma)(-1)^{\sigma[M](\gamma)}\;.
\end{equation}
Physically, $\lagr$ corresponds to the \emph{Lagrangian} of the model, and $S$ is known as the \emph{action}.
As shown, the Lagrangian gets a $\pm1$ prefactor depending on the orientation of the 3-cell $\gamma$.
The product of weights is the exponential of this summation,
\begin{equation}
\prod_{\gamma\in S_3[M]} \omega[A](\gamma)^{(-1)^{\sigma[M](\gamma)}} = e^{2\pi i S[A]}\;.
\end{equation}
Finally, the \emph{partition function} $Z$ is the number obtained by evaluating the state sum, that is, by summing the weights of all 1-cocycles $A$,
\begin{equation}
Z=\frac{1}{|G|^{|S_0[M]|}} \sum_{A\in Z^1(M,G)} e^{2\pi i S[A]}\;.
\end{equation}
Here, $Z^1(M,G)$ denotes the set of $G$-valued 1-cocycles on $M$.
The purpose of the normalization is to make the partition function independent on the cellulation, accounting for the larger number of 1-cocycles $A$ for larger cellulations.
Sometimes, we will write $Z_M$ or $S_M$ make the cellulation $M$ explicit.
We can also consider $S$ or $Z$ on cellulations $M$ with (spatial) boundary $\partial M$.
In this case, $Z$ depends on a 1-cocycle $a$ on $\partial M$, and is given by
\begin{equation}
Z[a]=\frac{1}{|G|^{|S_0[M]\setminus S_0[\partial M]|}} \sum_{A\in Z^1(M,G): \partial A=a} e^{2\pi i S[A]}\;.
\end{equation}
Here, $\partial A$ denotes $A$ restricted to $\partial M$.

The central property of $S$ is its gauge invariance,
\begin{equation}
S[A+d\alpha]=S[A]\;,
\end{equation}
for an arbitrary 0-cochain $\alpha$.
This holds on closed manifolds as well as manifolds with boundary if $\alpha$ is only non-zero on the interior vertices.
The condition that makes $\omega$ a group 3-cocycle implies this gauge invariance.
In the following, we will show that gauge invariance implies that the path integral fulfills the zero-correlation length condition in Definition~\ref{def:fixed_point_path_integral}.
To this end, we evaluate $Z[a]$ on a cellulation of topology $S_2\times [0,1]$.
We write $a=a^0+a^1$, where $a^0$ denotes $a$ restricted to the inside boundary $S_2\times 0$, and $a^1$ denotes $a$ restricted to the outside boundary $S_2\times 1$.
We then interpret $Z_{S_2\times [0,1]}[a^0+ a^1]$ as a bipartite state (or matrix), whose two degrees of freedom (or indices) are given by $a^0$ and $a^1$.
It remains to show that this state is a product state (or the matrix is rank-1), which we do below.
More precisely, we will show that this is the case as long as the cellulation is non-degenerate, meaning that the inside and outside boundaries do not share any vertices, edges, or faces.
To this end, we close off $S_2\times [0,1]$ to a 3-sphere $S_3$ by gluing a cellulated 3-ball $B_3^0$ to the inside boundary and $B_3^1$ to the outside boundary.
For every $a^0$ we choose a fixed 1-cocycle $A_+^0$ in $B_3^0$ with $\partial A_+^0=a^0$, and for every $a^1$ we choose $A_+^1$ inside $B_3^1$ with $\partial A_+^1=a^1$.
Since $S_3$ has trivial 1-cohomology, gauge invariance implies
\begin{equation}
\label{eq:gauge_to_robustness1}
\begin{multlined}
0=S_{S_3}[A+A_+^0+A_+^1]\\
= S_{S_2\times [0,1]}[A]+ S_{B_3^0}[A_+^0] + S_{B_3^1}[A_+^1]\;,
\end{multlined}
\end{equation}
for any $A$ with $\partial A=a^0+a^1$.
Using this, we find,
\begin{equation}
\label{eq:gauge_to_robustness2}
\begin{multlined}
Z_{S_2\times[0,1]}[a^0+a^1]
= \frac{1}{|G|^{|S_0[S_2\times [0,1]]|-|S_0[\partial(S_2\times [0,1])]|}}\\\cdot \sum_{\substack{A\in Z^1[S_2\times[0,1]]:\\\partial A=a^0+a^1}} e^{2\pi S_{S_2\times [0,1]}[A]}
\\
= |G| \cdot e^{-2\pi i S_{B_3^0}[A_+^0]} \cdot e^{-2\pi i S_{B_3^1}[A_+^1]}
\;.
\end{multlined}
\end{equation}
So $Z_{S_2\times[0,1]}[a^0+a^1]$ is indeed a product state, or a rank-1 matrix that is an exterior product of a vector only depending on $a^0$ and a vector only depending on $a^1$.
\footnote{The factor $|G|$ comes from the non-trivial 1-cohomology of $S_2\times [0,1]$.}

In this paper, we will specifically consider Dijkgraaf-Witten theories where $G$ is abelian and $\omega$ consists of so-called type-I and type-II 3-cocycles \cite{Chen2011}.
In this case, the action $S$ can indeed be written down as a simple formula involving a cup product, as we will see in Sections~\ref{sec:double_semion_pathintegral} and \ref{sec:general_path_integral}.

\subsection{1-form symmetries}
\label{sec:1form_symmetry}
As argued in Ref.~\cite{path_integral_qec}, we need to enhance our path integrals with defects if we want to turn them into fault-tolerant circuits.
In this paper, these defects will be \emph{(projective) 1-form symmetries}.
More generally, we can consider path integrals with $i$-form symmetries, which we define below.
\begin{mydef}
\label{def:homological_integral}
A \emph{$i$-form symmetric fixed-point path integral} is a fixed-point path integral (c.f.~Definition~\ref{def:fixed_point_path_integral}) defined on $d$-dimensional spacetime cellulations $M$ that are additionally equipped with a $K$-valued $d-i-1$-chain $s$.
Here, $K$ is an abelian group called the \emph{1-form symmetry group}.
Each weight (tensor) at a cell $m$ is allowed to depend on the restriction of $s$ to $m^{+\kappa}$ in addition to the combinatorics of the cellulation.
The following four conditions are required to hold, where the fourth contains Definition~\ref{def:fixed_point_path_integral}.
\begin{enumerate}
\item Consider a $d-i-2$-cell $\gamma$, and any possible cellulation and $d-i-1$-chain $s$ inside $\gamma^{+\chi+\kappa}$.
If $(\delta s)(\gamma)\neq 0$, then the evaluation of the path integral on $\gamma^{+\chi}$ yields $0$.
\item Consider a $d-i$-cell $\beta$, and any possible cellulation and $d-i-1$-cycle $s$ inside $\beta^{+\chi+\kappa}$.
Then the evaluated path integrals with $s$ and $s+\delta\beta$ are equal up to a complex phase.
\footnote{
This complex phase is what makes the 1-form symmetry projective.
An analogous condition might be called \emph{anomalous 1-form symmetry} in the literature.
}
\item Consider any sub-cellulation $M'\subset M$ whose boundary is a disjoint union $\partial M'=m_0\sqcup m_1$, such that the minimum number of edges connecting $m_0$ and $m_1$ is $\chi$.
Consider any possible cellulation and $d-i-1$-cycle $s$ on ${M'}^{+\kappa}$.
If the $m_0$-component of $\partial s$ is homologically non-trivial, then the path integral on $M'$ evaluates to zero.
\item Consider a sub-cellulation $M'\subset M$ of topology $S_{d-1}\times [0,1]$ of width $\chi$, and any possible cellulation and $d-i-1$-cycle $s$ inside ${M'}^{+\kappa}$.
If $s$ has trivial homology, then the evaluated path integral is a rank-1 matrix.
\end{enumerate}
Note that the first and second condition can be regarded as local versions of the third and fourth condition.
\end{mydef}
In this paper, we will consider $2+1$-dimensional path integrals with 1-form symmetries.
The path integrals will represent non-trivial topological phases, and the symmetry defects $s$ form a 1-cycle representing a pattern of \emph{anyon worldlines}.
The group $K$ then describes the fusion of anyons, so we see that this can only work if the anyons are \emph{abelian}.
The first condition in Definition~\ref{def:homological_integral} ensures that the anyon fusion rules are fulfilled.
The second condition implies that deforming or braiding the anyon worldlines only gives rise to phase factors.
The third condition states that the total charge of an anyon configuration must always be $0$.
The fourth condition states that the (ground-)state space on a sphere for any configuration of anyons (with trivial total charge) is $1$-dimensional.

An example for how 1-form symmetries can be implemented in a tensor-network path integral is described in Ref.~\cite{path_integral_qec}.
There, $m$ anyon worldlines are introduced by simply replacing the tensors located along a $\zz_2$-valued 2-cocycle $b$ by ``charged'' versions thereof, and the same holds for $e$ anyons along a $\zz_2$-valued 1-cycle $c$.
So we see that our actual examples do not strictly match Definition~\ref{def:homological_integral}, since the symmetry defects are represented by pair of 2-cocycle and 1-cycle $(b,c)$ instead of a single $K=\zz_2\times \zz_2$-valued 1-cycle $s$.
This is fine, since pairs $(b,c)$ still represent degree-1 homology classes, just in a microscopically different way.
For the general abelian Dijkgraaf-Witten models that we consider in this paper, symmetry defects still consist of a 2-cocycle $b$ valued in the gauge group $G$, as well as a $G$-valued 1-chain $c$ that is not a cycle but has a fixed boundary depending on $b$.
$b$ corresponds to the ``flux-like'' anyons, and $c$ to the ``charge-like'' anyons.
$b$ will modify the path integral in that we do not sum over 1-cocycles $A$ but over 1-cochains with $dA=b$, and both $b$ and $c$ lead to a modified action $S[A,b,c]$, which we write down in Sections~\ref{sec:semion_1form} and \ref{sec:general_1form}.
Again, the configurations $(b,c)$ represent degree-1 homology valued in some 1-form symmetry group $K$, just in a microscopically different way.
Due to the interaction between $b$ and $c$, $K$ is not in general a direct product $G\times G$, but instead a central group extension of $G$ by $G$ twisted by a group 2-cocycle.
We are assuming that the symmetry defects consist of only a $K$-valued 1-cycle $s$ in Definition~\ref{def:homological_integral} and for the fault-tolerance proof in Section~\ref{sec:fault_tolerance}, in order to not unnecessarily complicate things.
The generalization to other microscopic representations of symmetry defects is straight-forward and sketched at the end of Section~\ref{sec:fault_tolerance}.

\subsection{Fixed-point circuits}
\label{sec:dynamical_codes_from_integrals}
We are now ready to define fault-tolerant circuits based on 1-form symmetric fixed-point path integrals.
\begin{mydef}
\label{def:path_integral_code}
A \emph{1-form symmetric fixed-point circuit} (in $2+1$ dimensions) is a circuit that is equal to a ($2+1$-dimensional) 1-form symmetric fixed-point path integral.
By circuit, we mean a uniform geometrically local quantum circuit consisting of unitaries, classically controlled unitaries, and projective measurements, acting on qudits distributed over a $2$-dimensional space.
By equal, we mean that the circuit is a product of linear operators and therefore a path integral:
As a state sum, the variable configurations are the spacetime histories of qudit configurations, and the weights are the amplitudes of the linear operators between input and output configurations.
As a tensor network, the circuit diagram is also a tensor-network diagram, and the linear operators are tensors.
The operators in the circuit depend on the configuration of the classical controls and measurement outcomes, which corresponds to the configuration of 1-form symmetry defects in the path integral.
To this end, each control or outcome is valued in $K$ and associated with a nearby edge of the spacetime cellulation $M$.
The value of the control or outcome equals the value of the symmetry defect 1-chain $s$ on the associated edge.
\footnote{
Note that vice versa, \emph{not} every edge needs to be associated with a measurement outcome or control.
}
We can decompose $s=s_c+s_s$ into the 1-chain $s_s$ corresponding to the outcomes (``s'' for ``syndrome''), and the 1-chain $s_c$ corresponding to the controls (``c'' for ``corrections'').
\footnote{
The correspondence between controls and measurement outcomes and 1-form symmetry defect configurations can be generalized:
Each configuration of each individual control or measurement outcome is mapped to a 1-chain on $M$, supported only within a constant-size neighborhood of the position of the control or outcome.
$s$ is then obtained by summing over all these 1-chains.
We will not need this more general correspondence in this paper, but it is necessary, for example, for the Hastings-Haah honeycomb Floquet code as described in Ref.~\cite{path_integral_qec}.
}
\end{mydef}
Note that the controls here are much less important than the measurement outcomes, and only come into play when hypothetically interfacing the circuit with other types of fault-tolerant circuits as discussed below.
The usual QEC interpretation of the measurements is error syndrome extraction, and the usual interpretation of the controlled operations is performing corrections.

As such, 1-form symmetric fixed-point circuits do not yet define a complete fault-tolerant procedure.
We also need to specify a classical decoder whose task it is to choose the classical controls based on the measurement outcomes.
In other words, we need to choose the corrections based on the syndrome.
\footnote{Often, what is called ``syndrome'' in the literature would correspond to $\delta s_s$ rather than $s_s$, but we will use the term for the full configuration of measurement outcomes.}
Ideally, we would like the decoder to be a local classical circuit (in other words, a cellular automaton) as well, such that combining the quantum and classical circuits yields a self-correcting circuit of quantum channels.
However, it is unknown whether and how this is possible in physical spacetime dimensions smaller than 5 \cite{Kubica2018}.
This is why we only construct the quantum circuit and leave the classical decoder open in Definition~\ref{def:path_integral_code}.
There are many classical decoders that will work, but they are in general non-local algorithms.

Before we describe how the classical decoders work, let us discuss where we need to perform corrections.
General fixed-point circuits might require corrections throughout the circuit, such as for the non-abelian example described in Section~\ref{sec:z23_twisted}.
However, for 1-form symmetric fixed-point circuits, it suffices to perform corrections only ``at the very end'' of the circuit.
More precisely, we might need to perform corrections near spacetime interfaces of the circuit with other types of fault-tolerant circuits.
When interfacing with other 1-form symmetric fixed-point circuits, or at spacetime boundaries of such circuits, corrections are still not necessary.
For example, certain logical measurements correspond to temporal boundaries of the circuit.
The measurement outcome can be inferred from the syndrome alone, without having to perform any corrections near the boundary.
However, it is necessary to perform corrections at interfaces with non-abelian fault-tolerant circuits, because non-abelian phases need different decoding strategies that require a corrected ground state prepared at their spacetime boundary.
Since non-abelian phases seem to be necessary to perform purely topological universal quantum computation, performing corrections is unavoidable at some stage.
In this paper, we merely focus on the fault-tolerant storage of logical qudits via a circuit on an $L\times L$ torus by simulating an abelian topological phase.
We do not explicitly describe how to decode non-abelian phases (except for Section~\ref{sec:z23_twisted}) or how to interface them, so the above discussion is somewhat hypothetical.
Let us nonetheless consider a situation where we transfer our logical information into a non-abelian phase after storing it in an abelian phase for a time $T$, by a temporal interface with a hypothetical non-abelian circuit, and therefore need to perform corrections near time $T$.
A $1+1$-dimensional toy picture for the overall protocol is as follows:
\begin{equation}
\label{eq:qec_circuit}
\begin{tikzpicture}
\begin{scope}
\draw[mark={arr,e},mark={arr,s},mark={slab=$T$}] (0.3,0.9)--(0.3,5.4);
\draw[mark={arr,e},mark={arr,s},mark={slab=$L$,r}] (0.6,0.6)--(6.4,0.6);
\clip (0.6,0.9) rectangle (6.4,5.4);
\foreach \y in {0,1,2,3}{
\foreach \x in {0,1,2,3,4,5,6}{
\atoms{square,yscale=0.7,xscale=2.2}{{a\x\y/p={\x*1.4,\y*1.2}}}
\draw[classical,black!50] (a\x\y-t)edge[mark={three dots,a}]++(90:0.15);
}};
\foreach \y in {0,1,2,3}{
\foreach \x in {0,1,2,3,4,5,6}{
\atoms{square,yscale=0.7,xscale=2.2}{{b\x\y/p={\x*1.4+0.7,\y*1.2+0.6}}}
\draw[classical,black!50] (b\x\y-t)edge[mark={three dots,a}]++(90:0.15);
}};
\foreach[count=\yy from 1] \y in {0,1,2,3}{
\foreach[count=\xx from 0] \x in {1,2,3,4,5,6}{
\draw[quantum] ([sx=0.25]a\x\y-t)--([sx=-0.25]b\x\y-b);
\draw[quantum] ([sx=-0.25]a\x\y-t)--([sx=0.25]b\xx\y-b);
}};
\foreach[count=\yy from 1] \y in {0,1,2}{
\foreach[count=\xx from 0] \x in {1,2,3,4,5,6}{
\draw[quantum] ([sx=-0.25]b\xx\y-t)--([sx=0.25]a\xx\yy-b);
\draw[quantum] ([sx=0.25]b\xx\y-t)--([sx=-0.25]a\x\yy-b);
}};
\foreach \x in {0,1,2,3,4,5}{
\atoms{square}{a\x/p={\x*1.4+0.35,5}, b\x/p={\x*1.4+1.05,5}}
\draw[classical,black!50] ([sx=-0.1]a\x-b)edge[mark={three dots,a}]++(-100:0.15) ([sx=0.1]b\x-b)edge[mark={three dots,a}]++(-80:0.15);
\draw[quantum] ([sx=-0.25]b\x3-t)--(a\x-b) ([sx=0.25]b\x3-t)--(b\x-b) (a\x-t)--++(90:0.25) (b\x-t)--++(90:0.25);
};
\end{scope}
\atoms{square,xscale=4,yscale=1.7,lab={t=$D$,p={0,0}}}{c/p={7.4,3.2}}
\draw[classical,white] (b31-t)edge[mark={three dots,a}]++(90:0.15) (a41-t)edge[mark={three dots,a}]++(90:0.15) (b41-t)edge[mark={three dots,a}]++(90:0.15) ([sx=-0.1]a4-b)edge[mark={three dots,a}]++(-100:0.15) ([sx=0.1]b3-b)edge[mark={three dots,a}]++(-80:0.15);
\draw[classical,opacity=0.5,rc] ([sx=0.2]c-b)--++(-90:0.3)-|(b31-t) ([sx=0.4]c-b)--++(-90:0.5)-|(a41-t) ([sx=0.6]c-b)--++(-90:0.7)-|(b41-t) ([sx=-0.1]a4-b)--++(-100:0.2)-|([sx=0.6]c-t) ([sx=0.1]b3-b)--++(-80:0.4)-|([sx=0.4]c-t);
\path (c-b)--++(-0.3,-0.15)node{$\ldots$};
\path (c-t)--++(-0.3,0.15)node{$\ldots$};
\end{tikzpicture}\;.
\end{equation}
The double lines are qudits (ket and bra layer), the gray lines are classical degrees of freedom, the boxes are measurements or controlled operations, and the large box labeled $D$ is the classical decoder.

For the above layout, the input to the decoder is the 1-chain $s_s$ on the $L\times L\times T$ spacetime cellulation $M$.
The output is given by the 1-chain $s_c$ supported only at the final time $T$, or in a constant-time vicinity thereof.
To start with, we imagine running the circuit starting from a ground state and without noise, recording the syndrome $s_s$.
By construction in Definition~\ref{def:path_integral_code}, the probability of a configuration of measurement results is the squared absolute value of the path integral with the corresponding 1-chain $s_s$.
Thus, due to the first condition in Definition~\ref{def:homological_integral}, the probability of a measurement outcome is $0$ if $\delta s_s\neq 0$ except for at the boundary at time $T$ where $s_s$ is allowed to terminate at a 0-cycle $\partial s_s$.
Physically, $\partial s_s$ defines an anyon pattern in the resulting state at time $T$.
In this case, all the decoder needs to do is to choose $s_c$ arbitrarily such that $\delta s_c=\partial s_s$, and $s=s_s+s_c$ is homologically trivial.
In other words, we close off $s_s$ in a homologically trivial way at time $T$.
Since $s_s+s_c$ is homologically trivial, the path integral with $s=s_s+s_c$ equals the path integral with $s=0$, due to the second condition in Definition~\ref{def:homological_integral}.
Thus, the circuit acts as the identity within the ground state space.

Now, if we add perturb the circuit by weak noise, then the first condition in Definition~\ref{def:homological_integral} only holds approximately, and the syndrome $s_s$ will be slightly broken, that is, $\delta s_s\neq 0$.
This causes two problems for the decoder above:
First, $\partial s_s$ is undefined, since, for example, $s_s$ might terminate just before rather than on the boundary.
Second, it is undefined whether $s_s+s_c$ is homologically trivial, since $\delta s_s\neq 0$.
However, if we choose a 1-chain $\fix(\delta s_s)$ that fulfills
\begin{equation}
\label{eq:min_weight_perfect_matching}
\delta\fix(\delta s_s)=-\delta s_s\;,
\end{equation}
then we can proceed as in the noiseless case using $s_s+\fix(\delta s_s)$ instead of $s_s$.
Note that in the equation above, the boundary is closed at time $0$, but open at time $T$, that is, the 1-cycle $s_s+\fix(\delta s_s)$ can freely terminate at time $T$.
The following toy picture illustrates the situation for a $1+1$-dimensional spacetime,
\begin{equation}
\label{eq:decoder_toy_picture}
\begin{gathered}
\begin{tikzpicture}
\fill[black!10] (0,0)rectangle(7,5.5);
\draw[dotted] (0,0)edge[mark={lab=$0$,b},mark={slab=$x$,r},mark={arr,e}](7,0) (0,0)edge[mark={lab=$0$,b},mark={slab=$t$},mark={arr,e}] (0,5.5) (0,5)edge[mark={lab=$T$,b}]++(7,0) (7,0)edge[mark={lab=$L\sim 0$,b}]++(0,5.5);
\atoms{vertex}{ds0/p={1.25,2.25}, ds1/p={1.75,2.75}, ds2/p={5.25,2.25}, ds3/p={4.75,1.25}, ds4/p={4.25,1.75}, ds5/p={5.25,0.75}, ds6/p={5.75,3.75}, ds7/p={5.75,4.25}, ds8/p={1,4.7}, ds9/p={2.5,4.7}}
\draw (ds0)to[out=120,in=-90](0.5,5) (ds1)to[bend left](ds6) (ds7)--(5.75,5) (ds2)to[out=-30,in=180](7,1.5) (ds4)to[out=-170,in=0](0,1.5) (ds3)to[out=0,in=0,looseness=3](ds5) (2.5,0.8)circle(0.4) (3,5)to[bend right=60](4.5,5) (ds8)to[bend right](ds9);
\draw[red] (ds0)--(ds1) (ds2)--(ds4) (ds3)--(ds5) (ds6)--(ds7) (ds8)--(1,5) (ds9)--(2.5,5);
\draw[line width=2] (3,5)--(4.5,5) (0,5)--(0.5,5) (5.75,5)--(7,5) (1,5)--(2.5,5);
\end{tikzpicture}\\
\begin{tikzpicture}
\draw (0,0)edge[mark={lab=$s_s$,b}]++(180:0.4);
\path[mark={lab=$\delta s_s$,b}] (0,-0.4)--++(180:0.4);
\atoms{vertex}{ds/p={-0.2,-0.4}}
\draw[red] (3,0)edge[mark={lab=$\fix(\delta s_s)$,b}]++(180:0.4);
\draw[line width=2] (3,-0.4)edge[mark={lab=$s_c$,b}]++(180:0.4);
\end{tikzpicture}\;.
\end{gathered}
\end{equation}
Here, for $K=\zz_2$, 0-cycles are illustrated as superpositions of points, and 1-cycles as superpositions of black or red lines.
Of course, we still need to describe how to choose $\fix(\delta s_s)$.
To this end, we make use of the intuition that the syndrome $s_s$ is only slightly broken.
Thus, we can guess the ``real'' syndrome by choosing $\fix(\delta s)$ such that it is ``small''.
To make this slightly more formal, we can expand the noise-perturbed circuit as a sum over error configurations.
An isolated error has no effect on the circuit due to the fourth condition in Definition~\ref{def:homological_integral}, since it is surrounded by error-free path integral of topology $S_2\times [0,1]$.
Error configurations that contain non-contractable loops are a problem, but these are large and their probability decreases exponentially like $\epsilon^L$ where $\epsilon$ is the noise strength.
On the other hand, the number of problematic error configurations scales like $\epsilon_0^L$ for a fixed $\epsilon_0$, so if $\epsilon<\epsilon_0$, the effect of problematic error configurations vanishes exponentially in $L$.
Furthermore, due to the first condition in Definition~\ref{def:homological_integral}, $\delta s_s=0$ holds everywhere except for in the vicinity of the errors.
Again, individual isolated errors will break the syndrome only by a small amount such that $\fix(\delta s_s)$ will fix $s_s$ in a way that is homologically equivalent to the error that broke it.
For error configurations taking up a large enough portion of a non-contractible loop, it can happen that $\mwpm(\delta s_s)$ is homologically different from the error, leading to a logical error after correction.
Again, such configurations are exponentially unlikely and their overall contribution vanishes exponentially in $L$ if $\epsilon$ is smaller than some threshold.

There are many possible choices of decoders that lead to a fault-tolerant threshold.
A rather obvious choice is to take the 1-chain of minimum weight,
\begin{equation}
\label{eq:min_weight_matching}
\mwpm(\delta s_s)\coloneqq \operatorname{argmin}_{s_f\in K^{S_1[M]}: \delta s_f=-\delta s_s}(|\sup(s_f)|)\;,
\end{equation}
where $\sup(x)$ denotes the support of $x$, that is, the set of 1-cells $e$ with $x(e)\neq 0$.
If $K=\zz_2$, then the problem of finding $\mwpm(\delta s_s)$ is known as \emph{minimum weight perfect matching} of $\delta s_s$ \cite{Dennis2001}, and algorithms for solving it in polynomial runtime exist \cite{Edmonds1965, Higgott2023}.
However, it is important to note that for other abelian groups, efficient algorithms for finding the minimum-support fix do not exist to date.
In these cases, other computationally efficient decoders, such as clustering or RG decoders \cite{Bravyi2011}, still produce thresholds \cite{Duclos2013,Watson2014}.

We will formalize and prove the following proposition in Section~\ref{sec:fault_tolerance}.
\begin{myprop}
\label{prop:fault_tolerance}
A 1-form symmetric fixed-point circuit, combined with the minimum-weight decoder defined in Eq.~\eqref{eq:min_weight_matching}, fault-tolerantly performs the identity channel on the ground state space of the path integral.
That is, the process has a fault-tolerant threshold with respect to arbitrary local circuit-level noise.
\end{myprop}
Note that this proposition might not be exactly relevant to all of the circuits constructed in this paper, for two reasons:
First, as mentioned above, minimum-weight matching might not be efficient for $K\neq \zz_2^n$, and in practice we might have use an RG decoder or similar.
Second, as discussed at the end of Section~\ref{sec:1form_symmetry}, the symmetry defects will be represented by pairs $(b,c)$ instead of a single 1-chain $s$.
We will comment on how the proof generalizes to both situations at the end of Section~\ref{sec:general_proof}.

\section{Double-semion model}
\label{sec:double_semion}
In this section, we will focus on the simplest non-trivial example of the Dijkgraaf-Witten state sum introduced in Section~\ref{sec:fixed_points}, namely for gauge group $G=\zz_2$ and the non-trivial group 3-cocycle $\omega\in Z^1(BG,U(1))$.
The corresponding Hamiltonian model is known as the \emph{double-semion model} \cite{Levin2004, Hu2012}, and so its topological phase is called double-semion phase.
We will first write down the path integral, then equip it with 1-form symmetries, and finally turn it into a fault-tolerant circuit.

\subsection{Path integral}
\label{sec:double_semion_pathintegral}
Let us start by introducing the double-semion Dijkgraaf-Witten state sum.
We will introduce this state sum on an arbitrary 3-dimensional cellulation, in contrast to the original version defined on triangulations.
As discussed, the state sum is a sum over all configurations of $\zz_2$-elements on the edges, that is, over all $\zz_2$-valued 1-chains $A$.
There are two types of weights.
First, for every face whose edges have variables $a_0,a_1,\ldots$, we have a weight
\begin{equation}
\delta_{a_0+a_1+\ldots}\;,
\end{equation}
using a $\delta$ function that evaluates to $1$ if its argument is $0$, and to $0$ otherwise.
These weights force $A$ to be a 1-cocycle.
With only these weights, we have defined untwisted $\zz_2$ discrete gauge theory, the state-sum version of the toric code \cite{path_integral_qec}.
The second type of weight is given by the Lagrangian,
\begin{equation}
\label{eq:semion_lagrangian}
\lagr[A] = \frac14 \Abar\cup d\Abar\;.
\end{equation}
Here, $\ovl x$ denotes some fixed lifting map $\zz_2\rightarrow\zz$ applied to $x$, such that $\ovl x\mmod 2=x$.
An obvious choice is $\ovl0=0$ and $\ovl1=1$, with $\zz_2$-elements on the left and $\zz$-elements on the right.
More precisely, $\Abar$ denotes this lifting map applied element-wise to the 1-cocycle $A$.
$d\Abar$ defines a $\zz$-valued 2-cocycle, and hence the cup product between $\Abar$ and $d\Abar$ defines a $\zz$-valued 3-cochain.
The prefactor $\frac14$ maps the $\zz$-valued cochain to a $\rr/\zz$-valued one.

We notice that both the cup product and the coboundary in Eq.~\eqref{eq:semion_lagrangian}, as well as the summation in Eq.~\eqref{eq:semion_action} are $\zz$-linear.
We will exploit this to make our notation more compact.
Namely, we view $i$-chains as vectors in $\zz^{S_i[M]}$.
Then $d$ as a $\zz$-linear map
\begin{equation}
d: \zz^{S_i[M]}\rightarrow \zz^{S_{i+1}[M]}\;,
\end{equation}
represented by a $\zz$-valued $S_i[M]\times S_{i+1}[M]$ matrix.
We have
\begin{equation}
d^2=\delta^2=0\;,\qquad \delta=d^T\;,
\end{equation}
where $\bullet^T$ denotes transposition.
$\cup_x$ is interpreted as a 3-index $\zz$-valued tensor of dimensions $S_a[M]\times S_b[M]\times S_{a+b-x}[M]$, and the \emph{fundamental class} $(-1)^{\sigma[M](\gamma)}$ defines a $\zz$-valued vector of dimension $S_n[M]$.
In Eq.~\eqref{eq:semion_lagrangian}, $\Abar$ is thus a vector in $\zz^{S_1[M]}$, $d$ a $S_1[M]\times S_2[M]$ matrix, and $\cup$ a $S_1[M]\times S_2[M]\times S_3[M]$ tensor.
The summation in Eq.~\eqref{eq:semion_action} corresponds to contracting the third index of $\cup$ with the vector $(-1)^{\sigma[M](\gamma)}$, yielding a matrix that we will also denote by $\cup$.
After this, the action can be denoted as a product of $\zz$-valued vectors and matrices,
\begin{equation}
\label{eq:semion_action_linear}
S[A] = \frac14 \Abar^T\cup d\Abar\;.
\end{equation}
Sometimes we will write $\Sbar[\Abar]$ for the same function with $\Abar$ instead of $A$ as argument.
On a tetrahedron, using the cup product formula in Eq.~\eqref{eq:cup120_tetra}, the state-sum weight is given by the group 3-cocycle $\omega$,
\begin{equation}
\label{eq:semion_group_3cocycle}
\begin{multlined}
\omega(A_{01},A_{12},A_{23})=e^{2\pi iS[A]}\\
= i^{\ovl{A_{01}}\big(\ovl{A_{12}}+\ovl{A_{23}}-\ovl{(A_{12}+A_{23})}\big)} = (-1)^{\ovl{A_{01}}\ovl{A_{12}}\ovl{A_{23}}}\;.
\end{multlined}
\end{equation}
Here, $A_{01}$, $A_{12}$, and $A_{23}$, are the values of $A$ on the $01$, $12$, and $23$ edges of the tetrahedron, respectively.

The most important property of the action is its gauge invariance, that is, invariance under
\begin{equation}
A' = A+d\alpha\;.
\end{equation}
To obtain the lift $\ovl{A'}$ of $A'$, we realize that for every $x,y\in \zz_2$, there is $s_{x,y}\in\zz$ such that
\footnote{$s_{x,y}$ can be understood as the carry of $\mmod 2$ addition, or as the group 2-cocycle in $Z^2(B\zz_2,\zz)$ corresponding to the central extension $\zz\xrightarrow{\cdot 2}\zz\xrightarrow{\mmod 2}\zz_2$.}
\begin{equation}
\ovl{x+y}=\ovl x+\ovl y+2s_{x,y}\;.
\end{equation}
Since the coboundary is just a sum of group elements, for every $\zz_2$-valued $i$-chain $x$ there exists a $\zz$-valued $i+1$-chain $v_x$ such that
\footnote{If $x$ is an $i$-cocycle, the left-hand side of the following equation disappears, and $v_x$ is the inverse of the Bockstein homomorphism for the central extension mentioned in the previous footnote.}
\begin{equation}
\ovl{dx} = d\ovl x + 2v_x\;.
\end{equation}
Combining both, we find
\begin{equation}
\ovl{A'}=\ovl{A+d\alpha} = \Abar+d\ovl\alpha+2(s_{A,d\alpha}+v_\alpha)\eqqcolon \Abar+d\ovl\alpha+2x
\;.
\end{equation}
We first show that the action is invariant under adding $2x$,
\begin{equation}
\label{eq:semion_lift_invariance}
\begin{multlined}
\Sbar[\Abar+2x]-\Sbar[\Abar]\\
= \frac14 (\Abar+2x)^T\cup d(\Abar+2x) - \frac14 \Abar^T\cup d\Abar\\
= \frac14 \big(2x^T\cup d\Abar + \Abar^T\cup 2dx+ 2x^T \cup 2dx\big)\\
= \frac12 \big(x^T\cup d\Abar + (d\Abar)^T\cup x\big)\\
= -x^T\cup v_A - v_A^T\cup x = 0\;.
\end{multlined}
\end{equation}
Here we have used that
\begin{equation}
dA=0\quad\Rightarrow\quad d\Abar=-2v_A\;.
\end{equation}
We have also used that the action is valued in $\rr/\zz$, and thus terms with integer prefactor are zero.
Furthermore, we have used Eq.~\eqref{eq:cup_product_boundary}, which for the chosen dimensions and in our matrix notation becomes
\begin{equation}
\label{eq:cup0_relation_linear}
\cup d = \delta \cup\;.
\end{equation}
To this end, note that we can drop terms of the form $d(\ldots)$, since the matrix $\cup$ contains an implicit summation over the resulting 3-cocycle, and for any $2$-cochain $x$, we have
\begin{equation}
\sum_{\alpha\in S_3[M]} (dx)(\alpha) (-1)^{\sigma[M](\alpha)} = 0\;.
\end{equation}
Next, the action is also invariant under adding $d\ovl\alpha$,
\begin{equation}
\label{eq:semion_gauge_invariance}
\begin{multlined}
\Sbar[\Abar+d\ovl\alpha]-\Sbar[\Abar]\\
=\frac14 (\Abar+d\ovl\alpha)^T\cup d(\Abar+d\ovl\alpha) - \frac14 \Abar\cup d\Abar\\
= \frac14 (d\ovl\alpha)^T\cup d\Abar
= \frac14 \ovl\alpha^T\cup dd\Abar=0
\;.
\end{multlined}
\end{equation}
Again we have used Eq.~\eqref{eq:cup0_relation_linear}.
So in total, we find that the action is gauge invariant,
\begin{equation}
S[A']-S[A] = 0\;.
\end{equation}
As we argued at the end of Section~\ref{sec:fixed_points}, gauge invariance implies that the path integral fulfills the zero-correlation length condition in Definition~\ref{def:fixed_point_path_integral}.

\subsection{1-form symmetries}
\label{sec:semion_1form}
To turn the path integral into a fault-tolerant circuit, we need to equip the state sum with a set of defects whose absence or presence we measure.
In our case, these defects are the anyon worldlines of the model, which we view as projective 1-form symmetries.
The 1-form symmetry group is $K=\zz_2\times \zz_2$.
However, instead of representing a symmetry defect as a $K$-valued 1-chain $s$, we will slightly deviate from Definition~\ref{def:homological_integral}, and instead use a $\zz_2$-valued 2-cochain $b$ together with a $\zz_2$-valued 1-chain $c$.
Note that for turning the path integral into a circuit, we usually put it on cubic lattices.
In this case, we can simply shift $b$ by $(\frac12,\frac12,\frac12)$ to turn it from a 2-cocycle to a 1-cycle, and obtain an ordinary 1-chain $s$ as in Definition~\ref{def:homological_integral}.
On an arbitrary cellulation, however, there is no bijective mapping between 2-cocycles and 1-cycles, even though 2-cohomology and 1-homology are equivalent.
In this case we need to replace the four conditions in Definition~\ref{def:homological_integral} by analogous ones.
For example, the path integral being non-zero only if $\delta s=0$ via the first condition is turned into $db=0$ and $\delta c=0$.

In the presence of defects the path integral is modified in two ways.
First, the summation is not over 1-cocycles $A$, but over 1-cochains with fixed boundary $b$,
\begin{equation}
\label{eq:semion_boundary_constraint}
dA=b\quad\Rightarrow\quad
d\Abar=\bbar-2v_A\;.
\end{equation}
The second modification is to add terms to the action,
\begin{equation}
\label{eq:semion_1form_action}
S[A,b,c] = \frac14 \Abar^T\cup d\Abar + \frac14 \bbar^T\cup_1 d\Abar + \frac12 \Abar^T \cbar\;.
\end{equation}
Note that the term $\frac12\Abar^T\cbar$ is the direct product of two 1-chains and does not correspond to a summation over 3-cells but over 1-cells.
In the literature, this might be called a \emph{cap product} and denoted by $\frac12 \Abar\cap\cbar$.

Let us verify that the action is still gauge invariant in the presence of defects.
The invariance under adding $2x$ as in Eq.~\eqref{eq:semion_lift_invariance} becomes,
\begin{equation}
\label{eq:semion_a_periodicity}
\begin{multlined}
\Sbar[\Abar+2x,\bbar,\cbar]-\Sbar[\Abar,\bbar,\cbar]\\
 = \frac14 2x^T\cup (\bbar-2v_A) + \frac14 \Abar^T\cup 2dx + \frac14 2x^T\cup 2dx\\
+\frac14 \bbar^T\cup_1 2dx + \frac12 2x^T \cbar\\
= \frac14 2x^T\cup \bbar + \frac14 2(d\Abar)^T\cup x + \frac14 2 \bbar^T\cup_1 dx\\
= \frac12\big(x^T\cup \bbar+\bbar^T\cup x+\bbar^T\cup_1dx\big)\\
= \frac12 \bbar^T (\cup+\cup^T+\cup_1d ) x= \frac12 \bbar^Td^T\cup_1 x\\
= \frac12 2(dv_A)^T \cup_1 x = 0\;.
\end{multlined}
\end{equation}
Here we have used the coboundary $d\bbar=2dv_A$ of Eq.~\eqref{eq:semion_boundary_constraint}, as well as Eq.~\eqref{eq:cup_product_defining_equation}, which in matrix notation and for degrees $i=2$, $j=1$, $x=1$ becomes
\begin{equation}
\label{eq:firstcup21_identity}
0=\delta\cup_1+\cup_1d+\cup-\cup^T\;.
\end{equation}
Note that for the above usage the $\pm$ signs do not matter since the equation appears with a $\frac12$ prefactor and terms with integer prefactor vanish, but we still spelled out the correct signs for later reference.
Invariance under adding $d\ovl\alpha$ as in Eq.~\eqref{eq:semion_gauge_invariance} becomes,
\begin{equation}
\label{eq:semion_gauge_variance}
\begin{multlined}
\Sbar[\Abar+d\ovl\alpha,\bbar,\cbar]-\Sbar[\Abar,\bbar,\cbar]
=\frac14 (d\alpha)^T\cup d\Abar +\frac12 (d\alpha)^T\cbar\\
= \frac14 \alpha^T\cup dd\Abar+ \frac12 \alpha^T\delta \cbar
= \frac12 \alpha^T\delta \cbar\;.
\end{multlined}
\end{equation}
So the variance is zero if $c$ is a cycle, so $\delta c=0$ implies
\begin{equation}
S[A',b,c]-S[A,b,c]=0\;.
\end{equation}

Let us now argue why the above indeed defines a 1-form symmetric path integral obeying the four conditions of Definition~\ref{def:homological_integral}.
More precisely, we show that analogous conditions hold as discussed above.
To show the first condition, we consider Eq.~\eqref{eq:semion_gauge_variance} for a general $1$-chain $c$.
Consider a 3-ball sub-cellulation consisting of a vertex $v$ and all the 3-cells adjacent to it, and a fixed 1-cocycle $a$ inside its boundary.
After choosing a fixed $A_0$ in the interior with $\partial A_0=a$, the path integral is given by
\begin{equation}
\label{eq:semion_ccycle_derivation}
\begin{multlined}
Z[a,b,c] = \frac12 \sum_{\alpha(v)\in \zz_2} e^{2\pi i S[A_0+d\alpha]}\\
= \frac12 e^{2\pi i S[A_0]} \sum_{\alpha(v)\in \zz_2} (-1)^{\alpha(v)\delta c(v)}
= e^{2\pi i S[A_0]} \delta_{\delta c(v),0}\;.
\end{multlined}
\end{equation}
We see that the partition function on a 3-ball cellulation evaluates to zero if $\delta c\neq 0$ on any vertex in the interior.
So the first condition holds for a very small $\chi$, namely as long as the vertex is not part of the boundary of the 3-ball.
Also, by definition, the partition function on any 3-ball cellulation is zero if $db\neq 0$ at any 3-cell, since then the set of configurations $A$ with $dA=b$ that we sum over is empty.
So the 2-cochain analogue of the third condition also holds for $b$.

Next, let us look at the second condition of Definition~\ref{def:homological_integral}.
To this end, we show that for a single edge $\beta$ in the interior of a cellulation of $B_3$, we have
\begin{equation}
Z[a, b+d\beta, c] = e^{2\pi i\kappa(b,c,\beta)} \cdot Z[a,b,c]\;.
\end{equation}
To see this, we complete the cellulated 3-ball (that we simply refer to as $B_3$) by another cellulated 3-ball (that we refer to as $B_3^+$) into a 3-sphere.
We complete $a$, $\partial b$, and $\partial c$ with (co-)cycles $A^+$, $b^+$, and $c^+$ inside $B_3^+$.
We then have
\begin{equation}
\label{eq:bc_gauge_invariance}
\begin{gathered}
\begin{multlined}
S_{S_3}[A+A^+,b+b^+,c+c^+]\\=S_{B_3}[A,b,c]+S_{B_3^+}[A^+,b^+,c^+]\;,
\end{multlined}
\\
\begin{multlined}
S_{S_3}[A+\beta+A^+,b+d\beta+b^+,c+c^+]\\=S_{B_3}[A+\beta,b+d\beta,c]+S_{B_3^+}[A^+,b^+,c^+]
\end{multlined}
\\
\Rightarrow S_{B_3}[A+\beta,b+d\beta,c] = S_{B_3}[A,b,c]+\kappa(b,c,\beta)\;,
\end{gathered}
\end{equation}
with
\begin{equation}
\begin{multlined}
\kappa(b,c,\beta)\coloneqq
S_{S_3}[A+A^+,b+b^+,c+c^+]\\
-S_{S_3}[A+\beta+A^+,b+d\beta+b^+,c+c^+]\;.
\end{multlined}
\end{equation}
Here we have used that due to gauge invariance and since $S_3$ has trivial 1-cohomology, $S_{S_3}$ does not depend on the 1-cochain $A$ in its first argument.
Also note that we changed $A$ to $A+\beta$ in order to keep the constraint $dA=b$, but we could have taken any 1-cochain $A'$ such that $dA'=b+d\beta$.
Due to gauge invariance, this equation carries over to the partition functions,
\begin{equation}
Z_{B_3}[a,b+d\beta,c] = e^{2\pi i\kappa(b,c,\beta)}\cdot Z_{B_3}[a,b,c]\;.
\end{equation}
Analogously, one can show that for a single face $\gamma$ in the interior of a $B_3$ cellulation, we have
\begin{equation}
Z_{B_3}[a, b, c+\delta\gamma] = e^{2\pi i \lambda(b,c,\gamma)} \cdot Z_{B_3}[a,b,c]\;.
\end{equation}
The argument is completely analogous to Eq.~\eqref{eq:bc_gauge_invariance} after exchanging $b$ and $c$, and the coboundary with the boundary.
A slight difference is that we do not have to add $\beta$ to $A$, which only makes the argument simpler.
So we find that the second condition of Definition~\ref{def:homological_integral} (and its 2-cocycle analogue for $b$) holds for a very small $\chi$.
Even though this follows directly from the $A$ gauge invariance, we will explicitly compute the coefficients $\kappa$ and $\lambda$ in Appendix~\eqref{sec:cyww}, and find that they are indeed independent of $A$.
Physically, they are related to the braiding and topological twist of the abelian anyons.

Next, let us look at the third condition in Definition~\ref{def:homological_integral}.
Consider a sub-cellulation $M'$ with $\partial M=m_0\sqcup m_1$, equipped with a 2-cocycle $b$ and 1-cycle $c$.
If $\partial b$ restricted to $m_0$ is a cohomologically non-trivial 2-cocycle, then there exists no $A$ such that $dA=b$, so we sum over an empty set and the path integral evaluates to $0$.
If $\partial b$ is cohomologically trivial and $\partial c$ on $m_0$ is a homologically non-trivial 1-cycle, then there must be a connected component of $m_0$ such that the sum of $\partial c$ over all vertices of the connected component is $1$.
Consider the 1-cocycle $A_x$ consisting of all edges that are adjacent to a vertex in this connected component but not part of it.
By construction, we have $A_x^Tc=1$, so due to gauge invariance we find
\begin{equation}
\label{eq:semion_condition3_nontrivialc}
\begin{multlined}
Z_{S_2\times[0,1]}[a,b,c]=e^{2\pi iS[A_0,b,c]}+e^{2\pi iS[A_0+A_x,b,c]}\\
= e^{2\pi iS[A_0,b,c]} (1+(-1)^{A_x^Tc}) = 0\;,
\end{multlined}
\end{equation}
for some fixed $A_0$ with $\partial A_0=a$ and $dA_0=b$.
Here we have used that $S$ only depends on $c$ via the term $\frac12 A^Tc$, and the other terms yield the same for $A_0$ and $A_0+A_x$ because $A_x$ is cohomologically trivial after filling the interior of $M'$ at $m_0$, and $b$ can be completed to a 2-cocycle on this interior.

Finally, let us argue that the path integral also fulfills the fourth condition in Definition~\ref{def:homological_integral}.
The argument is a straight-forward generalization of the one we used at the end of Section~\ref{sec:fixed_points} to show that the path integral without 1-form symmetries fulfills Definition~\ref{def:fixed_point_path_integral}.
If $b$ and $c$ on $S_2\times [0,1]$ have trivial (co-)homology, then we can find extensions $b_+^0$ and $c_+^0$ on $B_3^0$, and $b_+^1$ and $c_+^1$ on $B_3^1$.
Eq.~\eqref{eq:gauge_to_robustness1} then generalizes to
\begin{equation}
\begin{multlined}
\kappa(b,c) \coloneqq S_{S_3}[A+A_+^0+A_+^1,b+b_+^0+b_+^1,c+c_+^0+c_+^1]\\
=S_{S_2\times [0,1]}[A,b,c]+ S_{B_3^0}[A_+^0,b_+^0,c_+^0]\\ + S_{B_3^1}[A_+^1,b_+^1,c_+^1]\;.
\end{multlined}
\end{equation}
Using this, we find analogous to \eqref{eq:gauge_to_robustness2},
\begin{equation}
\begin{multlined}
Z_{S_2\times [0,1]}[a,b,c]\\
=
|G|\cdot e^{2\pi i \kappa(b,c)} e^{-2\pi iS_{B_3^0}[A_+^0,b_+^0,c_+^0]} e^{-2\pi iS_{B_3^1}[A_+^1,b_+^1,c_+^1]}\;.
\end{multlined}
\end{equation}
So again, $Z_{S_2\times [0,1]}$ is an exterior product of a vector only depending on $a_0$ and another vector only depending on $a_1$, and the fourth condition in Definition~\ref{def:homological_integral} is fulfilled.

\subsection{Fault tolerant circuit}
\label{sec:double_semion_code}
In this section, we will turn the double-semion path integral from the previous section into a fault-tolerant circuit, related via Definition~\ref{def:path_integral_code}.
We start by choosing a spacetime cellulation and a time direction in which we traverse this cellulation.
We do this such that the resulting circuit contains the gates of the syndrome-extraction protocol for the stabilizer toric code in Section VII.B of Ref.~\cite{Dennis2001}.
We will look at other cellulations and time directions later in Section~\ref{sec:measurement_floquet}.
The resulting circuit will differ from the toric code only by some controlled-phase operations that implement the group cocycle twist.
The spacetime cellulation is a sheared cubic lattice.
We start with an ordinary cubic lattice and refer to the unit vectors as $x$, $y$, and $t$, where $t$ becomes the time direction of the circuit.
Then each volume of the sheared lattice is spanned by the new unit vectors $x_2\coloneqq -t$, $x_1\coloneqq y$, and $x_0\coloneqq t-x$.
For illustration, consider the following patch consisting of two cubes of the original $(x,y,t)$ lattice, containing one cube (drawn with thick lines) of the sheared lattice,
\begin{equation}
\label{eq:semion_spacetime_cellulation}
\begin{gathered}
\begin{tikzpicture}
\draw (0,0)edge[mark={arr,e},ind=$x$]++(0:0.35) (0,0)edge[mark={arr,e},ind=$y$]++(30:0.25) (0,0)edge[mark={arr,e},ind=$t$]++(90:0.35);
\end{tikzpicture}\\
\begin{tikzpicture}
\draw (0,0)edge[mark={arr,e},ind=$x_2$]++(-90:0.35) (0,0)edge[mark={arr,e},ind=$x_1$]++(30:0.35) (0,0)edge[mark={arr,e},ind=$x_0$]++(135:0.5);
\end{tikzpicture}
\end{gathered}
\quad
\begin{tikzpicture}
\atoms{vertex}{100/p={1.5,0}, 010/p={0,1.5}, {001/p={1.1,0.7}, astyle=gray}}
\atoms{vertex,astyle=gray}{000/, 120/p={$2*(010)+(100)$}, 121/p={$(100)+2*(010)+(001)$}}
\atoms{vertex}{110/p={$(100)+(010)$}, 011/p={$(010)+(001)$}, 101/p={$(100)+(001)$}, 111/p={$(100)+(010)+(001)$}, 020/p={$2*(010)$}, 021/p={$2*(010)+(001)$}}
\draw[line width=1.3] (100)edge[mark=arr](010) (110)edge[mark=arr](020) (100)edge[mark=arr](101) (110)edge[mark=arr](111) (110)edge[mark=arr](100) (020)edge[mark=arr](010) (111)edge[mark=arr](101) (020)edge[mark=arr](021);
\draw[gray] (100)edge[mark=arr](000) (010)edge[mark=arr](000) (120)edge[mark=arr](110) (120)edge[mark=arr](020) (120)edge[mark=arr](121) (121)edge[mark=arr](021) (121)edge[mark=arr](111) (110)edge[mark=arr](121) (110)edge[mark=arr](010) (100)edge[mark=arr](111);
\draw[dashed,line width=1.3] (010)edge[mark=arr](011) (101)edge[mark=arr](011) (111)edge[mark=arr](021) (021)edge[mark=arr](011);
\draw[gray,dashed] (000)edge[mark=arr](001) (101)edge[mark=arr](001) (011)edge[mark=arr](001) (000)edge[mark=arr](011) (111)edge[mark=arr](011) (010)edge[mark=arr](021);
\end{tikzpicture}
\;.
\end{equation}
The $\pm$ prefactors of the unit vectors reflect directions of the edges, which define their orientation $\sigma$ needed to define the (co-)boundary and higher-order cup products, see Section~\ref{sec:cohomology}.
We have also added edges that divide each $t$-like face into two triangles, which is helpful since the state-sum variables on these edges will correspond to an intermediate state of a qubit in the circuit.

Let us next look at the concrete variables, constraints, and weights of the $S[A,b,c]$ path integral on this lattice.
In the following picture, we have labeled the variables assigned to all the edges and faces on a patch of the cellulation,
\begin{equation}
\label{eq:sheared_cubic_labels}
\begin{tikzpicture}
\atoms{vertex}{100/p={3,0}, 010/p={0,3}, {001/p={2.2,1.4}, astyle=gray}}
\atoms{vertex,astyle=gray}{000/, 120/p={$2*(010)+(100)$}, 121/p={$(100)+2*(010)+(001)$}}
\atoms{vertex}{110/p={$(100)+(010)$}, 011/p={$(010)+(001)$}, 101/p={$(100)+(001)$}, 111/p={$(100)+(010)+(001)$}, 020/p={$2*(010)$}, 021/p={$2*(010)+(001)$}}
\draw[] (100)edge[mark=arr,varlab={A_{dx}^{000}}{r}](010) (110)edge[mark=arr,varlab={A_{dx}^{001}}{r}](020) (100)edge[mark=arr,varlab={A_y^{100}}{r}](101) (110)edge[mark=arr,varlab={A_y^{101}}{}](111) (110)edge[mark=arr,varlab={A_t^{100}}{r,p=0.4},varlab={c^{100}}{p=0.4}](100) (020)edge[mark=arr,varlab={A_t^{001}}{r},varlab={c^{001}}{}](010) (110)edge[mark=arr,varlab={A_x^{001}}{r,p=0.37}](010) (111)edge[mark=arr,varlab={A_t^{110}}{r},varlab={c^{110}}{}](101) (020)edge[mark=arr,varlab={A_y^{002}}{}](021) (100)edge[mark=arr,varlab={A_{dy}^{100}}{}](111);
\draw[] (100)edge[mark=arr,varlab={A_x^{000}}{}](000) (010)edge[mark=arr,varlab={A_t^{000}}{r},varlab={c^{000}}{}](000) (120)edge[mark=arr,varlab={A_t^{101}}{r,p=0.4},varlab={c^{101}}{p=0.4}](110) (120)edge[mark=arr,varlab={A_x^{002}}{r,p=0.37}](020) (120)edge[mark=arr,varlab={A_y^{102}}{}](121) (121)edge[mark=arr,varlab={A_x^{012}}{r}](021) (121)edge[mark=arr,varlab={A_t^{111}}{r},varlab={c^{111}}{}](111) (110)edge[mark=arr,varlab={A_{dy}^{101}}{}](121);
\draw[dashed] (010)edge[mark=arr,varlab={A_y^{001}}{r}](011) (101)edge[mark=arr,varlab={A_{dx}^{010}}{r}](011) (111)edge[mark=arr,varlab={A_{dx}^{011}}{r}](021) (021)edge[mark=arr,varlab={A_t^{011}}{r,p=0.6},varlab={c^{011}}{p=0.6}](011) (111)edge[mark=arr,varlab={A_x^{011}}{p=0.63}](011) (010)edge[mark=arr,varlab={A_{dy}^{001}}{p=0.55}](021);
\draw[dashed] (000)edge[mark=arr,varlab={A_y^{000}}{}](001) (101)edge[mark=arr,varlab={A_x^{010}}{p=0.63}](001) (011)edge[mark=arr,varlab={A_t^{010}}{r,p=0.6},varlab={c^{010}}{p=0.6}](001) (000)edge[mark=arr,varlab={A_{dy}^{000}}{p=0.55}](011);
\node at (2.6,2.2){$b^{000}$};
\node at (2.6,5.2){$b^{001}$};
\end{tikzpicture}
\;.
\end{equation}
The label name and its subscripts correspond to the different variables within one unit cell.
The superscripts of a variable named like $L^{abc}$ specify that the variable is part of the unit cell whose $x$, $y$, and $z$ coordinate is given by $a$, $b$, and $c$.
As shown, $b$ is only supported on the $(y,t-x)$-faces, and $c$ is only supported on the $-t$-edges.
In the resulting circuit, each measurement result will correspond to the value of $b$ on a $(y,t-x)$-face, or the value of $c$ on a $-t$-edge.
While it would be possible to perform measurements for the values of $b$ and $c$ on other faces and edges, this is unnecessary to turn the path integral into a circuit.

Next, let us look the constraints.
Namely, the constraint $dA=b$ of the state sum translates into the following constraints for the variables,
\begin{equation}
\label{eq:semion_lattice_constraints}
\begin{gathered}
A_{dx}^{000}+A_t^{000}-A_x^{000}=0\;,\\
A_t^{100}+A_{dx}^{000}-A_x^{001}=0\;,\\
A_{dy}^{000}+A_t^{010}-A_y^{000}=0\;,\\
A_t^{000}+A_{dy}^{000}-A_y^{001}=0\;,\\
A_{dx}^{000}+A_y^{001}-A_{dx}^{010}-A_y^{100}=b^{000}\;.
\end{gathered}
\end{equation}
Here, we have written down each constraint only for one particular unit cell, and shifting all the superscripts by a fixed $(x,y,t)$-vector yields the same constraint shifted to another unit cell.
Each of the first four constraints corresponds to one of $t$-like triangles inside the unit cell.
The last constraint corresponds to the $(y,t-x)$-face.
The $\pm$ signs do not matter here since the gauge group is $G=\zz_2$, but we show them for later reference when $G$ is an arbitrary abelian group.
These signs depend on the orientations $\sigma$, which we choose as for the cubic lattice in Appendix~\ref{sec:cup_product}, where we identify the basis vectors $x_0$, $x_1$, and $x_2$ in Eqs.~\eqref{eq:semion_spacetime_cellulation} and \eqref{eq:cup_product_cube_drawing}.
Note that all these constraints can be interpreted as weight, for example $a+b-c=0$ can be interpreted as a weight $\delta_{a+b-c}$.

Next, we consider how the terms of the action in \eqref{eq:semion_1form_action} yield concrete state-sum weights depending on the variables.
Let us start with the term $\frac14\Abar^T\cup d\Abar$.
We will take the formula for the cup product on a cube from Eq.~\eqref{eq:cup120} in Appendix~\ref{sec:cup_product}, after identifying $x_0$, $x_1$, and $x_2$ in Eqs.~\eqref{eq:semion_spacetime_cellulation} and \eqref{eq:cup_product_cube_drawing},
\begin{equation}
\label{eq:slanted_cubic_cup0}
\begin{tikzpicture}
\drawshearedcube
\fill[bichain2f] (020-c)--(021-c)--(011-c)--(010-c)--cycle;
\draw[bichain1] (110)--(020);
\end{tikzpicture}
-
\begin{tikzpicture}
\drawshearedcube
\fill[bichain2f] (111-c)--(101-c)--(011-c)--(021-c)--cycle;
\draw[bichain1] (110)--(111);
\end{tikzpicture}
+
\begin{tikzpicture}
\drawshearedcube
\fill[bichain2f] (100-c)--(101-c)--(011-c)--(010-c)--cycle;
\draw[bichain1] (110)--(100);
\end{tikzpicture}
\;.
\end{equation}
That is, the value of the term on the slanted-cube 3-cell is given by the product of $\Abar$ and $d\Abar$, summed over three different pairs of edges (marked red dashed) and faces (shaded blue).
Keeping in mind that the value of $d\Abar$ on a face is the sum over $\Abar$ on its edges, we obtain
\begin{equation}
\label{eq:semion_variable_weights0}
\begin{gathered}
\frac14 \ovl{A_{dx}^{001}}(\ovl{A_y^{002}}+\ovl{A_t^{011}}-\ovl{A_t^{001}}-\ovl{A_y^{001}})\\
-\frac14 \ovl{A_y^{101}}(\ovl{A_{dx}^{011}}+\ovl{A_t^{011}}-\ovl{A_t^{110}}-\ovl{A_{dx}^{010}})\\
+\frac14 \ovl{A_t^{100}}(\ovl{A_{dx}^{000}}+\ovl{A_y^{001}}-\ovl{A_y^{100}}-\ovl{A_{dx}^{010}})
\;.
\end{gathered}
\end{equation}
Eq.~\eqref{eq:semion_variable_weights0} denotes the term of the action for one single unit cell, and we have the same terms shifted to other unit cells.
That is, we have the same terms with the superscripts shifted by arbitrary $(x,y,t)$-coordinates.
Remember that the state-sum weights are the exponential of the action or Lagrangian, so each summand $X$ of the action above yields a weight $e^{2\pi X}$ of the state sum.
The next term is given by $\frac12 \Abar^T\cbar$.
In terms of the concrete lattice variables, this yields a term
\begin{equation}
\label{eq:semion_variable_weights2}
\frac12 \ovl{A_t^{000}}\ovl{c^{000}}\;,
\end{equation}
which we again shift to arbitrary unit cells.
Last, we look at the term $\frac14 \bbar^T\cup_1 d\Abar$.
We use the formula for the $\cup_1$ product on a cube from Eq.~\eqref{eq:cup221} in Appendix~\ref{sec:cup_product}, identifying $x_0$, $x_1$, and $x_2$ in Eqs.~\eqref{eq:semion_spacetime_cellulation} and \eqref{eq:cup_product_cube_drawing}.
However, since $b$ (and thus $\bbar$) is only non-zero on the $(y,t-x)$-faces, only two products of $\bbar$ and $d\Abar$ on two pairs of faces are non-zero.
The following drawing of the cube shows the first face of both pairs striped in red and the two possible second faces (front and left) shaded in blue,
\begin{equation}
-
\begin{tikzpicture}
\drawshearedcube
\fill[bichain1f] (100-c)--(101-c)--(011-c)--(010-c)--cycle;
\fill[bichain2f] (100-c)--(110-c)--(020-c)--(010-c)--cycle;
\fill[bichain2f] (020-c)--(021-c)--(011-c)--(010-c)--cycle;
\end{tikzpicture}
\;.
\end{equation}
Thus, in terms of the concrete variables in Eq.~\eqref{eq:sheared_cubic_labels}, the action inside one unit cell becomes
\begin{equation}
\label{eq:semion_variable_weights1}
\begin{gathered}
-\frac14 \ovl{b^{000}}(\ovl{A_{dx}^{001}}+\ovl{A_t^{001}}-\ovl{A_t^{100}}-\ovl{A_{dx}^{000}})\\
-\frac14 \ovl{b^{000}}(\ovl{A_y^{002}}+\ovl{A_t^{011}}-\ovl{A_t^{001}}-\ovl{A_y^{001}})\\
=
-\frac14 \ovl{b^{000}}(\ovl{A_{dx}^{001}}+\ovl{A_y^{002}}+\ovl{A_t^{011}}-\ovl{A_t^{100}}-\ovl{A_{dx}^{000}}-\ovl{A_y^{001}})
\;.
\end{gathered}
\end{equation}

Having spelled out the microscopic definition of the state sum, we will now turn it into a fault-tolerant circuit.
We will first describe this circuit and show that it implements the state sum, and later share some insights about how we have arrived at this circuit.
The computational-basis configuration of each qubit at each time in the circuit corresponds to the value of one state-sum variable, or at least some simple function of the variables.
The gates of the circuit correspond to the different weights of the state sum.
Roughly speaking, constraints correspond to controlled-$X$ gates, and terms of the action correspond to controlled-phase gates.

The circuit has qubits located at all edges, faces, and vertices of a square lattice.
The qubits at the edges are labeled $A_x$ or $A_y$, the qubits at the faces $b$, and the qubits at the vertices $c$.
Each qubit label $X$ also carries superscripts $X^{ij}$, indicating that it belongs to the unit cell with $x$-coordinate $i$ and $y$-coordinate $j$.
The following shows a patch of the lattice consisting of four unit cells, with the according qubit labels,
\begin{equation}
\label{eq:semion_code_qubits}
\begin{tikzpicture}
\draw (0,0)edge[mark={arr,e},ind=$x$]++(0:0.35) (0,0)edge[mark={arr,e},ind=$y$]++(90:0.35);
\end{tikzpicture}
\quad
\begin{tikzpicture}
\atoms{vertex}{{00/lab={t=$c^{00}$,p=-135:0.4}}}
\atoms{vertex,astyle=gray}{{10/p={2,0},lab={t=$c^{10}$,p=-135:0.35}}, {01/p={0,2},lab={t=$c^{01}$,p=-135:0.4}}, {11/p={2,2},lab={t=$c^{11}$,p=-135:0.4}}}
\draw (00)edge[mark={slab=$A_x^{00}$,r}](10) (00)edge[mark={slab=$A_y^{00}$}](01);
\draw[gray] (01)edge[mark={slab=$A_x^{01}$,r}](11) (10)edge[mark={slab=$A_y^{10}$}](11) (10)edge[mark={slab=$A_x^{10}$,r,e}]++(0:1) (11)edge[mark={slab=$A_x^{11}$,r,e}]++(0:1) (01)edge[mark={slab=$A_y^{01}$,e}]++(90:1) (11)edge[mark={slab=$A_y^{11}$,e}]++(90:1);
\draw[gray] (00)--++(-90:0.4) (00)--++(180:0.4) (10)--++(-90:0.4) (01)--++(180:0.4);
\node at (1,1){$b^{00}$};
\node[gray] at (3,1){$b^{10}$};
\node[gray] at (1,3){$b^{01}$};
\node[gray] at (3,3){$b^{11}$};
\end{tikzpicture}
\;.
\end{equation}
In addition to these qubits, we have classical bits associated to the $-t$-edges and to the $(y,t-x)$-faces of the spacetime cellulation.
These are the results from measuring the value of $c$ and $b$ at these spacetime edges and faces, and remain unchanged after the measurement.
They are later fed into the classical decoder to determine corrections that need to be performed at spacetime interfaces with other types of fault-tolerant circuits needed to perform universal quantum computation, as discussed in Section~\ref{eq:semion_decoding}.
We will also use the most recently read-out classical bits as classical controls for some of the unitary gates of the fault-tolerant circuit.
We will denote these classical bits by $\hat b$ and $\hat c$, with according superscripts to indicate their spacetime unit cell.

The intermediate computational-basis configurations of the qubits in the fault-tolerant circuit are labeled by (1) a spatial unit cell, (2) one out of four qubits per spatial unit cell, (3) the time \emph{period}, and (4) one out of 8 different \emph{stages} in the circuit.
Within the $(0,0)$ spatial unit cell and the $0$th time period, the qubit configurations are given by the following state-sum variables,
\begin{equation}
\label{eq:semion_code_stages}
\begin{tabular}{l|l|l|l|l|l}
Stage & $A_x^{00}$ & $A_y^{00}$ & $b^{00}$ & $c^{00}$ & meas.\\
\hline
\tcr{8'} & $\tcr{A_x^{000}}$ & $\tcr{A_y^{000}}$ & $\tcr{0}$ & $\tcr{c^{000}}$ & $\tcr{\hat c^{00(-1)}}$\\
0 & \ditto & \ditto & \ditto & $A_t^{000}$ & $\varnothing$\\
1 & $A_{dx}^{000}$ & \ditto & $A_y^{100}$ & \ditto &  $\varnothing$\\
2 & \ditto & $A_{dy}^{000}$ & $\scriptstyle{A_y^{100}+A_{dx}^{000}}$ & \ditto & $\varnothing$\\
3 & \ditto & $A_y^{001}$ & $\scriptstyle{A_y^{100}+A_{dx}^{000}+A_{dx}^{010}}$ & \ditto & $\varnothing$\\
4 & \ditto & \ditto & $b^{000}$ & \ditto & $\varnothing$\\
5 & \ditto & \ditto & \ditto & \ditto & $\hat b^{000}$\\
6 & $A_x^{001}$ & \ditto & 0 & \ditto & $\varnothing$\\
7 & \ditto & \ditto & \ditto & $c^{000}$ & $\varnothing$\\
8 & \ditto & \ditto & \ditto & \ditto & $\hat c^{000}$\\
\tcr{0'} & \gditto & \gditto & \gditto & $\tcr{A_t^{001}}$ & \tcr{$\varnothing$}\\
\tcr{1'} & $\tcr{A_{dx}^{001}}$ & \gditto & $\tcr{A_y^{101}}$ & \gditto &  \tcr{$\varnothing$}\\
$\tcr{\ldots}$ & $\tcr{\ldots}$ & $\tcr{\ldots}$ & $\tcr{\ldots}$ & $\tcr{\ldots}$ & $\tcr{\ldots}$
\end{tabular}
\;.
\end{equation}
The column labeled ``meas.'' contains the classical measurement outcomes that are newly available at the respective stage, in the $(0,0)$ spatial unit cell and $0$th time period.
In other words, these are the measurements made when entering the respective stage.
$\varnothing$ denotes that no new measurement results become available, and '' denotes that the qubit value remains unchanged from the proceeding stage.
The configuration in the $(i,j)$ unit cell and $k$th time period is obtained by shifting the $x$ and $y$ superscripts of the variables by $i$ and $j$, and the $t$ superscript by $k$.
The analogous holds for the newly available measurement outcomes.

The circuit consists of layers of unitary gates and measurements that mutually commute.
There are two types of such layers related to the stages above:
For each stage $i$, there are multiple layers $i$, $i'$, $i''$, $\ldots$, that apply controlled-phase gates to the qubits at that stage, which implement most of the terms in the action.
Additionally, there is a layer $i^+$ that maps from stage $i$ to stage $i+1$, implementing the constraints, summations, and also one particular term in the action.

Let us start with the layers $i^+$ implementing the constraints of the state sum.
If two qubits are in configurations $a,b$ at stage $i$ and $a,c$ at stage $i+1$, then a constraint of the form $a+b-c=0$ yields a weight $\delta_{a+b-c}$.
This weight can be implemented by a controlled-$X$ ($CX$, also known as CNOT) gate in the $i^+$ layer:
The amplitude of the $CX$ gate between configurations $a,b$ and $a',c$ is given by $\delta_{a-a'}\delta_{a+b-c}$, which ensures both that the first qubit remains in configuration $a$, and the constraint $a+b=c$ is fulfilled.
Thus, the constraints in Eq.~\eqref{eq:semion_lattice_constraints} yield the following gates within the $(0,0)$ unit cell and the $0$th time period,
\begin{equation}
\label{eq:semion_constraint_gates}
\begin{tabular}{l|l|l}
Weight & Gate & Layer\\
\hline
$\varnothing$ & $CX[A_y^{10}, b^{00}]$ & $0^+$\\
$\varnothing$ & $CX[A_x^{00}, b^{00}]$ & $1^+$\\
$\varnothing$ & $CX[A_x^{01}, b^{00}]$ & $2^+$\\
$\delta_{A_{dx}^{000}+A_y^{001}-A_{dx}^{010}-A_y^{100}-b^{000}}$ & $CX[A_y^{00}, b^{00}]$ & $3^+$\\
$\delta_{A_{dx}^{000}+A_t^{000}-A_x^{000}}$ & $CX[c^{00}, A_x^{00}]$ & $0^+$\\
$\delta_{A_t^{100}+A_{dx}^{000}-A_x^{001}}$ & $CX[c^{01}, A_y^{00}]$ & $1^+$\\
$\delta_{A_{dy}^{000}+A_t^{010}-A_y^{000}}$ & $CX[c^{00}, A_y^{00}]$ & $2^+$\\
$\delta_{A_t^{000}+A_{dy}^{000}-A_y^{001}}$ & $CX[c^{10}, A_x^{00}]$ & $5^+$
\end{tabular}
\;.
\end{equation}
Here $\varnothing$ denotes that the gate does not correspond to a constraint in the state sum, but results from the fact that some of the intermediate-qubit values in Eq.~\eqref{eq:semion_code_stages} are already expressed as sums of variables.
Specifically, this is the case for $b^{00}$ in stages $2$ and $3$.
The gates for other spatial unit cells are obtained by shifting the $x$ and $y$ superscripts of the qubit labels, and the gates for other time periods are the same.

Next, we also need to actually measure the values $b$ and $c$ and record them as classical information.
To this end, we simply perform a measurement on the $b$ and $c$ qubits in the computational basis at a stage where their value equals the value of the variable $b$ or $c$, respectively.
Thus, we can measure $b$ after stage $4$ and $c$ after stage $7$.
This is why $\hat b^{000}$ and $\hat c^{000}$ become available at stages $5$ and $8$ in Eq.~\eqref{eq:semion_code_stages}, respectively.

Next, the value $A_t^{000}$ of the qubit $c^{00}$ at stage $0$ is not determined by the other qubit values through any constraints.
Since the state sum is over all values of $A$ on any edge, subject to the constraints, we need to prepare $A$ in an equal-weight superposition of both computational basis states.
Thus, we need to erase the qubit $c^{00}$ after stage $8$ and prepare it in the state $\ket+$.
Similarly, we need to erase $b^{00}$ after stage $5$ and prepare it in the state $\ket0$.
Finally, the action term $\frac12 \ovl{A_t^{000}} \ovl{c^{000}}$ in Eq.~\eqref{eq:semion_variable_weights2} yields a weight $(-1)^{\ovl{A_t^{000}}\ovl{c^{000}}}$, which cannot be implemented by a controlled-phase gate since $A_t^{000}$ and $c^{000}$ are not simultaneously represented by qubit configurations at any stage.
However, the variables are the values of the qubit $c^{00}$ at two consecutive stages $5$ and $6$.
Now, looking at the weights,
\begin{equation}
\begin{tabular}{r|l|l}
\diagbox{$\scriptstyle{c^{000}}$}{$\scriptstyle{A_t^{000}}$}& $0$ & $1$\\
\hline
$0$ & $1$ & $1$\\
\hline
$1$ & $1$ & $-1$
\end{tabular}
\;,
\end{equation}
we see that this weight is proportional to a unitary matrix, and can be implemented by applying a Hadamard ($H$) gate on the $c^{00}$ qubit in layer $5^+$.
So taking into account all these considerations, we get the following additional gates in the $(0,0)$ unit cell and the $0$th time period,
\begin{equation}
\label{eq:semion_measurement_gates}
\begin{tabular}{l|l|l}
Weight & Gate & Layer\\
\hline
$\varnothing$ & $M_Z[b^{00},\hat b^{000}]$ & $4^+$\\
$\varnothing$ & $M_Z[c^{00},\hat c^{000}]$ & $7^+$\\
$\varnothing$ & $P_{\ket+}[c^{00}]$ & $8^+$\\
$\varnothing$ & $P_{\ket0}[b^{00}]$ & $5^+$\\
$(-1)^{\ovl{A_t^{000}}\ovl{c^{000}}}$ & $H[c^{00}]$ & $6^+$
\end{tabular}
\;.
\end{equation}
Here, $M_Z[a,b]$ denotes a computational-basis (or Pauli-$Z$ basis) measurement on qubit $a$ with the result being stored in the classical bit $b$.
Note that the measurement result is interpreted as a value in $\{0,1\}$, and not in $\{+1,-1\}$.
$P_{\ket\phi}$ denotes erasure and preparation of a qubit in state $\ket\phi$.
The gates for other spatial unit cells and time periods are obtained by shifting the $x$ and $y$ superscripts of qubits and measurement results, and also the $t$ superscripts of the measurement results.

As mentioned in the beginning of this section, the circuit consisting of the gates introduced so far coincides with the syndrome-extraction protocol for the stabilizer toric code in Section VII.B of Ref.~\cite{Dennis2001}.
To be precise, we have slightly ``stretched out'' the timing in our circuit, which makes it easier to include the terms of the action later.
In the following illustrative picture, we have labeled different places in the cellulation by the circuit layers that they are associated with,
\begin{equation}
\begin{tikzpicture}
\tikzset{numl/.style={mark={lab={$\scriptstyle #1$},a}}}
\atoms{vertex}{100/p={$1.4*(1.5,0)$}, 010/p={$1.4*(0,1.5)$}, {001/p={$1.4*(1.1,0.6)$}, astyle=gray}}
\atoms{void}{sz/p={$0.1*(001)$}, sx/p={$-0.07*(010)+0.07*(100)$}, sxx/p={$0.1*(100)$}}
\atoms{vertex,astyle=gray}{000/, 120/p={$2*(010)+(100)$}, 121/p={$(100)+2*(010)+(001)$}}
\atoms{vertex}{110/p={$(100)+(010)$}, 011/p={$(010)+(001)$}, 101/p={$(100)+(001)$}, 111/p={$(100)+(010)+(001)$}, 020/p={$2*(010)$}, 021/p={$2*(010)+(001)$}}
\draw[line width=1.3] (100)edge[](010) (110)edge[](020) (100)edge[](101) (110)edge[](111) (110)edge[](100) (020)edge[](010) (111)edge[](101) (020)edge[](021);
\draw[gray] (100)edge[](000) (010)edge[](000) (120)edge[](110) (120)edge[](020) (120)edge[](121) (121)edge[](021) (121)edge[](111) (110)edge[](121) (100)edge[](111) (110)edge[](010);
\draw[dashed,line width=1.3] (010)edge[](011) (101)edge[](011) (111)edge[](021) (021)edge[](011);
\draw[gray,dashed] (000)edge[](001) (101)edge[](001) (011)edge[](001) (021)edge[](011) (000)edge[](011) (111)edge[](011) (010)edge[](021);
\draw[red] ($(100)!0.5!(010)$) edge[numl=1]++(sz) ($(100)!0.5!(101)$)edge[numl=0]++($-1*(sx)$) ($(000)!0.5!(010)$) edge[numl=0]++(sxx) ($(000)!0.5!(010)$) edge[numl=2]++(sz) ($(100)!0.5!(110)$) edge[numl=3]++($-1*(sxx)$) ($(100)!0.5!(110)$) edge[numl=2]++(sz) ($(101)!0.5!(111)$) edge[numl=1]++($-1*(sz)$) ($(101)!0.5!(111)$) edge[numl=3]++($-1*(sxx)$) ($(001)!0.5!(011)$) edge[numl=0]++(sxx) ($(001)!0.5!(011)$) edge[numl=1]++($-1*(sz)$) ($(101)!0.5!(011)$) edge[numl=2]++($-1*(sz)$) ($(010)!0.5!(011)$) edge[numl=3]++(sx);
\tikzset{numl/.style={mark={lab={$\scriptstyle #1'$},a}}}
\draw[red] ($(110)!0.5!(020)$) edge[numl=1]++(sz) ($(110)!0.5!(111)$)edge[numl=0]++($-1*(sx)$) ($(010)!0.5!(020)$) edge[numl=0]++(sxx) ($(010)!0.5!(020)$) edge[numl=2]++(sz) ($(110)!0.5!(120)$) edge[numl=3]++($-1*(sxx)$) ($(110)!0.5!(120)$) edge[numl=2]++(sz) ($(111)!0.5!(121)$) edge[numl=1]++($-1*(sz)$) ($(111)!0.5!(121)$) edge[numl=3]++($-1*(sxx)$) ($(011)!0.5!(021)$) edge[numl=0]++(sxx) ($(011)!0.5!(021)$) edge[numl=1]++($-1*(sz)$) ($(111)!0.5!(021)$) edge[numl=2]++($-1*(sz)$) ($(020)!0.5!(021)$) edge[numl=3]++(sx);
\tikzset{numcirc/.style={red,draw,circle,inner sep=0.02cm,fill=white}}
\node[numcirc] at ($(010)!.5!(101)$) {$\scriptstyle 4$};
\node[numcirc] at ($(000)!.75!(010)$) {$\scriptstyle 6$};
\node[numcirc] at ($(100)!.75!(110)$) {$\scriptstyle 6$};
\node[numcirc] at ($(001)!.75!(011)$) {$\scriptstyle 6$};
\node[numcirc] at ($(101)!.75!(111)$) {$\scriptstyle 6$};
\node[numcirc] at ($(020)!.5!(111)$) {$\scriptstyle 4$};
\node[numcirc] at ($(010)!.75!(020)$) {$\scriptstyle 6$};
\node[numcirc] at ($(110)!.75!(120)$) {$\scriptstyle 6$};
\node[numcirc] at ($(011)!.75!(021)$) {$\scriptstyle 6$};
\node[numcirc] at ($(111)!.75!(121)$) {$\scriptstyle 6$};
\end{tikzpicture}
\;.
\end{equation}
The layer labels are in red, omitting the $+$ superscripts.
Each $CX$ gate corresponds to either a pair of a $-t$-edge and an adjacent triangle, or a pair of a $(y,x-t)$-face and an adjacent $y$-edge or $x-t$-edge.
We have marked these face-edge pairs by short red lines starting from the center of the edge pointing towards the adjacent face.
Each $M_z[b]$ measurement in layer $4^+$ and $P_{\ket0}$ preparation in layer $5^+$ corresponds to a $(y,x-t)$-face, which we have labeled accordingly by a red $4$.
Each $H$ gate in layer $6^+$, $M_Z[c]$ measurement in layer $7^+$, and $P_{\ket+}$ preparation in layer $8^+$ corresponds to a $-t$-edge, which we have labeled accordingly by a red $6$.
The reader may imagine successively building up the spacetime cellulation in $t$ direction from the bottom to the top:
In the steps labeled $0$ to $3$, we add an adjacent triangle for each $-t$-edge and an adjacent edges for each $(y,x-t)$-face.
In the steps $4$ and $6$, we ``finalize'' a $(y,x-t)$-face, or a $-t$-edge, respectively.

Next, translate the remaining terms of the action into gates of the circuit.
A term can be implemented at stage $i$ if for every involved variable there is a qubit at stage $i$ with the same value.
To this end, we apply the gate that is diagonal in the computational basis, and whose diagonal is given by corresponding the state-sum weight.
In other words, we implement the state-sum weights as controlled-phase gates.
In particular, a term in the action like $\frac14 \ovl a\ovl b$ corresponds to a weight $i^{\ovl a\ovl b}$, and would be implemented by a controlled-$S$ ($CS$) gate acting on the two qubits whose value is $a$ and $b$ at a given stage.
If there is a is a classical bit instead of a qubit that holds the value $a$ at stage $i$, we apply the analogous gate that leaves the value of the classical bit unchanged.
For a weight $i^{\ovl a\ovl b}$, this would be a classically controlled $S$ gate, which we denote by $cS$.
The following table shows the different state-sum weights from Eqs.~\eqref{eq:semion_variable_weights0} and \eqref{eq:semion_variable_weights1}, and the according gates of the circuit:
\begin{equation}
\label{eq:semion_phase_gates}
\begin{tabular}{l|l|l}
Weight & Gate & Stage\\
\hline
$i^{\ovl{A_{dx}^{001}}\ovl{A_y^{002}}}$ & $CS[A_x^{00},A_y^{00}]$ & \tcr{$3_1$,$4_1$,}$5_1'$\\
$i^{\ovl{A_{dx}^{001}}\ovl{A_t^{011}}}$ & $CS[A_x^{00},c^{01}]$ & $1_1''$\tcr{,$2_1$,$3_1$,$4_1$,$5_1$}\\
$(-i)^{\ovl{A_{dx}^{001}}\ovl{A_t^{001}}}$ & $\ovl{CS}[A_x^{00},c^{00}]$ & \tcr{$1_1$,$2_1$,}$3_1^+$\tcr{,$4_1$,$5_1$}\\
$(-i)^{\ovl{A_{dx}^{001}}\ovl{A_y^{001}}}$ & $\ovl{CS}[A_x^{00},A_y^{00}]$ & $1_1'$\\
 & \tcr{$\ovl{CS}[A_x^{00},b^{(-1)0}]$} & \tcr{$1_1$}\\
$(-i)^{\ovl{A_y^{101}}\ovl{A_{dx}^{011}}}$ & $\ovl{CS}[A_y^{10},A_x^{01}]$ & $1_1$\\
 & \tcr{$\ovl{CS}[b^{00},A_x^{01}]$} & \tcr{$1_1$}\\
$(-i)^{\ovl{A_y^{101}}\ovl{A_t^{011}}}$ & \tcr{$\ovl{CS}[A_y^{10},c^{01}]$} & \tcr{$0_1$,$1_1$}\\
 & $\ovl{CS}[b^{00},c^{01}]$ & $1_1$\\
$i^{\ovl{A_y^{101}}\ovl{A_t^{110}}}$ & $CS[A_y^{10},c^{11}]$ & \tcr{3,4,}5\tcr{,6}\\
$i^{\ovl{A_y^{101}}\ovl{A_{dx}^{010}}}$ & $CS[A_y^{10},A_x^{01}]$ & \tcr{3,}4\tcr{,5}\\
$i^{\ovl{A_t^{100}}\ovl{A_{dx}^{000}}}$ & $CS[c^{10},A_x^{00}]$ & \tcr{1,2,}3\tcr{,4,5}\\
$i^{\ovl{A_t^{100}}\ovl{A_y^{001}}}$ & $CS[c^{10},A_y^{00}]$ & \tcr{3,}$4'$\tcr{,5,6}\\
$(-i)^{\ovl{A_t^{100}}\ovl{A_y^{100}}}$ & \tcr{$\ovl{CS}[c^{10}, A_y^{10}]$} & \tcr{0,1}\\
 & $\ovl{CS}[c^{10}, b^{00}]$ & $1'$\\
$(-i)^{\ovl{A_t^{100}}\ovl{A_{dx}^{010}}}$ & $\ovl{CS}[c^{10},A_x^{01}]$ & \tcr{1,2,3,}$4^+$\tcr{,5}\\
$(-i)^{\ovl{b^{000}}\ovl{A_{dx}^{001}}}$ & $\ovl{cS}[\hat b^{000},A_x^{00}]$ & \tcr{$1_1$,$2_1$,$3_1$,}$4_1'$\tcr{,$5_1$}\\
$(-i)^{\ovl{b^{000}}\ovl{A_y^{002}}}$ & $\ovl{cS}[\hat b^{000},A_y^{00}]$ & $3_1$\tcr{,$4_1$--$8_1$,$0_2$,$1_2$}\\
$(-i)^{\ovl{b^{000}}\ovl{A_t^{011}}}$ & $\ovl{cS}[\hat b^{000},c^{01}]$ & \tcr{$0_1$--$3_1$,}$4_1$\tcr{,$5_1$,$6_1$}\\
$i^{\ovl{b^{000}}\ovl{A_t^{100}}}$ & $cS[\hat b^{000},c^{10}]$ & $5'$\tcr{,6}\\
 & \tcr{$CS[b^{00},c^{10}]$} & \tcr{4,5}\\
$i^{\ovl{b^{000}}\ovl{A_{dx}^{000}}}$ & $cS[\hat b^{000},A_x^{00}]$ & 5\\
& \tcr{$CS[b^{00},A_x^{00}]$} & \tcr{4,5}\\
$i^{\ovl{b^{000}}\ovl{A_y^{001}}}$ & $cS[\hat b^{000},A_y^{00}]$ & \tcr{5--8,$0_1$,}$1_1''$\\
 & \tcr{$cS[\hat b^{000},b^{(-1)0}]$} & \tcr{$1_1$}\\
 & \tcr{$CS[b^{00},A_y^{00}]$} & \tcr{4,5}
\end{tabular}
\;.
\end{equation}
A state-sum variable might be represented by multiple different qubits or measurement results in the circuit, at multiple stages.
So a weight can be implemented by different gates acting on different subsets of qubits at different stages.
Above, we have written down all different gates on qubit subsets, and all possible stages, for each weight.
We have chosen a subset and stage for each weight, and grayed out all others.
While any choice of subset and stage is fine, we have made our choice such that as many gates as possible can be applied in parallel acting on disjoint subsets of qubits.
This ensures that the time period of the circuit is as short as possible, and there are no unnecessary time steps with idle qubits.
The subscript of the stage is the time period in which we perform the gate, and no subscript means the $0$th time period.
The time period is only important since we need to shift the $t$ superscript of the classical bits when collecting the gates within the $0$th time period below.
Gates acting only on qubits (not classically controlled by previous measurement outcomes) are the same in each time period.
We have also divided the stages into layers of gates that can be performed in parallel, and indicated these layers by adding primes (like $1$, $1'$, and $1''$) to the stage.
In two cases, phase gates can be performed in parallel with the $CX$ gates or measurements of Eqs.~\eqref{eq:semion_constraint_gates} and \eqref{eq:semion_measurement_gates} in the layers $3^+$ and $4^+$.

For pedagogical reasons, we have translated the smallest possible features of the state sum into smallest possible gates of the circuit.
We can group some of these gates into larger gates.
In particular, we can combine the measurement and state preparation on the $b$ qubit into a single operation $P_{\ket0}\circ M_Z$, and we erase $b$ right after it is measured.
We can do this because $b$ is not required in any following phase gates, but only the measurement result $\hat b$.
Also, can combine the Hadamard gate, $Z$ measurement, and state preparation on the $c$ qubit into one single operation $P_{\ket+}\circ M_X$, where $M_X$ denotes an measurement in the $X$ basis.
After this, the layers $7^+$ and $8^+$ are empty, and the fault-tolerant circuit within the $(0,0)$ spatial unit cell and $0$th time period is given by
\begin{equation}
\begin{tabular}{l|l}
Layer & Gates\\
\hline
\tcr{$6^+_{-1}$} & \tcr{$P_{\ket+}\circ M_X[c^{00},\hat c^{00(-1)}]$}\\
\hline
$0^+$ & $CX[A_y^{10}, b^{00}]$, $CX[c^{00}, A_x^{00}]$\\
\hline
$1$ & $\ovl{CS}[A_y^{10},A_x^{01}]$, $\ovl{CS}[b^{00},c^{01}]$\\
$1'$ & $\ovl{CS}[A_x^{00},A_y^{00}]$, $\ovl{CS}[c^{10},b^{00}]$\\
$1''$ & $cS[\hat b^{00(-1)},A_y^{00}]$, $CS[A_x^{00},c^{01}]$\\
\hline
$1^+$ & $CX[A_x^{00}, b^{00}]$, $CX[c^{01}, A_y^{00}]$\\
\hline
$2^+$ & $CX[A_x^{01}, b^{00}]$, $CX[c^{00}, A_y^{00}]$\\
\hline
$3$ & $\ovl{cS}[\hat b^{00(-1)},A_y^{00}]$, $CS[c^{10},A_x^{00}]$\\
\hline
$3^+$ & $CX[A_y^{00}, b^{00}]$, $\ovl{CS}[A_x^{00},c^{00}]$\\
\hline
$4$ & $CS[A_y^{10},A_x^{01}]$, $\ovl{cS}[\hat b^{00(-1)},c^{01}]$\\
$4'$ & $\ovl{cS}[\hat b^{00(-1)},A_x^{00}]$, $CS[c^{10},A_y^{00}]$\\
\hline
$4^+$ & $P_{\ket0}\circ M_Z[b^{00},\hat b^{000}]$, $CS[c^{10},A_x^{01}]$\\
\hline
$5$ & $cS[\hat b^{000}, A_x^{00}]$, $CS[A_y^{10},c^{11}]$\\
$5'$ & $cS[\hat b^{000},c^{10}]$, $CS[A_x^{00},A_y^{00}]$\\
\hline
$5^+$ & $CX[c^{10}, A_x^{00}]$\\
\hline
$6^+$ & $P_{\ket+}\circ M_X[c^{00},\hat c^{000}]$\\
\hline
$\tcr{0^+_1}$ & $\tcr{CX[A_y^{10}, b^{00}]}$, $\tcr{CX[c^{00}, A_x^{00}]}$\\
\hline
$\tcr{\ldots}$ & $\tcr{\ldots}$
\end{tabular}
\;.
\end{equation}
The following drawing illustrate the different gate layers of the circuit:
\begin{equation}
\begin{tikzpicture}
\node[inner sep=0] (pic0) at (0,0){
\begin{tikzpicture}
\clip (-0.3,-0.3)rectangle(1.3,1.3);
\draw[orange] (-1,0)--++(3,0) (-1,1)--++(3,0) (0,-1)--++(0,3) (1,-1)--++(0,3);
\foreach \x/\y in {0/0, 2/0, 4/0, 1/1, 3/1, 0/2, 2/2, 4/2, 1/3, 3/3, 0/4, 2/4, 4/4}{
\atoms{vertex}{\x\y/p={0.5*\x-0.5,0.5*\y-0.5}}
}
\foreach \x/\y in {1/0, 3/0, 0/1, 2/1, 4/1, 1/2, 3/2, 0/3, 2/3, 4/3, 1/4, 3/4}{
\atoms{vertex,astyle=purple}{\x\y/p={0.5*\x-0.5,0.5*\y-0.5}}
}
\draw[line width=0.3cm,cyan,opacity=0.5,line cap=round] (22-c)--(32-c) (11-c)--(21-c) (13-c)--(23-c) (31-c)--(41-c) (33-c)--(43-c) (12-c)--(02-c);
\path (22-c)--node[midway,yshift=0.2cm]{$\scriptstyle{CX}$} (32-c);
\path (11-c)--node[midway,yshift=0.2cm]{$\scriptstyle{CX}$} (21-c);
\path (13-c)--node[midway,yshift=0.2cm]{$\scriptstyle{CX}$} (23-c);
\end{tikzpicture}
};
\node[inner sep=0] (pic1) at (2,0){
\begin{tikzpicture}
\clip (-0.3,-0.3)rectangle(1.3,1.3);
\draw[orange] (-1,0)--++(3,0) (-1,1)--++(3,0) (0,-1)--++(0,3) (1,-1)--++(0,3);
\foreach \x/\y in {0/0, 2/0, 4/0, 1/1, 3/1, 0/2, 2/2, 4/2, 1/3, 3/3, 0/4, 2/4, 4/4}{
\atoms{vertex}{\x\y/p={0.5*\x-0.5,0.5*\y-0.5}}
}
\foreach \x/\y in {1/0, 3/0, 0/1, 2/1, 4/1, 1/2, 3/2, 0/3, 2/3, 4/3, 1/4, 3/4}{
\atoms{vertex,astyle=purple}{\x\y/p={0.5*\x-0.5,0.5*\y-0.5}}
}
\draw[line width=0.3cm,cyan,opacity=0.5,line cap=round] (23-c)--(32-c) (13-c)--(22-c) (21-c)--(30-c) (03-c)--(12-c) (31-c)--(40-c) (33-c)--(42-c) (11-c)--(20-c);
\path (23-c)--node[midway]{$\scriptstyle{\textcolor{red}{\overline{CS}}}$} (32-c);
\path (13-c)--node[midway]{$\scriptstyle{\textcolor{red}{\overline{CS}}}$} (22-c);
\end{tikzpicture}
};
\node[inner sep=0] (pic2) at (4,0){
\begin{tikzpicture}
\clip (-0.3,-0.3)rectangle(1.3,1.3);
\draw[orange] (-1,0)--++(3,0) (-1,1)--++(3,0) (0,-1)--++(0,3) (1,-1)--++(0,3);
\foreach \x/\y in {0/0, 2/0, 4/0, 1/1, 3/1, 0/2, 2/2, 4/2, 1/3, 3/3, 0/4, 2/4, 4/4}{
\atoms{vertex}{\x\y/p={0.5*\x-0.5,0.5*\y-0.5}}
}
\foreach \x/\y in {1/0, 3/0, 0/1, 2/1, 4/1, 1/2, 3/2, 0/3, 2/3, 4/3, 1/4, 3/4}{
\atoms{vertex,astyle=purple}{\x\y/p={0.5*\x-0.5,0.5*\y-0.5}}
}
\draw[line width=0.3cm,cyan,opacity=0.5,line cap=round] (02-c)--(11-c) (12-c)--(21-c) (22-c)--(31-c) (32-c)--(41-c) (04-c)--(13-c) (14-c)--(23-c) (24-c)--(33-c) (34-c)--(43-c);
\path (12-c)--node[midway]{$\scriptstyle{\textcolor{red}{\overline{CS}}}$} (21-c);
\path (22-c)--node[midway]{$\scriptstyle{\textcolor{red}{\overline{CS}}}$} (31-c);
\end{tikzpicture}
};
\node[inner sep=0] (pic5) at (0,-2){
\begin{tikzpicture}
\clip (-0.3,-0.3)rectangle(1.3,1.3);
\draw[orange] (-1,0)--++(3,0) (-1,1)--++(3,0) (0,-1)--++(0,3) (1,-1)--++(0,3);
\foreach \x/\y in {0/0, 2/0, 4/0, 1/1, 3/1, 0/2, 2/2, 4/2, 1/3, 3/3, 0/4, 2/4, 4/4}{
\atoms{vertex}{\x\y/p={0.5*\x-0.5,0.5*\y-0.5}}
}
\foreach \x/\y in {1/0, 3/0, 0/1, 2/1, 4/1, 1/2, 3/2, 0/3, 2/3, 4/3, 1/4, 3/4}{
\atoms{vertex,astyle=purple}{\x\y/p={0.5*\x-0.5,0.5*\y-0.5}}
}
\draw[line width=0.3cm,cyan,opacity=0.5,line cap=round] (23-c)--(22-c) (21-c)--(20-c) (12-c)--(11-c) (32-c)--(31-c) (13-c)--(14-c) (33-c)--(34-c);
\path (22-c)--node[midway]{$\scriptstyle{CX}$} (23-c);
\path (11-c)--node[midway]{$\scriptstyle{CX}$} (12-c);
\path (31-c)--node[midway]{$\scriptstyle{CX}$} (32-c);
\end{tikzpicture}
};
\node[inner sep=0] (pic4) at (2,-2){
\begin{tikzpicture}
\clip (-0.3,-0.3)rectangle(1.3,1.3);
\draw[orange] (-1,0)--++(3,0) (-1,1)--++(3,0) (0,-1)--++(0,3) (1,-1)--++(0,3);
\foreach \x/\y in {0/0, 2/0, 4/0, 1/1, 3/1, 0/2, 2/2, 4/2, 1/3, 3/3, 0/4, 2/4, 4/4}{
\atoms{vertex}{\x\y/p={0.5*\x-0.5,0.5*\y-0.5}}
}
\foreach \x/\y in {1/0, 3/0, 0/1, 2/1, 4/1, 1/2, 3/2, 0/3, 2/3, 4/3, 1/4, 3/4}{
\atoms{vertex,astyle=purple}{\x\y/p={0.5*\x-0.5,0.5*\y-0.5}}
}
\draw[line width=0.3cm,cyan,opacity=0.5,line cap=round] (21-c)--(22-c) (23-c)--(24-c) (12-c)--(13-c) (32-c)--(33-c) (11-c)--(10-c) (31-c)--(30-c);
\path (21-c)--node[midway]{$\scriptstyle{CX}$} (22-c);
\path (12-c)--node[midway]{$\scriptstyle{CX}$} (13-c);
\path (32-c)--node[midway]{$\scriptstyle{CX}$} (33-c);
\end{tikzpicture}
};
\node[inner sep=0] (pic3) at (4,-2){
\begin{tikzpicture}
\clip (-0.3,-0.3)rectangle(1.3,1.3);
\draw[orange] (-1,0)--++(3,0) (-1,1)--++(3,0) (0,-1)--++(0,3) (1,-1)--++(0,3);
\foreach \x/\y in {0/0, 2/0, 4/0, 1/1, 3/1, 0/2, 2/2, 4/2, 1/3, 3/3, 0/4, 2/4, 4/4}{
\atoms{vertex}{\x\y/p={0.5*\x-0.5,0.5*\y-0.5}}
}
\foreach \x/\y in {1/0, 3/0, 0/1, 2/1, 4/1, 1/2, 3/2, 0/3, 2/3, 4/3, 1/4, 3/4}{
\atoms{vertex,astyle=purple}{\x\y/p={0.5*\x-0.5,0.5*\y-0.5}}
}
\draw[line width=0.3cm,cyan,opacity=0.5,line cap=round] (13-c)to[bend left=20](21-c) (33-c)to[bend left=20](41-c) (23-c)to[bend right=20]++(-0.5,1) (11-c)to[bend left=20]++(0.5,-1) (31-c)to[bend left=20]++(0.5,-1);
\fill[cyan,opacity=0.5] (12)circle(0.15) (32)circle(0.15);
\path (13-c)--node[pos=0.3,xshift=0.1cm]{$\scriptstyle{\textcolor{red}{CS}}$} (21-c);
\path (12-c)++(-90:0.15) node{$\scriptstyle{\textcolor{red}{cS}}$};
\path (32-c)++(-90:0.15) node{$\scriptstyle{\textcolor{red}{cS}}$};
\end{tikzpicture}
};
\node[inner sep=0] (pic6) at (0,-4){
\begin{tikzpicture}
\clip (-0.3,-0.3)rectangle(1.3,1.3);
\draw[orange] (-1,0)--++(3,0) (-1,1)--++(3,0) (0,-1)--++(0,3) (1,-1)--++(0,3);
\foreach \x/\y in {0/0, 2/0, 4/0, 1/1, 3/1, 0/2, 2/2, 4/2, 1/3, 3/3, 0/4, 2/4, 4/4}{
\atoms{vertex}{\x\y/p={0.5*\x-0.5,0.5*\y-0.5}}
}
\foreach \x/\y in {1/0, 3/0, 0/1, 2/1, 4/1, 1/2, 3/2, 0/3, 2/3, 4/3, 1/4, 3/4}{
\atoms{vertex,astyle=purple}{\x\y/p={0.5*\x-0.5,0.5*\y-0.5}}
}
\draw[line width=0.3cm,cyan,opacity=0.5,line cap=round] (21-c)--(31-c) (23-c)--(33-c) (01-c)--(11-c) (03-c)--(13-c);
\fill[cyan,opacity=0.5] (12)circle(0.15) (32)circle(0.15);
\path (21-c)--node[midway,yshift=0.1cm]{$\scriptstyle{\textcolor{red}{CS}}$} (31-c);
\path (23-c)--node[midway,yshift=0.1cm]{$\scriptstyle{\textcolor{red}{CS}}$} (33-c);
\path (12-c)++(180:0.25) node{$\scriptstyle{\textcolor{red}{c\overline S}}$};
\path (32-c)++(180:0.25) node{$\scriptstyle{\textcolor{red}{c\overline S}}$};
\end{tikzpicture}
};
\node[inner sep=0] (pic7) at (2,-4){
\begin{tikzpicture}
\clip (-0.3,-0.3)rectangle(1.3,1.3);
\draw[orange] (-1,0)--++(3,0) (-1,1)--++(3,0) (0,-1)--++(0,3) (1,-1)--++(0,3);
\foreach \x/\y in {0/0, 2/0, 4/0, 1/1, 3/1, 0/2, 2/2, 4/2, 1/3, 3/3, 0/4, 2/4, 4/4}{
\atoms{vertex}{\x\y/p={0.5*\x-0.5,0.5*\y-0.5}}
}
\foreach \x/\y in {1/0, 3/0, 0/1, 2/1, 4/1, 1/2, 3/2, 0/3, 2/3, 4/3, 1/4, 3/4}{
\atoms{vertex,astyle=purple}{\x\y/p={0.5*\x-0.5,0.5*\y-0.5}}
}
\draw[line width=0.3cm,cyan,opacity=0.5,line cap=round] (12-c)--(22-c) (32-c)--(42-c) (11-c)--(21-c) (31-c)--(41-c) (13-c)--(23-c) (33-c)--(43-c);
\path (11-c)--node[midway,yshift=0.1cm]{$\scriptstyle{\textcolor{red}{\overline{CS}}}$} (21-c);
\path (13-c)--node[midway,yshift=0.1cm]{$\scriptstyle{\textcolor{red}{\overline{CS}}}$} (23-c);
\path (12-c)--node[midway,yshift=0.1cm]{$\scriptstyle{CX}$} (22-c);
\end{tikzpicture}
};
\node[inner sep=0] (pic8) at (4,-4){
\begin{tikzpicture}
\clip (-0.3,-0.3)rectangle(1.3,1.3);
\draw[orange] (-1,0)--++(3,0) (-1,1)--++(3,0) (0,-1)--++(0,3) (1,-1)--++(0,3);
\foreach \x/\y in {0/0, 2/0, 4/0, 1/1, 3/1, 0/2, 2/2, 4/2, 1/3, 3/3, 0/4, 2/4, 4/4}{
\atoms{vertex}{\x\y/p={0.5*\x-0.5,0.5*\y-0.5}}
}
\foreach \x/\y in {1/0, 3/0, 0/1, 2/1, 4/1, 1/2, 3/2, 0/3, 2/3, 4/3, 1/4, 3/4}{
\atoms{vertex,astyle=purple}{\x\y/p={0.5*\x-0.5,0.5*\y-0.5}}
}
\draw[line width=0.3cm,cyan,opacity=0.5,line cap=round] (23-c)--(32-c) (21-c)--(30-c) (03-c)--(12-c);
\path (23-c)--node[midway]{$\scriptstyle{\textcolor{red}{CS}}$} (32-c);
\fill[cyan,opacity=0.5] (11)circle(0.15) (31)circle(0.15) (13)circle(0.15) (33)circle(0.15);
\path (11-c)++(90:0.2) node{$\scriptstyle{\textcolor{red}{c\overline S}}$};
\path (31-c)++(90:0.2) node{$\scriptstyle{\textcolor{red}{c\overline S}}$};
\path (13-c)++(90:0.2) node{$\scriptstyle{\textcolor{red}{c\overline S}}$};
\path (33-c)++(90:0.2) node{$\scriptstyle{\textcolor{red}{c\overline S}}$};
\end{tikzpicture}
};
\node[inner sep=0] (pic11) at (0,-6){
\begin{tikzpicture}
\clip (-0.3,-0.3)rectangle(1.3,1.3);
\draw[orange] (-1,0)--++(3,0) (-1,1)--++(3,0) (0,-1)--++(0,3) (1,-1)--++(0,3);
\foreach \x/\y in {0/0, 2/0, 4/0, 1/1, 3/1, 0/2, 2/2, 4/2, 1/3, 3/3, 0/4, 2/4, 4/4}{
\atoms{vertex}{\x\y/p={0.5*\x-0.5,0.5*\y-0.5}}
}
\foreach \x/\y in {1/0, 3/0, 0/1, 2/1, 4/1, 1/2, 3/2, 0/3, 2/3, 4/3, 1/4, 3/4}{
\atoms{vertex,astyle=purple}{\x\y/p={0.5*\x-0.5,0.5*\y-0.5}}
}
\draw[line width=0.3cm,cyan,opacity=0.5,line cap=round] (12-c)--(13-c) (32-c)--(33-c) (11-c)--(10-c) (31-c)--(30-c);
\path (12-c)--node[midway]{$\scriptstyle{\textcolor{red}{CS}}$} (13-c);
\path (32-c)--node[midway]{$\scriptstyle{\textcolor{red}{CS}}$} (33-c);
\fill[cyan,opacity=0.5] (21)circle(0.15) (23)circle(0.15);
\path (21-c)++(90:0.2) node{$\scriptstyle{\textcolor{red}{cS}}$};
\path (23-c)++(90:0.2) node{$\scriptstyle{\textcolor{red}{cS}}$};
\end{tikzpicture}
};
\node[inner sep=0] (pic10) at (2,-6){
\begin{tikzpicture}
\clip (-0.3,-0.3)rectangle(1.3,1.3);
\draw[orange] (-1,0)--++(3,0) (-1,1)--++(3,0) (0,-1)--++(0,3) (1,-1)--++(0,3);
\foreach \x/\y in {0/0, 2/0, 4/0, 1/1, 3/1, 0/2, 2/2, 4/2, 1/3, 3/3, 0/4, 2/4, 4/4}{
\atoms{vertex}{\x\y/p={0.5*\x-0.5,0.5*\y-0.5}}
}
\foreach \x/\y in {1/0, 3/0, 0/1, 2/1, 4/1, 1/2, 3/2, 0/3, 2/3, 4/3, 1/4, 3/4}{
\atoms{vertex,astyle=purple}{\x\y/p={0.5*\x-0.5,0.5*\y-0.5}}
}
\draw[line width=0.3cm,cyan,opacity=0.5,line cap=round] (23-c)to[bend left=20](31-c) (21-c)to[bend left=20]++(0.5,-1) (11-c)to[bend right=20]++(-0.5,1) (13-c)to[bend right=20]++(-0.5,1) (33-c)to[bend right=20]++(-0.5,1);
\fill[cyan,opacity=0.5] (22)circle(0.15);
\path (23-c)--node[pos=0.3,xshift=0.1cm]{$\scriptstyle{\textcolor{red}{CS}}$} (31-c);
\path (22-c)++(-90:0.2) node{$\scriptstyle{M_Z}$};
\end{tikzpicture}
};
\node[inner sep=0] (pic9) at (4,-6){
\begin{tikzpicture}
\clip (-0.3,-0.3)rectangle(1.3,1.3);
\draw[orange] (-1,0)--++(3,0) (-1,1)--++(3,0) (0,-1)--++(0,3) (1,-1)--++(0,3);
\foreach \x/\y in {0/0, 2/0, 4/0, 1/1, 3/1, 0/2, 2/2, 4/2, 1/3, 3/3, 0/4, 2/4, 4/4}{
\atoms{vertex}{\x\y/p={0.5*\x-0.5,0.5*\y-0.5}}
}
\foreach \x/\y in {1/0, 3/0, 0/1, 2/1, 4/1, 1/2, 3/2, 0/3, 2/3, 4/3, 1/4, 3/4}{
\atoms{vertex,astyle=purple}{\x\y/p={0.5*\x-0.5,0.5*\y-0.5}}
}
\draw[line width=0.3cm,cyan,opacity=0.5,line cap=round] (12-c)to[bend left=20](31-c) (33-c)to[bend right=20]++(-1,0.5) (13-c)to[bend right=20]++(-1,0.5) (11-c)to[bend right=20]++(-1,0.5) (32-c)to[bend left=20]++(1,-0.5);
\fill[cyan,opacity=0.5] (21)circle(0.15) (23)circle(0.15);
\path (12-c)--node[pos=0.7,yshift=0.1cm]{$\scriptstyle{\textcolor{red}{CS}}$} (31-c);
\path (21-c)++(-90:0.2) node{$\scriptstyle{\textcolor{red}{c\overline S}}$};
\path (23-c)++(-90:0.2) node{$\scriptstyle{\textcolor{red}{c\overline S}}$};
\end{tikzpicture}
};
\node[inner sep=0] (pic12) at (0,-8){
\begin{tikzpicture}
\clip (-0.3,-0.3)rectangle(1.3,1.3);
\draw[orange] (-1,0)--++(3,0) (-1,1)--++(3,0) (0,-1)--++(0,3) (1,-1)--++(0,3);
\foreach \x/\y in {0/0, 2/0, 4/0, 1/1, 3/1, 0/2, 2/2, 4/2, 1/3, 3/3, 0/4, 2/4, 4/4}{
\atoms{vertex}{\x\y/p={0.5*\x-0.5,0.5*\y-0.5}}
}
\foreach \x/\y in {1/0, 3/0, 0/1, 2/1, 4/1, 1/2, 3/2, 0/3, 2/3, 4/3, 1/4, 3/4}{
\atoms{vertex,astyle=purple}{\x\y/p={0.5*\x-0.5,0.5*\y-0.5}}
}
\draw[line width=0.3cm,cyan,opacity=0.5,line cap=round] (12-c)--(21-c) (23-c)--(14-c) (32-c)--(41-c);
\path (12-c)--node[midway]{$\scriptstyle{\textcolor{red}{CS}}$} (21-c);
\fill[cyan,opacity=0.5] (11)circle(0.15) (31)circle(0.15) (13)circle(0.15) (33)circle(0.15);
\path (11-c)++(-90:0.2) node{$\scriptstyle{\textcolor{red}{cS}}$};
\path (31-c)++(-90:0.2) node{$\scriptstyle{\textcolor{red}{cS}}$};
\path (13-c)++(-90:0.2) node{$\scriptstyle{\textcolor{red}{cS}}$};
\path (33-c)++(-90:0.2) node{$\scriptstyle{\textcolor{red}{cS}}$};
\end{tikzpicture}
};
\node[inner sep=0] (pic13) at (2,-8){
\begin{tikzpicture}
\clip (-0.3,-0.3)rectangle(1.3,1.3);
\draw[orange] (-1,0)--++(3,0) (-1,1)--++(3,0) (0,-1)--++(0,3) (1,-1)--++(0,3);
\foreach \x/\y in {0/0, 2/0, 4/0, 1/1, 3/1, 0/2, 2/2, 4/2, 1/3, 3/3, 0/4, 2/4, 4/4}{
\atoms{vertex}{\x\y/p={0.5*\x-0.5,0.5*\y-0.5}}
}
\foreach \x/\y in {1/0, 3/0, 0/1, 2/1, 4/1, 1/2, 3/2, 0/3, 2/3, 4/3, 1/4, 3/4}{
\atoms{vertex,astyle=purple}{\x\y/p={0.5*\x-0.5,0.5*\y-0.5}}
}
\draw[line width=0.3cm,cyan,opacity=0.5,line cap=round] (21-c)--(31-c) (23-c)--(33-c) (03-c)--(13-c) (01-c)--(11-c);
\path (21-c)--node[midway,yshift=0.1cm]{$\scriptstyle{CX}$} (31-c);
\path (23-c)--node[midway,yshift=0.1cm]{$\scriptstyle{CX}$} (33-c);
\end{tikzpicture}
};
\node[inner sep=0] (pic14) at (4,-8){
\begin{tikzpicture}
\clip (-0.3,-0.3)rectangle(1.3,1.3);
\draw[orange] (-1,0)--++(3,0) (-1,1)--++(3,0) (0,-1)--++(0,3) (1,-1)--++(0,3);
\foreach \x/\y in {0/0, 2/0, 4/0, 1/1, 3/1, 0/2, 2/2, 4/2, 1/3, 3/3, 0/4, 2/4, 4/4}{
\atoms{vertex}{\x\y/p={0.5*\x-0.5,0.5*\y-0.5}}
}
\foreach \x/\y in {1/0, 3/0, 0/1, 2/1, 4/1, 1/2, 3/2, 0/3, 2/3, 4/3, 1/4, 3/4}{
\atoms{vertex,astyle=purple}{\x\y/p={0.5*\x-0.5,0.5*\y-0.5}}
}
\fill[cyan,opacity=0.5] (11)circle(0.15) (31)circle(0.15) (13)circle(0.15) (33)circle(0.15);
\path (11-c)++(-90:0.2) node{$\scriptstyle{M_X}$};
\path (31-c)++(-90:0.2) node{$\scriptstyle{M_X}$};
\path (13-c)++(-90:0.2) node{$\scriptstyle{M_X}$};
\path (33-c)++(-90:0.2) node{$\scriptstyle{M_X}$};
\end{tikzpicture}
};
\draw[very thick] (pic0.east)edge[->](pic1.west) (pic1.east)edge[->](pic2.west) (pic2.south)edge[->](pic3.north) (pic3.west)edge[->](pic4.east) (pic4.west)edge[->](pic5.east) (pic5.south)edge[->](pic6.north) (pic6.east)edge[->](pic7.west) (pic7.east)edge[->](pic8.west) (pic8.south)edge[->](pic9.north) (pic9.west)edge[->](pic10.east) (pic10.west)edge[->](pic11.east) (pic11.south)edge[->](pic12.north) (pic12.east)edge[->](pic13.west) (pic13.east)edge[->](pic14.west);
\draw[very thick,->] (pic14.south)--++(-90:0.4)-|($(pic0.west)+(180:0.4)$)--(pic0.west);
\end{tikzpicture}
\;.
\end{equation}
Here, $A$ qubits correspond to purple dots on the square lattice (orange lines), and $b$ can $c$ qubits are black dots.
Each gate is indicated by shading the involved qubits in blue with an according label.
$cS$ represents a single-qubit $S$ gate that may or may not be performed depending on previous measurement outcomes.

Finally, after having shown how the 1-form symmetric state sum can be turned into a circuit, let us share some insights into how we have found this circuit.
We have started with the toric-code syndrome-extraction protocol in Section VII.B of Ref.~\cite{Dennis2001}, and turned this protocol into a cellulation.
An instructive intermediate step for this might be to turn the $+1$ postselected circuit into a ZX diagram, and then replace every $X$-type by face and every $Z$-type tensor by an edge, see Ref.~\cite{path_integral_qec}.
Then we have constructed the higher-cup product formulas on a cubic lattice in Appendix~\ref{sec:cup_product}.
We have tried different ways to choose the edge directions in Eq.~\eqref{eq:semion_spacetime_cellulation}, and to identify the basis vectors of the sheared cubic lattice with $x_0$, $x_1$, and $x_2$ in Eq.~\eqref{eq:cup_product_cube_drawing}.
We have found choices such that for every term in the action, there is a stage and a subset of qubits that represent all the involved state-sum variables.
To achieve this goal, we have also ``stretched out'' the timing of the circuit slightly:
If we were only implementing the toric code with trivial action, we could update the qubit $A_x^{00}$ from $A_{dx}^{000}$ to $A_x^{001}$ via the gate $CX[c^{10},A_x^{00}]$ already in layer $3^+$ instead of $5^+$.
However, to implement the action term $\frac14 \ovl{b^{000}}\ovl{A_{dx}^{000}}$, $b^{000}$ and $A_{dx}^{000}$ need to be simultaneously represented by qubits at some stage.
To achieve this, we need to delay the update of the qubit $A_x^{00}$ by one stage.
In fact, we have chosen to delay it even by two stages.
Then, we can implement the term via a classically controlled $S$ gate via the measurement outcome $\hat b^{000}$ which is then available, instead of via a coherent $CS$ gate.

All in all, it turned out that we are lucky and it is possible to implement all terms in the action by making appropriate choices of the microscopic definition of the action, as well as delaying some of the gates in the untwisted (trivial-action) circuit.
If we had been less lucky, there would have been two less preferable options for implementing the terms in the action.
The first option is to express the state-sum variables that are not represented by a qubit at a chosen stage through the available qubit values.
For example, looking at Eq.~\eqref{eq:semion_code_stages}, $A_{dy}^{000}$ is not represented at stage $1$, but equals the sum of qubit values $A_y^{00}+c^{00}$.
Then, a weight involving $A_{dy}^{000}$ could be implemented by a gate involving both qubits $A_y^{00}$ and $c^{00}$ instead of a single qubits.
So this method is less preferable since it will in general lead to gates acting on more qubits, for example 3-qubit gates instead of 2-qubit gates.
The second option is to copy the computational-basis configuration of a qubit to an auxiliary qubit, such that its value at one stage remains available also at later stages.
When refreshing the value of this auxiliary qubit in the next round, we need to measure it in the $X$ basis before erasure.
The measurement result then corresponds to the value of $c$ on the edge whose state-sum variable the qubit was representing.
This method is less preferable simply because it requires additional qubits, copy operations, and measurements.

\subsection{Decoding and correction}
\label{eq:semion_decoding}
In this section we will show how to specialize the general method for decoding and correction given in Section~\ref{sec:double_semion_code} to the present fault-tolerant circuit.
We assume that we need to perform corrections at a time $T$ because we want to transfer the logical information into a non-abelian phase via a temporal interface with a hypothetical non-abelian circuit.

Let us start by showing how to implement the corrections in the microscopic circuit.
To this end, we insert classically controlled unitaries into the circuit, with one $\zz_2$-valued control at every $t-x$ and $y$-edge, and every $(-t,y)$ and $(-t,t-x)$-faces.
A configuration of controls thus corresponds to a 1-chain $c_c$ supported on these edges, and a 2-cochain $b_c$ supported on these faces.
For a fixed configuration of controls $(b_c,c_c)$ and measurement outcomes $(b_s,c_s)$, the circuit is equal to the path integral with 1-form symmetry defects given $b_s+b_c$ and $c_s+c_c$.
Note that it does not make sense add controls that insert defects at the $-t$-edges or $(y,t-x)$-faces, since this is equivalent to simply adding the controls to the measurement results at these edges and faces in classical post-processing.
The additional $b$ and $c$ variables in the microscopic state sum will be labeled as follows,
\begin{equation}
\label{eq:semion_correction_defects_variables}
\begin{tikzpicture}
\atoms{vertex}{100/p={3,0}, 010/p={0,3}, {001/p={2.2,1.4}, astyle=gray}}
\atoms{vertex,astyle=gray}{000/, 120/p={$2*(010)+(100)$}, 121/p={$(100)+2*(010)+(001)$}}
\atoms{vertex}{110/p={$(100)+(010)$}, 011/p={$(010)+(001)$}, 101/p={$(100)+(001)$}, 111/p={$(100)+(010)+(001)$}, 020/p={$2*(010)$}, 021/p={$2*(010)+(001)$}}
\draw[] (100)edge[mark=arr,varlab={c_{dx}^{000}}{r}](010) (110)edge[mark=arr,varlab={c_{dx}^{001}}{r}](020) (100)edge[mark=arr,varlab={c_y^{100}}{r}](101) (110)edge[mark=arr,varlab={c_y^{101}}{}](111) (110)edge[mark=arr](100) (020)edge[mark=arr](010) (111)edge[mark=arr](101) (020)edge[mark=arr,varlab={c_y^{002}}{}](021);
\draw[] (010)edge[mark=arr](000) (120)edge[mark=arr](110) (120)edge[mark=arr,varlab={c_y^{102}}{}](121) (121)edge[mark=arr](111);
\draw[dashed] (010)edge[mark=arr,varlab={c_y^{001}}{r}](011) (101)edge[mark=arr,varlab={c_{dx}^{010}}{r}](011) (111)edge[mark=arr,varlab={c_{dx}^{011}}{r}](021) (021)edge[mark=arr](011);
\draw[dashed] (000)edge[mark=arr,varlab={c_y^{000}}{}](001) (011)edge[mark=arr](001);
\node at ($(010)!0.5!(110)$) {$b_x^{000}$};
\node at ($(011)!0.5!(111)$) {$b_x^{010}$};
\node at ($(000)!0.5!(011)$) {$b_y^{000}$};
\node at ($(010)!0.5!(021)$) {$b_y^{001}$};
\node at ($(100)!0.5!(111)$) {$b_y^{100}$};
\node at ($(110)!0.5!(121)$) {$b_y^{101}$};
\end{tikzpicture}
\;.
\end{equation}
Here we put each face label at the center of the corresponding face.
Let us start by discussing how the additional support of $b$ and $c$ changes the state sum microscopically.
First of all, the first and third constraint in Eq.~\eqref{eq:semion_lattice_constraints} are replaced by
\begin{equation}
\begin{gathered}
A_{dx}^{001}+A_t^{001}-A_x^{000}=b_x^{000}\;,\\
A_y^{001}-A_t^{000}-A_{dy}^{000}=b_y^{000}\;.
\end{gathered}
\end{equation}
When taking these constraints literally, $b_x^{000}$ and $b_y^{000}$ correspond to the value of $b$ not on the 4-gon faces as indicated in Eq.~\eqref{eq:semion_correction_defects_variables}, but rather on one of its triangles.
These two things are the same though, since we define the action on the sheared cubic lattice without the diagonal edges, independent of the ``intermediate'' variables $A_x$ and $A_{dy}$.
The action term $\frac12 A^Tc$ yields additional terms
\begin{equation}
\label{eq:correction_cweight}
\frac12 \ovl{c_{dx}^{000}}\ovl{A_{dx}^{000}}+\frac12 \ovl{c_y^{000}}\ovl{A_y^{000}}\;.
\end{equation}
Finally, consider the term $\frac14 \bbar^T\cup_1 d\Abar$.
Taking the cup product formula from Eq.~\eqref{eq:cup221} after identifying $x_0$, $x_1$, and $x_2$ in Eqs.~\eqref{eq:semion_spacetime_cellulation} and \eqref{eq:cup_product_cube_drawing}, and taking into account that the first argument ($\bbar$) restricted to the $(-t,y)$ or $(-t,t-x)$-faces, we get
\begin{equation}
+
\begin{tikzpicture}
\drawshearedcube
\fill[bichain1f] (100-c)--(101-c)--(111-c)--(110-c)--cycle;
\fill[bichain2f] (101-c)--(111-c)--(021-c)--(011-c)--cycle;
\fill[bichain2f] (110-c)--(111-c)--(021-c)--(020-c)--cycle;
\end{tikzpicture}
-
\begin{tikzpicture}
\drawshearedcube
\fill[bichain1f] (100-c)--(110-c)--(020-c)--(010-c)--cycle;
\fill[bichain2f] (020-c)--(021-c)--(011-c)--(010-c)--cycle;
\end{tikzpicture}
+
\begin{tikzpicture}
\drawshearedcube
\fill[bichain1f] (101-c)--(111-c)--(021-c)--(011-c)--cycle;
\fill[bichain2f] (110-c)--(111-c)--(021-c)--(020-c)--cycle;
\end{tikzpicture}
\;.
\end{equation}
So the term of the action expressed in terms of the microscopic state-sum variables is given by,
\begin{equation}
\label{eq:b_correction_weights}
\begin{gathered}
\frac14 \ovl{b_y^{100}}(\ovl{A_{dx}^{001}}+\ovl{A_y^{002}}+\ovl{A_t^{011}}-\ovl{A_{y}^{101}}-\ovl{A_{t}^{110}}-\ovl{A_{dx}^{010}})\\
-\frac14 \ovl{b_x^{000}}(\ovl{A_y^{002}}+\ovl{A_t^{011}}-\ovl{A_t^{001}}-\ovl{A_y^{001}})\\
+\frac14 \ovl{b_x^{010}}(\ovl{A_{dx}^{001}}+\ovl{A_y^{002}}-\ovl{A_y^{101}}-\ovl{A_{dx}^{011}})
\;.
\end{gathered}
\end{equation}
Let us reorganize the weights in Eqs.~\eqref{eq:correction_cweight} and \eqref{eq:b_correction_weights}, such that there is one weight for each $A$-variable.
When doing this, we also redistribute weights between different unit cells.
After this, we get the following weight per unit cell,
\begin{equation}
\frac14 (\ovl{A_{dx}^{000}} \ovl{f_{dx}^{000}} + \ovl{A_y^{000}} \ovl{f_y^{000}} + \ovl{A_t^{000}} \ovl{f_t^{000}})\;,\\
\end{equation}
with
\begin{equation}
\begin{gathered}
\begin{multlined}
f_{dx}^{000} \coloneqq 2\ovl{c_{dx}^{000}}+\ovl{b_y^{10(-1)}}-\ovl{b_y^{1(-1)0}}+\ovl{b_x^{01{-1}}}\\-b_x^{00(-1)}\mod 4\;,
\end{multlined}
\\
\begin{multlined}
f_y^{000}\coloneqq 2\ovl{c_y^{000}}+\ovl{b_y^{10(-2)}}-\ovl{b_y^{00(-1)}}-\ovl{b_x^{00{-2}}}\\+\ovl{b_x^{00(-1)}}+\ovl{b_x^{01(-1)}}-\ovl{b_x^{(-1)1(-1)}}\mod 4\;,
\end{multlined}
\\
\begin{multlined}
f_t^{000}\coloneqq \ovl{b_y^{1(-1)(-1)}} - \ovl{b_y^{0(-1)0}} - \ovl{b_x^{0(-1)(-1)}}\\ + \ovl{b_x^{00(-1)}}\mod 4\;.
\end{multlined}
\end{gathered}
\end{equation}
Here, $f_{dx}$, $f_y$, and $f_t$ are valued in $\zz_4$, and, e.g., $\ovl{f_{dx}}$ denotes a lift of $f_{dx}$ from $\zz_4$ to $\zz$.

Let us now implement the state-sum weights above as gates in the circuit.
Weights coming from constraints are implemented by classically controlled $X$ gates, and weights coming from the action as classically controlled phase gates.
To this end, we introduce a new gate $c^4S[a,b]$, with $a\in\zz_4$ as classical input, acting on the qubit $b$ by $S^a$, that is, by applying an $S$ gate $a$ times.
A term $\frac14 \ovl a \ovl f$ giving rise to a weight $i^{\ovl a\ovl f}$ can then be implemented by the gate $c^4S[f,a]$, acting at a stage where $f\in\zz_4$ is represented by a classical degree of freedom with the same name, and $a\in\zz_2$ is represented by a qubit with the same name.
With this, the gates that we have to apply at specific stages are as follows,
\begin{equation}
\begin{tabular}{l|l|l}
Weight & Gate & Stage\\
\hline
$\delta_{\substack{A_{dx}^{001}+A_t^{001}\\-A_x^{000}-b_x^{000}}}$ & $cX[\hat b_x^{000}, A_x^{00}]$ & $0_1^{++}$\\
$\delta_{\substack{A_y^{001}-A_t^{000}\\-A_{dy}^{000}-b_y^{000}}}$ & $cX[\hat b_y^{000}, A_y^{00}]$ & $2^{++}$\\
$i^{\ovl{A_{dx}^{000}} \ovl{f_{dx}^{000}}}$ & $c^4S[f_{dx}^{000},A_x^{00}]$ & $1'''$\tcr{,2,3,4,5}\\
$i^{\ovl{A_y^{000}} \ovl{f_y^{000}}}$ & $c^4S[f_y^{000},A_y^{00}]$ & \tcr{$4_{-1}$--$8_{-1}$,0,}$1'''$ \\
$i^{\ovl{A_t^{000}} \ovl{f_t^{000}}}$ & $c^4S[f_t^{000},A_y^{00}]$ & \tcr{0,}$1'''$\tcr{,2--6}
\end{tabular}
\;.
\end{equation}
Here, $i^{++}$ denotes an additional layer in the protocol that is executed right after the layer $i^+$, and $1'''$ is a new layer executed after layer $1''$.
Each of action weights can be implemented by multiple gates acting on multiple stages, and we have decided to perform all gates at stage $1$.
When given the classical inputs consisting on the values of $b_x$, $b_y$, $c_y$, and $c_{dx}$, we proceed by computing the $\zz_4$ values of $f_{dx}$, $f_y$, and $f_t$ purely classically.
Then we apply the identity, $S$ gate, $Z$ gate, or $\ovl S$ gate depending on these values.
So corrections can be implemented by only 5 single-qubit gates per unit cell, introducing arbitrary $b$ and $c$ defects on all edges and faces where we do not measure $b$ and $c$.

After having discussed how to perform corrections, let us review how the classical decoder described in Section~\ref{sec:dynamical_codes_from_integrals} specializes to the present fault-tolerant circuit.
The input to the classical decoder at time $T$ is the syndrome recorded up to that point, consisting of a $b_s$ 2-chain supported on the $(y,t-x)$-faces, and a $c_s$ 1-chain supported on the $-t$-edges on the $L\times L\times T$ spacetime lattice.
The decoder proceeds by determining a (minimum-weight) fix of $b_s$ and $c_s$ on the sheared cubic lattice, with open boundary conditions at time $T$.
Then we close off the fixed $b_s+\mwpm(db_s)$ and $c_s+\mwpm(\delta c_s)$ in a homologically trivial way at time $T$ via $\widetilde b_c$ and $\widetilde c_c$.
Note that there is no such closure supported on the $(-t,y)$ and $(-t,t-x)$-faces, and $y$ and $t-x$-edges, since some of these move anyons ``diagonally'' instead of perpendicularly to the $t$ direction.
This is not a problem:
The controls $b_c$ and $c_c$ are chosen to be the restrictions of $\widetilde b_c$ and $\widetilde c_c$ to the corresponding edges and faces.
Even though these controls do not directly close off $b_s+\mwpm(db_s)$ and $c_s+\mwpm(\delta c_s)$, the following measurement results at the $-t$-edges and $(y,t-x)$-faces will automatically close off the syndrome.
At least this is the case if there is no noise from time $T$ onward, since the measurement results of the noiseless circuit deterministically form a $c$ 1-cycle and $b$ 2-cocycle.
Even if there is noise during the corrections at time $T$, the resulting $b$ and $c$ fail to be closed only by a small amount, which is good enough for the hypothetical non-abelian circuit to take over.

\section{General abelian twisted quantum doubles}
\label{sec:general_doubles}
In this section, we show how to generalize the methods from Section~\ref{sec:double_semion} to arbitrary abelian Dijkgraaf-Witten theories, or equivalently twisted quantum doubles.
This allows us to construct dynamic fault-tolerant protocols for arbitrary non-chiral abelian topological phases.

\subsection{Path integral}
\label{sec:general_path_integral}
Let us start by defining abelian Dijkgraaf-Witten (or twisted quantum double) path integrals in a way that makes them particularly accessible for explicit computation.
By abelian, we mean that the anyon theory describing these models is abelian.
This implies that the gauge group $G$ needs to be abelian, but also that the group 3-cocycle $\omega\in Z^3(BG,U(1))$ needs to be composed of type-I and type-II cocycles.

Any abelian group is isomorphic to a product of cyclic factors,
\begin{equation}
G=\bigoplus_{0\leq i<k} \zz_{l_i}\;.
\end{equation}
After putting the orders on the diagonal of a diagonal matrix $f$,
\begin{equation}
f_{ij}=\delta_{ij} l_i\;,
\end{equation}
we have
\begin{equation}
G=\zz_f\coloneqq f\zz^k\backslash\zz^k\;.
\end{equation}
That is, $\zz_f$ denotes the finite abelian group formed by equivalence classes of $\zz^k$-elements under $a\simeq a+fv$.
This notation still makes sense for arbitrary, non-diagonal invertible integer matrices
\begin{equation}
f\in  \zz^{k\times k}\cap GL(k,\rr)\;,
\end{equation}
and all expressions derived in this section will also apply to non-diagonal $f$.

The group 3-cocycle, or equivalently the action of the model, is determined by a $\rr/\zz$-valued matrix
\begin{equation}
F\in (\rr/\zz)^{k\times k}\;,
\end{equation}
such that
\begin{equation}
\label{eq:fFcondition}
f^TFf= 0\;,
\end{equation}
as an equation valued in $\rr/\zz$.

The Dijkgraaf-Witten state sum is a sum over all $G$-valued 1-cocycles $A$, as discussed in Section~\ref{sec:fixed_points}.
The action is given by
\begin{equation}
\label{eq:general_action}
S[A]\coloneqq \Abar^T F \cup d\Abar\;,
\end{equation}
where $\ovl x\in \zz^k$ is a fixed lift of $x\in \zz_f$ such that $f\backslash \ovl x=x$, and $\Abar$ denotes this lift applied to $A$ element-wise.
Also, $\Abar$ is now interpreted as a vector in $\zz^{k\times S_1[M]}$.
Accordingly, $\cup$ and $d$ are linear maps acting only on the $S_1[M]$ tensor factor and $F$ acts only on the $k$ tensor factor, and we have
\begin{equation}
\label{eq:tensor_factors_commute}
F\cup=\cup F\;,\quad Fd=dF\;.
\end{equation}
The according 3-cocycle $\omega\in Z^3(BG,U(1))$ in its usual form can be obtained by evaluating $S$ on a tetrahedron, analogous to Eq.~\eqref{eq:semion_group_3cocycle},
\begin{equation}
\begin{multlined}
\omega(A_{01},A_{12},A_{23})=e^{2\pi iS[A]}\\
= e^{2\pi i \ovl{A_{01}}^T F\big(\ovl{A_{12}}+\ovl{A_{23}}-\ovl{(A_{12}+A_{23})}\big)}\;.
\end{multlined}
\end{equation}

The central property is still the invariance of the action under the gauge transformation,
\begin{equation}
A'=A+d\alpha\;.
\end{equation}
Analogous to Section~\ref{sec:double_semion_pathintegral}, we have
\footnote{
Below, $s$ is the group 2-cocycle in $Z^2(B\zz_f,\zz^k)$ corresponding to the central extension $\zz^k\xrightarrow{f\cdot}\zz^k\xrightarrow{f\backslash\bullet} \zz_f$.
Applied to $i$-cocycles $x$, $v_x$ is the Bockstein homomorphism corresponding to the same short exact sequence.
Note that for $i=1$, the Bockstein homomorphism again corresponds to a group 2-cocycle in $Z^2(B\zz_f,\zz^k)$, which is the same as $s_{x,y}$ above.}
\begin{equation}
\ovl{x+y}=\ovl x+\ovl y + fs_{x,y}\;,
\end{equation}
for $x,y\in \zz_f$, and
\begin{equation}
\label{eq:general_bockstein}
\ovl{dx} = d\ovl x+fv_x\;,
\end{equation}
for a $\zz_f$-valued $i$-cochain $x$ and some $\zz^k$-valued $i+1$-cochain $v_x$.
So we have
\begin{equation}
\ovl{A'}=\ovl{A+d\alpha} = \Abar+d\ovl\alpha+f(s_{A,d\alpha}+v_\alpha)\eqqcolon \Abar+d\ovl\alpha+fx
\;.
\end{equation}
So to show gauge invariance, we first compute the variance under adding $fx$ to $\ovl A$,
\begin{equation}
\label{eq:a_periodicity}
\begin{multlined}
\Sbar[\Abar+fx]-\Sbar[\Abar]\\
=(\Abar+fx)^T F\cup d(\Abar+fx) - \Abar^TF\cup d\Abar\\
= x^Tf^T F\cup d\Abar + \Abar^T F f\cup dx + x^T f^T F f\cup dx\\
= -x^Tf^TFf\cup v_A + (d\Abar)^TFf\cup x = 0\;.
\end{multlined}
\end{equation}
We have used Eqs.~\eqref{eq:fFcondition}, \eqref{eq:tensor_factors_commute}, \eqref{eq:general_bockstein}, and \eqref{eq:cup0_relation_linear}.
Next, we show invariance under adding $d\ovl\alpha$,
\begin{equation}
\begin{multlined}
\Sbar[\Abar+\ovl\alpha]-\Sbar[\Abar]\\=
(\Abar+d\alpha)^T F\cup d(\Abar+d\alpha) - \Abar^TF\cup \Abar\\
= (d\alpha)^T F\cup d\Abar = (dd\alpha)^T F\cup \Abar = 0\;.
\end{multlined}
\end{equation}
Thus, the action is gauge invariant,
\begin{equation}
S[A']-S[A]=0\;.
\end{equation}

Not all different choices of $f$ and $F$ yield different models.
For example, indicating the matrix $F$ as a superscript, we have
\begin{equation}
\begin{multlined}
S_{F^T}[A]=
\Abar^TF\cup d\Abar = (d\Abar)^TF\cup \Abar\\ = \Abar^TF^T\cup^T d\Abar
= \Abar^TF^T\cup d\Abar + (d\Abar)^TF^T\cup_1 d\Abar\\ = \Abar^TF^T\cup d\Abar + v_A^Tf^TF^Tf\cup_1 v_A\\
= \Abar^TF^T\cup d\Abar = S_F[A]\;,
\end{multlined}
\end{equation}
using Eq.~\eqref{eq:firstcup21_identity}.
Thus, we have
\begin{equation}
S_{F+Y-Y^T}[A]=S_F[A]\;.
\end{equation}
Also it is easy to see that
\begin{equation}
S_F[A]=S_{F+X}[A]
\end{equation}
for any $X$ such that $Xf=0$ or $f^TX=0$.
Further, if $f'=fg$, then we have $f\zz=f'\zz$ and $\zz_f=\zz_{f'}$.
In general, note that many different choices of $f$ give isomorphic $\zz_f$, even in the diagonal case.

Let us give a few examples of twisted quantum doubles and how they are phrased in our language.
For the double-semion model discussed in Section~\ref{sec:double_semion_pathintegral}, we have
\begin{equation}
f=\begin{pmatrix}2\end{pmatrix},\quad F=\begin{pmatrix}\frac14\end{pmatrix}\;.
\end{equation}
All twisted quantum doubles with group $G=\zz_n$ are represented by
\begin{equation}
\label{eq:type1_example}
f=\begin{pmatrix}n\end{pmatrix},\quad F=\begin{pmatrix}\frac a{n^2}\end{pmatrix}\;,\quad 0\leq a<n\;,
\end{equation}
corresponding to a type-I 3-cocycle.
A type-II 3-cocycle on $\zz_n\times \zz_n$ corresponds to
\begin{equation}
\label{eq:type2_twist}
f=\begin{pmatrix}n&0\\0&n\end{pmatrix},\quad F=\begin{pmatrix} 0 & \frac{a}{n^2}\\0&0\end{pmatrix}\;.
\end{equation}
As a slightly more involved example, the so-called 6-semion model consists of a type-2 twist, and a type-1 twist on each factor of $\zz_2\times \zz_2$,
\begin{equation}
l=\begin{pmatrix} 2&0\\0&2\end{pmatrix},\quad F=\begin{pmatrix} \frac14&\frac14\\0&\frac14\end{pmatrix}\;.
\end{equation}

\subsection{1-form symmetries}
\label{sec:general_1form}
Next, we equip the twisted quantum double path integral with projective 1-form symmetries.
The 1-form symmetry group is $K=\zz_m\coloneqq m\zz^{2k}\backslash\zz^{2k}$ with
\footnote{
In other words, $K=\zz_m$ is a central extension $\zz_{f^T}\rightarrow \zz_m\rightarrow \zz_f$ determined by the 2-cocycle $\psi\in H^2(B\zz_f,\zz_{f^T})$ given by $\psi(a,b)=f^T\backslash\big(f^T(F+F^T)(\ovl a+\ovl b-\ovl{a+b})\big)$.
}
\begin{equation}
\label{eq:general_anyon_fusion_group}
m \coloneqq \begin{pmatrix}f&0\\f^T(F+F^T)f & f^T\end{pmatrix}\;.
\end{equation}
However, instead of a $K$-valued 1-chain $s$, we will deviate from Definition~\ref{def:homological_integral} and use a $\zz_f$-valued 2-cochain $b$ together with a $\zz_{f^T}$-valued 1-chain $c$.
Instead of $\delta s=0$ for a non-zero path integral due to the first condition in Definition~\ref{def:homological_integral}, $b$ and $c$ have to obey the following,
\begin{equation}
\label{eq:bcycle_constraint}
db=0 \quad\Rightarrow\quad d\bbar=-fv_b\;,
\end{equation}
and
\begin{equation}
\label{eq:ccycle_constraint}
\begin{gathered}
\delta c = f^T\backslash \big(\cup f^T (F+F^T) d\bbar\big)\\
\Rightarrow\quad \delta \cbar = -\cup f^T (F+F^T) fv_b -f^T v_c
\;.
\end{gathered}
\end{equation}
Here, $F$ is interpreted as having entries in $\rr$ and not $\rr/\zz$, such that $f^T(F+F^T)f$ is not zero but has entries valued in $\zz$.
Combining both into one equation yields
\begin{equation}
\label{eq:cycle_cocycle_constraints}
\begin{pmatrix}
db\\\delta c
\end{pmatrix}
=
\begin{pmatrix}f&0\\\cup f^T(F+F^T)f & f^T\end{pmatrix}
\begin{pmatrix}
-v_b\\-v_c
\end{pmatrix}
\;.
\end{equation}
Noting that the effect of the $\zz$-linear operator $\cup$ is to ``shift'' an $i$-cocycle to a homologically equivalent $d-i$-cycle, we can read off Eq.~\eqref{eq:general_anyon_fusion_group}.
Also note that on a cubic lattice, $\cup$ can be chosen such that it is literally a shift in the $(\frac12,\frac12,\frac12)$-direction.
In this case, the combination $s\coloneqq f^T\backslash (\cup b,c)^T$ defines an ordinary $K$-valued 1-cycle.
However, on arbitrary cellulations, we have to slightly modify Definition~\ref{def:homological_integral}.

The path integral is a sum over all $\zz_f$ 1-cochains $A$ such that
\begin{equation}
\label{eq:1form_boundary_costraint}
dA=b
\quad\Rightarrow\quad
d \Abar=\bbar-f v_A\;.
\end{equation}
The action is given by
\begin{equation}
\label{eq:action_1form_general}
\begin{multlined}
S[A,b,c] = \Abar^T F \cup d\Abar + \Abar^T(f^{-1})^T \cbar\\ - \Abar^T (F+F^T)\cup \bbar + \bbar^TF\cup_1d\Abar\;.
\end{multlined}
\end{equation}
This action is still gauge invariant under
\begin{equation}
A'=A+d\alpha\;,
\end{equation}
even in the presence of defects $b$ and $c$ as we show below.
As in Section~\ref{sec:general_path_integral}, we start by showing invariance under adding $fx$,
\begin{equation}
\label{eq:a_periodicity_1form}
\begin{multlined}
\Sbar[\Abar+fx,\bbar,\cbar]-\Sbar[\Abar,\bbar,\cbar]\\
= (fx)^TF\cup d\Abar + \Abar^TF\cup dfx + (fx)^TF\cup dfx\\
+ (fx)^T (f^{-1})^T \cbar
- (fx)^T(F+F^T)\cup \bbar
+ \bbar^TF\cup_1 dfx\\
= 
(fx)^TF\cup d\Abar-(fx)^TF\cup \bbar\\
+ \Abar^T F\cup dfx-(fx)^TF^T\cup \bbar + \bbar^TF\cup_1 dfx\\
= x^Tf^TF\cup(d\Abar-\bbar) + (d\Abar)^TF\cup fx\\
+ \bbar^TF(-\cup^T+\cup_1d)fx\\
\overset{\eqref{eq:firstcup21_identity}}{=} -x^Tf^TFf\cup v_A+(d\Abar)^TF\cup fx\\
- \bbar^TF\cup fx-(d\bbar)^TF\cup_1fx\\
= (d\Abar-\bbar)^TF\cup fx + v_b^Tf^TFf\cup_1x\\
= -v_A^Tf^T Ff\cup x = 0\;.
\end{multlined}
\end{equation}
Next, we consider the variance under adding $d\ovl\alpha$,
\begin{equation}
\label{eq:1form_gauge_variance}
\begin{multlined}
\Sbar[\Abar+d\ovl\alpha,\bbar,\cbar]-\Sbar[\Abar,\bbar,\cbar]\\
=(d\ovl\alpha)^T F\cup d\Abar + (d\ovl\alpha)^T(f^{-1})^T \cbar\\ -(d\ovl\alpha)^T(F+F^T)\cup \bbar\\
= \ovl\alpha^T (f^{-1})^T\delta \cbar -\ovl\alpha^T\cup (F+F^T)d\bbar\\
= \ovl\alpha^T(f^{-1})^T\big(\delta \cbar -\cup f^T (F+F^T)d\bbar\big)\;.
\end{multlined}
\end{equation}
Thus, if Eq.~\eqref{eq:ccycle_constraint} holds, then we indeed have
\begin{equation}
S[A',b,c]-S[A,b,c]=0\;.
\end{equation}

Let us now discuss how the four conditions of Definition~\ref{def:homological_integral} hold for the presented 1-form symmetries, or better, the analogous conditions since we have pairs $(b,c)$ instead of 1-cycles $s$.
The arguments are analogous to the discussion towards the end of Section~\ref{sec:semion_1form}, and we will not repeat them to full extent.
To show that the first condition holds, we note that the path integral on a 3-ball evaluates to zero by construction unless Eq.~\eqref{eq:bcycle_constraint} is fulfilled, since otherwise the set of 1-cochains $A$ with $dA=b$ that we sum over is empty.
Further, let us consider Eq.~\eqref{eq:1form_gauge_variance} for an arbitrary 1-chain $c$ that does not need to fulfill Eq.~\eqref{eq:ccycle_constraint}.
Analogous to Eq.~\eqref{eq:semion_ccycle_derivation}, $Z[a,b,c]$ on the set of 3-cells surrounding a vertex $v$ is given by
\begin{equation}
\label{eq:ccycle_derivation}
\begin{multlined}
Z[a,b,c] = \frac12 \sum_{\alpha(v)\in \zz_f} e^{2\pi i S[A_0+d\alpha]}\\
= \frac{1}{|\zz_f|} e^{2\pi i S[A_0]} \sum_{\alpha(v)\in \zz_f} e^{2\pi i (\delta c-\cup f^T(F+F^T)d\bbar)(v)^T f^{-1} \ovl\alpha(v)}\\
= e^{2\pi i S[A_0]} \delta_{(\delta c-\cup f^T(F+F^T)d\bbar)(v)}\;.
\end{multlined}
\end{equation}
Here, we have used that the $\zz_f$ discrete Fourier transform of the constant function is the $\delta$-function,
\begin{equation}
\label{eq:discrete_fourier_constant}
\sum_{y\in\zz_f} e^{2\pi i x^T f^{-1} y} = |\zz_f| \delta_x\;.
\end{equation}
So the path integral evaluates to zero on a 3-ball if either Eq.~\eqref{eq:bcycle_constraint} or Eq.~\eqref{eq:ccycle_constraint} is violated at any volume or vertex in the interior, which implies the first condition in Definition~\ref{def:homological_integral} holds for a minimal $\chi$.

For the second condition, we note that when adding $d\beta$ to $b$ for a single edge $\beta$, we also need to change $A$ and $c$ in order to preserve Eqs.~\eqref{eq:1form_boundary_costraint} and \eqref{eq:ccycle_constraint},
\begin{equation}
\label{eq:b_gauge}
\begin{gathered}
A'=A+\beta\;,\\
b'=b+d\beta\;,\\
c'=c+f^T\backslash \cup f^T(F+F^T)fy\;,
\end{gathered}
\end{equation}
with
\begin{equation}
y\coloneqq s_{b,d\beta}+v_\beta\;.
\end{equation}
Indeed, we find that this transformation preserves Eq.~\eqref{eq:ccycle_constraint},
\begin{equation}
\begin{multlined}
\delta c' = \delta \big(c+\cup f^T(F+F^T)f y\big)\\
= \delta c+\delta\cup f^T(F+F^T)f (\ovl{b'}-\bbar-d\ovl\beta)\\
= \delta c+\cup f^T(F+F^T) d (\ovl{b'}-\bbar)\\
= \cup f^T(F+F^T) d\ovl{b'}\;.
\end{multlined}
\end{equation}
Apart from this, the proof of the second condition in Definition~\ref{def:homological_integral} works as in Section~\ref{sec:semion_1form}, by extending $b$ and $c$ with $b^+$ and $c^+$ inside a 3-ball $B_3^+$.

For the proof of the third condition in Definition~\ref{def:homological_integral}, we note if $\partial c$ is homologically non-trivial, then there is a connected component of $m_0$ on which $\partial c$ sums to $0\neq y\in \zz_f$.
For every $x\in \zz_f$, consider the 1-cocycle $A_x$ that associates $\pm x$ to all edges of $M'$ adjacent to but not inside the chosen connected component.
With this, we have $\ovl{A_x}^T (f^{-1})^T c=\ovl x^T(f^{-1})^T \ovl y$.
To show that the evaluation $Z_{M'}$ is zero, we then use Eq.~\eqref{eq:discrete_fourier_constant} inside the analog of Eq.~\eqref{eq:semion_condition3_nontrivialc}.
Finally, also the proof of the fourth condition in Definition~\ref{def:homological_integral} works the same way as in Section~\ref{sec:semion_1form}, by extending $S_2\times [0,1]$ to $S_3$ by gluing $B_+^0$ and $B_+^1$.

We conclude this section by noting that if we spell out the action in Eq.~\eqref{eq:action_1form_general} for the double-semion model (Eq.~\eqref{eq:type1_example} with $n=2$), we get
\begin{equation}
\begin{gathered}
\frac14 \Abar^T\cup d\Abar+ \frac12 \Abar^T\cbar + \frac12 \Abar^T\cup\bbar +\frac14 \bbar^T\cup_1 d\Abar\\
= \frac14 \Abar^T\cup d\Abar+ \frac12 \Abar^T(\ovl{c+\cup b}) +\frac14 \bbar^T\cup_1 d\Abar\;.
\end{gathered}
\end{equation}
So the action equals that Eq.~\eqref{eq:semion_1form_action} after we replace $c$ by $c+\cup b$

\subsection{Fault-tolerant circuit}
\label{sec:codes_from_arbitrary_1form}
In this section, we generalize the fault-tolerant circuit constructed from the double-semion path integral in Section~\ref{sec:double_semion_code} to arbitrary abelian twisted quantum double path integrals.
For the underlying spacetime cellulation, we choose the same as in Section~\ref{sec:double_semion_code}, depicted in Eq.~\eqref{eq:semion_spacetime_cellulation}.
The labels for the state-sum variables are again the same as in Eq.~\eqref{eq:sheared_cubic_labels}, just that now the labels take values in $\zz_f$ (or $\zz_{f^T}$) instead of $\zz_2$.
Also the constraints in Eq.~\eqref{eq:semion_lattice_constraints} are the same.

The first visible difference arises when we resolve the action in terms of the microscopic state-sum variables.
Let we start with the first term of Eq.~\eqref{eq:action_1form_general}, $\Abar^T F \cup d\Abar$.
Using the cup product formula in Eq.~\eqref{eq:slanted_cubic_cup0}, we obtain similar to Eq.~\eqref{eq:semion_variable_weights0},
\begin{equation}
\label{eq:variable_weights0}
\begin{gathered}
\ovl{A_{dx}^{001}}^TF(\ovl{A_y^{002}}+\ovl{A_t^{011}}-\ovl{A_t^{001}}-\ovl{A_y^{001}})\\
-\ovl{A_y^{101}}^TF(\ovl{A_{dx}^{011}}+\ovl{A_t^{011}}-\ovl{A_t^{110}}-\ovl{A_{dx}^{010}})\\
+\ovl{A_t^{100}}^TF(\ovl{A_{dx}^{000}}+\ovl{A_y^{001}}-\ovl{A_y^{100}}-\ovl{A_{dx}^{010}})
\;.
\end{gathered}
\end{equation}
The next term in the action is $A^T(f^{-1})^T c$, which becomes
\begin{equation}
\ovl{A_t^{000}}^T(f^{-1})^T c^{000}\;.
\end{equation}
The next term is $\bbar^TF\cup_1d\Abar$.
Analogous to Eq.~\eqref{eq:semion_variable_weights1}, we find
\begin{equation}
\begin{gathered}
-\ovl{b^{000}}^TF(\ovl{A_{dx}^{001}}+\ovl{A_t^{001}}-\ovl{A_t^{100}}-\ovl{A_{dx}^{000}})\\
-\ovl{b^{000}}^TF(\ovl{A_y^{002}}+\ovl{A_t^{011}}-\ovl{A_t^{001}}-\ovl{A_y^{001}})\\
=-\ovl{b^{000}}^TF(\ovl{A_{dx}^{001}}+\ovl{A_y^{002}}+\ovl{A_t^{011}}-\underbrace{\ovl{A_t^{100}}}_{\text{cancel}}-\ovl{A_{dx}^{000}}-\ovl{A_y^{001}})\;.
\end{gathered}
\end{equation}
The term marked by ``cancel'' will cancel with another term of the action below.
Finally, we have the additional term $-A^T (F+F^T)\cup b$ that was not present for the double-semion model, as explained at the end of Section~\ref{sec:general_1form}.
For this we still use the cup product formula shown in Eq.~\eqref{eq:slanted_cubic_cup0}.
Since $b$ is only supported on the $(y,t-x)$-faces, we only get one term,
\begin{equation}
-\ovl{A_t^{100}}^T(F+\underbrace{F^T}_{\text{cancel}})\ovl{b^{000}}\;.
\end{equation}

After describing the microscopics of the state sum on the chosen spacetime cellulation, it is now time to construct the fault-tolerant circuit.
The circuit has degrees of freedom $A_x^{ij}$ and $A_y^{ij}$ located on the edges, $b^{ij}$ on the faces, and $c^{ij}$ on the vertices of a square lattice, just as for the double-semion circuit in Eq.~\eqref{eq:semion_code_qubits}.
The only difference is that the degrees of freedom are not qubits but $|\zz_f|$-dimensional qudits, whose basis vectors are labeled by elements of $\zz_f$.
Additionally, the circuit generates classical measurement results $\hat b$ valued in $\zz_f$ at every $(y,t-x)$-face of the spacetime lattice, and $\hat c$ taking values in $\zz_{f^T}$ at every $-t$-edge.
The stages of the circuit are the same as for the double-semion case in Eq.~\eqref{eq:semion_code_stages}.

To get from stage $i$ to stage $i+1$, we update the values of the qubits using the constraints in Eq.~\eqref{eq:semion_lattice_constraints} and below, which yields the same gates as in Eq.~\eqref{eq:semion_constraint_gates}.
The only difference is that instead of the $CX$ gate on qubits, we use the $\zz_f$ analogue of the gate,
\begin{equation}
CX_f\ket{a,b} \coloneqq \ket{a,b+a}\;.
\end{equation}
The measurements in Eq.~\eqref{eq:semion_measurement_gates} are completely analogous again.
We replace $M_Z$ by the measurement $M_{Zf}$ in the computational basis yielding the $\zz_f$-value of the qubit.
We replace $\ket+$ by the equal-weight superposition $\ket{+_f}$ of $\zz_f$-elements,
\begin{equation}
\label{eq:general_plus_state}
\ket{+_f}\coloneqq \frac{1}{\sqrt{|\zz_f|}}\sum_{a\in \zz_f} \ket a\;.
\end{equation}
The state $\ket0$ is replaced by $\ket{0_f}$, the computational basis state corresponding to the identity element of $\zz_f$.
The Hadamard gate is replaced by the discrete Fourier transform for the group $\zz_f$,
\begin{equation}
H_f\ket a = \sum_{b\in \zz_{f^T}} e^{2\pi i \ovl a^Tf^{-1}\ovl b} \ket b\;.
\end{equation}
Unitarity of this operator follows from Eq.~\eqref{eq:discrete_fourier_constant},
\begin{equation}
\begin{gathered}
\bra a H_f^\dagger H_f \ket c = \sum_{b\in \zz_{f^T}} e^{2\pi i \ovl a^T f^{-1}\ovl b} e^{-2\pi i \ovl c^T f^{-1}\ovl b}\\
=\sum_{b\in \zz_{f^T}} e^{2\pi i (\ovl a-\ovl c)^T f^{-1}\ovl b} = \delta_{a,c}\;.
\end{gathered}
\end{equation}

The next step is to implement the state-sum weights as controlled-phase gates, diagonal in the computational basis.
Most importantly, an action term like $\ovl a^T F \ovl b$ yields a weight $e^{2\pi i\ovl a^TF\ovl b}$, which is implemented by applying the gate
\begin{equation}
CS_F\ket{a,b} = e^{2\pi i \ovl a^TF\ovl b} \ket{a,b}\;.
\end{equation}
If $a$ is represented by a measurement outcome instead of a qubit, the weight can be implemented by an analogous classically controlled gate $cS_F$.
With this, the weights can be implemented as the following gates acting on the qubits, analogous to Eq.~\eqref{eq:semion_phase_gates},
\begin{equation}
\begin{tabular}{l|l|l}
Weight & Gate & Stage\\
\hline
$e^{2\pi i\ovl{A_{dx}^{001}}^TF\ovl{A_y^{002}}}$ & $CS_{F}[A_x^{00},A_y^{00}]$ & \tcr{$3_1$,$4_1$,}$5_1'$\\
$e^{2\pi i\ovl{A_{dx}^{001}}^TF\ovl{A_t^{011}}}$ & $CS_{F}[A_x^{00},c^{01}]$ & $1_1''$\tcr{,$2_1$,$3_1$,$4_1$,$5_1$}\\
$e^{-2\pi i\ovl{A_{dx}^{001}}^TF\ovl{A_t^{001}}}$ & $CS_{-F}[A_x^{00},c^{00}]$ & \tcr{$1_1$,$2_1$,}$3_1^+$\tcr{,$4_1$,$5_1$}\\
$e^{-2\pi i\ovl{A_{dx}^{001}}^TF\ovl{A_y^{001}}}$ & $CS_{-F}[A_x^{00},A_y^{00}]$ & $1_1'$\\
 & \tcr{$CS_{-F}[A_x^{00},b^{(-1)0}]$} & \tcr{$1_1$}\\
$e^{-2\pi i\ovl{A_y^{101}}^TF\ovl{A_{dx}^{011}}}$ & $CS_{-F}[A_y^{10},A_x^{01}]$ & $1_1$\\
 & \tcr{$CS_{-F}[b^{00},A_x^{01}]$} & \tcr{$1_1$}\\
$e^{-2\pi i\ovl{A_y^{101}}^TF\ovl{A_t^{011}}}$ & $CS_{-F}[A_y^{10},c^{01}]$ & \tcr{$0_1$,$1_1$}\\
 & $CS_{-F}[b^{00},c^{01}]$ & $1_1$\\
$e^{2\pi i\ovl{A_y^{101}}^TF\ovl{A_t^{110}}}$ & $CS_F[A_y^{10},c^{11}]$ & \tcr{3,4,}5\tcr{,6}\\
$e^{2\pi i\ovl{A_y^{101}}^TF\ovl{A_{dx}^{010}}}$ & $CS_F[A_y^{10},A_x^{01}]$ & \tcr{3,}4\tcr{,5}\\
$e^{2\pi i\ovl{A_t^{100}}^TF\ovl{A_{dx}^{000}}}$ & $CS_F[c^{10},A_x^{00}]$ & \tcr{1,2,}3\tcr{,4,5}\\
$e^{2\pi i\ovl{A_t^{100}}^TF\ovl{A_y^{001}}}$ & $CS_F[c^{10},A_y^{00}]$ & \tcr{3,}$4'$\tcr{,5,6}\\
$e^{-2\pi i\ovl{A_t^{100}}^TF\ovl{A_y^{100}}}$ & $CS_{-F}[c^{10}, A_y^{10}]$ & \tcr{0,1}\\
 & $CS_{-F}[c^{10}, b^{00}]$ & $1'$\\
$e^{-2\pi i\ovl{A_t^{100}}^TF\ovl{A_{dx}^{010}}}$ & $CS_{-F}[c^{10},A_x^{01}]$ & \tcr{1,2,3,}$4^+$\tcr{,5}\\
$e^{-2\pi i\ovl{b^{000}}^TF\ovl{A_{dx}^{001}}}$ & $cS_{-F}[\hat b^{000},A_x^{00}]$ & \tcr{$1_1$,$2_1$,$3_1$,}$4_1'$\tcr{,$5_1$}\\
$e^{-2\pi i\ovl{b^{000}}^TF\ovl{A_y^{002}}}$ & $cS_{-F}[\hat b^{000},A_y^{00}]$ & $3_1$\tcr{,$4_1$--$8_1$,$0_2$,$1_2$}\\
$e^{-2\pi i\ovl{b^{000}}^TF\ovl{A_t^{011}}}$ & $cS_{-F}[\hat b^{000},c^{01}]$ & \tcr{$0_1$--$3_1$,}$4_1$\tcr{,$5_1$,$6_1$}\\
$e^{-2\pi i\ovl{A_t^{100}}^TF\ovl{b^{000}}}$ & $cS_{-F^T}[\hat b^{000},c^{10}]$ & $5'$\tcr{,6}\\
 & \tcr{$CS_{-F^T}[b^{00},c^{10}]$} & \tcr{4,5}\\
$e^{2\pi i\ovl{b^{000}}^TF\ovl{A_{dx}^{000}}}$ & $cS_{F}[\hat b^{000},A_x^{00}]$ & 5\\
& \tcr{$CS_{F}[b^{00},A_x^{00}]$} & \tcr{4,5}\\
$e^{2\pi i\ovl{b^{000}}^TF\ovl{A_y^{001}}}$ & $cS_{F}[\hat b^{000},A_y^{00}]$ & \tcr{5--8,$0_1$,}$1_1''$\\
 & \tcr{$cS_{F}[\hat b^{000},b^{(-1)0}]$} & \tcr{$1_1$}\\
 & \tcr{$CS_{F}[b^{00},A_y^{00}]$} & \tcr{4,5}
\end{tabular}
\;.
\end{equation}
Taking everything together, the circuit is given by
\begin{equation}
\begin{tabular}{l|l}
Step & Gates\\
\hline
\tcr{$6_{-1}^+$} & \tcr{$P_{\ket+}\circ M_{Xf}[c^{00},\hat c^{00(-1)}]$}\\
\hline
$0^+$ & $CX_f[A_y^{10}, b^{00}]$, $CX_f[c^{00}, A_x^{00}]$\\
\hline
$1$ & $CS_{-F}[A_y^{10},A_x^{01}]$, $CS_{-F}[b^{00},c^{10}]$\\
$1'$ & $CS_{-F}[A_x^{00},A_y^{00}]$, $CS_{-F}[c^{10},b^{00}]$\\
$1''$ & $cS_{F}[\hat b^{00(-1)},A_y^{00}]$, $CS_{-F}[A_x^{00},c^{01}]$\\
\hline
$1^+$ & $CX_f[A_x^{00}, b^{00}]$, $CX_f[c^{01}, A_y^{00}]$\\
\hline
$2^+$ & $CX_f[A_x^{01}, b^{00}]$, $CX_f[c^{00}, A_y^{00}]$\\
\hline
$3$ & $cS_{-F}[\hat b^{00(-1)},A_y^{00}]$, $CS_F[c^{10},A_x^{00}]$\\
\hline
$3^+$ & $CX_f[A_y^{00}, b^{00}]$, $CS_F[A_x^{00},c^{00}]$\\
\hline
$4$ & $CS_F[A_y^{10},A_x^{01}]$, $cS_{-F}[\hat b^{00(-1)},c^{01}]$\\
$4'$ & $cS_{-F}[\hat b^{00(-1)},A_x^{00}]$, $CS_F[c^{10},A_y^{00}]$\\
\hline
$4^+$ & $P_{\ket0}\circ M_{Zf}[b^{00},\hat b^{000}]$, $CS_{-F}[c^{10},A_x^{01}]$\\
\hline
$5$ & $cS_{F}[\hat b^{000}, A_x^{00}]$, $CS_F[A_y^{10},c^{11}]$\\
$5'$ & $cS_{-F^T}[\hat b^{000},c^{10}]$, $CS_{-F}[A_x^{00},A_y^{00}]$\\
\hline
$5^+$ & $CX_f[c^{10}, A_x^{00}]$\\
\hline
$6^+$ & $P_{\ket+}\circ M_{Xf}[c^{00},\hat c^{000}]$\\
\hline
$\tcr{0_1^+}$ & $\tcr{CX_f[A_y^{10}, b^{00}]}$, $\tcr{CX_f[c^{00}, A_x^{00}]}$\\
\hline
$\tcr{\ldots}$ & $\tcr{\ldots}$
\end{tabular}
\;.
\end{equation}
That is, in addition to the qudit toric code, we end up with $12$ generalized $CS$ gates, and depending on the classical measurement outcomes up to $6$ generalized $S$ gates.
Here, we again combined $H_f$ and $M_{Zf}$ into a generalized $X$ measurement $M_{Xf}$.

\subsection{Decoding and correction}
\label{sec:general_decoding}
In order to incorporate corrections, we generalize the double-semion case in Section~\ref{eq:semion_decoding}, just as we generalized Section~\ref{sec:double_semion_code} to Section~\ref{sec:codes_from_arbitrary_1form}.
Since this would be rather repetitive, and is anyways slightly hypothetical, we omit this.

However, we want to stress one important qualitative difference to the double-semion case, which comes into play when we look at the classical decoder:
We cannot anymore fix $b$ and $c$ separately, since for a noise-free syndrome, $c$ is now a 1-chain whose boundary depends on $b$ through Eq.~\eqref{eq:ccycle_constraint}.
We thus suggest the following decoding procedure:
We first find a fix $\fix(db)$ for $b$ such that $b+\fix(db)$ is a $\zz_f$-valued 2-cocycle.
Then we fix $c$ with
\begin{equation}
\fix(\delta c - f^T\backslash\cup f^T (F+F^T) d\ovl{(b+\fix(db))}
\;.
\end{equation}
As usual, we choose open boundary conditions at time $T$ in the calculation of $\fix$, that is both 2-cocycles like $b$ and 1-cycles like $c$ are allowed to freely terminate at the time-$T$ boundary.
Also, in general, computing the minimum-weight fix $\mwpm$ is computationally inefficient, so we might have to use another decoder such as an RG decoder.

\section{Measurement-based and Floquet architectures}
\label{sec:measurement_floquet}
The fault-tolerant circuits of Sections~\ref{sec:double_semion_code} and \ref{sec:codes_from_arbitrary_1form} were based on a standard syndrome extraction circuit for the stabilizer (qudit) toric code, into which we included phase gates implementing the twists.
In Ref.~\cite{path_integral_qec}, we have shown that many other fault-tolerant topological protocols are equivalent to the toric code:
They are based on the same toric-code path integral, on different spacetime cellulations traversed in different time directions.
In this section, we will show that also the other protocols corresponding to other spacetime cellulations or time directions can be equipped with phase gates.
In particular, we will look at measurement-based topological quantum computation as described in \cite{Raussendorf2007}, and the CSS honeycomb Floquet code defined in Refs.~\cite{Kesselring2022,Davydova2022,Aasen2022}.

\subsection{Measurement-based topological quantum computation}
\label{sec:measurement_based}
Let us start by measurement-based topological quantum computation.
As we have discussed at the end of Section~\ref{sec:double_semion_code}, the major challenge when turning the path integral into a fault-tolerant circuit is to ensure that for every weight in the action, there is a stage in the circuit where each involved state-sum variable is represented by a qubit.
We have shown that for a standard syndrome-extraction circuit of the stabilizer toric code, this can be achieved by choosing an according local formula for the higher order cup products.
For different spacetime cellulations, different time directions, or different cup product formulas, it is not so clear that the weights in the action can be realized by phase gates inserted into the ``untwisted'' circuit.
As argued at the end of Section~\ref{sec:double_semion_code}, a way to solve this problem is to copy a state-sum variable stored in one qubit to an auxiliary qubit, such that this variable remains accessible also at later stages.
After usage, these auxiliary qubits must be measured in a generalized $X$ basis, such that the measurement outcome corresponds to the presence of a $c$ defect at the according edge.
If we take this method to an extreme, we store every state-sum variable in a separate auxiliary qubit and only perform the $X$-type measurement after it is not needed anymore.
Of course, the large number of auxiliary qubits means that these circuits have a larger overhead compared to the one constructed in Section~\ref{sec:double_semion}, but it guarantees that we can straight-forwardly implement any sort of twists in an abelian gauge theory.
Even more extremely, we could consider keeping all of the qubits, and performing all measurements at the very end, including the $Z$-type measurements corresponding to the $b$ defects.
After doing this, we can rotate the protocol from a $2+1$-dimensional to a $3+0$-dimensional circuit.
That is, the protocol now consists of a constant-time circuit preparing a resource state in $3$ spatial dimensions, which we then measure.
We have arrived at a measurement-based quantum computation protocol.

Topological measurement-based quantum computation has been originally established for a so-called cluster state as resource state \cite{Raussendorf2007}.
This cluster state is obtained from preparing the qubits of a $3+1$-dimensional toric code in the $\ket+$ state, and then preparing the ancillas whose measurement would yield the values of all $Z$-type stabilizers.
As was pointed out in Refs.~\cite{Roberts2020,Williamson2021}, this resource state is a ground state of a modular Crane-Yetter-Walker-Wang (CYWW) model \cite{Crane1993,Walker2011}, see also Ref.~\cite{Williamson2024}.
More precisely, modular CYWW models are defined for any anyon theory, and the resource state is the CYWW ground state for the toric code anyon model.
CYWW models for non-chiral anyon theories are trivial and can be prepared or disentangled with constant-depth circuits \cite{walker_wang_boundaries}.
\footnote{
Since ``the CYWW ground state'' is usually interpreted up to local unitary equivalence, the equivalence between the cluster resource state and the CYWW ground state as such is trivial, at least in the non-chiral case.
It does become a non-trivial statement, however, if we view the CYWW ground state as an SPT state protected by a 1-form symmetry \cite{Roberts2020,Williamson2021}.
}
The circuit that prepares the non-chiral modular CYWW ground state in constant time corresponds to a tensor-network representation (or PEPS) of a ground state, which at the same time defines a boundary for the $3+1$-dimensional CYWW path integral.
\footnote{
Perhaps unexpectedly, this boundary is not the usual one that hosts the topological order of the input anyon model, but rather the invertible domain wall to vacuum that only exists in the non-chiral case \cite{walker_wang_boundaries}.
}
It has been suggested in Ref.~\cite{Williamson2021} that other abelian CYWW ground states could be used for measurement-based quantum computation as well, but how precisely this works has not been spelled out in the literature so far.
Ref.~\cite{Roberts2020} discusses a computation scheme based on the 3-fermion CYWW model, but does not describe how to prepare the CYWW ground state to obtain a fully explicit $3+0$ or $2+1$-dimensional protocol.
Due to its chirality, the ground state cannot be prepared with a constant-depth circuit like that of the toric-code CYWW.
We can instead measure the CYWW stabilizers and apply corrections, which indeed yields a $3+0$ or $2+1$-dimensional circuit.
However, the phase of the corresponding path integral will then be the double of the 3-fermion phase, which is equal to two copies of the toric code.

When viewing the $3+0$-dimensional measurement-based circuit of Ref.~\cite{Raussendorf2007} as a path integral using the methods in Ref.~\cite{path_integral_qec}, we precisely get the toric code path integral.
This way, one can see that the stabilizer toric code and measurement-based quantum computation are in fact equivalent, as was also pointed out in Ref.~\cite{Bombin2023}.
In this section, we construct $3+0$-dimensional measurement-based quantum computation circuits based on our non-chiral abelian $2+1$-dimensional 1-form symmetric path integrals from Section~\ref{sec:general_1form}.
Equivalently, we realize measurement-based quantum computation for arbitrary non-chiral abelian CYWW models, as suggested in Ref.~\cite{Williamson2021}.
\footnote{
We prefer thinking in terms of $2+1$-dimensional path integrals, since CYWW models describe $3+1$-dimensional physics, but our circuits are really $2+1$ or $3+0$-dimensional.
Viewing the $2+1$-dimensional path integral as a boundary of the $3+1$-dimensional CYWW path integral does neatly describe the anomaly of the (projective) 1-form symmetries of the former, that is, the phases we get via the second condition in Definition~\ref{def:homological_integral}.
However, this anomaly is irrelevant for fault tolerance.
}

Let us now describe the resource state $\ket\psi$ for our protocol.
While this state could be defined on any 3-cellulation, we would like to demonstrate a specific task, namely ``teleporting'' logical information from one side of the cellulation to the other.
This is the measurement-based analogue of fault-tolerant storage.
To be concrete, we consider a 3-cellulation of topology $S_1\times S_1\times [0,1]$.
The resource state has $\zz_f$-valued qudits on all faces and all interior edges.
Configurations of the qubits thus consist of a 2-chain $b$ and a 1-chain $c$.
The amplitude of the resource state simply equals our 1-form symmetric path integral derived in Section~\ref{sec:general_1form} with symmetry defects $b$ and $c$, and $A=0$ at the boundary,
\begin{equation}
\braket{b,c|\psi} = Z[0,b,c]\;.
\end{equation}
We will now construct the finite-depth circuit that prepares the state $\ket\psi$ above in the bulk, discussing the boundaries at $S_1\times S_1\times 0$ and $S_1\times S_1\times 1$ later.
We start by preparing every $c$ qudit in the state $\ket{+_f}$, defined in Eq.~\eqref{eq:general_plus_state}, and every $b$ qudit in the state $\ket{0_f}$.
The value of these $c$ qudits represent the value of $A$, and the fact that we prepare them in the $\ket{+_f}$ state corresponds to the fact that we sum over all $A$-configurations.
The constraint $dA=b$ is then implemented by applying a $CX_f$ (or $CX_f^T$) gate, controlled by the $c$ qudit and acting on the $b$ qubit, for every pair of adjacent edge and face.
Next, we implement the action terms depending on the $A$-variables and $b$ variables as (controlled-)phase gates acting on the corresponding $b$ and $c$ qudits.
Finally, we implement the weight $A^T (f^{-1})^T c$ by applying a $H_f$ gate at every $c$ qudit.
After this, the $c$ qudits represent the values of $c$ instead of $A$.

Let us look at the microscopics of this preparation circuit on an ordinary cubic lattice.
We will name the state-sum variables within a unit cell as follows,
\begin{equation}
\label{eq:cubic_labels}
\begin{tikzpicture}
\draw (0,0)edge[mark={arr,e},ind=$x$]++(0:0.35) (0,0)edge[mark={arr,e},ind=$y$]++(30:0.25) (0,0)edge[mark={arr,e},ind=$z$]++(90:0.35);
\end{tikzpicture}
\quad
\begin{tikzpicture}
\tikzset{varlab/.style 2 args={mark={slab=$\scriptstyle{#1}$,#2}}}
\atoms{vertex}{100/p={3,0}, 010/p={0,3}, {001/p={2.2,1.4}}}
\atoms{vertex}{000/}
\atoms{vertex}{110/p={$(100)+(010)$}, 011/p={$(010)+(001)$}, 101/p={$(100)+(001)$}, 111/p={$(100)+(010)+(001)$}}
\draw (100)edge[mark=arr,varlab={A_y^{100}}{p=0.4},varlab={c_y^{100}}{r,p=0.4}](101) (110)edge[mark=arr,varlab={A_y^{101}}{},varlab={c_y^{101}}{r}](111) (110)edge[mark={arr,-},varlab={A_z^{100}}{p=0.4},varlab={c_z^{100}}{r,p=0.4}](100) (110)edge[mark={arr,-},varlab={A_x^{001}}{},varlab={c_x^{001}}{r}](010) (111)edge[mark={arr,-},varlab={A_z^{110}}{},varlab={c_z^{110}}{r}](101);
\draw (100)edge[mark={arr,-},varlab={A_x^{000}}{},varlab={c_x^{000}}{r}](000) (010)edge[mark={arr,-},varlab={A_z^{000}}{},varlab={c_z^{000}}{r}](000) (010)edge[mark=arr,varlab={A_y^{001}}{},varlab={c_y^{001}}{r}](011) (111)edge[mark={arr,-},varlab={A_x^{011}}{p=0.6},varlab={c_x^{011}}{r,p=0.6}](011);
\draw[dashed] (000)edge[mark=arr,varlab={A_y^{000}}{},varlab={c_y^{000}}{r}](001) (011)edge[mark={arr,-},varlab={A_z^{010}}{p=0.6},varlab={c_z^{010}}{r,p=0.6}](001) (101)edge[mark={arr,-},varlab={A_x^{010}}{},varlab={c_x^{010}}{r}](001);
\node at (1.5,1.5){$b_y^{000}$};
\node at (2.6,0.7){\tcr{$b_z^{000}$}};
\node at (1.1,2.2){\tcr{$b_x^{000}$}};
\node at (3.7,2.9){\tcr{$b_y^{010}$}};
\node at (2.6,3.7){$b_z^{001}$};
\node at (4.1,2.2){$b_x^{100}$};
\end{tikzpicture}
\;.
\end{equation}
The constraint $dA=b$ yields three constraints of the state-sum variables per unit cell,
\begin{equation}
\begin{gathered}
A_z^{000}+A_y^{001}-A_y^{000}-A_z^{010}=b_x^{000}\;,\\
A_x^{000}+A_z^{100}-A_z^{000}-A_x^{001}=b_y^{000}\;,\\
A_x^{000}+A_y^{100}-A_y^{000}-A_x^{010}=b_z^{000}
\;.
\end{gathered}
\end{equation}
Next, let us look at the terms of the action.
To express the term $\Abar^T F \cup d\Abar$ in terms of the microscopic state-sum variables, we use the cup product formula in Eq.~\eqref{eq:cup120}.
Thereby, we identify the cube in Eq.~\eqref{eq:cubic_labels} with the one in Eq.~\eqref{eq:cup_product_cube_drawing} such that the drawings coincide, and $x_0,x_1,x_2$ are matched with $x,z,y$.
This yields the following weight per unit cell,
\begin{equation}
\begin{gathered}
\ovl{A_x^{000}}^T F (\ovl{A_z^{100}}+\ovl{A_y^{101}}-\ovl{A_y^{100}}-\ovl{A_z^{110}})\\
-\ovl{A_z^{000}}^T F (\ovl{A_x^{001}}+\ovl{A_y^{101}}-\ovl{A_y^{001}}-\ovl{A_x^{011}})\\
+\ovl{A_y^{000}}^T F (\ovl{A_x^{010}}+\ovl{A_z^{110}}-\ovl{A_z^{010}}-\ovl{A_x^{011}})
\;.
\end{gathered}
\end{equation}
The term $\Abar^T(f^{-1})^T \cbar$ yields
\begin{equation}
\ovl{A_x^{000}}^T(f^{-1})^T \ovl{c_x^{000}}
+\ovl{A_y^{000}}^T(f^{-1})^T \ovl{c_y^{000}}
+\ovl{A_z^{000}}^T(f^{-1})^T \ovl{c_z^{000}}
\;.
\end{equation}
The term $-\Abar^T (F+F^T)\cup \bbar$ yields, again using Eq.~\eqref{eq:cup120},
\begin{equation}
\label{eq:measurementbased_weights1}
\begin{gathered}
-\ovl{A_x^{000}}^T (F+F^T) \ovl{b_x^{100}}
+\ovl{A_z^{000}}^T (F+\underbrace{F^T}_{\text{cancel}}) \ovl{b_z^{001}}\\
-\ovl{A_y^{000}}^T (F+\underbrace{F^T}_{\text{cancel}}) \ovl{b_y^{010}}
\;.
\end{gathered}
\end{equation}
For the term $\bbar^TF\cup_1d\Abar$, we use the $\cup_1$ product formula in Eq.~\eqref{eq:cup221}, identifying the cubes in Eqs.~\eqref{eq:cubic_labels} and \eqref{eq:cup_product_cube_drawing} in the same way as above.
We get
\begin{equation}
\label{eq:measurementbased_weights2}
\begin{gathered}
\ovl{b_x^{000}}^TF(\ovl{A_x^{000}}+\ovl{A_z^{100}}+\ovl{A_y^{101}}-\ovl{A_z^{000}}-\ovl{A_y^{001}}-\ovl{A_x^{011}})\\
-\ovl{b_y^{010}}^TF(\ovl{A_x^{000}}+\ovl{A_z^{100}}+\ovl{A_y^{101}}-\underbrace{\ovl{A_y^{000}}}_{\text{cancel}}-\ovl{A_x^{010}}-\ovl{A_z^{110}})\\
-\ovl{b_z^{000}}^TF(\ovl{A_z^{100}}+\ovl{A_y^{101}}-\ovl{A_y^{100}}-\ovl{A_z^{110}})\\
+\ovl{b_z^{001}}^TF (\ovl{A_x^{000}}+\ovl{A_z^{100}}-\underbrace{\ovl{A_z^{000}}}_{\text{cancel}}-\ovl{A_x^{001}})
\;.
\end{gathered}
\end{equation}

After preparing the resource state with the $3+0$-dimensional circuit, we measure all $c$ qudits and $b$ qudits in the computational basis.
We can combine the $H_f$ gates with the $c$-qudit measurements, yielding $c$-qubit measurements in the generalized $X$ basis.
Furthermore, we can measure the $b$ qubits before implementing the weights in Eqs.~\eqref{eq:measurementbased_weights1} and \eqref{eq:measurementbased_weights2}.
After this, the value of $b$ is represented by classical degrees of freedom, so we can implement these weights as classically controlled instead of coherently controlled unitaries.
All in all, we get the following $3+0$-dimensional circuit in the bulk,
\begin{equation}
\begin{tabular}{l|l}
Step & Gates\\
\hline
0 & $P_{\ket+}[c_x^{000}]$, $P_{\ket+}[c_y^{000}]$, $P_{\ket+}[c_z^{000}]$,\\
& $P_{\ket0}[b_x^{000}]$, $P_{\ket0}[b_y^{000}]$, $P_{\ket0}[b_z^{000}]$\\
\hline
1 & $CX_f[c_z^{000},b_x^{000}]$, $CX_f[c_y^{001},b_x^{000}]$, $CX_f^T[c_y^{000},b_x^{000}]$,\\
& $CX_f^T[c_z^{010},b_x^{000}]$, $CX_f[c_x^{000},b_y^{000}]$, $CX_f[c_z^{100},b_y^{000}]$,\\
& $CX_f^T[c_z^{000},b_y^{000}]$, $CX_f^T[c_x^{001},b_y^{000}]$, $CX_f[c_x^{000},b_z^{000}]$,\\
& $CX_f[c_y^{100},b_z^{000}]$, $CX_f^T[c_y^{000},b_z^{000}]$, $CX_f^T[c_x^{010},b_z^{000}]$\\
\hline
2 & $CS_F[c_x^{000},c_z^{100}]$, $CS_F[c_x^{000},c_y^{101}]$, $CS_{-F}[c_x^{000},c_y^{100}]$,\\
& $CS_{-F}[c_x^{000},c_z^{110}]$, $CS_{-F}[c_z^{000},c_x^{001}]$, $CS_{-F}[c_z^{000},c_y^{101}]$,\\
& $CS_F[c_z^{000},c_y^{001}]$, $CS_F[c_z^{000},c_x^{011}]$, $CS_F[c_y^{000},c_x^{010}]$,\\
& $CS_F[c_y^{000},c_z^{110}]$, $CS_{-F}[c_y^{000},c_z^{010}]$, $CS_{-F}[c_y^{000},c_x^{011}]$\\
\hline
3 & $M_{Zf}[b_x^{000},\hat b_x^{000}]$, $M_{Zf}[b_y^{000},\hat b_y^{000}]$, $M_{Zf}[b_z^{000},\hat b_z^{000}]$\\
\hline
4 & $cS_{-F}[\hat b_x^{100},c_x^{000}]$, $cS_{-F^T}[\hat b_x^{100},c_x^{000}]$, $cS_{F^T}[\hat b_z^{001},c_z^{000}]$,\\
& $cS_{-F^T}[\hat b_y^{010},c_y^{000}]$, $cS_{F}[\hat b_x^{000},c_x^{000}]$, $cS_{F}[\hat b_x^{000},c_z^{100}]$,\\
& $cS_{F}[\hat b_x^{000},c_y^{101}]$, $cS_{-F}[\hat b_x^{000},c_z^{000}]$, $cS_{-F}[\hat b_x^{000},c_y^{001}]$,\\
& $cS_{-F}[\hat b_x^{000},c_x^{011}]$, $cS_{-F}[\hat b_y^{010},c_x^{000}]$, $cS_{-F}[\hat b_y^{010},c_z^{100}]$,\\
& $cS_{-F}[\hat b_y^{010},c_y^{101}]$, $cS_{F}[\hat b_y^{010},c_x^{010}]$, $cS_{F}[\hat b_y^{010},c_z^{110}]$,\\
& $cS_{-F}[\hat b_z^{000},c_z^{100}]$, $cS_{-F}[\hat b_z^{000},c_y^{101}]$, $cS_{F}[\hat b_z^{000},c_y^{100}]$,\\
& $cS_{F}[\hat b_z^{000},c_z^{110}]$, $cS_F[\hat b_z^{001},c_x^{000}]$, $cS_{F}[\hat b_z^{001},c_z^{100}]$\\
& $cS_{-F}[\hat b_z^{001},c_x^{001}]$,\\
\hline
5 & $M_{Xf}[c_x^{000},\hat c_x^{000}]$, $M_{Xf}[c_y^{000},\hat c_y^{000}]$, $M_{Xf}[c_z^{000},\hat c_z^{000}]$\\
\end{tabular}
\end{equation}
If we wanted to move the $b$ measurements to the end, we would have to replace all the classically controlled $cS$ gates by coherently controlled $CS$ gates.

Let us briefly discuss how to use this protocol to fault-tolerantly ``teleport'' logical information from the space boundary at $S_1\times S_1\times 0$ to $S_1\times S_1\times 1$.
We assume that the input logical information is encoded as a toric-code ground space for the space cellulation at $S_1\times S_1\times 0$, and the output is encoded in the cellulation at $S_1\times S_1\times 1$.
In stage $0$, we use the input toric-code ground state as the state of the $c$ qubits inside the $S_1\times S_1\times 0$ boundary, while all $b$ qubits are prepared in state $\ket0$, and all other $c$ qubits in state $\ket+$.
We then continue with steps $1$, $2$, $3$, and $4$ as above.
In step $5$, we measure all qubits as described except for the ones inside the $S_1\times S_1\times 1$ boundary.
These qubits hold the teleported state.
However, we need to apply corrections to this teleported state, depending on the measurement outcomes in the bulk.
These measurement outcomes consist of a 2-cochain $b$ and a 1-chain $c$.
In contrast to Sections~\ref{eq:semion_decoding} and \ref{sec:general_decoding}, $b$ and $c$ are supported on all edges and faces of the 3-dimensional cellulation.
This is not a fundamental problem, but it is related to the larger overhead per unit cell of the 3-cellulation for the measurement-based approach compared to the stabilizer-syndrome-extraction approach.
Like in Section~\ref{sec:general_decoding}, we perform minimum-weight matching (or any other suitable decoder) with open boundary conditions at $S_1\times S_1\times 1$, and then close off the matched $b$ and $c$ near $S_1\times S_1\times 1$.
However, since all faces and edges in the bulk are already used for measurements, there are none left for controlled insertion of defect segments to close off the fixed $b$ and $c$.
Instead, we append a round of stabilizer-syndrome-extraction-based $2+1$-dimensional error correction similar to Section~\ref{sec:general_decoding}, into which we can insert additional defects.
Alternatively, we could look at the restriction of the matched $b$ and $c$ to the 2-cellulation at $S_1\times S_1\times 1$, yielding a $0$-cycle for $c$ and a 2-cocycle for $b$.
This restricted $(b,c)$ configuration defines an anyon pattern in the teleported state, which can be fixed using the string operators of these anyons.
We will not describe the microscopics of these corrections here, since in practice whether and how we perform them depends on what logical gate we next want to apply to our encoded information.

\subsection{CSS honeycomb Floquet code}
\label{sec:floquet}
Next, we look at the Floquet architecture.
We will not show how to implement arbitrary twisted quantum doubles, but only the one for $G=\zz_2\times \zz_2$ with a type-2 twist as shown in Eq.~\eqref{eq:type2_twist} for $n=2$.
We restrict ourselves to this case because the twist can be implemented without any auxiliary qudits.
We will write $A=(A_0,A_1)$, $b=(b_0,b_1)$, and $c=(c_0,c_1)$, explicitly spelling out the two components of the gauge field and symmetry defects.
Instead of the general action in Eq.~\eqref{eq:action_1form_general}, we choose a simpler alternative of equipping the path integral with 1-form symmetries:
$b_0$ and $b_1$ are $\zz_2$ 2-cocycles, and $c_0$ and $c_1$ are 1-chains such that
\begin{equation}
\label{eq:type2_defects}
\delta c_0=(\frac12\cup d\ovl{b_1})\mmod 2\;,\quad
\delta c_1=(\frac12\cup d\ovl{b_0})\mmod 2\;.
\end{equation}
Note that analogous to Eq.~\eqref{eq:general_anyon_fusion_group}, we have
\begin{equation}
m=\begin{pmatrix}2&0&0 & 0\\0&2&0&0\\0&1&2&0\\1&0&0&2\end{pmatrix}\;,
\end{equation}
so the effective 1-form symmetry group $K=Z_m$ is isomorphic to $\zz_4\times \zz_4$.
In fact, this twisted quantum double is in the same phase as the $G=\zz_4$ untwisted quantum double, in other words, the $\zz_4$ toric code.
Thus, our circuit provides a way to implement the $\zz_4$ toric code phase in a Floquet way using only qubits.
\footnote{
Surely, this could also be implemented by representing a 4-dimensional qudit as two qubits.
We would then have to implement the $\zz_4$ generalized $CX$ gate in terms of qubit pairs, or equivalently, $\mmod 4$ addition of two binary 2-digit numbers.
This involves a Toffoli ($CCX$) gate to implement the carry, whose decomposition into simple 2-qubit gates seems to yield a larger overhead than our circuits.
}
The state sum is a sum over two 2-cochains $A=(A_0,A_1)$ with
\begin{equation}
dA_0=b_0\;,\quad dA_1=b_1\;.
\end{equation}
The action in Eq.~\eqref{eq:general_action} can be equipped with 1-form symmetries as
\begin{equation}
\label{eq:type2_action}
\begin{multlined}
S[A,b,c] = \frac14 \big(\ovl{A_0}^T\cup d \ovl{A_1} - \ovl{b_0}^T\cup \ovl{A_1} - \ovl{A_0}^T\cup \ovl{b_1}\big)\\
+\frac12 (\ovl{c_0}^T\ovl{A_0}+\ovl{c_1}^T\ovl{A_1})\;.
\end{multlined}
\end{equation}
Since this way of introducing 1-form symmetries deviates from the general case in Eq.~\eqref{eq:action_1form_general}, we need to show invariance under the gauge transformation
\begin{equation}
\begin{gathered}
A_0'=A_0+d\alpha\\
\Rightarrow\quad \ovl{A_0'}=\ovl{A_0}+d\ovl\alpha+2(s_{A_0,d\alpha}+v_\alpha)\coloneqq \ovl{A_0}+d\ovl\alpha+2x\;.
\end{gathered}
\end{equation}
The action is invariant under adding $2x$,
\begin{equation}
\begin{gathered}
\Sbar[(\ovl{A_0}+2x,\ovl{A_1}),\bbar,\cbar]-\Sbar[(\ovl{A_0},\ovl{A_1}),\bbar,\cbar]\\
= \frac14 2x^T\cup (d\Abar_1-\bbar_1) = -x^T\cup v_{A_1} = 0\;.
\end{gathered}
\end{equation}
The variance under adding $d\ovl\alpha$ is
\begin{equation}
\begin{gathered}
\Sbar[(\ovl{A_0}+d\ovl\alpha,\ovl{A_1}),\bbar,\cbar]-\Sbar[(\ovl{A_0},\ovl{A_1}),\bbar,\cbar]\\
= \frac14 (d\ovl\alpha)^T \cup (d\Abar_1-\bbar_1)+\frac12 c_0^Td\ovl\alpha\\
= \frac12 \ovl\alpha^T (-\frac12\cup d\bbar_1+\delta c_0)\;.
\end{gathered}
\end{equation}
The exact same derivation holds for gauge invariance of $A_1$, after exchanging the subscripts $0$ and $1$ everywhere.
So we find that due to Eq.~\eqref{eq:type2_defects}, the action is gauge invariant.
Using arguments analogous to these towards the end of Section~\ref{sec:semion_1form}, we find that the four conditions of Definition~\ref{def:homological_integral} hold.

As described in Section~III.C of Ref.~\cite{path_integral_qec}, we consider a cubic lattice, and let the time direction be the $t=x+y+z$ direction.
We modify this lattice by splitting every edge into two edges separated by a 2-gon, and every face into two triangles separated by a diagonal 2-valent edge.
The 2-cochains $b_0$ and $b_1$ are only supported on the 2-gon faces, and the 2-chains $c_0$ and $c_1$ are only supported on the diagonal edges.
We will label the variables at the faces and edges of one spacetime volume as follows,
\begin{equation}
\label{eq:modified_cubic_variables}
\begin{tikzpicture}
\draw (-0.8,0.3)edge[mark={arr,e},ind=$x$]++(0:0.35) (-0.8,0.3)edge[mark={arr,e},ind=$y$]++(25:0.25) (-0.8,0.3)edge[mark={arr,e},ind=$z$]++(90:0.35) (-0.8,0.3)edge[gray,mark={arr,e},ind=$t$]++(50:0.5);
\end{tikzpicture}
\begin{tikzpicture}
\atoms{vertex}{{0/}, {1/p={4,0}}, {2/p={2,1.2}}, {3/p={6,1.2}}}
\atoms{vertex}{{0x/p={0,4}}, {1x/p={4,4}}, {2x/p={2,5.2}}, {3x/p={6,5.2}}}
\draw[dashed] (0)edge[bend left=20,varlabo={A_{y-}^{000}}{p=0.8,sideoff=0.05}](2) (0)edge[bend right=20,varlabo={A_{y+}^{000}}{p=0.8}](2) (2)edge[varlabo={A_{z-}^{010}}{}] (2x) (2)edge[varlabo={A_{x+}^{010}}{p=0.4}](3) (3)edge[bend left=30,varlabo={A_{z+}^{110}}{p=0.4}](3x) (2x)edge[bend right=30,varlabo={A_{x-}^{011}}{p=0.45}](3x);
\draw[dashed] (0)edge[varlabo={c_z^{000}}{p=0.6},varlabo={A_z^{000}}{p=0.8}](3) (0)edge[varlabo={c_x^{000}}{p=0.55},varlabo={A_x^{000}}{p=0.65}](2x) (2)edge[varlabo={c_y^{010}}{p=0.4},varlabo={A_y^{010}}{p=0.3}](3x);
\draw (0)edge[bend left=5,varlabo={A_{x-}^{000}}{}](1) (0)edge[bend left=30,varlabo={A_{x+}^{000}}{p=0.55}](1) (0)edge[bend right=5,varlabo={A_{z+}^{000}}{p=0.4}](0x);
\draw (0)edge[bend right=30,varlabo={A_{z-}^{000}}{p=0.6}](0x) (0x)edge[varlabo={A_{y+}^{001}}{}](2x) (0x)edge[varlabo={A_{x-}^{001}}{p=0.6}](1x);
\draw (1)edge[varlabo={A_{z+}^{100}}{}](1x) (1)edge[varlabo={A_{y-}^{100}}{}](3) (3)edge[bend left=5,varlabo={A_{z-}^{110}}{p=0.6}](3x) (2x)edge[bend right=5,varlabo={A_{x+}^{011}}{}](3x) (1x)edge[bend left=20,varlabo={A_{y-}^{101}}{p=0.2}](3x) (1x)edge[bend right=20,varlabo={A_{y+}^{101}}{p=0.2,sideoff=-0.05}](3x);
\draw (0)edge[varlabo={c_y^{000}}{p=0.6},varlabo={A_y^{000}}{p=0.7}](1x) (0x)edge[varlabo={c_z^{001}}{p=0.4},varlabo={A_z^{001}}{p=0.2}](3x) (1)edge[varlabo={c_x^{100}}{p=0.45},varlabo={A_x^{100}}{p=0.35}](3x);
\node at ($(0)!0.5!(2)$){$\scriptstyle{b_y^{000}}$};
\node at ($(1x)!0.5!(3x)$){$\scriptstyle{b_y^{101}}$};
\node at ($(0)!0.65!(1)+(90:0.3)$){$\scriptstyle{b_x^{000}}$};
\node at ($(2x)!0.35!(3x)+(-90:0.3)$){$\scriptstyle{b_x^{011}}$};
\node at ($0.5*(0)+0.5*(0x)+(0:0.35)$){$\scriptstyle{b_z^{000}}$};
\node at ($0.5*(3)+0.5*(3x)+(180:0.35)$){$\scriptstyle{b_z^{110}}$};
\end{tikzpicture}\;.
\end{equation}
Here, $A$, $b$, and $c$ denote $\zz_2\times \zz_2$-variables.
The corresponding $\zz_2$-variables have an additional subscript $0$ or $1$ in front.
There are each $6$ $A_0$ and $A_1$-variables, $3$ $b_0$ and $b_1$-variables, and $3$ $c_0$ and $c_1$ variables per unit cell.
As shown, we indicate the unit cell by superscripts like $A^{ijk}$, where $i,j,k$ are the $x,y,z$-coordinates of the unit cell.
The $\pm$ subscript indicates whether the edge is positively ($-$) or negatively ($+$) oriented inside the adjacent 2-gon face.

Let us now write the action in terms of the microscopic state-sum variables, as labeled above.
First, we need to construct a formula for the cup product on the modified cubic lattice using the method in Appendix~\ref{sec:cup_product}.
Thereby, we can neglect the diagonal edges, since $c$ is not involved in any cup products.
To define the cup product and also the (co-)boundary, we need to choose the orientation $\sigma$ for all subcells.
For this, we need to identify all cells with a standard representative.
We choose a single standard edge representative, and choose the identification such that the positively oriented vertex of the edge has the larger $t=x+y+z$ coordinate, which can be indicated in the drawing by equipping all edges with directions,
\begin{equation}
\label{eq:modified_cube_edge_directions}
\begin{tikzpicture}
\atoms{vertex}{{0/}, {1/p={1.5,0}}, {2/p={0.75,0.45}}, {3/p={2.25,0.45}}}
\atoms{vertex}{{0x/p={0,1.5}}, {1x/p={1.5,1.5}}, {2x/p={0.75,1.95}}, {3x/p={2.25,1.95}}}
\draw[dashed] (0)edge[bend left=20,mark=arr](2) (0)edge[bend right=20,mark=arr](2) (2x)edge[mark={arr,-}](2) (2)edge[mark=arr](3) (3)edge[bend left=30,mark=arr](3x) (2x)edge[bend right=30,mark=arr](3x);
\draw (0)edge[bend left=5,mark=arr](1) (0)edge[bend left=30,mark=arr](1) (0)edge[bend right=5,mark=arr](0x) (0)edge[bend right=30,mark=arr](0x) (2x)edge[mark={arr,-}](0x) (0x)edge[mark=arr](1x) (1x)edge[mark={arr,-}](1) (1)edge[mark=arr](3) (3)edge[bend left=5,mark=arr](3x) (2x)edge[bend right=5,mark=arr](3x) (1x)edge[bend left=20,mark=arr](3x) (1x)edge[bend right=20,mark=arr](3x);
\end{tikzpicture}
\;.
\end{equation}
There are two 2-cell representatives,
\begin{equation}
\label{eq:floquet_standard_representatives}
\begin{tikzpicture}
\atoms{vertex}{0/, 1/p={1,0}}
\draw (0)edge[bend left=30,mark=arr](1) (0)edge[bend right=30,mark=arr](1);
\end{tikzpicture}
\;,\quad
\begin{tikzpicture}
\atoms{vertex}{0/, 1/p={1,0}, 2/p={0,1}, 3/p={1,1}}
\draw (0)edge[mark=arr](1) (1)edge[mark=arr](3) (0)edge[mark=arr](2) (2)edge[mark=arr](3);
\end{tikzpicture}
\;.
\end{equation}
The counter-clockwise pointing edges of these 2-cell representatives are positively oriented and the clockwise ones negatively.
The identification with these standard representatives is such that the vector pointing into the drawing plane in Eq.~\eqref{eq:floquet_standard_representatives} has positive overlap with the $t$ direction in Eq.~\eqref{eq:modified_cube_edge_directions}.
Finally, the orientation of the faces in the cube in Eq.~\eqref{eq:modified_cube_edge_directions} is positive for the three 2-gons and three squares with the smaller $t$ coordinates, and negative for the larger $t$ coordinates.
We have chosen the orientations symmetric under the $\zz_3$ symmetry of the cubic lattice generated by the $\frac{2\pi}{3}$ rotation around the $t$ axis.
The same will be true for the formulas for $\cupsymb_0$ which we choose below.
For the edges and square faces, we use the formulas from Appendix~\ref{sec:cup_product} in Eqs.~\eqref{eq:cup010}, \eqref{eq:cup100}, \eqref{eq:cubic_cup_020}, and \eqref{eq:cubic_cup_110}.
The first new cell representative is the 2-gon, on which $\cupsymb_0^{02}$ can be chosen
\begin{equation}
\begin{tikzpicture}
\atoms{vertex}{0/, 1/p={1,0}}
\draw (0)to[bend left=30](1) (0)to[bend right=30](1);
\fill[bichain2f] (0-c)to[bend left=30](1-c)to[bend left=30]cycle;
\path (0)pic{bichain1p};
\end{tikzpicture}
\;.
\end{equation}
$\cupsymb_0^{11}$ is zero, and $\cupsymb_0^{20}$ given by
\begin{equation}
\begin{tikzpicture}
\atoms{vertex}{0/, 1/p={1,0}}
\draw (0)to[bend left=30](1) (0)to[bend right=30](1);
\path (1)pic{bichain2p};
\fill[bichain1f] (0-c)to[bend left=30](1-c)to[bend left=30]cycle;
\draw[bichainhelp] (0)edge[bend left=30,mark={slab=$-$}](1) (0)to[bend right=30](1);
\end{tikzpicture}
\;.
\end{equation}
Now, we are ready to define $\cupsymb_0$ for the modified-cubic 3-cell in Eq.~\eqref{eq:modified_cubic_variables}.
We start by $\cupsymb_0^{03}$,
\begin{equation}
\begin{tikzpicture}
\drawmodcube
\fill[bichain2ff] (0-c)--(1-c)--(3-c)to[bend left=5](3x-c)to[bend left=5](2x-c)--(0x-c)to[bend left=5]cycle;
\path (0)pic{bichain1p};
\end{tikzpicture}
\;.
\end{equation}
For $\cupsymb_0^{12}$, we get
\begin{equation}
\label{eq:modified_cube_cup012}
-
\begin{tikzpicture}
\drawmodcube
\fill[bichain2f] (0x-c)--(1x-c)to[bend left=30](3x-c)to[bend left=5](2x-c)--cycle;
\path (0)pic{bichainhelpp} (0x)pic{bichainhelpp} node[\bichaincolhelp,left]{$-$};
\draw[bichain1] (0)to[bend right=30](0x);
\end{tikzpicture}
-
\begin{tikzpicture}
\drawmodcube
\fill[bichain2f] (2x-c)to[bend right=30](3x-c)to[bend left=5]cycle;
\path (0)pic{bichainhelpp} (2x)pic{bichainhelpp} node[\bichaincolhelp,above]{$-$};
\draw[bichain1] (0)to[bend right=30](0x)--(2x);
\end{tikzpicture}
+\text{rot}
\;.
\end{equation}
Here, ``rot'' symbolizes that we have to include two more two copies of the terms above, rotated by $\pm \frac{2\pi}{3}$ around the $t$ axis.
For $\cupsymb_0^{21}$, we get
\begin{equation}
\label{eq:modified_cube_cup021}
-
\begin{tikzpicture}
\drawmodcube
\draw[bichain2] (1x-c)to[bend left=20](3x-c);
\draw[bichainhelp] (0-c)edge[bend left=5,mark={slab=$-$,r}](1-c) (1-c)edge[mark={slab=$-$,r}](1x-c) (0-c)to[bend right=30](0x-c) (0x-c)--(1x-c);
\fill[bichain1f] (0-c)to[bend left=5](1-c)--(1x-c)--(0x-c)to[bend left=30]cycle;
\end{tikzpicture}
-
\begin{tikzpicture}
\drawmodcube
\draw[bichain2] (1-c)--(1x-c);
\draw[bichainhelp] (0-c)edge[bend left=5,mark={slab=$-$,r}](1-c) (1-c)to[bend right=30](0-c);
\fill[bichain1f] (0-c)to[bend left=5](1-c)to[bend right=30]cycle;
\end{tikzpicture}
+\text{rot}
\;.
\end{equation}
With this, we are ready to translate the action in Eq.~\eqref{eq:type2_action} into microscopic terms.
For $\frac14 \ovl{A_0}^T\cup d \ovl{A_1}$, we get
\begin{equation}
\label{eq:floquet_weight0}
\begin{gathered}
-\frac14 \ovl{A_{0z-}^{000}}(\ovl{A_{1y+}^{001}}+\ovl{A_{1x-}^{011}}-\ovl{A_{1x-}^{001}}-\ovl{A_{1y-}^{101}})\\
-\frac14 \ovl{A_{0y+}^{001}}(\ovl{A_{1x-}^{011}}-\ovl{A_{1x+}^{011}}) +\text{rot}\;,
\end{gathered}
\end{equation}
using Eq.~\eqref{eq:modified_cube_cup012}.
For $-\frac14\ovl{b_0}^T\cup \ovl{A_1}$, we get
\begin{equation}
\label{eq:floquet_weight1}
\frac14 \ovl{b_{0x}^{000}} (\ovl{A_{1z+}^{100}}+\ovl{A_{1y-}^{101}})+\text{rot}\;,
\end{equation}
using Eq.~\eqref{eq:modified_cube_cup021}.
For $-\frac14\ovl{A_0}^T\cup \ovl{b_1}$, we get
\begin{equation}
\label{eq:floquet_weight2}
\frac14 (\ovl{A_{0z-}^{000}}+\ovl{A_{0y+}^{001}}) \ovl{b_{1x}^{011}} +\text{rot}\;,
\end{equation}
using Eq.~\eqref{eq:modified_cube_cup012}.
Finally, the terms $\frac12 (\ovl{c_0}^T\ovl{A_0}+\ovl{c_1}^T\ovl{A_1})$ yield
\begin{equation}
\frac12(\ovl{c_{0y}^{000}}\ovl{A_{0y}^{000}} + \ovl{c_{1y}^{000}}\ovl{A_{1y}^{000}}) + \text{rot}\;.
\end{equation}

Having constructed the microscopics of the state sum, we now turn it into a circuit of unitaries and measurements.
The time direction in which we choose to traverse the path integral is $t=x+y+z$ as drawn in Eq.~\eqref{eq:modified_cubic_variables}.
That is, the configuration of qubits at one fixed time represents the state-sum variables with a fixed value of the coordinate $t$.
One time period of the circuit corresponds to a translation by $t=x+y+z$.
The $t$ coordinates of the different vertices in the cubic lattice are $n$, $n+\frac13$, or $n+\frac23$ for $n\in\zz$, and we will color them red (r), green (g), or blue (b), respectively.
For example, the cube whose vertex with the smallest $(x,y,z)$-coordinates is $(0,0,0)$ has vertices colored as follows,
\begin{equation}
\begin{tikzpicture}
\draw (-0.8,0.3)edge[mark={arr,e},ind=$x$]++(0:0.35) (-0.8,0.3)edge[mark={arr,e},ind=$z$]++(25:0.25) (-0.8,0.3)edge[mark={arr,e},ind=$y$]++(90:0.35) (-0.8,0.3)edge[gray,mark={arr,e},ind=$t$]++(50:0.5);
\atoms{vertex}{{0/,vca}, {1/p={1,0},vcb}, {2/p={0.5,0.3},vcb}, {3/p={1.5,0.3},vcc}}
\atoms{vertex}{{0x/p={0,1},vcb}, {1x/p={1,1},vcc}, {2x/p={0.5,1.3},vcc}, {3x/p={1.5,1.3},vca}}
\draw[dashed] (0)--(2) (2x)--(2)--(3);
\draw (0)--(1) (0)--(0x) (0x)--(1x)--(1)--(3) (3)--(3x) (2x)--(3x) (1x)--(3x) (0x)--(2x);
\end{tikzpicture}
\;.
\end{equation}
Let us first consider the case of trivial twist, corresponding to two separate CSS honeycomb Floquet codes.
Let us focus on the circuit formed by the $A_0$-variables.
In accordance with the three colors, one period of this circuit consists of three different stages.
In the $i$th stage, the qubits represent all the $A_0$-variables at the edges connecting vertices with $t$-coordinates $n+\frac{i}{3}$ and $n+\frac{i+1}{3}$.
A natural spatial lattice on which we place these qubits is the projection of the cubic lattice along the $t$ direction.
This projection is a regular triangular lattice with rgb-colored vertices.
On this lattice, there is one $A_0$-qubit for each triangle.
We label them as follows,
\begin{equation}
\label{eq:triangular_projection}
\begin{tikzpicture}
\clip (2.2,0.7)rectangle (5.7,3.4);
\foreach \su in {0,3,6,9}{
\foreach \x in {0,...,9}{
\atoms{vertex,vca}{\x\su/p={$(1.2*\x,0)+{\su-\x}*(120:1.2)$}}
}
}
\foreach \su in {1,4,7,10}{
\foreach \x in {0,...,9}{
\atoms{vertex,vcc}{\x\su/p={$(1.2*\x,0)+{\su-\x}*(120:1.2)$}}
}
}
\foreach \su in {2,5,8,11}{
\foreach \x in {0,...,9}{
\atoms{vertex,vcb}{\x\su/p={$(1.2*\x,0)+{\su-\x}*(120:1.2)$}}
}
}
\draw (46)edge[mark={arr,e},mark={lab=$\bar x$,a}]++($(0:1.2)+(60:1.2)$);
\draw (46)edge[mark={arr,e},mark={lab=$\bar y$,a}]++($(0:1.2)+(-60:1.2)$);
\draw (46)edge[mark={arr,e},mark={lab=$x$,a}]++(60:1.2);
\draw (46)edge[mark={arr,e},mark={lab=$y$,a}]++(-60:1.2);
\draw (46)edge[mark={arr,e},mark={lab=$z$,a}]++(180:1.2);
\end{tikzpicture}
\begin{tikzpicture}
\clip (1.4,0.8)rectangle (5.7,4.3);
\foreach \y in {0,...,6}{
\foreach \x in {0,...,9}{
\draw ($\x*(0:1.2)+\y*(120:1.2)$)--(${\x+1}*(0:1.2)+\y*(120:1.2)$);
\draw ($\x*(0:1.2)+\y*(120:1.2)$)--(${\x+1}*(0:1.2)+{\y+1}*(120:1.2)$);
\draw (${\x+1}*(0:1.2)+\y*(120:1.2)$)--(${\x+1}*(0:1.2)+{\y+1}*(120:1.2)$);
}
}
\foreach \su in {0,3,6,9}{
\foreach \x in {0,...,9}{
\atoms{vertex,vca}{\x\su/p={$(1.2*\x,0)+{\su-\x}*(120:1.2)$}}
}
}
\foreach \su in {1,4,7,10}{
\foreach \x in {0,...,9}{
\atoms{vertex,vcc}{\x\su/p={$(1.2*\x,0)+{\su-\x}*(120:1.2)$}}
}
}
\foreach \su in {2,5,8,11}{
\foreach \x in {0,...,9}{
\atoms{vertex,vcb}{\x\su/p={$(1.2*\x,0)+{\su-\x}*(120:1.2)$}}
}
}
\node at ($1/3*(46)+1/3*(57)+1/3*(45)$){$\scriptstyle{A_{y+}^{00}}$};
\node at ($1/3*(34)+1/3*(46)+1/3*(45)$){$\scriptstyle{A_{y-}^{00}}$};
\node at ($1/3*(46)+1/3*(34)+1/3*(35)$){$\scriptstyle{A_{z+}^{00}}$};
\node at ($1/3*(46)+1/3*(35)+1/3*(47)$){$\scriptstyle{A_{z-}^{00}}$};
\node at ($1/3*(46)+1/3*(47)+1/3*(58)$){$\scriptstyle{A_{x+}^{00}}$};
\node at ($1/3*(46)+1/3*(57)+1/3*(58)$){$\scriptstyle{A_{x-}^{00}}$};
\node at ($1/3*(56)+1/3*(45)+1/3*(57)$){$\scriptstyle{A_{z-}^{01}}$};
\node at ($1/3*(56)+1/3*(57)+1/3*(68)$){$\scriptstyle{A_{x+}^{01}}$};
\node at ($1/3*(69)+1/3*(58)+1/3*(57)$){$\scriptstyle{A_{z+}^{10}}$};
\node at ($1/3*(69)+1/3*(57)+1/3*(68)$){$\scriptstyle{A_{y-}^{10}}$};
\node at ($1/3*(58)+1/3*(610)+1/3*(69)$){$\scriptstyle{A_{z-}^{10}}$};
\node at ($1/3*(69)+1/3*(610)+1/3*(711)$){$\scriptstyle{A_{x+}^{10}}$};
\node at ($1/3*(59)+1/3*(58)+1/3*(610)$){$\scriptstyle{A_{y+}^{1(\text{-}1)}}$};
\node at ($1/3*(47)+1/3*(58)+1/3*(59)$){$\scriptstyle{A_{y-}^{1(\text{-}1)}}$};
\node at ($1/3*(47)+1/3*(48)+1/3*(59)$){$\scriptstyle{A_{z+}^{1(\text{-}1)}}$};
\node at ($1/3*(36)+1/3*(47)+1/3*(48)$){$\scriptstyle{A_{x-}^{0(\text{-}1)}}$};
\node at ($1/3*(36)+1/3*(37)+1/3*(48)$){$\scriptstyle{A_{x+}^{0(\text{-}1)}}$};
\node at ($1/3*(35)+1/3*(36)+1/3*(47)$){$\scriptstyle{A_{y+}^{0(\text{-}1)}}$};
\node at ($1/3*(24)+1/3*(35)+1/3*(36)$){$\scriptstyle{A_{y-}^{0(\text{-}1)}}$};
\node at ($1/3*(23)+1/3*(24)+1/3*(35)$){$\scriptstyle{A_{x+}^{(\text{-}1)0}}$};
\node at ($1/3*(23)+1/3*(34)+1/3*(35)$){$\scriptstyle{A_{x-}^{(\text{-}1)0}}$};
\end{tikzpicture}\;.
\end{equation}
The superscripts refer to the $\bar x$ and $\bar y$ coordinates of red vertices, as shown on the left.
We have also shown the projections of the $x$, $y$, and $z$ basis vectors onto the $t=0$ plane.
The subscripts specify the triangle within each unit cell consisting of the $6$ triangle surrounding each red vertex.
The $\pm$ subscripts of the qubits coincide with these of the state-sum variables that they represent in Eq.~\eqref{eq:modified_cubic_variables}, but the $x/y/z$ subscripts do not.

Let us now add back the $A_1$ variables and the twist.
For every $A_0$ qubit, there will also be an $A_1$ qubit, and we distinguish them by an additional subscript $0$ or $1$ in front.
When traversing the path integral in $t$ direction, we notice that the $A_1$ and $b_1$-variables involved in the weights in Eqs.~\eqref{eq:floquet_weight0}, \eqref{eq:floquet_weight1}, and \eqref{eq:floquet_weight2} have larger $t$ coordinates than the $A_0$ and $b_0$-variables.
Since we want to implement the weights through phase gates acting at a fixed time, we let the $A_1$-qubits represent $A_1$-variables whose time coordinate is by $\frac12 t$ larger on average than the $A_0$-variables represented at the same stage.
In other words, we delay the circuit evolving the $A_0$ qubits by half a period.
This way we can ensure that for each weight there is a time where all involved state-sum variables are represented by qubits.
In addition, we also need to introduce additional sub-stages, such that between any two stages we either update the values of the $A_0$-qubits, or these of the $A_1$-qubits, or we perform a $b_0$ or $b_1$ measurement.
After this, we end up with the following $12$ stages,
\begin{equation}
\begin{tabular}{l|l|l|l|l|l}
Stage & $A_{0x+}^{00}$ & $A_{0x-}^{00}$ & $A_{1x+}^{00}$ & $A_{1x-}^{00}$ & meas.\\
\hline
\tcr{$11_{-1}$} & \tcr{$A_{0x+}^{0(-1)0}$} & \tcr{$A_{0x-}^{00(-1)}$} & \tcr{$A_{1z+}^{100}$} & \tcr{$A_{1y-}^{100}$} & \tcr{$\hat b_{1x}^{100}$}\\
\hline
$0$ & $A_{0x+}^{000}$ & $A_{0x-}^{000}$ & \ditto & \ditto & $\hat c_{0x}^{-100}$\\
\hline
$1$ & \ditto & \ditto & \ditto & \ditto & $\hat b_{0x}^{000}$\\
\hline
$2$ & \ditto & \ditto & $A_{1y+}^{101}$ & $A_{1z-}^{110}$ & $\hat c_{1x}^{100}$\\
\hline
$3$ & \ditto & \ditto & \ditto & \ditto & $\hat b_{1x}^{101}$\\
\hline
$4$ & $A_{0z+}^{100}$ & $A_{0y-}^{100}$ & \ditto & \ditto & $\hat c_{0x}^{000}$\\
\hline
$5$ & \ditto & \ditto & \ditto & \ditto & $\hat b_{0x}^{100}$\\
\hline
$6$ & \ditto & \ditto & $A_{1x+}^{111}$ & $A_{1x-}^{111}$ & $\hat c_{1x}^{101}$\\
\hline
$7$ & \ditto & \ditto & \ditto & \ditto & $\hat b_{1x}^{111}$\\
\hline
$8$ & $A_{0y+}^{101}$ & $A_{0z-}^{110}$ & \ditto & \ditto & $\hat c_{0x}^{100}$\\
\hline
$9$ & \ditto & \ditto & \ditto & \ditto & $\hat b_{0x}^{101}$\\
\hline
$10$ & \ditto & \ditto & $A_{1z+}^{211}$ & $A_{1y-}^{211}$ & $\hat c_{1x}^{111}$\\
\hline
$11$ & \ditto & \ditto & \ditto & \ditto & $\hat b_{1x}^{211}$
\end{tabular}
\;.
\end{equation}
The above table only lists the values for the $0$th time period and the spatial unit cell near $(0,0)$, and also for only $4$ out of $12$ qubits within the unit cell.
The values for other periods, unit cells, and qubits within the unit cell can be inferred from the values given above by the two following symmetries of the state sum and circuit:
First, they are translation symmetric under shifts in the $\bar x, \bar y, t$ coordinate system.
So when the value of the qubit $A_{\ldots}^{00}$ in period $0$ and stage $x$ equals the state-sum variable $A_{\ldots}^{abc}$, then $A_{\ldots}^{ij}$ in period $t$ and stage $x$ takes the value $A_{\ldots}^{(a+i+t)(b+j+t)(c-i-j+t)}$, for any choice of subscripts $\ldots$.
Also, when we measure $\hat f_x^{abc}$ in period $0$ and stage $x$, then we also measure $\hat f_x^{(a+i+t)(b+j+t)(c-i-j+t)}$ in period $t$ and stage $x$, for all $i$ and $j$, and $f\in \{b,c\}$.
Second, they are symmetric under $\frac{2\pi}{3}$ rotations around the $t$ axis.
So when the value of the qubit $A_{\eta\alpha\epsilon}^{00}$ in period $0$ stage $x$ equals the state-sum variable $A_{\eta\beta\epsilon}^{abc}$, then the qubit $A_{\eta\alpha'\epsilon}^{00}$ in period $0$ stage $x$ equals the state-sum variable $A_{\eta\beta'\epsilon}^{cab}$.
Here, $\eta\in \{0,1\}$, $\epsilon\in\{+,-\}$, $\alpha,\beta\in \{x,y,z\}$, and $x'=y$, $y'=z$, $z'=x$.
Also, when we measure $\hat f_x^{abc}$, then we also measure $\hat f_y^{abc}$ and $\hat f_z^{abc}$.

In fact, the $(x,y,z)$ unit cell of the state sum is smaller than the $(\bar x,\bar y,t)$ unit cell of the circuit.
So the translation invariance of the state sum yields a third symmetry of the circuit:
The circuit is invariant under a spatial translation by the projections of $x$, $y$, or $z$ shown in Eq.~\eqref{eq:triangular_projection}, together with a shift by a third period $\frac{t}{3}$ or $4$ stages.
To formally spell out the change of qubit sub and superscripts when shifting by $x$, $y$, or $z$ is tedious, so we will not do so and instead just have a look at Eq.~\eqref{eq:triangular_projection}.
Note that this shift will cycle through the roles of the red, green, and blue sub-lattices in Eq.~\eqref{eq:triangular_projection}.

Let us now translate the weights of the state sum into gates in the circuit.
We reduce the number of weights we need to write down by making use of all three symmetries discussed above.
We obtain
\begin{equation}
\begin{tabular}{l|l|l}
Weight & Gate & Step\\
\hline
$\delta_{A_{0x+}^{000}+A_{0x-}^{000}-b_{0x}^{000}}$ & $M_{ZZ}[A_{0x+}^{00},A_{0x-}^{00}, \hat b_{0x}^{000}]$ & $0^+$\\
\hline
\makecell{$\delta_{A_{0x+}^{000}+A_{0z+}^{100}-A_{0y}^{000}}$,\\$\delta_{A_{0z-}^{000}+A_{0x-}^{001}-A_{0y}^{000}}$,\\$(-1)^{\ovl{A_{0y}^{000}}\ovl{c_{0y}^{000}}}$} & $M_{XX}[A_{0x+}^{00},A_{0z-}^{00}, \hat c_{0y}^{000}]$ & $3^+$\\
\hline
$\delta_{A_{1x+}^{000}+A_{1x-}^{000}-b_{1x}^{000}}$ & $M_{ZZ}[A_{1x+}^{00},A_{1x-}^{00}, \hat b_{1x}^{000}]$ & $6^+_{-1}$\\
\hline
\makecell{$\delta_{A_{1x+}^{000}+A_{1z+}^{100}-A_{1y}^{000}}$,\\$\delta_{A_{1z-}^{000}+A_{1x-}^{001}-A_{1y}^{000}}$,\\$(-1)^{\ovl{A_{1y}^{000}}\ovl{c_{1y}^{000}}}$} & $M_{XX}[A_{1x+}^{00},A_{1z-}^{00}, \hat c_{1y}^{000}]$ & $9^+_{-1}$\\
\hline
$(-i)^{\ovl{A_{0z-}^{000}}\ovl{A_{1y+}^{001}}}$ & $\ovl{CS}[A_{0z-}^{00},A_{1z+}^{00}]$ & \tcr{0,}1\\
\hline
$(-i)^{\ovl{A_{0z-}^{000}}\ovl{A_{1x-}^{011}}}$ & $\ovl{CS}[A_{0z-}^{00},A_{1y-}^{00}]$ & 2\tcr{,3}\\
\hline
$i^{\ovl{A_{0z-}^{000}}\ovl{A_{1x-}^{001}}}$ & $CS[A_{0z-}^{00},A_{1z-}^{00}]$ & \tcr{0,}$1'$\\
\hline
$i^{\ovl{A_{0z-}^{000}}\ovl{A_{1y-}^{101}}}$ & $CS[A_{0z-}^{00},A_{1z-}^{00}]$ & $2'$\tcr{,3}\\
\hline
$(-i)^{\ovl{A_{0y+}^{001}}\ovl{A_{1x-}^{011}}}$ & $\ovl{CS}[A_{0z+}^{00},A_{1y-}^{00}]$ & \tcr{4,}5\\
\hline
$i^{\ovl{A_{0y+}^{001}}\ovl{A_{1x+}^{011}}}$ & $CS[A_{0z+}^{00},A_{1z+}^{00}]$ & \tcr{4,}$5'$\\
\hline
$i^{\ovl{b_{0x}^{000}} \ovl{A_{1z+}^{100}}}$ & $cS[\hat b_{0x}^{000},A_{1x+}^{00}]$ & $1''$\\
\hline
$i^{\ovl{b_{0x}^{000}}\ovl{A_{1y-}^{101}}}$ & $cS[\hat b_{0x}^{000},A_{1z-}^{00}]$ & \tcr{2,}3\tcr{,4,5}\\
\hline
$i^{\ovl{A_{0z-}^{000}} \ovl{b_{1x}^{011}}}$ & $cS[\hat b_{1x}^{011},A_{0z-}^{00}]$ & 3\\
\hline
$i^{\ovl{A_{0y+}^{001}} \ovl{b_{1x}^{011}}}$ & $cS[\hat b_{1x}^{011},A_{0z+}^{00}]$ & \tcr{4,}$5''$\tcr{,6,7}\\
\end{tabular}
\end{equation}
Here, $M_{ZZ}[a,b,c]$ denotes a $Z_0Z_1$-measurement performed on the qubits $a$ and $b$, whose value is stored in the classical bit $c$.
Analogously, $M_{XX}$ is a $X_0X_1$-measurement.
Including the $t$-axis rotation symmetry discussed above, there are $3$ times as many state-sum weights per $(x,y,z)$ unit cell, and due to the $(x,y,z)$-translation symmetry there are $9$ times as many gates per $(\bar x, \bar y, t)$-unit cell of the circuit.

The probably most human-readable way to spell out the circuit is by referring to the triangular lattice in Eq.~\eqref{eq:triangular_projection}, and specifying the qubits that the gates act on with respect to the rgb-coloring of the vertices.
Then, a $\frac13$ period of the circuit is given by
\begin{equation}
\begin{tabular}{l|l}
Step & Gate\\
\hline
$0^+$ & $M_{ZZ}^{gr}[0-,0+]$\\
\hline
$1$ & $\overline{CS^{gr}}[0-,1+]$, $\overline{CS^{rb}}[0+,1-]$\\
\hline
$1'$ & $CS^+[0,1]$, $CS^-[0,1]$\\
\hline
$1''$ & $cS^+[0br,1]$, $cS^+[1rg,0]$\\
\hline
$1^+$ & $M_{XX}^{gb}[1-,1+]$\\
\hline
$2$ & $\overline{CS^{gr}}[0-,1-']$\\
\hline
$2'$ & $CS^-[0,1]$\\
\hline
$2^+$ & $M_{ZZ}^{gr}[1-,1+]$\\
\hline
$3$ & $cS^-[1rb',0]$, $cS^-[0rg',1]$\\
\hline
$3^+$ & $M_{XX}^{rg}[0-,0+]$
\end{tabular}
\;.
\end{equation}
The superscript of each gate specifies at which places each gate or measurement acts, and the arguments in square brackets specify which qubits or bits it acts on.
For example, $M_{ZZ}^{gr}[0-,0+]$ denotes a $ZZ$ measurement performed at every green-red ($gr$) edge, acting on the $A_0$ qubit of the adjacent negative triangle ($0-$) and the $A_0$ qubit of the adjacent positive triangle ($0+$).
Similarly, $CS^{gr}[0-,1+]$ denotes a controlled-$S$ gate performed at every $gr$ edge, acting on the $A_0$ qubit on the adjacent negative triangle ($0-$), and the $A_1$ qubit on the adjacent positive triangle ($1+$).
$CS^{gr}[0-,1-']$ acts on the $A_0$ qubit on the adjacent negative triangle ($0-$), and the $A_1$ qubit on the negative triangle ($1-'$) sharing the $r$ vertex with the $gr$ edge, and not adjacent but otherwise closest to the edge.
$CS^+[0,1]$ denotes a $CS$ gate performed at every positive triangle, acting on the $A_0$ and $A_1$ qubit at this triangle.
$cS^+[0gr,1]$ denotes an $S$ gate acting on the $A_1$ qubit at every positive triangle, conditioned on the outcome of the last $ZZ$ measurement on the $A_0$ qubits at the green-red edge of the triangle.
$cS^+[0gr',1]$ denotes the same operation except that the classical control is the last measurement outcome at the $gr$ edge that shares the $r$ vertex with the triangle, and is not adjacent but otherwise closest to the triangle.
The other two third periods are obtained from the above by a cyclic permutation of the labels $r\rightarrow g\rightarrow b\rightarrow r$.
The following drawings illustrate the circuit:
\begin{equation}
\begin{tikzpicture}
\node[inner sep=0] (pic0) at (4.2,0){
\includegraphics{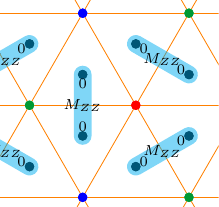}
};
\node[inner sep=0] (pic1) at (4.2,-4){
\includegraphics{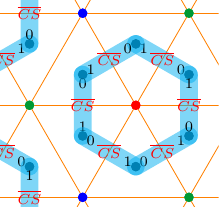}
};
\node[inner sep=0] (pic2) at (4.2,-8){
\includegraphics{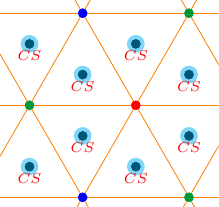}
};
\node[inner sep=0] (pic3) at (4.2,-12){
\includegraphics{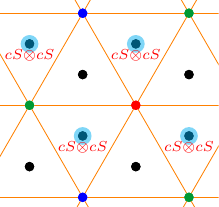}
};
\node[inner sep=0] (pic4) at (4.2,-16){
\includegraphics{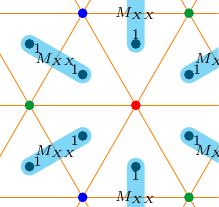}
};
\node[inner sep=0] (pic5) at (0,-16){
\includegraphics{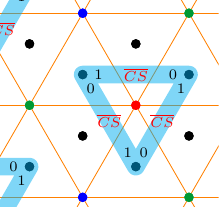}
};
\node[inner sep=0] (pic6) at (0,-12){
\includegraphics{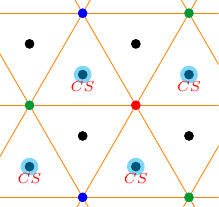}
};
\node[inner sep=0] (pic7) at (0,-8){
\includegraphics{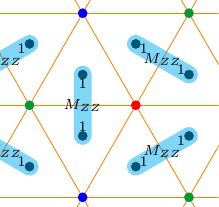}
};
\node[inner sep=0] (pic8) at (0,-4){
\includegraphics{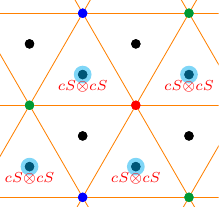}
};
\node[inner sep=0] (pic9) at (0,0){
\includegraphics{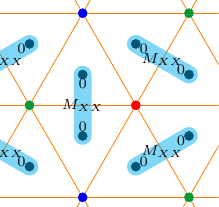}
};
\draw[very thick] (pic0.south)edge[->](pic1.north) (pic1.south)edge[->](pic2.north) (pic2.south)edge[->](pic3.north) (pic3.south)edge[->](pic4.north) (pic4.west)edge[->](pic5.east) (pic5.north)edge[->](pic6.south) (pic6.north)edge[->](pic7.south) (pic7.north)edge[->](pic8.south) (pic8.north)edge[->](pic9.south);
\draw[very thick,->] (pic9.north)--++(90:0.4)--++(0:4.2) node[midway, inner sep=0,above]{
\begin{tikzpicture}
\node[red,circle,inner sep=0] (r) at (0:0.4){$r$};
\node[green,circle,inner sep=0] (g) at (120:0.4){$g$};
\node[blue,circle,inner sep=0] (b) at (-120:0.4){$b$};
\draw (r)edge[out=90,in=30,->](g) (g)edge[out=-150,in=150,->](b) (b)edge[out=-30,in=-90,->](r);
\end{tikzpicture}
}
--(pic0.north);
\end{tikzpicture}
\;.
\end{equation}

\section{Fault tolerance for 1-form symmetric fixed-point circuits}
\label{sec:fault_tolerance}
In this section, we will show the existence of a fault-tolerant threshold for homological fault-tolerant circuits.
That is, we prove Proposition~\ref{prop:fault_tolerance}, after making it more formal.
\subsection{Definition of local fault tolerance}
\label{sec:fault_tolerance_definition}
In this section we will define the notion of fault tolerance which we will prove in the next section.
Let us start by discussing the general notion of local noise in a geometrically local circuit of unitaries and measurements for which we later prove robustness.
Roughly speaking, local noise is implemented by changing all the instruments/channels in the circuit by a small perturbation.
\begin{mydef}
Consider a uniform circuit of channels and instruments $M_0,\ldots, M_n$, each of which occur once in every unit cell.
A \emph{perturbation} is a set of alternative channels/instruments $\widetilde M_0,\ldots, \widetilde M_n$ with the same indices and bond dimensions.
Using $E_i=\widetilde M_i-M_i$, we can define the \emph{perturbation strength} as $\|E\|=\max_i \|E_i\|$.
The norm one the right is the 2-norm of tensors, which is the Frobenius norm if we view instruments/channels as linear operators.
\footnote{Note that since all classical and quantum degrees of freedom are assumed to be finite-dimensional, the choice of norm actually does not matter.}
The \emph{incoherent perturbation strength} is the smallest $\epsilon>0$ such that
\begin{equation}
F_i\coloneqq M_i+\frac1\epsilon E_i
\end{equation}
is a valid (completely positive) channel/instrument for all $i$.
\end{mydef}
We will only prove fault tolerance with respect to the \emph{incoherent} perturbation strength.
Let us comment on the difference between perturbation strength and incoherent perturbation strength.
The space of all classical/quantum hybrid channels (including instruments, POVMs, quantum channels, classical stochastic maps, etc.) with fixed input/output dimensions is a convex set, since the complete positivity condition is robust under convex combination of channels.
The following shows two toy pictures for such convex sets,
\begin{equation}
\begin{gathered}
\begin{tikzpicture}
\draw[cyan,line width=1.5,fill=gray] (0,0)circle(1);
\begin{scope}
\clip (0,0)circle(1);
\draw[green] (120:1)circle(0.5);
\draw[red] (120:0.6)circle(0.4);
\draw[green] (-60:0.5)circle(0.4);
\draw[red] (-60:0.3)circle(0.4);
\end{scope}
\path (120:1)pic{channelpoint} (-60:0.5)pic{channelpoint};
\end{tikzpicture}\qquad\;,
\begin{tikzpicture}
\fill[gray] (-120:2)--(0,0)--(-60:2)--cycle;
\atoms{vertex,astyle=cyan}{0/, 1/p=-120:2, 2/p=-60:2}
\begin{scope}
\clip (-120:2)--(0,0)--(-60:2)--cycle;
\draw[green] (0,0)circle(0.6);
\draw[red] (-120:0.5)--(-60:0.5);
\draw[green] (-60:1.2)circle(0.4);
\draw[red] (-60:1.4)--++(180:0.5)--(-60:0.9);
\end{scope}
\path (0,0)pic{channelpoint} (-60:1.2)pic{channelpoint};
\end{tikzpicture}\\
\begin{tikzpicture}
\draw[gray,line width=7] (0,0)edge[mark={lab=Channels/Instruments,b}]++(180:0.4);
\path[mark={lab=Unperturbed,b}] (0,-0.4)--++(180:0.4) pic[midway]{channelpoint};
\draw[cyan,line width=1.5] (0,-0.8)edge[mark={lab=Extremal,b}]++(180:0.4);
\draw[green] (0,-1.2)edge[mark={lab=Constant strength,b}]++(180:0.4);
\draw[red] (0,-1.6)edge[mark={lab=Incoherent,b}]++(180:0.4);
\end{tikzpicture}\;.
\end{gathered}
\end{equation}
The sets of channels/instruments themselves are in gray.
Extremal points, that is, points which are not convex combinations of any other points, are in blue.
In each set we have marked two selected points, as well as the perturbations around these points of some constant (incoherent) strength in green (red).
The convex sets of hybrid classical/quantum channels with multiple input/output bits/qubits are high dimensional and much more complicated, but the above toy pictures capture the essence.
For fixed-point circuits, the channels/instruments are typically (controlled) unitaries or projective measurements, and thus extremal points.
The convex sets of purely classical channels have \emph{isolated} extremal points without any other extremal points inside some open neighborhood, as shown on the right.
Channels with at least one quantum input/output only have \emph{non-isolated} extremal points with other extremal points arbitrarily close, as shown on the left.
Such nearby extremal points can be obtained by applying a small unitary $e^{i\epsilon H}$ to a quantum input or output.
For all non-extremal points or isolated extremal points, coherent and incoherent perturbation strengths are equivalent in the sense that their quotient is bounded by a constant factor.
For non-isolated extremal points however, the nearby extremal points have arbitrarily small perturbation strength but an incoherent perturbation strength of $1$.

Note that, in contrast to most fault-tolerance proofs, the notion of noise we consider is agnostic towards the Pauli basis.
That is, from the very beginning, we consider noise that is not a convex combination of Pauli, or even Clifford unitaries.
A toy picture to illustrate this point is the following:
\begin{equation}
\begin{tikzpicture}
\draw[cyan, line width=1.5,fill=gray] (0,0)circle(1);
\atoms{vertex,style=red}{0/p=0:1, 1/p=120:1, 2/p=-120:1}
\fill[red!30] (0-c)--(1-c)--(2-c)--cycle;
\path (0:1)pic{channelpoint};
\end{tikzpicture}
\begin{tikzpicture}
\draw[gray,line width=7] (0,-0.8)edge[mark={lab=Incoherent non-Clifford noise,b}]++(180:0.4);
\path[mark={lab=Unperturbed,b}] (0,0)--++(180:0.4) pic[midway]{channelpoint};
\draw[red!30,line width=7] (0,-0.4)edge[mark={lab=Clifford/Pauli noise,b}]++(180:0.4);
\draw[cyan,line width=1.5] (0,-1.2)edge[mark={lab=Coherent noise,b}]++(180:0.4);
\end{tikzpicture}
\;.
\end{equation}
The set of Clifford channels is discrete, so it spans a polytope inside the set of all channels.
There are incoherent perturbations that are not inside this polytope.

Arbitrarily changing whole channels or instruments in the circuit already seems like a very general notion of noise, and includes, for example, any single-qubit noise before the channel is applied.
However, these perturbations have a fixed locality, and we would like to show fault-tolerance under noise of geometric distance $k$, for $k$ constant but arbitrarily large.
This is covered by our proof since fault tolerance holds not only for the original circuit, but also after we apply local circuit identities to rewrite the circuit.
This includes combining two nearby channels of the circuit and then perturbing the combined channel.
For example, in a $1+1$-dimensional brick-layer as in Eq.~\eqref{eq:qec_circuit}, we could combine every even-layer channel with the channel at its top right,
\begin{equation}
\label{eq:gate_combination}
\begin{tikzpicture}
\atoms{square,xscale=8}{0/}
\draw[quantum] ([sx=-1]0-b)--++(-90:0.4) (0-b)--++(-90:0.4) ([sx=1]0-b)--++(-90:0.4) ([sx=-1]0-t)--++(90:0.4) (0-t)--++(90:0.4) ([sx=1]0-t)--++(90:0.4);
\draw[classical] ([sx=-0.5]0-t)--++(90:0.4) ([sx=0.5]0-t)--++(90:0.4);
\end{tikzpicture}
\coloneqq
\begin{tikzpicture}
\atoms{square,xscale=4}{0/, 1/p={1,0.8}}
\draw[quantum] ([sx=-0.5]0-b)--++(-90:0.4) ([sx=0.5]0-b)--++(-90:0.4) ([sx=-0.5]0-t)--++(90:1.2) ([sx=0.5]0-t)--([sx=-0.5]1-b) ([sx=0.5]1-b)--++(-90:1.2) ([sx=-0.5]1-t)--++(90:0.4) ([sx=0.5]1-t)--++(90:0.4);
\draw[classical] (0-t)--++(90:0.4) (1-t)--++(90:0.4);
\end{tikzpicture}
\;.
\end{equation}
Perturbation of this combined channel will then have a larger distance $k$ than perturbations of the circuit in its original form.
In Appendix~\ref{sec:noise}, we discuss how common types of noise, and also weakly correlated noise, can be implemented as circuit perturbations after locally rewriting the circuit.

Having discussed the kind of noise we consider, we will now define what it means for the circuit to be fault tolerant under noise.
Intuitively, fault tolerance means that the memory lifetime of the protected logical information in the circuit on an $L\times L$ torus increases exponentially with $L$, after applying any perturbation of incoherent strength $\epsilon$ below some threshold $\epsilon_0$.
In other words, the error probability of the circuit on a $L\times L\times T$ spacetime scales like $\propto T (\frac{\epsilon}{\epsilon_0})^{\frac{L}{L_0}}$.
Since the perturbed logical subspace itself is hard to describe, we define fault tolerance with respect to the logical subspace of the unperturbed circuit.
That is, we prepare an unperturbed logical state, run the perturbed circuit, and then perform unperturbed corrections to map the evolved state back to the unperturbed logical subspace.
The assumption that we prepare and extract the unperturbed logical state is only made to test whether the logical information is preserved.
If we would define fault-tolerance for a circuit that performs a computation by moving around boundaries, anyons, twist defects, or interfacing with other topological phases, then we would make these assumptions only at the very beginning and very end of the computation.

More precisely, we consider the following setup.
Let $C$ denote the instrument describing the complete evolution of a $L\times L\times T$ 1-form symmetric fixed-point circuit.
Let $\widetilde C$ denote the instrument of the perturbed circuit.
So $C$ and $\widetilde C$ are linear operators from the space of density matrices of all qubits to that same space tensored with the vector space of probability distributions over all classical measurement outcomes $s_s$.
That is, as tensors, $C$ and $\widetilde C$ have two input and output (ket and bra) indices for every qubit, and an additional single output index for each classical measurement outcome.

Let $V=L^2T$ denote the spacetime volume of the circuit.
Let $G_b$ ($G_e$) be an operator that prepares a ground state of the fixed-point path integral at the initial (final) slice at time $0$ ($T$).
So $G_b$ ($G_e$) is a linear map from some abstract logical vector space to the vector space of density matrices associated to the initial (final) slice.
Let $\overline D$ be the quantum channel consisting of (1) applying the unperturbed circuit for a constant time independent of $L$, (2) applying the classical minimum-weight matching decoder $D$ to the measurement outcomes, and (3) applying the unperturbed corrections (see Eq.~\eqref{eq:qec_circuit}).
So $\overline D$ is a linear map from the (density-matrix) space of qubits and classical measurement outcomes to the space of qubits, that is, acting in the opposite direction compared to $C$ and $\widetilde C$.
With these notations, we can now state our fault-tolerance theorem:
\begin{mythm}
\label{thm:fault_tolerance}
For any 1-form symmetric fixed-point circuit, there exist $\epsilon_0>0$, $L_0>0$, and $\alpha>0$ such that for any perturbation $\widetilde M_i$ of incoherent perturbation strength $\epsilon<\epsilon_0$, we have
\footnote{The choice of norm on the left-hand side does not matter, since the logical Hilbert space dimension is independent of $L$, and all norms in finite vector spaces are equivalent.
For concreteness, we will pick the 2-norm as a tensor, which is the Frobenius norm as an operator.}
\begin{equation}
\label{eq:fault_tolerance_definition}
\|G_e^\dagger \overline D \widetilde C G_b  - \onelog\|
\leq \alpha V (\frac{\epsilon}{\epsilon_0})^{\frac{L}{L_0}}\;,
\end{equation}
where $\onelog$ is the identity channel on the logical subspace.
\end{mythm}
Note that $\epsilon_0$, $L_0$ and $\alpha$ are constants, but may change when we rewrite the circuit using local identities, in particular, when we regroup what the elementary operations in $\{M_i\}$ are.
For example, when we combine two neighboring gates as shown in Eq.~\eqref{eq:gate_combination}, the threshold for perturbations on the combined gate will in general be smaller than for perturbing the individual gates.

\subsection{Proof of fault tolerance}
\label{sec:general_proof}
In this section, we prove Theorem~\ref{thm:fault_tolerance}.
The outline of the proof is as follows:
First we expand the perturbed circuit $\widetilde C$ into a sum over different error configurations of unperturbed gates or perturbations at every site.
Then we formulate a criterion for error configurations to be correctable, and show that correctable configurations indeed do not affect the logical subspace.
Finally, we estimate the number and weight of non-correctable error configurations and find that their overall effect vanishes exponentially quickly in the system size $L$.

Let us start by introducing a few basic geometric notations.
We will denote by $S$ the set of \emph{sites} in the circuit, that is, the set of individual gate applications, or the set of boxes in the circuit diagram.
By Definition~\ref{def:path_integral_code}, these are also the weights or tensors in the underlying 1-form symmetric fixed-point path integral.
Recall from Definition~\ref{def:fixed_point_path_integral} that each site $r\in S$ is associated with a nearby volume $\widetilde r\in S_3[M]$.
Vice versa, for any volume $c\in S_3[M]$, define $\widetilde c\coloneqq \{r\in S: \widetilde r=c\}$ as the set of associated sites.
Also, define $\widetilde X\coloneqq\bigcup_{x\in X} \widetilde x\subset S$ for $X\subset S_3[M]$, and $\widetilde R\coloneqq \bigcup_{r\in R}\{\widetilde r\}\subset S_3[M]$ for $R\subset S$.
Further, write $x^+$ as short-hand for $x^{+2\chi}$, as used in Definitions~\ref{def:fixed_point_path_integral} and \ref{def:homological_integral}.
Let $V_\chi$ be an integer such that $|\widetilde{r^+}|\leq V_\chi$ for any edge, vertex, or site $r$.
Further, we consider a cellulation $M$ of topology $S_1\times S_1\times [0,1]$, where $L$ is the length of the shortest non-contractable loop around either of the $S_1$.
The canonical example is a $L\times L\times T$ cubic lattice with periodic boundary conditions in $L$.
Also choose the constant time for which we apply the unperturbed circuit in $\ovl D$ as discussed before Definition~\ref{thm:fault_tolerance}, such that all gates that are associated with volumes of distance less than $2\chi$ to the boundary at time $T$ are unperturbed.
Similarly, since we are applying the perturbed circuit to the ground-state subspace, we can extend the circuit before time $0$ by a distance-$2\chi$ unperturbed circuit with $s=0$.

With this, let us start by writing down the expansion of the perturbed circuit.
Given a perturbation of incoherent strength $\epsilon$, there exist channels/instruments $F_i$ such that
\begin{equation}
\widetilde M_i=(1-\epsilon)M_i+\epsilon F_i\;.
\end{equation}
Define an \emph{error configuration} as a subset of sites $R\subset S$.
Let $C_R$ denote the overall instrument obtained from the circuit diagram by taking the instrument/channel $F_i$ at every site in $R$, and $M_i$ at every site in the complement $\overline R$ of $R$.
We can now expand the perturbed circuit $\widetilde C$ as a sum
\begin{equation}
\label{eq:error_expansion}
\widetilde C=
\sum_{R\subset S} \epsilon^{|R|} (1-\epsilon)^{|\bar R|} C_R\;.
\end{equation}

Next, we will write down a criterion that is supposed to tell us when an error configuration can be corrected.
Define a \emph{logical loop} $\Lambda\subset S_1[M]$ as a connected subset of the edges of $M$ such that there is a homologically non-trivial 1-cycle $\alpha$ with $\sup(\alpha)\subset \Lambda$.
Here, $\sup(\alpha)$ denotes the support of $\alpha$, that is, the set of edges $e$ with $\alpha(e)\neq 0$.
We call an error configuration $R\subset S$ \emph{correctable} if for every logical loop $\Lambda$, we have
\begin{equation}
\label{eq:correctability_condition}
|\widetilde R^+\cap \Lambda|<\frac{|\Lambda|}2\;.
\end{equation}
Here and at some other places below, we abuse the notation $\widetilde R^+$ to mean the set of edges adjacent to $\widetilde R^+$.
We will denote the set of correctable error configurations by $\mathcal R_{\text{cor}}$.

The next step is to show that correctable error configurations can in fact be corrected.
Consider a correctable error configuration $R\in \mathcal R_{\text{cor}}$, and the according channel $C^s_R$, where we denote the measurement outcome 1-chain $s$ by an explicit superscript.
By Definition~\ref{def:path_integral_code}, $C^s_R$ on $\overline{ \widetilde R}$ is equal to the underlying fixed-point path integral with symmetry defect 1-chain $s$.
Thus, due to the first condition in Definition~\ref{def:homological_integral}, $C^s_R=0$ unless $\delta s|_{\ovl{\widetilde R^+}}=0$, in other words, the anyon fusion rules are deterministically fulfilled at each vertex outside of $\widetilde R^+$.
Due to the third condition in Definition~\ref{def:homological_integral}, $C_R^s=0$ unless $s$ restricted to the boundary of $\widetilde R^{+\chi}$ is homologically trivial.
Thus, $\delta s$ can be fixed within $\widetilde R^+$.
That is, there is a 1-chain $\fix_R(\delta s)$ such that
\begin{equation}
\label{eq:fault_tolerance_fixr}
\sup(\fix_R(\delta s))\subset \widetilde R^+\;,\qquad \delta(\fix_R(\delta s))=-\delta s\;,
\end{equation}
such that $s+\fix_R(\delta s)$ is a 1-cycle with open boundary conditions at time $T$.
That $R$ is correctable implies that $\widetilde R^+=\widetilde R^{+2\chi}$ is contained in a sub-cellulation $X$ of 3-ball topology.
We know that $\delta s=0$ outside of $\widetilde R^{+\chi}$ unless $C_R^s=0$, and that the distance between $\widetilde R^{+\chi}$ and the boundary of $X$ is at least $\chi$.
Thus, $X\setminus R^{+\chi}$ contains a sub-cellulation of topology $S_2\times [0,1]$ and width $\chi$, on which $s$ is a homologically trivial 1-cycle.
Due to the fourth condition in Definition~\ref{def:homological_integral}, the evaluation of $C^s_R$ on this $S_1\times [0,1]$ sub-cellulation (where all tensors are $M_i$) is a rank-1 operator for any $s$.
Thus we find
\begin{equation}
\label{eq:fixed_point_proportionality}
C^s_R \propto C^{s+\fix_R(\delta s)}\;.
\end{equation}
Note that the proportionality would also hold with $s$ instead of $s+\fix_R(\delta s)$ on the right-hand side, however, this right-hand side would then be $0$.

Now, consider the minimum-weight 1-chain $\mwpm(\delta s)$ defined around Eq.~\eqref{eq:min_weight_perfect_matching}.
We will now show that for $R\in \mathcal R_{\text{cor}}$, $\fix_R(\delta s)-\mwpm(\delta s)$ is a homologically trivial 1-cycle.
To this end, assume the opposite.
Then,
\begin{equation}
\label{eq:logical_loop_contradiction}
\Lambda\coloneqq \sup(\fix_R(\delta s))\cup \sup(\mwpm(\delta s))
\end{equation}
is a logical loop.
Using Eq.~\eqref{eq:correctability_condition}, we find
\begin{equation}
\begin{gathered}
\begin{multlined}
|\sup(\fix_R(\delta s))| + |\sup(\mwpm(\delta s))| \overset{\eqref{eq:logical_loop_contradiction}}{\geq} |\Lambda|\\\overset{\eqref{eq:correctability_condition}}{>} 2|\Lambda\cap \widetilde R^+|\overset{\eqref{eq:fault_tolerance_fixr}}{\geq} 2|\sup(\fix_R(\delta s))|
\end{multlined}\\
\Rightarrow |\sup(\mwpm(\delta s))| > |\sup(\fix_R(\delta s))|\;,
\end{gathered}
\end{equation}
a contradiction to the minimum-weight condition of $\mwpm(\delta s)$.

The following toy picture in $1+1$ spacetime dimensions (left and right identified) illustrates the situation,
\begin{equation}
\begin{gathered}
\begin{tikzpicture}
\fill[black!10] (0,0)rectangle(7,5);
\draw[dotted] (0,0)edge[mark={lab=$0$,b},mark={slab=$x$,r},mark={arr,e}](7,0) (0,0)edge[mark={lab=$0$,b},mark={slab=$t$},mark={arr,e}] (0,5) (0,5)edge[mark={lab=$T$,b}]++(7,0) (7,0)edge[mark={lab=$L\sim 0$,b}]++(0,5);
\atoms{circ,scale=0.25,all,nobd,astyle=blue}{r0/p={1.5,2.5}, r1/p={1.5,3}, r2/p={2,2.5}, r3/p={4.5,1}, r4/p={5,1}, r5/p={4.5,1.5}, r6/p={5,1.5}, r7/p={4.5,2}, r8/p={5,2}, r9/p={6,4}}
\fill[rc,blue,fill opacity=0.3] (1,2)--++(1.5,0)--++(0,1)--++(-0.5,0)--++(0,0.5)--++(180:1)--cycle (4,0.5)--++(0:1.5)--++(90:2)--++(180:1.5)--cycle (5.5,3.5)rectangle(6.5,4.5);
\atoms{vertex}{ds0/p={1.25,2.25}, ds1/p={1.75,2.75}, ds2/p={5.25,2.25}, ds3/p={4.75,1.25}, ds4/p={4.25,1.75}, ds5/p={5.25,0.75}, ds6/p={5.75,3.75}, ds7/p={5.75,4.25}}
\draw (ds0)to[out=120,in=-90](0.5,5) (ds1)to[bend left](ds6) (ds7)--(5.75,5) (ds2)to[out=-30,in=180](7,1.5) (ds4)to[out=-170,in=0](0,1.5) (ds3)to[out=0,in=0,looseness=3](ds5) (2.5,0.8)circle(0.4) (3,5)to[bend right=60](4.5,5);
\draw[red] (ds0)--(ds1) (ds2)--(ds4) (ds3)--(ds5) (ds6)--(ds7);
\draw[blue,rc] (ds0)-|(ds1) (ds2)--(ds5) (ds3)-|(ds4) (ds6)--++(0:0.5)|-(ds7);
\draw[green,dashed] (ds1)--(ds4) (ds0)--(0,2.25) (ds3)to[out=30,in=180](7,2.25) (ds5)to[bend right](ds2) (ds6)to[bend left=45](ds7);
\end{tikzpicture}\\
\begin{tikzpicture}
\path[mark={lab=$R$,b},blue] (0,0)--++(180:0.4);
\atoms{circ,scale=0.25,all,nobd,astyle=blue}{r/p={-0.2,0}}
\draw[blue!30,line width=10] (0,-0.4)edge[mark={lab=$\textcolor{blue}{R^+}$,b}]++(180:0.4);
\draw (0,-0.8)edge[mark={lab=$s$,b}]++(180:0.4);
\path[mark={lab=$\delta s$,b}] (0,-1.2)--++(180:0.4);
\atoms{vertex}{ds/p={-0.2,-1.2}}
\draw[blue] (3,0)edge[mark={lab=$\fix_R(\delta s)$,b}]++(180:0.4);
\draw[red] (3,-0.4)edge[mark={lab={Real $\mwpm(\delta s)$},b}]++(180:0.4);
\draw[green,dashed] (3,-0.8)edge[mark={lab={Wrong $\mwpm(\delta s)$},b}]++(180:0.4);
\end{tikzpicture}
\end{gathered}\;.
\end{equation}
Since $s+\fix_R(\delta s)$ and $s+\mwpm(\delta s)$ are in the same homology class, we have
\begin{equation}
\label{eq:correctable_error_condition}
C^{s+\fix_R(\delta s)} G_b = C^{s+\mwpm(\delta s)} G_b\;,
\end{equation}
due to the second condition in Definition~\ref{def:homological_integral}.
Combined with Eq.~\eqref{eq:fixed_point_proportionality}, we find
\begin{equation}
\label{eq:correctable_error}
G_e^\dagger \overline D C_R G_b = G_e^\dagger \overline D C G_b = \onelog\;,
\end{equation}
for a correctable error configuration $R\in \mathcal R_{\text{cor}}$.
Note that the proportionality in Eq.~\eqref{eq:fixed_point_proportionality} becomes an equality since we know that both sides are correctly normalized (trace preserving) channels.

Having shown that correctable error configurations have no effect on the logical subspace, we can now ignore any correctable errors in the expansion in Eq.~\eqref{eq:error_expansion}, when calculating the logical error in Eq.~\eqref{eq:fault_tolerance_definition}.
An additional minor technical result that we will use is that the Frobenius norm of any quantum channel $X$ is bounded by the Hilbert space dimension $d_{\text{log}}$.
To show this, we use that the channel defines a positive semi-definite matrix $\mathcal X$ after blocking the input and output ket indices, and the input and output bra indices.
Then we use that the Frobenius norm of a positive semi-definite matrix is bounded by its trace,
\begin{equation}
\label{eq:trace_inequality}
\|X\|=\sqrt{\operatorname{Tr}(\mathcal X^2)}\leq \operatorname{Tr}(\mathcal X)=
\begin{tikzpicture}
\atoms{square,xscale=1.5,lab={t=$X$,p=0:0}}{x/}
\draw[rc] ([sx=-0.15]x-b)--++(-90:0.4)-|([sx=0.15]x-b) ([sx=-0.15]x-t)--++(90:0.4)-|([sx=0.15]x-t);
\end{tikzpicture}
=
\begin{tikzpicture}
\draw[rc] (0,0)rectangle(0.3,0.3);
\end{tikzpicture}
=
d_{\text{log}}\;.
\end{equation}

Putting all of the above together, we can upper bound the error in Eq.~\eqref{eq:fault_tolerance_definition} by
\begin{equation}
\label{eq:error_expansion_bound}
\begin{gathered}
\|G_e^\dagger \overline D \widetilde C G_b-\onelog\|\\
\overset{\text{\eqref{eq:error_expansion}}}{=} \Bigl\|\sum_{R\subset C} \epsilon^{|R|} (1-\epsilon)^{|\ovl R|} (G_e^\dagger \overline D C_R G-\onelog)\Bigr\|\\
= \Bigl\|\sum_{R\in \mathcal R_{\text{cor}}} \epsilon^{|R|} (1-\epsilon)^{|\ovl R|} (G_e^\dagger \overline D C_R G_b-\onelog)\\
+ \sum_{R\in \overline{\mathcal R_{\text{cor}}}} \epsilon^{|R|} (1-\epsilon)^{|\ovl R|} (G_e^\dagger \overline D C_R G_b -\onelog)\Bigr\|\\
\overset{\text{\eqref{eq:correctable_error}}}{=} \Bigl\|\sum_{R\in \overline{\mathcal R_{\text{cor}}}} \epsilon^{|R|} (1-\epsilon)^{|\ovl R|} (G_e^\dagger \overline D C_R G_b -\onelog)\Bigr\|\\
\leq \sum_{R\in \overline{\mathcal R_{\text{cor}}}} \epsilon^{|R|} (1-\epsilon)^{|\ovl R|} \|G_e^\dagger \overline D C_R G_b -\onelog\|\\
\overset{\text{\eqref{eq:trace_inequality}}}{\leq} 2d_{\text{log}} \sum_{R\in \overline{\mathcal R_{\text{cor}}}} \epsilon^{|R|} (1-\epsilon)^{|\ovl R|}\;,
\end{gathered}
\end{equation}
where $\overline{\mathcal R_{\text{cor}}}$ is the set of error configurations that are not correctable.

All remains is to count non-correctable error configurations, weighted by the above expression.
For each non-correctable error configuration $R\in \overline{\mathcal R_{\text{cor}}}$, per definition there exists a logical loop $\Lambda$ of length $\lambda\coloneqq |\Lambda|\geq L$ such that $|\widetilde R^+\cap \Lambda|>\frac\lambda2$.
Define $R_\Lambda\coloneqq R\cap\widetilde{\Lambda^+}$ and $R_{\overline\Lambda}\coloneqq R\cap\overline{\widetilde{\Lambda^+}}$, such that $R=R_\Lambda\cup R_{\ovl\Lambda}$ and $|\widetilde R_\Lambda^+\cap \Lambda|>\frac\lambda2$.

The number of logical loops of length $|\Lambda|=\lambda$ is bounded by
\begin{equation}
\label{eq:loop_counting}
|\{\Lambda: |\Lambda|=\lambda\}|\leq V \eta^{2\lambda}\;,
\end{equation}
where $V=|S_0[M]|$ is the number of vertices of $M$, and $\eta$ is the maximal number of edges adjacent to a vertex in $M$.
To see this, we note that since $\Lambda$ is connected it can be viewed as a tree.
We can run through this tree by increasing depth before width in $2\lambda$ steps.
At every step there are $\eta$ possibilities to choose the next edge.
In addition, there are $V$ possible starting points of the tree.

By non-correctability, we have
\begin{equation}
\label{eq:error_size_bound}
|R_\Lambda|\geq \frac{|\widetilde R_\Lambda^+|}{V_\chi} \geq \frac{\lambda}{2\nu V_\chi}\;,
\end{equation}
where $\nu$ is the maximum number of edges of a volume.
On the other hand, the number of error configurations $R_\Lambda\subset \widetilde{\Lambda^+}$ can be upper bounded by
\begin{equation}
\label{eq:rlambda_counting}
|\{R_\Lambda: R_\Lambda\subset \widetilde{\Lambda^+}\}|\leq 2^{|\widetilde{\Lambda^{+}}|}\leq 2^{\lambda V_\chi}\;.
\end{equation}
Last, we will use two simple identities that hold for all $0<x<1$ and positive integers $L$ and $L_0$,
\begin{gather}
\label{eq:exp_summation}
\sum_{\lambda\geq L} x^\lambda = \frac{x^L}{1-x}\;,\\
\label{eq:inequality1}
1-x \leq (1-\frac{x}{L_0})^{L_0}
\Rightarrow \frac{1}{1-(1-x)^{1/L_0}}\leq \frac{L_0}{x}\;.
\end{gather}
We now rewrite the sum in Eq.~\eqref{eq:error_expansion_bound} as four nested sums over (1) $\lambda$, (2) $\Lambda$, (3) $R_\Lambda$, and (4) $R_{\overline\Lambda}$.
Note that for every non-correctable error configuration $R\in \overline{\mathcal R_{\text{cor}}}$, there will in general be many logical loops $\Lambda$.
However, since all summands are positive, we get an upper bound by the triangle inequality,
\begin{equation}
\label{eq:summation_bound}
\begin{gathered}
2d_{\text{log}} \sum_{R\in \overline{\mathcal R_{\text{cor}}}} \epsilon^{|R|} (1-\epsilon)^{|\bar R|}\\
\begin{multlined}
\overset{0\leq\epsilon\leq 1}{\leq} 2d_{\text{log}}\sum_{\lambda\geq L}\sum_{\Lambda: |\Lambda|=\lambda}\sum_{R_\Lambda\subset \Lambda^+: |R_\Lambda^+\cap \Lambda|\geq\frac\lambda2}\\ \sum_{R_{\bar\Lambda}\subset \overline{\Lambda^{+}}}
\epsilon^{|R_\Lambda\cup R_{\overline\Lambda}|} (1-\epsilon)^{|\overline{R_\Lambda\cup R_{\overline\Lambda}}|}
\end{multlined}
\\
=
2d_{\text{log}} \sum_{\lambda\geq L}\sum_{\Lambda: |\Lambda|=\lambda}\sum_{R_\Lambda\subset \Lambda^+: |R_\Lambda^+\cap \Lambda|\geq\frac\lambda2} \epsilon^{|R_\Lambda|} \\
\overset{\text{\eqref{eq:error_size_bound}}}{\leq}
2d_{\text{log}} \sum_{\lambda\geq L}\sum_{\Lambda: |\Lambda|=\lambda}\sum_{R_\Lambda\subset \Lambda^+} \epsilon^{\frac{\lambda}{2\nu V_\chi}}\\
\overset{\text{\eqref{eq:rlambda_counting}}}{\leq}
2d_{\text{log}} \sum_{\lambda\geq L} \sum_{\Lambda:|\Lambda|=\lambda} 2^{\lambda V_\chi} \epsilon^{\frac{\lambda}{2\nu V_\chi}}\\
\overset{\text{\eqref{eq:loop_counting}}}{\leq}
2d_{\text{log}} \sum_{\lambda\geq L} V \eta^{2\lambda} 2^{\lambda V_\chi} \epsilon^{\frac{\lambda}{2\nu V_\chi}}\\
\overset{\text{\eqref{eq:l0def}, \eqref{eq:epsilon0def}}}{=}
2d_{\text{log}} V \sum_{\lambda\geq L} (\frac{\epsilon}{2\epsilon_0})^{\frac{\lambda}{L_0}}\\
\overset{\text{\eqref{eq:exp_summation}}}{=}
2d_{\text{log}} V \frac{(\frac{\epsilon}{2\epsilon_0})^{\frac{L}{L_0}}}{1-(\frac{\epsilon}{2\epsilon_0})^{\frac{1}{L_0}}}\\
\overset{\text{\eqref{eq:inequality1}}}{\leq}
\frac{L_0}{1-\frac{\epsilon}{2\epsilon_0}} 2d_{\text{log}} V (\frac{\epsilon}{2\epsilon_0})^{\frac{L}{L_0}}\\
\overset{\epsilon\leq \epsilon_0}{\leq} 4 L_0 d_{\text{log}} V (\frac{\epsilon}{2\epsilon_0})^{\frac{L}{L_0}}\\
\overset{\text{\eqref{eq:alphadef}}}{\leq} \alpha V (\frac{\epsilon}{\epsilon_0})^{\frac{L}{L_0}}
\;,
\end{gathered}
\end{equation}
with
\begin{gather}
\label{eq:l0def}
L_0\coloneqq 2\nu V_\chi\;,\\
\label{eq:epsilon0def}
\epsilon_0\coloneqq \frac12 \eta^{-4\nu V_\chi}2^{-2\nu V_\chi^2}\;,\\
\label{eq:alphadef}
\alpha\coloneqq 4 L_0 d_{\text{log}}\;.
\end{gather}
Combining Eq.~\eqref{eq:summation_bound} and Eq.~\eqref{eq:error_expansion_bound} yields Eq.~\eqref{eq:fault_tolerance_definition} and proves Theorem~\ref{thm:fault_tolerance}.

Finally, let us briefly comment on how one might adapt this proof to apply precisely to the circuits in Sections~\ref{sec:double_semion}, \ref{sec:general_doubles}, and \ref{sec:measurement_floquet}.
First, the fact that 1-form symmetry defects form pairs $(b,c)$ instead of 1-chains $s$ can be incorporated with only minor changes.
For example, we need to define a logical loop as a connected subset of faces and edges that supports a (co-)homologically non-trivial $(b,c)$-configuration.
``Connected'' should be defined through sharing either an adjacent vertex or cube, and $L$ should be defined as the minimum weight of such a logical loop.
The counting of logical loops in Eq.~\eqref{eq:loop_counting} still holds, just that the coefficient $\eta$ might be different.

Second, if $K\neq \zz_2^n$, we might have to use an RG decoder instead for efficiency reasons.
To this end, note that the proof above is similar to the loop-counting arguments in Ref.~\cite{Dennis2001}, but starts from the four general conditions in Definition~\ref{def:homological_integral} instead of a specific stabilizer code.
In similar spirit, one could use the proof in Appendix~B of Ref.~\cite{Bravyi2011} starting from the four general conditions in Definition~\ref{def:homological_integral}.
Roughly, the proof relies on the fact that if an error is contained in a cubic box of length $x$ and there are no further errors inside the enlarged box of size $\alpha x$ (for large enough some constant $\alpha$), then the RG decoder will correctly fix $\delta s$ generated by the error.
This fact still holds in our general setting.
\section{A fault-tolerant circuit for a non-abelian phase}
\label{sec:z23_twisted}
In this section we will show how our methods can be adapted to construct fault-tolerant circuits for a particular non-abelian phase, namely the one represented by the twisted quantum double with gauge group $\zz_2^3$ and a type-III group 3-cocycle as twist.
We will first sketch how this can be achieved by using defects of a different kind than 1-form symmetries.
Then we argue that circuits like this are already part of the protocols for non-Clifford CCZ gates in Refs.~\cite{Bombin2018,Brown2019}, even though the connection to non-abelian phases has not been spelled out previously, see also Ref.~\cite{Davydova2024}.
Finally, we propose a concrete microscopic implementation of the circuit with a lower overhead than the protocols in Refs.~\cite{Bombin2018,Brown2019}, similar to the ones we use for abelian phases in Sections~\ref{sec:double_semion} and \ref{sec:general_doubles}.

As a state sum, the twisted quantum double we consider is a sum over three $\zz_2$ 1-cocycles $A\coloneqq (A_0,A_1,A_2)$, with a Lagrangian given by
\begin{equation}
\label{eq:type3_action}
\lagr[A]=\frac12 A_0\cup A_1\cup A_2\;.
\end{equation}
This twisted quantum double is known be in the same phase as the untwisted quantum double for the non-abelian dihedral group $D_4$, and hosts non-abelian anyons \cite{Propitius1995}.
We can still equip the path integral with ``charge'' 1-form symmetries located at three $\zz_2$-valued 1-chains $c=(c_0,c_1,c_2)$, by adding the terms $\frac12 A_i c_i$ to the action.
However, the ``bare fluxes'' located at three 2-cochains $b=(b_0,b_1,b_2)$ such that $dA_i=b_i$ do \emph{not} define 1-form symmetries due to the non-trivial action in Eq.~\eqref{eq:type3_action}.
In contrast to abelian quantum doubles, this cannot be fixed by additional terms in the action like in Section~\ref{sec:general_1form}, because the flux-type anyons we would like to place along $b$ are non-abelian in this model.
\footnote{
While it is possible to equip the path integral with non-abelian flux-type anyon worldlines invariant under topological deformation, their non-abelian nature prevents us from putting them on arbitrary 2-cocycles invariant under homological deformation.
The latter is necessary for error correction unless we are happy to deal with a much more complex protocol and decoder \cite{Schotte2020,Schotte2022}.
}
Since the path integral cannot be equipped with a ``complete'' set of 1-form symmetries, we cannot turn it into a circuit using the methods in Section~\eqref{sec:path_integral_codes}.
Instead, we will use other sorts of defects and other kinds of classical decoding.
Namely, we introduce three additional defects, forming three $\zz_2$ 1-chains $e=(e_0,e_1,e_2)$, such that $de_i=b_i$.
The overall action after adding the $e$ defects is
\begin{equation}
\label{eq:type3_defect_action}
\begin{multlined}
S[A,b,c,e] = \frac12 (A_0+e_0)\cup (A_1+e_1)\cup (A_2+e_2)\\
+ \frac12 (A_0 c_0+A_1c_1+A_2c_2)
\;.
\end{multlined}
\end{equation}
So the $e$ defects are a fairly trivial thing:
We just add $e$ to $A$.
\footnote{
Interestingly, $e$ defines a non-trivial topological domain wall when classified through a subgroup $H\subset G\times G$, and a group 2-cocycle in $Z^2(BH,U(1))$.
However, this domain wall appears to be in a trivial phase relative to the bulk, with $b$ defining the invertible twist defect that interfaces it with the trivial domain wall in an invertible manner.
}
Even though the defects are not purely 1-form symmetries, this path integral still has properties very similar to the ones in Definition~\ref{def:homological_integral}:
Similar to the first condition, the path integral evaluates to zero if $\delta c\neq 0$ or $db\neq 0$ at any vertex or volume.
Note that the path integral is not necessarily zero if $de\neq b$ though.
Similar to the second condition, the path integral is invariant under $c'=c+\delta \gamma$ for some 2-cell $\gamma$.
It is also invariant under $b'=b+d\beta$ together with $e'=e+\beta$ for some 1-cell $\beta$.
Similar to the third condition, on any patch of topology $S_2\times [0,1]$ and defects satisfying $db=0$, $\delta c=0$, $de=b$, and $c$ homologically trivial, the path integral is a rank-1 matrix.
Similar to the fourth condition, the path integral evaluates to zero on $M'$ with $\partial M'=m_0\sqcup m_1$ if $\partial c$ has non-trivial homology, or $\partial b$ has non-trivial cohomology.

Let us now describe how to turn the path integral into a circuit using the defects $b$, $e$, and $c$.
For each of the three $\zz_2$ components, we can use some toric-code phase circuit with $Z$ measurements corresponding to $b$ defects and $X$ measurements corresponding to $c$ defects.
The action in Eq.~\eqref{eq:type3_action} can be implemented by inserting $CCZ$ gates into the circuit.
At the end of this section we will describe how to do this explicitly for a circuit layout similar to Sections~\ref{sec:double_semion_code} and \ref{sec:codes_from_arbitrary_1form} 
Now, the $e$ defects correspond to classical controls rather than measurement results.
\footnote{
We may thus call these controlled operations ``corrections'', but it might be more appropriate to think of them as ``keeping the $b$ syndrome correctable'' rather than correcting it.
}
There are two ways in which the $e$ defects can be implemented as classically controlled unitaries in the circuit.
The first possibility is the following:
If a qubit at an edge is involved in a $CCZ$ gate that implements the action in Eq.~\eqref{eq:type3_action}, then $e_i=1$ on this edge can be implemented as an $X$ gate acting on the $A_i$ qubit before and after the $CCZ$ gate.
This will effectively add the value of $e$ to that of $A$ and transform the action in Eq.~\eqref{eq:type3_action} into that of Eq.~\eqref{eq:type3_defect_action}.
The second possibility is to decompose the action into individual terms like $\frac12 e_0\cup A_1\cup A_2$.
$e_0=1$ on an edge can then be implemented as a $CZ$ gate acting on an $A_1$ and an $A_2$-qubit at different edges.
We might also need to insert $b$ and $c$ defects as controlled unitaries in order to close these off near interfaces with other phases, as discussed for the abelian case in Sections~\ref{sec:dynamical_codes_from_integrals} and \ref{eq:semion_decoding}.
This can be done by $Z$ operators for $c$ and $X$ operators for $b$, similar to Section~\ref{eq:semion_decoding}.

Last, let us describe the classical decoder for the above circuit.
This classical decoder has two separate tasks.
The first task is to choose how to close off $b$ and $c$ near interfaces to other circuits.
To this end, we perform minimum-weight matching on $b$ and $c$, and then close them off in a (co-)homologically trivial way.
The second task is to choose $e_i$ along the way.
In the absence of noise, we simply need to choose any $e_i$ such that $de_i=b_i$.
However, if there is noise, then we need to first find a minimum-weight fix $\mwpm(db_i)$ for $b_i$ in spacetime and then choose $de_i=b_i+\mwpm(db_i)$.
Unfortunately, at the moment when we need to make a choice for the control $e$ at some edge, we do not have access to the full spacetime history of $b$ measurement outcomes but only these from the past.
Even though we cannot exactly know $\mwpm(db_i)$ at a moment in time without having access to future $db_i$, it is possible to make an estimate for $\mwpm(db_i)$ that still leads to a fault-tolerant threshold.
To this end, Ref.~\cite{Bombin2018} proposes a method called \emph{just-in-time decoding}, see also Refs.~\cite{Brown2019,Scruby2020}.

Since they do not purely rely on 1-form symmetries, the circuits described above are not covered by the fault-tolerance proof in Section~\ref{sec:fault_tolerance}.
Since the path integral with defects obeys conditions very similar to Definition~\ref{def:homological_integral}, one might be able to use similar proof techniques.
However, replacing minimum-weight matching by the just-in-time version thereof will require significant modifications.
To further support our proposed circuits, we will instead show that they are implicitly used by existing protocols in Refs.~\cite{Bombin2018,Brown2019}, for which fault tolerance was proven.
Of course, this is how we found the construction above in the first place.

Let us start by describing the protocols in Refs.~\cite{Bombin2018,Brown2019}.
Both protocols take existing $3+0$-dimensional protocols for non-Clifford gates, and turn them into $2+1$-dimensional circuits by applying them sequentially.
The $3+0$ dimensional protocol consists of (1) code switching from a $2+1$-dimensional code to a $3+1$-dimensional code, (2) applying a transversal logical non-Clifford gate on the $3+1$-dimensional code, and (3) switching back to a $2+1$-dimensional code.
In Ref.~\cite{Brown2019}, we switch from (some copies of) the $2+1$-dimensional toric code to three copies of the $3+1$-dimensional toric code, which allows for a transversal logical $CCZ$ gate \cite{Vasmer2018}.
In Ref.~\cite{Bombin2018}, we switch from a $2+1$-dimensional color code to a $3+1$-dimensional color code, which allows for a transversal $T$ gate \cite{Bombin2006a,Bombin2018a}.
Apart form the global layout, the two protocols are equivalent in the same way as the $3+1$-dimensional color code is equivalent to three copies of the toric code \cite{Kubica2015}.
We will thus on focus on Ref.~\cite{Brown2019} from now on since its microscopics are closer to twisted quantum doubles.
In order to switch into the $3+1$-dimensional toric code, we prepare its ground state in the $3$-dimensional bulk, by preparing each qubit in a $\ket+$ state and performing the $Z$-type plaquette measurements.
In the absence of noise, this yields three 2-cochains $b=(b_0,b_1,b_2)$, one for each of the three copies.
Before applying the $CCZ$ gate, we need to correct the outcome of these $Z$ measurements by applying $X$ operators everywhere at 1-cochains $e=(e_0,e_1,e_2)$ with $de_i=b_i$.
If there is noise, we need to fix $b$ by minimum-weight perfect matching (or another suitable decoder), before choosing $e$ where we apply $X$.
When switching back from the $3+1$-dimensional toric code to the $2$-dimensional toric code, we perform $X$ measurements on all qubits in the $3$-dimensional bulk.
The results form three 1-chains $c=(c_0,c_1,c_2)$ on the 3-dimensional cellulation.
To obtain the corrections for the $2+1$-dimensional toric code on the boundary, we need to perform minimum-weight matching of $c$, and close off the fixed $c$ inside this boundary.
We also need to close off $b$ for which we already performed matching.

Since the $3+1$-dimensional toric code only exists for a constant amount of time, the protocol is $3+0$-dimensional rather than $3+1$-dimensional.
Without the transversal non-Clifford $CCZ$ gate, we also do not have to perform the $X$ corrections along $e$.
Then, the $3+0$-dimensional part of the protocol is equivalent to three copies of measurement-based topological quantum computation as in Ref.~\cite{Raussendorf2007}.
This measurement-based protocol can be applied sequentially as a $2+1$-dimensional circuit, as also discussed in Section~\ref{sec:measurement_based}.
However, including the $CCZ$ gates along with the required $X$ corrections poses an obstacle to this:
At the moment where we have to apply $X$ and choose $e$, we only know the part of $b$ that lies in the past.
This is the reason why just-in-time decoding was developed in Ref.~\cite{Bombin2018}.

We hope that by now the reader is starting to see the equivalence of the protocol in Ref.~\cite{Bombin2018} and our construction above, but let us motivate this more explicitly.
For a fixed configuration $b,c,e$ of classical measurement outcomes and controls, the circuit equals the path integral with that same defect configuration.
As discussed in Section~\ref{sec:measurement_floquet}, $3+0$-dimensional measurement-based quantum computation corresponds to the toric-code path integral with $b$ and $c$ defects.
The transversal $CCZ$ gates in the protocol become additional weights in the path integral, and these weights implement the action in Eq.~\eqref{eq:type3_action}.
\footnote{
More precisely, the three $3+1$-dimensional toric code states of Ref.~\cite{Vasmer2018} used in Ref.~\cite{Brown2019} are defined on three different superimposed $3$-cellulations.
The $CCZ$ gates act on triples of qubits associated to the same place in the $3$-dimensional space, but to edges of three different cellulations.
One could equivalently use three toric codes on the same cellulation, and $CCZ$ gates acting on qubits on different edges.
The phases corresponding to these $CCZ$ gates are then given by $e^{2\pi i\mathcal L[A]}$ in Eq.~\eqref{eq:type3_action}, where $A$ is the qubit configuration of the three $3+1$-dimensional toric codes.
}
The $X$ operators controlled by $e$ are applied before the $CCZ$ gates, which implements adding $e$ to $A$ in Eq.~\eqref{eq:type3_defect_action}.
Note that we do not have to apply $X$ operators after the $CCZ$ gates, since we perform $X$ measurements on all qubits right after this.

As one can see, Ref.~\cite{Brown2019} corresponds to implementing our construction in a measurement-based style as described in Section~\ref{sec:measurement_based}.
This style makes it particularly easy to implement terms in the action as controlled-phase gates.
As we have demonstrated, however, our framework offers great flexibility and allows to freely move between a stabilizer-circuit approach and a measurement-based approach.
Therefore, our approach has potential for developing stabilizer-circuit-type realizations of fault-tolerant non-Clifford gates, which could greatly reduce the large overhead of the protocols proposed in Ref.~\cite{Bombin2018,Brown2019}.
To support this suggestion, let us now show explicitly how the action in Eq.~\eqref{eq:type3_action} can be inserted into three copies of the stabilizer syndrome-extraction circuit for the toric code.
This yields a circuit defined on the same qubits as three toric codes, and with $6$ $CCZ$ gates in addition to the $3\cdot 8$ $CX$ gates and $3\cdot 2$ measurements of the three toric codes per spacetime unit cell.

We start from the untwisted toric-code syndrome-extraction circuit that we also used as base for our abelian fault-tolerant circuits in Sections~\ref{sec:double_semion_code} and \ref{sec:codes_from_arbitrary_1form}.
That is, we put the path integral on the lattice shown in Eq.~\eqref{eq:semion_spacetime_cellulation}, and label the variables as in Eq.~\eqref{eq:sheared_cubic_labels}.
The variables are valued in $G=\zz_2^3$, and we will denote their $A_0$, $A_1$, and $A_2$ component by an additional subscript $0$, $1$, or $2$ in front.
To express the Lagrangian $\frac12 A_0\cup A_1\cup A_2$ in Eq.~\eqref{eq:type3_action} in terms of the microscopic state-sum variables, we combine the cup-product formulas Eqs.~\eqref{eq:cubic_cup_110} and \eqref{eq:cup120} from Appendix~\ref{sec:cup_product}.
Note that due to the $\frac12$ prefactor, the signs do not matter, and we have
\begin{equation}
\label{eq:type3_weights}
\begin{multlined}
\frac12\Big(A_{0t}^{100}(A_{1dx}^{000}A_{2y}^{001}+A_{1y}^{100}A_{2dx}^{010})\\
+A_{0dx}^{001}(A_{1y}^{002}A_{2t}^{011}+A_{1t}^{001}A_{2y}^{001})\\
+ A_{0y}^{101}(A_{1dx}^{011}A_{2t}^{011}+A_{1t}^{110}A_{2dx}^{010})\Big)
\;.
\end{multlined}
\end{equation}

For the quantum circuit, we use the qudits as in Eq.~\eqref{eq:semion_code_qubits}, and the stages as in Eq.~\eqref{eq:semion_code_stages}.
We split each $\zz_2^3$ qudit into three qubits with an additional $0$, $1$, or $2$ subscript in front.
We note that an action term of the form $\frac12 abc$ gives rise to a weight $(-1)^{abc}$, which can be implemented by a $CCZ$ gate acting on three qubits representing the variables $a$, $b$, and $c$ simultaneously at a stage.
The weights in Eq.~\eqref{eq:type3_weights} can thus be implemented as the following gates,
\begin{equation}
\label{eq:type2_ccz_gates}
\begin{tabular}{l|l|l}
Weight & Gate & Stage\\
\hline
$(-1)^{A_{0t}^{100}A_{1dx}^{000}A_{2y}^{001}}$ & $CCZ[c_0^{10},A_{1x}^{00},A_{2y}^{00}]$ & 3\tcr{,4,5}\\
$(-1)^{A_{0t}^{100}A_{1y}^{100}A_{2dx}^{010}}$ & $CCZ[c_0^{10},A_{1y}^{10},A_{2x}^{01}]$ & 1\\
$(-1)^{A_{0dx}^{001}A_{1y}^{002}A_{2t}^{011}}$ & $CCZ[A_{0x}^{00},A_{1y}^{00},c_2^{01}]$ & $3_1$\tcr{,$4_1$,$5_1$}\\
$(-1)^{A_{0dx}^{001}A_{1t}^{001}A_{2y}^{001}}$ & $CCZ[A_{0x}^{00},c_1^{00},A_{2y}^{00}]$ & $1_1$\\
$(-1)^{A_{0y}^{101}A_{1dx}^{011}A_{2t}^{011}}$ & $CCZ[A_{0y}^{10},A_{1x}^{01},c_2^{01}]$ & $1_1$\\
$(-1)^{A_{0y}^{101}A_{1t}^{110}A_{2dx}^{010}}$ & $CCZ[A_{0y}^{10},c_1^{11},A_{2x}^{01}]$ & 3\tcr{,4,5}
\end{tabular}
\;.
\end{equation}
As usual, there are sometimes multiple options for stages when the gate can be applied.
We have chosen to apply every gate either at stage $1$ or at stage $3$.
Neatly, the three gates in each of these two stages act on non-overlapping triples of qubits, so we can apply them all in parallel.
The action in Eq.~\eqref{eq:type3_action} can thus be implemented by only two additional layers of gates compared to the stabilizer-toric-code syndrome-extraction circuit.
The following drawing illustrates the different gate layers of the circuit:
\begin{equation}
\begin{tikzpicture}
\node (0) at (0,0){
\begin{tikzpicture}
\clip (-0.3,-0.3)rectangle(1.8,1.8);
\draw[orange] (-1,0)--++(4,0) (-1,1.5)--++(4,0) (0,-1)--++(0,4) (1.5,-1)--++(0,4);
\foreach \x/\y in {0/0, 2/0, 4/0, 1/1, 3/1, 0/2, 2/2, 4/2, 1/3, 3/3, 0/4, 2/4, 4/4}{
\atoms{vertex}{\x\y/p={0.75*\x-0.75,0.75*\y-0.75}}
}
\foreach \x/\y in {1/0, 3/0, 0/1, 2/1, 4/1, 1/2, 3/2, 0/3, 2/3, 4/3, 1/4, 3/4}{
\atoms{vertex,astyle=purple}{\x\y/p={0.75*\x-0.75,0.75*\y-0.75}}
}
\draw[line width=0.3cm,cyan,opacity=0.5,line cap=round] (22-c)--(32-c) (11-c)--(21-c) (13-c)--(23-c) (31-c)--(41-c) (33-c)--(43-c) (12-c)--(02-c);
\path (22-c)--node[midway,yshift=0.2cm]{$\scriptstyle{CX^{\otimes3}}$} (32-c);
\path (11-c)--node[midway,yshift=0.2cm]{$\scriptstyle{CX^{\otimes3}}$} (21-c);
\path (13-c)--node[midway,yshift=0.2cm]{$\scriptstyle{CX^{\otimes3}}$} (23-c);
\end{tikzpicture}
};
\node (1) at (2.8,0){
\begin{tikzpicture}
\clip (-0.3,-0.3)rectangle(1.8,1.8);
\draw[orange] (-1,0)--++(4,0) (-1,1.5)--++(4,0) (0,-1)--++(0,4) (1.5,-1)--++(0,4);
\foreach \x/\y in {0/0, 2/0, 4/0, 1/1, 3/1, 0/2, 2/2, 4/2, 1/3, 3/3, 0/4, 2/4, 4/4}{
\atoms{vertex}{\x\y/p={0.75*\x-0.75,0.75*\y-0.75}}
}
\foreach \x/\y in {1/0, 3/0, 0/1, 2/1, 4/1, 1/2, 3/2, 0/3, 2/3, 4/3, 1/4, 3/4}{
\atoms{vertex,astyle=purple}{\x\y/p={0.75*\x-0.75,0.75*\y-0.75}}
}
\draw[line width=0.2cm,cyan,opacity=0.5,line cap=round] ($(22)!0.5!(31)$)--(31-c) ($(22)!0.5!(31)$)--(32-c) ($(22)!0.5!(31)$)--(23-c);
\path[red] ($(22)!0.5!(31)$)--node[pos=0.65]{$\scriptstyle0$}(31-c) ($(22)!0.5!(31)$)--node[pos=0.65]{$\scriptstyle1$}(32-c) ($(22)!0.5!(31)$)--node[pos=0.85]{$\scriptstyle2$}(23-c);
\path[red] ($(22)!0.5!(31)$) node[]{$\scriptstyle{CCZ}$};
\draw[line width=0.2cm,cyan,opacity=0.5,line cap=round] ($(22)!0.5!(13)$)--(13-c) ($(22)!0.5!(13)$)--(23-c) ($(22)!0.5!(13)$)--(32-c);
\path[red] ($(22)!0.5!(13)$)--node[pos=0.65]{$\scriptstyle2$}(13-c) ($(22)!0.5!(13)$)--node[pos=0.65]{$\scriptstyle1$}(23-c) ($(22)!0.5!(13)$)--node[pos=0.85]{$\scriptstyle0$}(32-c);
\path[red] ($(22)!0.5!(13)$) node[]{$\scriptstyle{CCZ}$};
\draw[line width=0.2cm,cyan,opacity=0.5,line cap=round] ($(22)!0.5!(11)$)--(11-c) ($(22)!0.5!(11)$)--(12-c) ($(22)!0.5!(11)$)--(21-c);
\path[red] ($(22)!0.5!(11)$)--node[pos=0.65]{$\scriptstyle1$}(11-c) ($(22)!0.5!(11)$)--node[pos=0.65]{$\scriptstyle2$}(12-c) ($(22)!0.5!(11)$)--node[pos=0.65]{$\scriptstyle0$}(21-c);
\path[red] ($(22)!0.5!(11)$) node[]{$\scriptstyle{CCZ}$};
\draw[line width=0.2cm,cyan,opacity=0.5,line cap=round] (33-c)--(44-c) (31-c)--(42-c) (32-c)--(41-c) (13-c)--(24-c) (23-c)--(14-c) (12-c)--(01-c) (11-c)--(02-c) (33-c)--(24-c) (21-c)--++(0.375,-0.875) (33-c)--(42-c) (11-c)--(20-c) (21-c)--(10-c) (31-c)--(40-c) (12-c)--++(-0.875,0.375) (13-c)--(04-c);
\path[red] (33-c)--node[pos=0.175]{$\scriptstyle1$}(44-c) (31-c)--node[pos=0.175]{$\scriptstyle1$}(42-c) (32-c)--node[pos=0.175]{$\scriptstyle2$}(41-c) (13-c)--node[pos=0.175]{$\scriptstyle1$}(24-c) (23-c)--node[pos=0.175]{$\scriptstyle0$}(14-c) (12-c)--node[pos=0.175]{$\scriptstyle1$}(01-c) (11-c)--node[pos=0.175]{$\scriptstyle0$}(02-c) (33-c)--node[pos=0.175]{$\scriptstyle0$}(24-c) (21-c)--node[pos=0.15]{$\scriptstyle2$}++(0.375,-0.875) (33-c)--node[pos=0.175]{$\scriptstyle2$}(42-c) (11-c)--node[pos=0.175]{$\scriptstyle2$}(20-c) (21-c)--node[pos=0.175]{$\scriptstyle1$}(10-c) (31-c)--node[pos=0.175]{$\scriptstyle2$}(40-c) (12-c)--node[pos=0.15]{$\scriptstyle0$}++(-0.875,0.375) (13-c)--node[pos=0.175]{$\scriptstyle0$}(04-c);
\end{tikzpicture}
};
\node (2) at (5.6,0){
\begin{tikzpicture}
\clip (-0.3,-0.3)rectangle(1.8,1.8);
\draw[orange] (-1,0)--++(4,0) (-1,1.5)--++(4,0) (0,-1)--++(0,4) (1.5,-1)--++(0,4);
\foreach \x/\y in {0/0, 2/0, 4/0, 1/1, 3/1, 0/2, 2/2, 4/2, 1/3, 3/3, 0/4, 2/4, 4/4}{
\atoms{vertex}{\x\y/p={0.75*\x-0.75,0.75*\y-0.75}}
}
\foreach \x/\y in {1/0, 3/0, 0/1, 2/1, 4/1, 1/2, 3/2, 0/3, 2/3, 4/3, 1/4, 3/4}{
\atoms{vertex,astyle=purple}{\x\y/p={0.75*\x-0.75,0.75*\y-0.75}}
}
\draw[line width=0.3cm,cyan,opacity=0.5,line cap=round] (21-c)--(22-c) (23-c)--(24-c) (12-c)--(13-c) (32-c)--(33-c) (11-c)--(10-c) (31-c)--(30-c);
\path (21-c)--node[midway]{$\scriptstyle{CX^{\otimes3}}$} (22-c);
\path (12-c)--node[midway]{$\scriptstyle{CX^{\otimes3}}$} (13-c);
\path (32-c)--node[midway]{$\scriptstyle{CX^{\otimes3}}$} (33-c);
\end{tikzpicture}
};
\node(3) at (5.6,-2.8){
\begin{tikzpicture}
\clip (-0.3,-0.3)rectangle(1.8,1.8);
\draw[orange] (-1,0)--++(4,0) (-1,1.5)--++(4,0) (0,-1)--++(0,4) (1.5,-1)--++(0,4);
\foreach \x/\y in {0/0, 2/0, 4/0, 1/1, 3/1, 0/2, 2/2, 4/2, 1/3, 3/3, 0/4, 2/4, 4/4}{
\atoms{vertex}{\x\y/p={0.75*\x-0.75,0.75*\y-0.75}}
}
\foreach \x/\y in {1/0, 3/0, 0/1, 2/1, 4/1, 1/2, 3/2, 0/3, 2/3, 4/3, 1/4, 3/4}{
\atoms{vertex,astyle=purple}{\x\y/p={0.75*\x-0.75,0.75*\y-0.75}}
}
\draw[line width=0.3cm,cyan,opacity=0.5,line cap=round] (23-c)--(22-c) (21-c)--(20-c) (12-c)--(11-c) (32-c)--(31-c) (13-c)--(14-c) (33-c)--(34-c);
\path (22-c)--node[midway]{$\scriptstyle{CX^{\otimes3}}$} (23-c);
\path (11-c)--node[midway]{$\scriptstyle{CX^{\otimes3}}$} (12-c);
\path (31-c)--node[midway]{$\scriptstyle{CX^{\otimes3}}$} (32-c);
\end{tikzpicture}
};
\node (4) at (5.6,-5.6){
\begin{tikzpicture}
\clip (-0.3,-0.3)rectangle(1.8,1.8);
\draw[orange] (-1,0)--++(4,0) (-1,1.5)--++(4,0) (0,-1)--++(0,4) (1.5,-1)--++(0,4);
\foreach \x/\y in {0/0, 2/0, 4/0, 1/1, 3/1, 0/2, 2/2, 4/2, 1/3, 3/3, 0/4, 2/4, 4/4}{
\atoms{vertex}{\x\y/p={0.75*\x-0.75,0.75*\y-0.75}}
}
\foreach \x/\y in {1/0, 3/0, 0/1, 2/1, 4/1, 1/2, 3/2, 0/3, 2/3, 4/3, 1/4, 3/4}{
\atoms{vertex,astyle=purple}{\x\y/p={0.75*\x-0.75,0.75*\y-0.75}}
}
\draw[line width=0.2cm,cyan,opacity=0.5,line cap=round] ($(22)!0.5!(31)$)--(31-c) ($(22)!0.5!(31)$)--(21-c) ($(22)!0.5!(31)$)--(12-c);
\path[red] ($(22)!0.5!(31)$)--node[pos=0.65]{$\scriptstyle0$}(31-c) ($(22)!0.5!(31)$)--node[pos=0.65]{$\scriptstyle1$}(21-c) ($(22)!0.5!(31)$)--node[pos=0.85]{$\scriptstyle2$}(12-c);
\path[red] ($(22)!0.5!(31)$) node[]{$\scriptstyle{CCZ}$};
\draw[line width=0.2cm,cyan,opacity=0.5,line cap=round] ($(22)!0.5!(13)$)--(13-c) ($(22)!0.5!(13)$)--(12-c) ($(22)!0.5!(13)$)--(21-c);
\path[red] ($(22)!0.5!(13)$)--node[pos=0.65]{$\scriptstyle2$}(13-c) ($(22)!0.5!(13)$)--node[pos=0.65]{$\scriptstyle1$}(12-c) ($(22)!0.5!(13)$)--node[pos=0.85]{$\scriptstyle0$}(21-c);
\path[red] ($(22)!0.5!(13)$) node[]{$\scriptstyle{CCZ}$};
\draw[line width=0.2cm,cyan,opacity=0.5,line cap=round] ($(22)!0.5!(33)$)--(33-c) ($(22)!0.5!(33)$)--(23-c) ($(22)!0.5!(33)$)--(32-c);
\path[red] ($(22)!0.5!(33)$)--node[pos=0.65]{$\scriptstyle1$}(33-c) ($(22)!0.5!(33)$)--node[pos=0.65]{$\scriptstyle2$}(23-c) ($(22)!0.5!(33)$)--node[pos=0.65]{$\scriptstyle0$}(32-c);
\path[red] ($(22)!0.5!(33)$) node[]{$\scriptstyle{CCZ}$};
\draw[line width=0.2cm,cyan,opacity=0.5,line cap=round] (13-c)--(02-c) (12-c)--(03-c) (21-c)--(30-c) (31-c)--(20-c) (11-c)--(00-c) (33-c)--(42-c) (32-c)--(43-c) (11-c)--(20-c) (23-c)--++(-0.375,0.875) (33-c)--(24-c) (23-c)--(34-c) (11-c)--(02-c) (13-c)--(04-c) (32-c)--++(0.875,-0.375) (31-c)--(40-c);
\path[red] (13-c)--node[pos=0.175]{$\scriptstyle1$}(02-c) (12-c)--node[pos=0.175]{$\scriptstyle0$}(03-c) (21-c)--node[pos=0.175]{$\scriptstyle2$}(30-c) (31-c)--node[pos=0.175]{$\scriptstyle1$}(20-c) (11-c)--node[pos=0.175]{$\scriptstyle1$}(00-c) (33-c)--node[pos=0.175]{$\scriptstyle2$}(42-c) (32-c)--node[pos=0.175]{$\scriptstyle1$}(43-c) (11-c)--node[pos=0.175]{$\scriptstyle2$}(20-c) (23-c)--node[pos=0.15]{$\scriptstyle0$}++(-0.375,0.875) (33-c)--node[pos=0.175]{$\scriptstyle0$}(24-c) (23-c)--node[pos=0.175]{$\scriptstyle1$}(34-c) (11-c)--node[pos=0.175]{$\scriptstyle0$}(02-c) (13-c)--node[pos=0.175]{$\scriptstyle0$}(04-c) (32-c)--node[pos=0.15]{$\scriptstyle2$}++(0.875,-0.375) (31-c)--node[pos=0.175]{$\scriptstyle2$}(40-c);
\end{tikzpicture}
};
\node (5) at (2.8,-5.6){
\begin{tikzpicture}
\clip (-0.3,-0.3)rectangle(1.8,1.8);
\draw[orange] (-1,0)--++(4,0) (-1,1.5)--++(4,0) (0,-1)--++(0,4) (1.5,-1)--++(0,4);
\foreach \x/\y in {0/0, 2/0, 4/0, 1/1, 3/1, 0/2, 2/2, 4/2, 1/3, 3/3, 0/4, 2/4, 4/4}{
\atoms{vertex}{\x\y/p={0.75*\x-0.75,0.75*\y-0.75}}
}
\foreach \x/\y in {1/0, 3/0, 0/1, 2/1, 4/1, 1/2, 3/2, 0/3, 2/3, 4/3, 1/4, 3/4}{
\atoms{vertex,astyle=purple}{\x\y/p={0.75*\x-0.75,0.75*\y-0.75}}
}
\draw[line width=0.3cm,cyan,opacity=0.5,line cap=round] (23-c)--(33-c) (12-c)--(22-c) (21-c)--(31-c) (11-c)--(01-c) (32-c)--(42-c) (13-c)--(03-c);
\path (23-c)--node[midway,yshift=0.2cm]{$\scriptstyle{CX^{\otimes3}}$} (33-c);
\path (12-c)--node[midway,yshift=0.2cm]{$\scriptstyle{CX^{\otimes3}}$} (22-c);
\path (21-c)--node[midway,yshift=0.2cm]{$\scriptstyle{CX^{\otimes3}}$} (31-c);
\end{tikzpicture}
};
\node(6) at (0,-5.6){
\begin{tikzpicture}
\clip (-0.3,-0.3)rectangle(1.8,1.8);
\draw[orange] (-1,0)--++(4,0) (-1,1.5)--++(4,0) (0,-1)--++(0,4) (1.5,-1)--++(0,4);
\foreach \x/\y in {0/0, 2/0, 4/0, 1/1, 3/1, 0/2, 2/2, 4/2, 1/3, 3/3, 0/4, 2/4, 4/4}{
\atoms{vertex}{\x\y/p={0.75*\x-0.75,0.75*\y-0.75}}
}
\foreach \x/\y in {1/0, 3/0, 0/1, 2/1, 4/1, 1/2, 3/2, 0/3, 2/3, 4/3, 1/4, 3/4}{
\atoms{vertex,astyle=purple}{\x\y/p={0.75*\x-0.75,0.75*\y-0.75}}
}
\fill[cyan,opacity=0.5] (11)circle(0.15) (31)circle(0.15) (13)circle(0.15) (33)circle(0.15) (22)circle(0.15);
\path (11-c)++(-90:0.2) node{$\scriptstyle{M_X^{\otimes3}}$};
\path (31-c)++(-90:0.2) node{$\scriptstyle{M_X^{\otimes3}}$};
\path (13-c)++(-90:0.2) node{$\scriptstyle{M_X^{\otimes3}}$};
\path (33-c)++(-90:0.2) node{$\scriptstyle{M_X^{\otimes3}}$};
\path (22-c)++(-90:0.2) node{$\scriptstyle{M_Z^{\otimes3}}$};
\end{tikzpicture}
};
\draw[very thick] (0.east)edge[->](1.west) (1.east)edge[->](2.west) (2.south)edge[->](3.north) (3.south)edge[->](4.north) (4.west)edge[->](5.east) (5.west)edge[->](6.east) (6.north)edge[->](0.south);
\end{tikzpicture}
\;.
\end{equation}
Here, each purple and black dot represents three qubits that can be labelled by $0$, $1$, and $2$.
The $CCZ$ gates act on three qubits of three different dots, whose numbers are indicated in red.

We can also implement the full action in Eq.~\eqref{eq:type3_defect_action}, where $e_0$, $e_1$, and $e_2$ correspond to classical controls of the circuit.
To this end, we apply a $cX$ gate before and after each $CCZ$ gate.
The classical control of the $cX$ gate is the value of $e$ on the corresponding edge.
That is, we need to introduce four additional rounds of $cX$ gates in Eq.~\eqref{eq:type2_ccz_gates}, before and after stage $1$ and $3$, respectively.

One aspect that remains to be verified is whether the alternative low-overhead circuit we propose is still compatible with just-in-time decoding.
We must make our choice of $e$ at an edge before we apply a $cX$ gate controlled by this edge.
This might happen at a moment that is too early for us to predict $e$ from the measurement results collected so far.
We believe that just-in-time decoding works as long as in the absence of noise, we end up with a valid choice of $e$ for which $de=b$.
This can be achieved for the proposed circuit, since it is of a stabilizer syndrome-extraction type where in the absence of noise the $b$ syndrome is constant in time.
However, showing that this is indeed the case would require a detailed understanding and adaption of the just-in-time fault-tolerance proofs provided in Refs.~\cite{Bombin2018,Brown2019}.

\section{Conclusion and outlook}
\label{sec:outlook}
In this paper, we have presented a method to construct fault-tolerant quantum-local circuits for non-chiral abelian topological phases, represented by abelian twisted-quantum-double or Dijkgraaf-Witten path integrals.
The circuits are based on these of the untwisted qudit toric code phase, such as syndrome extraction for the stabilizer toric code, measurement-based quantum computation, or Floquet codes.
As we have shown in Ref.~\cite{path_integral_qec}, these untwisted codes are associated with the same toric-code path integral that is put on different spacetime cellulations and traversed in different time directions.
The non-trivial twist can be added to any of these circuits by inserting 2-qudit controlled-phase gates.
This way, we can implement exotic abelian phases at a moderate overhead relative to the qudit toric code.

To arrive at our goal, we had to solve two major technical challenges.
The first challenge was to find an explicit expression for the complete set of projective 1-form symmetries, whose defects correspond to anyon worldlines.
While it is well-known how to introduce anyons along individual worldlines in the path integral via a technique called \emph{tube algebra} \cite{Evand1995,Ocneanu2001,Bullivant2019,Lan2013,tensor_lattice,Magdalena2023}, we needed to be able to introduce anyons along arbitrary cellular (co-)cycles.
This is simple for the charge-type anyons, which are the same as in the untwisted case, but the flux-type anyons required additional terms in the action containing higher-order cup products.
In Section~\ref{sec:general_1form}, we have reverse-engineered these terms by demanding gauge invariance even in the presence of 1-form symmetry defects.
In addition, we have found that the overall configurations of 1-form symmetry defects are intricate combinations of a flux 2-cocycle $b$ together with a charge 1-chain $c$, both valued in the gauge group $G$.
The boundary of $c$ is obtained from a cohomology operation of $b$, which is the Bockstein homomorphism for the central extension of $G$ by itself, whose overall group is the 1-form symmetry group $K$.
To make practical use of the 1-form symmetries, we have come up with a way to construct explicit formulas for higher order cup products on arbitrary cellulations.

The second challenge was to turn the path integral into a circuit without requiring any auxiliary qudits compared to the untwisted versions.
This was simple in the case of measurement-based quantum computation in Section~\ref{sec:measurement_based} since all qudits are active at the same time representing all state-sum variables, and the twist weight can be implemented as a phase gate at this time.
However, for other architectures, the challenge was to have a stage in the circuit for every weight, where all state-sum variables involved in this weight are represented by qubits.
In Sections~\ref{sec:double_semion_code} and \ref{sec:codes_from_arbitrary_1form}, we have solved this problem for the stabilizer syndrome-extraction protocol by making appropriate choices for the orientations of cells and the local formulas for cup products.
For the Floquet architecture in Section~\ref{sec:floquet}, we have only shown how to solve this for a particular example, namely the type-II twisted $\zz_2\times \zz_2$ quantum double, which is in the same phase as the $\zz_4$ toric code.

In addition to constructing explicit fault-tolerant circuits, we have provided a general formalism for 1-form symmetric fixed-point path integrals and the corresponding fault-tolerant circuits which we have termed \emph{1-form symmetric fixed-point circuits} in Section~\ref{sec:dynamical_codes_from_integrals}.
We have proven fault tolerance of arbitrary 1-form symmetric fixed-point circuits under arbitrary incoherent local perturbations in Section~\ref{sec:fault_tolerance}.
To this end, we have used the basic properties of the path integral, namely zero correlation length, and the fact that symmetry defects must be closed and can be freely deformed.
Interestingly, the idea of explicitly ``detecting'' and ``correcting'' an error does neither appear in the general formalism nor in the fault-tolerance proof.
So our formalism provides a new perspective on topological error correction, where we focus on implementing a topological path integral, and errors are taken care of automatically in an implicit manner.
Further, our formalism is fully agnostic towards the Pauli basis, which is in striking contrast to much of the quantum-error-correction literature.

Finally, we have demonstrated that path integrals can also be used to construct fault-tolerant circuits for a specific non-abelian phase in Section~\ref{sec:z23_twisted}.
To this end, we have interpreted existing $2+1$-dimensional protocols for non-Clifford gates Refs.~\cite{Bombin2018,Brown2019} in terms of path integrals, which revealed that they represent a non-abelian phase.
We used the flexibility of our approach to propose an alternative low-overhead implementation of these protocols.

There are several interesting directions in which our method can be generalized or applied in other contexts.
The first direction is to enhance the circuits with boundaries and defects, as well as to go from mere information storage to fault-tolerant processing of logical information.
The geometric flexibility of the path-integral approach makes it particularly suited for this task.
For example, boundaries for the Dijkgraaf-Witten path integral are obtained by constraining the 1-cocycle $A$ to some subgroup $H\in G$ at the boundary.
All one needs to do is to extend the 1-form symmetries to the case of the boundary.
Thereby, of the overall 1-form symmetry group $K$ represented by the 2-cocycle $b$ and 1-chain $c$, only some subgroup of is allowed to terminate at the boundary, corresponding to the anyons that condense.

Another direction is to look at different microscopic representations of path integrals and the associated fixed-point circuits.
For example, in Ref.~\cite{liquid_intro}, we suggested a path-integral version of the color code.
It might be possible to equip this path integral, representing the $\zz_2\times \zz_2$ untwisted quantum double, with phase factors such that we end up with some twisted phase.
Working out the 1-form symmetries in this path integral would then lead to a dynamic twisted version of the color code.

A natural question is in how far our methods can be applied to non-abelian phases, in light of the fact that these seem to be necessary to achieve computational universality in $2+1$ dimensions.
While we have demonstrated that this is possible for a particular non-abelian phase in Section~\ref{sec:z23_twisted}, it is unclear how this transfers to other phases.
The generalization to other twisted quantum doubles with abelian gauge group $G$ (but non-abelian anyon theory) is straight-forward, but these are all just type-III twisted models and not much more interesting than the discussed $G=\zz_2^3$ case.
We note that the discussed type-III twisted quantum double can be obtained by gauging a $\zz_2$ symmetry in a $\zz_2\times \zz_2$ untwisted model, if we interpret $A_2$ in Eq.~\eqref{eq:type3_action} as symmetry defects and $(A_0,A_1)$ as the untwisted model.
This leaves hope that the methods might carry over to similar scenarios, for example the Ising string-net model, which is obtained by gauging the $\zz_2$ $e/m$ duality symmetry of the toric code.

\subsubsection*{Acknowledgments}
I would like to thank Julio Magdalena de la Fuente, Margarita Davydova, Ben Brown, and Tyler Ellison for stimulating conversations.
This work was supported by the DFG (CRC 183 project B01), the BMBF (RealistiQ, QSolid), the Munich Quantum Valley (K-8), the BMWK (PlanQK), the U. S. Army Research Laboratory and the U. S. Army Research Office under contract/grant number W911NF2310255, and by the U.S. Department of Energy, Office of Science, National Quantum Information Science Research Centers, and the Co-design Center for Quantum Advantage (C2QA) under contract number DE-SC0012704.

\bibliographystyle{quantum}
\bibliography{twisted_double_code_refs}{}

\appendix

\section{(Higher-order) cup products on arbitrary cellulations}
\label{sec:cup_product}
In this appendix, we define higher order cup products on arbitrary cellulations.
We recall the basic notions of cellular (co-)homology on a cellulation $M$ from Section~\ref{sec:cohomology}.
To define the higher order cup products, we introduce a new notion, namely ($\zz$-valued) \emph{$c$-bichains}.
A $c$-bichain is an $\zz$-valued function taking pairs of cells as argument, whose dimensions sum to $c$,
\begin{equation}
\bigcup_{a,b:a+b=c} S_a[M]\times S_b[M]\rightarrow \zz\;.
\end{equation}
The central operation on $c$-bichains is the $\zz$-linear \emph{boundary} map $\delta$ from the set of $c$-bichains to $c-1$-bichains, defined by its action on a basis element $(\alpha,\beta)\in S_a[M]\times S_b[M]$,
\begin{equation}
\delta(\alpha,\beta) =  (\delta \alpha, \beta) + (-1)^a (\alpha, \delta \beta)\;.
\end{equation}
It fulfills the common relation for a boundary map $\delta^2=0$,
\begin{equation}
\begin{gathered}
\delta^2(\alpha,\beta) = \delta\big((\delta \alpha,\beta)+(-1)^a (\alpha,\delta \beta)\big)\\
=(\delta^2 \alpha,\beta)+(-1)^{a-1} (\delta \alpha,\delta \beta)\\
+ (-1)^a (\delta \alpha,\delta \beta) + (-1)^{a+a} (\alpha,\delta^2 \beta) = 0\;.
\end{gathered}
\end{equation}
$c$-bichains $W$ with $\delta W=0$ are called \emph{$c$-bicycles}, and \emph{$c$-biboundaries} if $W=\delta V$ for some $c+1$-bichain $V$.
In fact, the homology of bichains and their boundary is directly equivalent to the cellular homology of $M\times M$.
Next, a \emph{$d$-dimensional $c$-bichain family} $X$ associates to every $d$-cell representative $\gamma$ a $c$-bichain $X[\gamma]$ on $\gamma$.
Thereby, $\gamma$ does not mean the $d-1$-dimensional boundary cellulation of $\gamma$, but also includes the $d$-cell $\gamma$ itself, whose boundary is given by the orientation,
\begin{equation}
\delta \gamma=(-1)^{\sigma[\gamma]}\;.
\end{equation}
In other words, we consider the homology of $\gamma$ as a $d$-cellulation with boundary.
A $d$-dimensional $c$-bichain family $X$ can be used to map an $a$-cocycle $A$ and a $b$-cocycle $B$ with $a+b=c$ to a $d$-cocycle $X(A,B)$,
\begin{equation}
X(A,B)(\gamma)\coloneqq \sum_{\alpha\in S_a[\gamma], \beta\in S_b[\gamma]} X[\gamma](\alpha, \beta) A(\alpha) B(\beta)\;,
\end{equation}
for $\gamma\in S_d[M]$.
The boundary map can also be applied to a bichain family, which means applying it for every representative.
With this, we have
\begin{equation}
\delta X(A,B) = X(dA,B) + (-1)^a X(A,dB)\;.
\end{equation}
Next, we define the \emph{transpose} $\bullet^T$ of bichains as a $\zz$-linear map by its action on basis elements,
\begin{equation}
(\alpha,\beta)^T=(-1)^{ab} (\beta,\alpha)\;.
\end{equation}
Taking the transpose commutes with taking the boundary,
\begin{equation}
\begin{gathered}
\delta (\alpha,\beta)^T = (-1)^{ab} \delta (\beta,\alpha)\\
= (-1)^{ab} ( (\delta \beta, \alpha)+(-1)^b (\beta, \delta \alpha))\\
=(-1)^{(a-1)b} (\beta,\delta \alpha) + (-1)^{a+a(b-1)} (\delta \beta, \alpha)\\
=((\delta \alpha, \beta) + (-1)^a (\alpha, \delta \beta))^T
= (\delta (\alpha,\beta))^T\;.
\end{gathered}
\end{equation}
We can also define a \emph{coboundary}, mapping a $d$-dimensional $c$-bichain family $X$ to a $d+1$-dimensional $c$-bichain family $dX$.
On a $d+1$-cell representative $\epsilon$, $dX$ is given by
\begin{equation}
\label{eq:bichainfamily_coboundary}
(dX)[\epsilon]\coloneqq \sum_{\gamma\in S_{d}[\epsilon]} (-1)^{\sigma[\epsilon](\gamma)} X[\gamma]\;.
\end{equation}
Note that this coboundary acts on bichain families and not bichains, and is not a Poincar\'e dual notion to the boundary defined above.
For analogue reasons to the standard boundary map of cochains, we find $d^2=0$.
It is also easy to see that $\delta d=d\delta$, and $(dX)^T=d(X^T)$.
For the according action on pairs of cochains, we find
\begin{equation}
(dX)(A,B) = d(X(A,B))\;.
\end{equation}

Equipped with these notions, we are now ready to define the higher-order cup products on arbitrary cellulations.
More precisely, we will define an $c-x$-dimensional $c$-bichain family $\cupsymb_x^c$, such that
\begin{equation}
\label{eq:cup_from_bichain}
A\cup_x B\coloneqq \cupsymb_x^c(A,B)\;,
\end{equation}
for every $a$-chain $A$ and $b$-chain $B$ with $a+b=c$.

$\cupsymb_x^c$ is defined inductively:
Given $\cupsymb_x^{c-1}$ and $\cupsymb_{x-1}^{c-1}$ as an induction hypothesis, $\cupsymb_x^c$ can be chosen arbitrarily such that
\begin{equation}
\label{eq:cup_any_cellulation_definition}
\delta \cupsymb_x^c
= d \cupsymb_x^{c-1} +(-1)^{c+x} \cupsymb_{x-1}^{c-1} + (-1)^{c} {\cupsymb_{x-1}^{c-1}}^T\;.
\end{equation}
The induction is terminated by the conventions
\begin{equation}
\label{eq:cup_product_induction_start}
\cupsymb_x^{x-1}=\varnothing
\;,\quad \cupsymb_{-1}^c[\gamma]= 0
\;,\quad \cupsymb_0^0[\text{pt}]= (\text{pt},\text{pt})
\;,
\end{equation}
where $\gamma$ is any $c+1$-cell representatives, and $\text{pt}$ denotes the point, which is the only 0-cell.
Any set of choices fulfilling Eq.~\eqref{eq:cup_any_cellulation_definition} yields a consistent definition of the cup product.

A solution to Eq.~\eqref{eq:cup_any_cellulation_definition} always exists since the right-hand side is a (bi-)boundary.
So see this, we first show that it is a bicycle,
\begin{equation}
\begin{gathered}
\delta\Big(d\cupsymb_x^{c-1}+(-1)^{c+x} \cupsymb_{x-1}^{c-1} +(-1)^{c} {\cupsymb_{x-1}^{c-1}}^T\Big)
\\
= d\Big(d\cupsymb_x^{c-2}+(-1)^{(c-1)+x} \cupsymb_{x-1}^{c-2}+ (-1)^{c-1} {\cupsymb_{x-1}^{c-2}}^T\Big)
\\
\begin{multlined}
+(-1)^{c+x}\Big(d \cupsymb_{x-1}^{c-2}+ (-1)^{(c-1)+(x-1)}\cupsymb_{x-2}^{c-2}\\
+(-1)^{c-1} {\cupsymb_{x-2}^{c-2}}^T\Big)
\end{multlined}
\\
\begin{multlined}
+ (-1)^{c}\Big(d \cupsymb_{x-1}^{c-2}+ (-1)^{(c-1)+(x-1)} \cupsymb_{x-2}^{c-2}\\
+ (-1)^{c-1} {\cupsymb_{x-2}^{c-2}}^T\big)^T
\end{multlined}
\\
=0
\;.
\end{gathered}
\end{equation}
Then we use that the homology of bicycles defined by $\delta$ is equal to the homology of $M\times M$ for a cellulation $M$.
Here the cellulation is the $c$-cell representative $\gamma$, whose topology is the $c$-ball $B_c$.
Since $B_c\times B_c$ has trivial homology (except for the $0$th degree), every (bi-)cycle is a (bi-)boundary.

Finally, we can use the recursive formula Eq.~\eqref{eq:cup_any_cellulation_definition} for the action of the cup product on an $a$-cochain $A$ and a $b$-cochain $B$ in Eq.~\eqref{eq:cup_from_bichain}.
This yields the familiar formula in Eq.~\eqref{eq:cup_product_defining_equation},
\begin{equation}
\begin{gathered}
d(A\cup_x B) = d(\cupsymb_x^c(A,B)) = (d\cupsymb_x^c)(A,B)\\
= \Big(\delta \cupsymb_x^{c+1} + (-1)^{c+x} \cupsymb_{x-1}^{c} + (-1)^{c} {\cupsymb_{x-1}^{c}}^T\Big)(A,B)\\
= \cupsymb_x^{c+1}(dA,B) + (-1)^a \cupsymb_x^{c+1}(A,dB)\\+ (-1)^{a+b+x} \cupsymb_{x-1}^c(A,B) + (-1)^{a+b+ab} \cupsymb_{x-1}^c(B,A)\\
= dA\cup_x B + (-1)^a A\cup_x dB\\+ (-1)^{a+b+x} A\cup_{x-1} B + (-1)^{a+b+ab} B\cup_{x-1} A\;,
\end{gathered}
\end{equation}
noting that $a+b=c$.

In constructing the bichains $\cupsymb$ defining the higher order cup product through Eq.~\eqref{eq:cup_any_cellulation_definition}, all we need to do is finding a bichain whose boundary is the given bicycle on the right-hand side.
Note that this involves finding the solution to a $\zz$-valued linear equation, which can computed efficiently.
Also note that when using cup products in the study of fixed-point models, the dimension of the cellulations, the number of different cell representatives, and the size of these cell representatives are small and do not scale.
Furthermore, as we have mentioned earlier, bichains on a cellulation $M$ are equivalent to chains on $M\times M$, so all we need to do is to find an ordinary chain with a given boundary.
Despite all this, it would be desirable to have a method to construct $\cupsymb$ by hand, that does not involve doubling the dimension of the cellulations.
This can be done as follows.
Assume we want to find a $c$-bichain $W$ on a $d$-cellulation with $d<c$, whose boundary is a fixed $c-1$-bichain $U$, 
Let $W^{c-i, i}$ denote the component of $W$ defined on $S_{c-i}\times S_i$, such that $\delta W=U$ becomes
\begin{equation}
\begin{multlined}
(\delta\otimes \idop) W^{c-i, i}+(-1)^{c-i-1} (\idop\otimes \delta) W^{c-i-1, i+1}\\= U^{c-i-1, i}\;.
\end{multlined}
\end{equation}
We can use this to compute $W^{c-i,i}$ inductively from $i=d$ to $i=c-d$ by
\begin{equation}
\begin{multlined}
(\delta\otimes \idop) W^{c-i, i}\\
= U^{c-i-1, i} + (-1)^{c-i} (\idop\otimes \delta) W^{c-i-1, i+1}\;,
\end{multlined}
\end{equation}
using the convention $W^{c-d-1, d+1}\coloneqq 0$ for the first step of the induction.
We solve this equation explicitly for every fixed $i$-cell $\lambda$ in the second component,
\begin{equation}
\label{eq:bichain_iterative_definition}
\begin{multlined}
\delta W^{c-i, i}(\bullet, \lambda)\\
= U^{c-i-1, i}(\bullet, \lambda) + (-1)^{c-i} W^{c-i-1, i+1}(\bullet, d\lambda)\;,
\end{multlined}
\end{equation}
where $\delta$ now denotes the boundary operator for ordinary cellular chains.
The following diagram illustrates the situation for $d=3$ and $c=4$,
\begin{equation}
\begin{tabular}{|c|c|c|c|}
\hline
$U^{03}$ & $W^{13}$ &&\\
\hline
& $U^{12}$ & $W^{22}$ &\\
\hline
&& $U^{21}$ & $W^{31}$\\
\hline
&&& $U^{30}$\\
\hline
\end{tabular}\;.
\end{equation}
Here we would successively compute $W^{13}$, then $W^{22}$, and then $W^{31}$.
We now apply this method to find $\cupsymb_x^c$ on some $c-x$-cell representative $\gamma$, whose component defined on $S_{c-i}[\gamma]\times S_i[\gamma]$ we denote by $\cupsymb_x^{c-i,i}$.
Plugging Eq.~\eqref{eq:cup_any_cellulation_definition} as $U$ into Eq.~\eqref{eq:bichain_iterative_definition}, and spelling out Eq.~\eqref{eq:bichainfamily_coboundary} yields
\begin{equation}
\label{eq:cup_product_induction}
\begin{multlined}
\delta \cupsymb_x^{c-i,i}(\bullet, \lambda)\\
=
\sum_{s\in S_{d-1}[\gamma]} (-1)^{\sigma[\gamma](s)} \cupsymb_x^{c-i,i}[s](\bullet, \lambda)\\
+ (-1)^{c-x}\cupsymb_{x-1}^{c-i-1,i}(\bullet, \lambda) + (-1)^{c+ic} \cupsymb_{x-1}^{i,c-i-1}(\lambda, \bullet)\\
+(-1)^{c-i} \cupsymb_x^{c-i-1,i+1}(\bullet, d\lambda)\;.
\end{multlined}
\end{equation}
Note that for the $\cup_0$ product, we have $d=c$.
We can still apply the method describe above, but we have to guess the 0-homology of
\begin{equation}
\label{eq:cup_product_induction_start0}
\cupsymb_0^{0,c}[\gamma](\bullet,\gamma)
\end{equation}
for $\lambda=\gamma$.
We find that choosing the generating homology class, where the sum over all vertices yields $1$, works.
In explicit constructions, we pick $\cupsymb_0^{0,c}[\gamma](\bullet,\gamma)$ to consist of a single vertex.

Let us explicitly construct higher order cup products for some examples of families of cellulations.
We will recover known formulas for cup products, but also stress that our method allows us to freely choose any compatible formula for higher cup products on any cellulation.
When constructing components $\cupsymb_x^{c-i,i}$ via the inductive formula in Eq.~\eqref{eq:cup_product_induction}, we will denote them as formal sums over pairs of cells $(\alpha,\beta)$, which we more compactly denote as $\alpha|\beta$.
We will also illustrate the cell pairs by marking them inside a drawing of the $c-x$-cell.
The first cell will be marked with a red cross for 0-cells, a red dotted line for 1-cells, red stripes for 2-cells, and a red dot pattern for 3-cells.
The second cell will be marked by a blue dot for 0-cells, a blue thick line for 1-cells, a light blue shading for 2-cells, and a stronger blue shading for 3-cells.
When constructing the cup product via Eq.~\eqref{eq:cup_product_induction}, we go through all possible choices of $\lambda$ which we mark in blue as described above.
We then consider the right-hand side of Eq.~\eqref{eq:cup_product_induction}.
We will also mark this right-hand side as a green crosses for 0-cells, and as green dotted lines for 1-cells (but not for 2-cells since this makes the drawings too cluttered).
We put a $-$ sign next to the corresponding cell if it has a $-1$ prefactor.
Then we mark our choice of $\cupsymb_x^{c-i,i}(\bullet, \lambda)$ on the left-hand side of Eq.~\eqref{eq:cup_product_induction} in red, as described above.
That is, the chain marked in red is chosen such that its boundary is the cycle marked in green (if drawn).
For cells $\lambda$ where the right-hand side of Eq.~\eqref{eq:cup_product_induction} is zero, we will choose $\cupsymb_x^{c-i,i}(\bullet, \lambda)=0$ and draw nothing.
Finally, for the special case of $x=0$ and $i=c$, we pick
\begin{equation}
\label{eq:0simplex_pick}
\cupsymb_0^{0,c}[\gamma](\bullet,\gamma)=v_\gamma\;,
\end{equation}
for some vertex $v_\gamma$ of the $c$-cell representative $\gamma$.

Let us start with the most restricted kind of cellulation, namely triangulations where all cells are simplices.
Each simplex can be identified with a canonical representative by equipping it with a \emph{branching structure}, that is, a direction for all the edges that is acyclic around each triangle.
This is equivalent to choosing an ordering of the vertices of a $d$-simplex by numbering them from $0$ to $d$.
Subsimplices can be labeled by ordered subsets of $0\ldots d$ corresponding to their vertices.
The orientation of the $d-1$-subsimplex labeled $0\ldots (i-1) (i+1)\ldots d$ is chosen to be
\begin{equation}
\sigma(0\ldots (i-1) (i+1)\ldots d)=i\mod 2\;.
\end{equation}

With this, let us inductively write down the formulas for $\cupsymb_x^c[\gamma]$ with $\gamma$ being the $c-x$-simplex, for a few values of $x$ and $c$.
We start with $\cupsymb_0$ on a 0-simplex.
Using the induction start in Eq.~\eqref{eq:cup_product_induction_start}, we have
\begin{equation}
\cupsymb^{00}_0 =
\begin{tikzpicture}
\atoms{vertex}{0/}
\path (0) pic{bichain2p} pic{bichain1p};
\end{tikzpicture}
= 0|0
\;.
\end{equation}

Next, we choose $\cupsymb_0$ on a 1-simplex.
Choosing the $0$-vertex for $v_\gamma$ in Eq.~\eqref{eq:0simplex_pick}, we get
\begin{equation}
\label{eq:cup010}
\cupsymb^{01}_0 =
\begin{tikzpicture}
\drawedge
\draw (0)edge[bichain2] (1);
\path (0) pic {bichain1p};
\end{tikzpicture}
= 0|01
\;.
\end{equation}
Next, we choose $\cupsymb_0^{10}$,
\begin{equation}
\label{eq:cup100}
\cupsymb^{10}_0 =
\begin{tikzpicture}
\drawedge
\draw (0)edge[bichain1] (1);
\path (1) pic{bichain2p};
\path (0) pic{bichainhelpp} (1) pic{bichainhelpp} (0) node[\bichaincolhelp, above]{$-$};
\end{tikzpicture}
= 01|1
\;.
\end{equation}
That is, for $\lambda$ (marked in blue) equal to the $1$-vertex, the right-hand side of Eq.~\eqref{eq:cup_product_induction} (marked in green) consists of both vertices, and the only choice for $\cupsymb^{10}_0(\bullet,\lambda)$ (in red) is the 1-simplex itself.
For $\lambda$ equal to the $0$-vertex, the right-hand side of Eq.~\eqref{eq:cup_product_induction} is zero, so we choose $\cupsymb^{10}_0(\bullet,\lambda)$ to be zero and do not draw anything.

Next is $\cupsymb_0$ on a 2-simplex.
Again, we choose $v_\gamma=0$ in Eq.~\eqref{eq:0simplex_pick},
\begin{equation}
\cupsymb^{02}_0 =
\begin{tikzpicture}
\drawtriangle
\fill[bichain2f] (0-c)--(1-c)--(2-c)--cycle;
\path (0) pic{bichain1p};
\end{tikzpicture}
=0|012
\;.
\end{equation}
Next, we find
\begin{equation}
\cupsymb^{11}_0 =
\begin{tikzpicture}
\drawtriangle
\draw[bichain2] (1)--(2);
\draw[bichain1] (0)--(1);
\path (0) pic{bichainhelpp} (1) pic{bichainhelpp} (0) node[\bichaincolhelp, left]{$-$};
\end{tikzpicture}
=01|12
\;.
\end{equation}
That is, for $\lambda$ (in blue) equal to the $12$ edge of the triangle, the right-hand side of Eq.~\eqref{eq:cup_product_induction} (in green) consists of the $0$ and $1$ vertices, and we choose $\cupsymb^{11}_0(\bullet,\lambda)$ (in red) equal to the $01$ edge.
We could also have chosen the formal sum $02-12$ instead.
For $\lambda$ equal to the $01$ or the $02$ edge, the right-hand side of Eq.~\eqref{eq:cup_product_induction} is zero.
Last, we have
\begin{equation}
\cupsymb^{20}_0 =
\begin{tikzpicture}
\drawtriangle
\draw[bichainhelp] (0)--(1)--(2) (2)--node[midway,shift=(-90:0.15),\bichaincolhelp]{$-$}(0);
\path (2) pic{bichain2p};
\fill[bichain1f] (0-c)--(1-c)--(2-c)--cycle;
\end{tikzpicture}
=012|2
\;.
\end{equation}
That is, for $\lambda$ (in blue) equal to the $2$ vertex, the right-hand side of Eq.~\eqref{eq:cup_product_induction} (in green) consists of all three edges, and the only choice for $\cupsymb^{20}_0(\bullet,\lambda)$ (in red) is the triangle $012$ itself.
For $\lambda$ equal to the $0$ or $1$ vertex, the right-hand side of Eq.~\eqref{eq:cup_product_induction} is zero.

Next, $\cupsymb_0$ on the 3-simplex is given by
\begin{equation}
\cupsymb^{03}_0 =
\begin{tikzpicture}
\drawtetrahedron
\fill[bichain2ff] (0-c)--(1-c)--(3-c)--cycle;
\path (0)pic{bichain1p};
\end{tikzpicture}
=0|0123
\;,
\end{equation}
\begin{equation}
\label{eq:cup120_tetra}
\cupsymb^{12}_0 =
\begin{tikzpicture}
\drawtetrahedron
\fill[bichain2f] (1-c)--(2-c)--(3-c)--cycle;
\draw[bichain1] (0)--(1);
\path (0)pic{bichainhelpp} node[\bichaincolhelp, left]{$-$} (1)pic{bichainhelpp};
\end{tikzpicture}
=01|123
\;,
\end{equation}
\begin{equation}
\cupsymb^{21}_0 =
\begin{tikzpicture}
\drawtetrahedron
\fill[bichain1f] (0-c)--(1-c)--(2-c)--cycle;
\draw[bichain2] (2)--(3);
\draw[bichainhelp] (0-c)--(1-c)--(2-c)--node[midway,shift=(-60:0.15),\bichaincolhelp]{$-$}(0-c);
\end{tikzpicture}
=012|23
\;,
\end{equation}
\begin{equation}
\cupsymb^{30}_0 =
\begin{tikzpicture}
\drawtetrahedron
\fill[bichain1ff] (0-c)--(1-c)--(3-c)--cycle;
\path (3)pic{bichain2p};
\end{tikzpicture}
=0123|3
\;,
\end{equation}
In the last picture, the right-hand side of Eq.~\eqref{eq:cup_product_induction} is not shown and consists of all four triangles.

Next, let us look at $\cupsymb_1$.
On the 1-simplex, we get
\begin{equation}
\cupsymb^{11}_1 =
\begin{tikzpicture}
\drawedge
\draw (0)edge[bichain2](1) (0)edge[bichain1](1);
\path (0) pic{bichainhelpp} (1) pic{bichainhelpp} (1) node[\bichaincolhelp, above]{$-$};
\end{tikzpicture}
=01|01
\;.
\end{equation}
The right-hand side of Eq.~\eqref{eq:cup_product_induction} is built from $\cupsymb_0$ and $\cupsymb_0^T$ on the 1-simplex, which we constructed in Eqs.~\eqref{eq:cup010} and \eqref{eq:cup100}.
Next, $\cupsymb_1$ on the 2-simplex is given by
\begin{equation}
\cupsymb^{12}_1 =
-
\begin{tikzpicture}
\drawtriangle
\fill[bichain2f] (0-c)--(1-c)--(2-c)--cycle;
\draw[bichain1] (0)--(2);
\path (0)pic{bichainhelpp} (2)pic{bichainhelpp} (2) node[\bichaincolhelp, below]{$-$};
\end{tikzpicture}
=-02|012
\;,
\end{equation}
\begin{equation}
\cupsymb^{21}_1 =
\begin{tikzpicture}
\drawtriangle
\fill[bichain1f] (0-c)--(1-c)--(2-c)--cycle;
\draw[bichainhelp] (0)--(1) (1)--(2) (0)edge[mark={slab=$-$,r}](2);
\draw[bichain2] (0)--(1);
\end{tikzpicture}
+
\begin{tikzpicture}
\drawtriangle
\fill[bichain1f] (0-c)--(1-c)--(2-c)--cycle;
\draw[bichainhelp] (0)--(1) (1)--(2) (0)edge[mark={slab=$-$,r}](2);
\draw[bichain2] (1)--(2);
\end{tikzpicture}
= 012|01 + 012|12
\;.
\end{equation}
On the 3-simplex, we get
\begin{equation}
\cupsymb^{13}_1 =
\begin{tikzpicture}
\drawtetrahedron
\fill[bichain2ff] (0-c)--(1-c)--(3-c)--cycle;
\draw[bichain1] (0)--(3);
\path (0)pic{bichainhelpp} (3)pic{bichainhelpp} (0) node[\bichaincolhelp, below]{$-$};
\end{tikzpicture}
= 03|0123
\;,
\end{equation}
\begin{equation}
\begin{multlined}
\cupsymb^{22}_1 =
-
\begin{tikzpicture}
\drawtetrahedron
\fill[bichain1f] (0-c)--(1-c)--(3-c)--cycle;
\fill[bichain2f] (1-c)--(2-c)--(3-c)--cycle;
\draw[bichainhelp] (0)edge[mark={slab=$-$}](1) (1)edge[mark={slab=$-$}](3) (0)--(3);
\end{tikzpicture}
+
\begin{tikzpicture}
\drawtetrahedron
\fill[bichain2f] (0-c)--(1-c)--(2-c)--cycle;
\draw[bichainhelp] (0)--(2) (2)--(3) (0)edge[mark={slab=$-$}](3);
\fill[bichain1f] (0-c)--(2-c)--(3-c)--cycle;
\end{tikzpicture}\\
= -013|123 + 023|012
\;,
\end{multlined}
\end{equation}
\begin{equation}
\begin{multlined}
\cupsymb^{13}_1 =
\begin{tikzpicture}
\drawtetrahedron
\fill[bichain1ff] (0-c)--(1-c)--(3-c)--cycle;
\draw[bichain2] (0)--(1);
\end{tikzpicture}
+
\begin{tikzpicture}
\drawtetrahedron
\fill[bichain1ff] (0-c)--(1-c)--(3-c)--cycle;
\draw[bichain2] (1)--(2);
\end{tikzpicture}
+
\begin{tikzpicture}
\drawtetrahedron
\fill[bichain1ff] (0-c)--(1-c)--(3-c)--cycle;
\draw[bichain2] (2)--(3);
\end{tikzpicture}
\\
= 0123|01 + 0123|12 + 0123|23
\;.
\end{multlined}
\end{equation}
In each of the last three pictures, the right-hand side of Eq.~\eqref{eq:cup_product_induction} is not shown and consists of all four triangles.

Finally, let us consider $\cupsymb_2$.
On the 2-simplex, we get
\begin{equation}
\cupsymb^{22}_2 =
\begin{tikzpicture}
\drawtriangle
\fill[bichain1f] (0-c)--(1-c)--(2-c)--cycle;
\fill[bichain2f] (0-c)--(1-c)--(2-c)--cycle;
\draw[bichainhelp] (0)--(1)--(2) (2)--node[midway,shift=(-90:0.15),\bichaincolhelp]{$-$}(0);
\end{tikzpicture}
=012|012
\;.
\end{equation}
On the 3-simplex, we have
\begin{equation}
\cupsymb^{23}_2 =
\begin{tikzpicture}
\drawtetrahedron
\fill[bichain2ff] (0-c)--(1-c)--(3-c)--cycle;
\fill[bichain1f] (0-c)--(1-c)--(2-c)--(3-c)--cycle;
\draw[bichainhelp] (0-c)--(1-c)--(2-c)--(3-c) (0)edge[mark={slab=$-$,r}](3);
\end{tikzpicture}
= 012|0123 + 023|0123
\;,
\end{equation}
\begin{equation}
\begin{multlined}
\cupsymb^{32}_2 =
\begin{tikzpicture}
\drawtetrahedron
\fill[bichain1ff] (0-c)--(1-c)--(3-c)--cycle;
\fill[bichain2f] (0-c)--(1-c)--(3-c)--cycle;
\end{tikzpicture}
+
\begin{tikzpicture}
\drawtetrahedron
\fill[bichain1ff] (0-c)--(1-c)--(3-c)--cycle;
\fill[bichain2f] (1-c)--(2-c)--(3-c)--cycle;
\end{tikzpicture}
\\
= 0123|013 + 0123|123
\;.
\end{multlined}
\end{equation}
In the original Ref.~\cite{Steenrod1947}, Steenrod defines higher order cup products on branching-structure triangulations via an explicit formula.
In our notation, this formula becomes
\begin{equation}
\begin{gathered}
\cupsymb_x^c = \sum_{\mathbf e:0<e_0<\ldots<e_x<e_{x+1}\coloneqq c-x} (-1)^{\sigma(\mathbf e)} s_0(\mathbf e)|s_1(\mathbf e)\;,\\
s_0(\mathbf e)\coloneqq [0e_0][e_1e_2]\ldots[e_{2 \left\lfloor\frac{x-1}{2}\right\rfloor+1}e_{2 \left\lfloor\frac{x-1}{2}\right\rfloor+2}]\;,\\
s_1(\mathbf e)\coloneqq [e_0e_1][e_2e_3]\ldots[e_{2 \left\lfloor\frac{x}{2}\right\rfloor}e_{2 \left\lfloor\frac{x}{2}\right\rfloor+1}]\;,
\end{gathered}
\end{equation}
where $[a,b]$ denotes the integer sequence $a(a+1)\ldots b$, and $\left\lfloor a\right\rfloor$ denotes the largest integer smaller than $a$.
In words, $s_0$ and $s_1$ are obtained by appending the intervals $[e_ie_{i+1}]$ to either $s_0$ or $s_1$ alternatingly until none are left.
$\sigma(\mathbf e)$ is the number of permutations of individual vertex-numbers needed to reorder the intervals in $s_0(\mathbf e)s_1(\mathbf e)$ as
\begin{equation}
[0e_0][e_0e_1]\ldots[e_xe_{x+1}]\;.
\end{equation}
We find that Steenrod's formulas agree with ours in all cases above.

As a next example, consider hypercubic lattices.
More generally, the resulting cup products can be applied to any cellulations where all cells are hypercubes.
Again we want to unambiguously identify each hypercube with its standard representative, which for triangulations was achieved by the branching structure.
Here, we choose an ``origin'' vertex and an ordering of the adjacent edges, numbering them from $0$ to $d-1$.
Using a coordinate system where the $i$th edge defines a basis vector $x_i$, the vertices of the hypercube can be labeled by bitstrings in $\{0,1\}^d$ corresponding to their coordinate.
Accordingly, the sub-hypercubes can be labeled by strings $\{0,1,x\}^d$, where $x$ at position $i$ means that $x_i$ is a spanning vector of the sub-hypercube.
A $d$-dimensional sub-hypercube is identified with the $d$-hypercube representative in the following way:
The $0^d$ vertex the $d$-hypercube is identified with the vertex obtained by setting all $x$s to $0$.
The ordering of basis vectors for the $d$-hypercube is identified with that of the containing hypercube, restricted to the sub-hypercube.
The orientation of the $d-1$-dimensional sub-hypercube $x^ijx^{d-i-1}$ for $j\in \{0,1\}$ is
\begin{equation}
\sigma(x^ijx^{d-i-1}) = i+1+j\mod 2\;.
\end{equation}

$0$ and $1$-dimensional hypercubes coincide with $0$-simplices and $1$-simplices, and we choose $\cupsymb$ for vertices and edges like in the case of triangulations.
For a square, we take the following ordering of basis vectors relative to the drawings,
\begin{equation}
\begin{tikzpicture}
\draw (0,0)edge[mark={arr,e}, mark={lab=$x_0$,a}]++(0:0.3) (0,0)edge[mark={arr,e}, mark={lab=$x_1$,a}]++(90:0.3);
\end{tikzpicture}
\qquad
\begin{tikzpicture}
\drawsquare
\end{tikzpicture}\;.
\end{equation}
Then $\cupsymb_0$ is given by
\begin{equation}
\label{eq:cubic_cup_020}
\cupsymb^{02}_0=
\begin{tikzpicture}
\drawsquare
\path (0)pic{bichain1p};
\fill[bichain2f] (0-c)--(1-c)--(3-c)--(2-c)--cycle;
\end{tikzpicture}
= 00|xx\;,
\end{equation}
picking $v_\gamma=00$ in Eq.~\eqref{eq:0simplex_pick}, and
\begin{equation}
\label{eq:cubic_cup_110}
\cupsymb^{11}_0=
\begin{tikzpicture}
\drawsquare
\draw[bichain1] (0)--(1);
\draw[bichain2] (1)--(3);
\path (0)pic{bichainhelpp} (1)pic{bichainhelpp} (0)node[\bichaincolhelp,below]{$-$};
\end{tikzpicture}
-
\begin{tikzpicture}
\drawsquare
\draw[bichain1] (0)--(2);
\draw[bichain2] (2)--(3);
\path (0)pic{bichainhelpp} (2)pic{bichainhelpp} (2)node[\bichaincolhelp,left]{$-$};
\end{tikzpicture}
= x0|1x - 0x|x1\;,
\end{equation}
\begin{equation}
\cupsymb^{20}_0=
\begin{tikzpicture}
\drawsquare
\path (3)pic{bichain2p};
\fill[bichain1f] (0-c)--(1-c)--(3-c)--(2-c)--cycle;
\draw[bichainhelp] (0)edge[mark={slab=$-$}](2) (2)edge[mark={slab=$-$}](3) (0)--(1)--(3);
\end{tikzpicture}
= xx|11\;.
\end{equation}

Next, for a cube, we take the following ordering of basis vectors relative to the drawings,
\begin{equation}
\label{eq:cup_product_cube_drawing}
\begin{tikzpicture}
\draw (0,0)edge[mark={arr,e}, mark={lab=$x_0$,a}]++(0:0.35) (0,0)edge[mark={arr,e}, mark={lab=$x_1$,a}]++(90:0.35) (0,0)edge[mark={arr,e}, mark={lab=$x_2$,a}]++(35:0.25);
\end{tikzpicture}
\qquad
\begin{tikzpicture}
\drawcube
\end{tikzpicture}\;.
\end{equation}
$\cupsymb_0$ is then given by
\begin{equation}
\cupsymb^{03}_0
=
\begin{tikzpicture}
\drawcube
\fill[bichain2ff] (000-c)--(100-c)--(101-c)--(111-c)--(011-c)--(010-c)--cycle;
\path (000)pic{bichain1p};
\end{tikzpicture}
=
000|xxx
\;,
\end{equation}
\begin{equation}
\label{eq:cup120}
\begin{gathered}
\cupsymb^{12}_0
=
\begin{tikzpicture}
\drawcube
\fill[bichain2f] (100-c)--(110-c)--(111-c)--(101-c)--cycle;
\draw[bichain1] (000)--(100);
\path (000)pic{bichainhelpp} (100)pic{bichainhelpp} (000)node[\bichaincolhelp,below]{$-$};
\end{tikzpicture}
-
\begin{tikzpicture}
\drawcube
\fill[bichain2f] (010-c)--(011-c)--(111-c)--(110-c)--cycle;
\draw[bichain1] (000)--(010);
\path (000)pic{bichainhelpp} (010)pic{bichainhelpp} (010)node[\bichaincolhelp,left]{$-$};
\end{tikzpicture}
+
\begin{tikzpicture}
\drawcube
\fill[bichain2f] (001-c)--(011-c)--(111-c)--(101-c)--cycle;
\draw[bichain1] (000)--(001);
\path (000)pic{bichainhelpp} (001)pic{bichainhelpp} (000)node[\bichaincolhelp,below]{$-$};
\end{tikzpicture}
\\
=
x00|1xx - 0x0|x1x + 00x|xx1\;,
\end{gathered}
\end{equation}
\begin{equation}
\begin{gathered}
\cupsymb^{21}_0
=
\begin{tikzpicture}
\drawcube
\draw[bichain2] (110)--(111);
\fill[bichain1f] (000-c)--(100-c)--(110-c)--(010-c)--cycle;
\draw[bichainhelp] (000)--(100)--(110) (000)edge[mark={slab=$-$}](010) (010)edge[mark={slab=$-$}](110);
\end{tikzpicture}
-
\begin{tikzpicture}
\drawcube
\draw[bichain2] (101)--(111);
\fill[bichain1f] (000-c)--(100-c)--(101-c)--(001-c)--cycle;
\draw[bichainhelp] (000)--(001)--(101) (000)edge[mark={slab=$-$,r}](100) (100)edge[mark={slab=$-$,r}](101);
\end{tikzpicture}
+
\begin{tikzpicture}
\drawcube
\draw[bichain2] (011)--(111);
\fill[bichain1f] (000-c)--(010-c)--(011-c)--(001-c)--cycle;
\draw[bichainhelp] (000)--(010)--(011) (000)edge[mark={slab=$-$,r,p=0.8}](001) (001)edge[mark={slab=$-$,r}](011);
\end{tikzpicture}
\\
=
xx0|11x - x0x|1x1 + 0xx|x11\;,
\end{gathered}
\end{equation}
\begin{equation}
\cupsymb^{30}_0
=
\begin{tikzpicture}
\drawcube
\fill[bichain1ff] (000-c)--(100-c)--(101-c)--(111-c)--(011-c)--(010-c)--cycle;
\path (111)pic{bichain2p};
\end{tikzpicture}
=
xxx|111\;.
\end{equation}
Next, let us look at $\cupsymb_1$.
On the square, we find
\begin{equation}
\cupsymb^{12}_1=
-
\begin{tikzpicture}
\drawsquare
\fill[bichain2f] (0-c)--(1-c)--(3-c)--(2-c)--cycle;
\draw[bichain1] (0)--(2)--(3);
\path (0)pic{bichainhelpp} (3)pic{bichainhelpp} (3)node[\bichaincolhelp,right]{$-$};
\end{tikzpicture}
= -0x|xx - x1|xx\;,
\end{equation}
\begin{equation}
\cupsymb^{21}_1=
\begin{tikzpicture}
\drawsquare
\draw[bichain2] (0)--(1);
\draw[bichainhelp] (0)--(1)--(3) (0)edge[mark={slab=$-$}](2) (2)edge[mark={slab=$-$}](3);
\fill[bichain1f] (0-c)--(1-c)--(3-c)--(2-c)--cycle;
\end{tikzpicture}
+
\begin{tikzpicture}
\drawsquare
\draw[bichain2] (1)--(3);
\draw[bichainhelp] (0)--(1)--(3) (0)edge[mark={slab=$-$}](2) (2)edge[mark={slab=$-$}](3);
\fill[bichain1f] (0-c)--(1-c)--(3-c)--(2-c)--cycle;
\end{tikzpicture}
= xx|x0 + xx|1x\;.
\end{equation}
On the cube, we have
\begin{equation}
\begin{multlined}
\cupsymb^{13}_1
=
\begin{tikzpicture}
\drawcube
\fill[bichain2ff] (000-c)--(100-c)--(101-c)--(111-c)--(011-c)--(010-c)--cycle;
\draw[bichain1] (000)--(001)--(011)--(111);
\path (000)pic{bichainhelpp} (111)pic{bichainhelpp} (000)node[\bichaincolhelp,left]{$-$};
\end{tikzpicture}
\\
=
00x|xxx + 0x1|xxx + x11|xxx\;,
\end{multlined}
\end{equation}
\begin{equation}
\label{eq:cup221}
\begin{gathered}
\cupsymb^{22}_1
=
-
\begin{tikzpicture}
\drawcube
\fill[bichain2f] (000-c)--(001-c)--(101-c)--(100-c)--cycle;
\fill[bichain1f] (001-c)--(101-c)--(111-c)--(011-c)--(001-c);
\draw[bichainhelp] (001)edge[mark={slab=$-$,r,p=0.55,sideoff=-0.05}](101) (101)edge[mark={slab=$-$,r}](111) (001)--(011) (011)--(111);
\end{tikzpicture}
-
\begin{tikzpicture}
\drawcube
\fill[bichain2f] (100-c)--(101-c)--(111-c)--(110-c)--cycle;
\fill[bichain1f] (000-c)--(001-c)--(011-c)--(111-c)--(101-c)--(100-c)--cycle;
\draw[bichainhelp] (000)--(001) (001)--(011) (011)--(111) (000)edge[mark={slab=$-$,r}](100) (100)edge[mark={slab=$-$,r}](101) (101)edge[mark={slab=$-$,r}](111);
\end{tikzpicture}
\\
+
\begin{tikzpicture}
\drawcube
\fill[bichain2f] (010-c)--(011-c)--(111-c)--(110-c)--cycle;
\fill[bichain1f] (000-c)--(001-c)--(011-c)--(010-c)--cycle;
\draw[bichainhelp] (000)edge[mark={slab=$-$,r,p=0.7,sideoff=-0.1}](001) (001)edge[mark={slab=$-$,r}](011) (000)--(010) (010)--(011);
\end{tikzpicture}
+
\begin{tikzpicture}
\drawcube
\fill[bichain2f] (000-c)--(100-c)--(110-c)--(010-c)--cycle;
\fill[bichain1f] (000-c)--(001-c)--(011-c)--(010-c)--cycle;
\fill[bichain1f] (010-c)--(011-c)--(111-c)--(110-c)--cycle;
\draw[bichainhelp] (000)edge[mark={slab=$-$,r,p=0.7,sideoff=-0.1}](001) (001)edge[mark={slab=$-$,r,p=0.4}](011) (011)edge[mark={slab=$-$}](111) (000)--(010) (010)--(110) (110)--(111);
\end{tikzpicture}
\\=
-xx1|x0x - x0x|1xx - xx1|1xx\\
+ 0xx|x1x + 0xx|xx0 + x1x|xx0\;,
\end{gathered}
\end{equation}
\begin{equation}
\label{eq:cup311}
\begin{gathered}
\cupsymb^{31}_1
=
\begin{tikzpicture}
\drawcube
\fill[bichain1ff] (000-c)--(100-c)--(101-c)--(111-c)--(011-c)--(010-c)--cycle;
\draw[bichain2] (110-c)--(111-c);
\end{tikzpicture}
+
\begin{tikzpicture}
\drawcube
\fill[bichain1ff] (000-c)--(100-c)--(101-c)--(111-c)--(011-c)--(010-c)--cycle;
\draw[bichain2] (100-c)--(110-c);
\end{tikzpicture}
+
\begin{tikzpicture}
\drawcube
\fill[bichain1ff] (000-c)--(100-c)--(101-c)--(111-c)--(011-c)--(010-c)--cycle;
\draw[bichain2] (000-c)--(100-c);
\end{tikzpicture}
\\=
xxx|11x + xxx|1x0 + xxx|x00
\;.
\end{gathered}
\end{equation}
Finally, let us look at $\cupsymb_2$.
On the square, we have
\begin{equation}
\cupsymb^{22}_2=
\begin{tikzpicture}
\drawsquare
\fill[bichain2f] (0-c)--(1-c)--(3-c)--(2-c)--cycle;
\fill[bichain1f] (0-c)--(1-c)--(3-c)--(2-c)--cycle;
\draw[bichainhelp] (0)--(1)--(3) (0)edge[mark={slab=$-$}](2) (2)edge[mark={slab=$-$}](3);
\end{tikzpicture}
= xx|xx\;.
\end{equation}
On a cube, we find
\begin{equation}
\cupsymb^{23}_2
=
\begin{tikzpicture}
\drawcube
\fill[bichain2ff] (000-c)--(100-c)--(101-c)--(111-c)--(011-c)--(010-c)--cycle;
\fill[bichain1f] (000-c)--(001-c)--(011-c)--(010-c)--cycle (000-c)--(100-c)--(110-c)--(111-c)--(011-c)--(010-c)--cycle;
\draw[bichainhelp] (000)edge[mark={slab=$-$}](001) (001)edge[mark={slab=$-$}](011) (011)edge[mark={slab=$-$}](111) (000)--(100)--(110)--(111);
\end{tikzpicture}
=
0xx|xxx + x1x|xxx + xx0|xxx\;,
\end{equation}
\begin{equation}
\begin{gathered}
\cupsymb^{32}_2
=
\begin{tikzpicture}
\drawcube
\fill[bichain2f] (000-c)--(100-c)--(101-c)--(001-c)--cycle;
\fill[bichain1ff] (000-c)--(100-c)--(101-c)--(111-c)--(011-c)--(010-c)--cycle;
\end{tikzpicture}
+
\begin{tikzpicture}
\drawcube
\fill[bichain2f] (100-c)--(101-c)--(111-c)--(110-c)--cycle;
\fill[bichain1ff] (000-c)--(100-c)--(101-c)--(111-c)--(011-c)--(010-c)--cycle;
\end{tikzpicture}
+
\begin{tikzpicture}
\drawcube
\fill[bichain2f] (001-c)--(101-c)--(111-c)--(011-c)--cycle;
\fill[bichain1ff] (000-c)--(100-c)--(101-c)--(111-c)--(011-c)--(010-c)--cycle;
\end{tikzpicture}
\\
=
xxx|x0x + xxx|1xx + xxx|xx1\;.
\end{gathered}
\end{equation}

\section{Projective 1-form symmetry and anyon models}
\label{sec:cyww}
In this appendix, we will relate the 1-form symmetry of the path integral to the abelian anyon model of the corresponding topological phase.
Abelian anyon models are known to be in one-to-one correspondence with \emph{metric groups} \cite{Wang2020, Galindo2016}.
A metric group \cite{Davydov2010,Drinfeld2009} is a (finite) abelian group $K$ (denoted additively) together with a function
\begin{equation}
q: K\rightarrow \rr/\zz
\end{equation}
such that
\begin{equation}
q(x)=q(-x)\;,
\end{equation}
and
\begin{equation}
\begin{aligned}
b: K\times K &\rightarrow \rr/\zz\;,\\
b(g,h)&=q(g+h)-q(g)-q(h)
\end{aligned}
\end{equation}
is a non-degenerate symmetric bilinear form.
Physically, the elements of $K$ are the anyons, and the group $K$ defines their fusion.
$q$ is the topological spin of each anyon such that a $2\pi$ rotation of the anyon in space yields a phase $\theta=e^{2\pi iq}$.
Similarly, $e^{2\pi i b(g,h)}$ is the phase obtained from braiding the anyons $g$ and $h$.

The data of a metric group can be organized by choosing $l$ independent generators $\{g_i\}_{0\leq i<l}$ of $K$ identifying $K=\bigoplus_{0\leq i<l} \zz_{m_i}$.
Then, the anyon model is fully specified the spins and braidings between all generators, which we can assemble to a matrix
\begin{equation}
(M_<)_{ij}=
\begin{cases}
0 & \text{if } i<j\\
b(g_i,g_j) & \text{if } i>j\\
q(g_i) & \text{if } i=j
\end{cases}
\;.
\end{equation}
After defining the diagonal matrix $m_{ij}\coloneqq \delta_{i,j} m_i$, the conditions on this data defining a metric group can be expressed as follows.
First, we have
\begin{equation}
\label{eq:anyon_model_condition1}
m^TM_<m =0\;,\quad Mm=0\;,
\end{equation}
using
\begin{equation}
M\coloneqq M_<+M_<^T\;,
\end{equation}
where we think of both equations as valued in $\rr/\zz$.
Further, we demand that there exists $\ovl M\in \rr^{l\times l}$ with $\ovl M/\zz=M$ and
\begin{equation}
\label{eq:anyon_model_condition2}
\det(\ovl Mm)=\pm 1\;,
\end{equation}
that is, $\ovl Mm$ is a unimodular matrix.

More generally, we can define an abelian anyon model by arbitrary matrices $m\in \zz^{l\times l}\cap GL(l)$ and $M_<\in (\rr/\zz)^{l\times l}$, such that the conditions in Eqs.~\ref{eq:anyon_model_condition1} and \ref{eq:anyon_model_condition2} hold.
The anyon fusion group is $K=\zz_l\coloneqq m\zz^{l}\backslash\zz^{l}$, where the $\zz_l$ unit vectors are now potentially dependent generators.
The spin and braiding of these generators are given by $q(g_i)=(M_<)_{ii}$ and $b(g_i,g_j)=(M_<)_{ij}+(M_<)_{ji}$.
So two anyon models with the same $m$ but different $M_<$ and $\widetilde{M_<}$ are equivalent if there is a $Y\in (\rr/\zz)^{l\times l}$ such that $M_<-\widetilde{M_<}=Y-Y^T$.
Also, the same fusion group and anyon model can be represented by different matrices $m$ (which is already true in the case of independent generators).
The generalization to non-diagonal $m$ might seem overkill, but it allows us to elegantly express the anyon theories (in other words, the \emph{Drinfeld center} \cite{Drinfeld2009}) of abelian twisted quantum doubles.
In this case, $m$ is given in Eq.~\eqref{eq:general_anyon_fusion_group}.

Let us consider a few examples for anyon models.
The toric code corresponds to
\begin{equation}
m=
\begin{pmatrix}
2&0\\0&2
\end{pmatrix}
\;,\quad
M_<=
\begin{pmatrix}
0&\frac12\\0&0
\end{pmatrix}
\;,
\end{equation}
if we use $e$ and $m$ as generators.
The double-semion model is represented by
\begin{equation}
m=
\begin{pmatrix}
2&0\\0&2
\end{pmatrix}
\;,\quad
M_<=
\begin{pmatrix}
\frac14&0\\0&-\frac14
\end{pmatrix}
\;,
\end{equation}
using $s$ and $\ovl s$ as generators.
The 3-fermion model on the other hand corresponds to
\begin{equation}
m=
\begin{pmatrix}
2&0\\0&2
\end{pmatrix}
\;,\quad
M_<=
\begin{pmatrix}
\frac12&\frac12\\0&\frac12
\end{pmatrix}
\;,
\end{equation}

In order to find the matrix $M_<$ determining the spin and braiding of the anyons in the twisted quantum doubles, we look at the variance of the path integral under gauge transformations acting on $b$ and $c$.
The phases factors arising from such gauge transformations due to the second condition in Definition~\ref{def:homological_integral} correspond to the twists and braiding of the anyons.
There are two such gauge transformations.
The first gauge transformation acts on $c$ only and is given by
\begin{equation}
c'=c+\delta\gamma\;.
\end{equation}
This transformation acts on the $\zz^k$ lifts as
\begin{equation}
\cbar'=\cbar+\delta\ovl\gamma+f(s_{c,d\gamma}+v_\gamma)\eqqcolon \cbar+\delta\ovl\gamma+fz\;.
\end{equation}
We first compute the variance under adding the $fy$ term,
\begin{equation}
\label{eq:c_periodicity}
\begin{multlined}
S[\Abar,\bbar,\cbar+fy]-S[\Abar,\bbar,\cbar]\\
= \Abar^T (f^{-1})^T\cup f^T z = \Abar^T\cup z = 0\;.
\end{multlined}
\end{equation}
Next, the variance under adding the $\delta\ovl\gamma$ term is given by
\begin{equation}
\begin{multlined}
S[\Abar,\bbar,\cbar+\delta\gamma]-S[\Abar,\bbar,\cbar]\\
= \Abar^T (f^{-1})^T\delta\ovl\gamma = (d\Abar)^T(f^{-1})^T\ovl\gamma = \bbar^T(f^{-1})^T\ovl\gamma\;.
\end{multlined}
\end{equation}
So overall we have
\begin{equation}
\label{eq:c_gauge_variance}
S[A,b,c']-S[A,b,c]= \bbar^T(f^{-1})^T\ovl\gamma\;.
\end{equation}

The second gauge transformation is given in Eq.~\eqref{eq:b_gauge}.
It acts on the $\zz^k$ lifts as follows,
\begin{equation}
\begin{gathered}
\ovl{A'}=\Abar+\ovl\beta+fx\;,\\
\ovl{b'}=\bbar+d\ovl\beta+fy\;,\\
\ovl{c'}=\cbar+\cup f^T(F+F^T)fy+f^Tz\;.
\end{gathered}
\end{equation}
We start by computing the variance under adding the two terms involving $\beta$,
\begin{equation}
\begin{multlined}
S[\Abar+\ovl\beta,\bbar+d\ovl\beta,\cbar]-S[\Abar,\bbar,\cbar]\\
= \textcolor{red}{\ovl\beta^TF\cup d\Abar} + \textcolor{blue}{\Abar^TF\cup d\ovl\beta} + \textcolor{green}{\ovl\beta^TF\cup d\ovl\beta}\\
+\ovl\beta^T(f^{-1})^T\cbar\\
- \ovl\beta^T(F+F^T)\cup\bbar - \Abar^T(\textcolor{blue}{F}+\textcolor{red}{F^T})\cup d\ovl\beta\\ - \ovl\beta^T(\textcolor{green}{F}+F^T)\cup d\ovl\beta\\
+ \textcolor{red}{(d\ovl\beta)^TF\cup_1 d\Abar} + \bbar^TF\cup_1 d\ovl\beta + (d\ovl\beta)^T F\cup_1 d\ovl\beta\\
=\ovl\beta^T(f^{-1})^T\cbar
- \ovl\beta^T(F+F^T)\cup\bbar - \ovl\beta^TF^T\cup d\ovl\beta\\
+ \bbar^TF\cup_1 d\ovl\beta + (d\ovl\beta)^T F\cup_1 d\ovl\beta
\;.
\end{multlined}
\end{equation}
Terms marked in the same color cancel.
Next, we consider the variance under adding the two terms containing $y$,
\begin{equation}
\begin{multlined}
S[\Abar,\bbar+fy,\cbar+\cup f^T(F+F^T)fy]-S[\Abar,\bbar,\cbar]\\
=\Abar^T\cup (F+F^T)fy - \Abar^T(F+F^T)\cup fy\\ + (fy)^TF\cup_1 d\Abar\\
=(fy)^TF\cup_1 d\Abar = y^Tf^TF\cup_1\bbar\;.
\end{multlined}
\end{equation}
The variance under adding the $x$ and $z$ terms is zero, as shown in Eqs.~\eqref{eq:a_periodicity_1form} and \eqref{eq:c_periodicity}.
Adding all four variances consecutively, we get
\begin{equation}
\label{eq:b_gauge_variance}
\begin{multlined}
S[A',b',c']-S[A,b,c]\\
=\ovl\beta^T(f^{-1})^T\cbar
- \ovl\beta^T(F+F^T)\cup\bbar - \ovl\beta^TF^T\cup d\ovl\beta\\
+ \bbar^TF\cup_1 d\ovl\beta + (d\ovl\beta)^T F\cup_1 d\ovl\beta
+y^Tf^TF\cup_1(\bbar+d\ovl\beta)\;.
\end{multlined}
\end{equation}
We can confirm that both gauge variances in Eqs.~\eqref{eq:c_gauge_variance} and \eqref{eq:b_gauge_variance} are independent of $A$, as we argued around Eq.~\eqref{eq:bc_gauge_invariance} for the double-semion case.

While we cannot give a detailed explanation for all the terms in Eqs.~\eqref{eq:c_gauge_variance} and \eqref{eq:b_gauge_variance}, we remark that terms corresponding to the braiding between $b$ fluxes and $c$ charges are bilinear in $b$ and $\gamma$, or in $\beta$ and $c$.
Terms corresponding to the topological twist of $b$ fluxes are bilinear in $\beta$ and $\beta$, and terms corresponding to the topological twist of $c$ charges are bilinear in $\gamma$ and $\gamma$ (and do not appear since the charges are always bosons).
Based on this, we suggest that the $M_<$ and $M$ are given by
\begin{equation}
M_<=\begin{pmatrix} -F & (f^{-1})^T\\ 0 & 0\end{pmatrix}\;,\quad M=\begin{pmatrix}-F-F^T & (f^{-1})^T\\ f^{-1} & 0\end{pmatrix}\;.
\end{equation}
To support this claim, we explicitly check the conditions in Eq.~\eqref{eq:anyon_model_condition1} and \eqref{eq:anyon_model_condition2} with $\ovl M=M$:
\begin{equation}
\begin{gathered}
m^T M_< m\\
\begin{multlined}
= 
\begin{pmatrix}f^T&f^T(F+F^T)f\\0 & f\end{pmatrix}
\begin{pmatrix}-F& (f^{-1})^T\\ 0 & 0\end{pmatrix}\\
\begin{pmatrix}f&0\\f^T(F+F^T)f & f^T\end{pmatrix}
\end{multlined}
\\
=
\begin{pmatrix}f^T&f^T(F+F^T)f\\0 & f\end{pmatrix}
\begin{pmatrix}F^Tf&\idop\\ 0 & 0\end{pmatrix}\\
=
\begin{pmatrix}f^TF^Tf & f^T\\ 0 & 0\end{pmatrix} =0
\;,
\end{gathered}
\end{equation}
\begin{equation}
\begin{multlined}
\ovl Mm = \begin{pmatrix}-F-F^T & (f^{-1})^T\\ f^{-1} & 0\end{pmatrix}\begin{pmatrix}f&0\\f^T(F+F^T)f & f^T\end{pmatrix}\\
=
\begin{pmatrix}0 & \idop \\\idop & 0\end{pmatrix}\;.
\end{multlined}
\end{equation}

\section{Common noise models in terms of circuit perturbation}
\label{sec:noise}
In this appendix, we discuss how the concept of perturbation of a circuit from Section~\ref{sec:fault_tolerance_definition} covers many common types of noise.
For illustration, we consider a $1+1$-dimensional brick-layer circuit of 2-qubit instruments as shown in Eq.~\eqref{eq:qec_circuit}.
As such, a perturbation would mean replacing each channel $M$,
\begin{equation}
\label{eq:whole_channel_perturbation}
\begin{tikzpicture}
\atoms{square,xscale=4}{0/}
\draw[quantum] ([sx=-0.5]0-b)--++(-90:0.4) ([sx=0.5]0-b)--++(-90:0.4) ([sx=-0.5]0-t)--++(90:0.4) ([sx=0.5]0-t)--++(90:0.4);
\draw[classical] (0-t)--++(90:0.6);
\draw[rc,gray,dashed] (-0.8,-0.3)rectangle (0.8,0.3);
\node[gray] at (1,0){$M$};
\end{tikzpicture}\;.
\end{equation}
by some other channel $\widetilde M$.
As mentioned in Section~\ref{sec:fault_tolerance_definition}, perturbations of a larger locality can be obtained, for example, by blocking neighboring channels into a single channel, and then perturbing the blocked channel.
We can also go the other direction, and decompose channels into smaller operations that we perturb individually.
For example, we can represent single-qubit errors by inserting identity channels at all corresponding places.
Then we use this identity channel as the channel $M$ that we perturb,
\begin{equation}
\begin{tikzpicture}
\atoms{square,xscale=4}{0/}
\draw[quantum] ([sx=-0.5]0-b)--++(-90:0.4) ([sx=0.5]0-b)--++(-90:0.8) ([sx=-0.5]0-t)--++(90:0.4) ([sx=0.5]0-t)--++(90:0.4);
\draw[rc,gray,dashed] (0.3,-0.4)rectangle (0.7,-0.8);
\node[gray] at (0.9,-0.6){$M$};
\draw[classical] (0-t)--++(90:0.6);
\end{tikzpicture}\;,\quad
M=
\begin{tikzpicture}
\draw[quantum] (0,0)--++(90:0.8);
\end{tikzpicture}\;.
\end{equation}
That is, we replace $M$ by a channel $\widetilde M$ that applies a unitary $U$ with a probability $p$,
\begin{equation}
\label{eq:single_incoherent_perturbation}
\widetilde M=
(1-p)\cdot
\begin{tikzpicture}
\draw[quantum] (0,0)--++(90:0.8);
\end{tikzpicture}+
p\cdot
\begin{tikzpicture}
\atoms{circ,lab={t=$U$,p=0:0}}{x0/p={-0.25,0}, x1/p={0.25,0}}
\draw (x0)--++(90:0.6) (x0)--++(-90:0.6) (x1)--++(90:0.6) (x1)--++(-90:0.6);
\end{tikzpicture}\;.
\end{equation}
This perturbation has incoherent perturbation strength $p$.
If $U$ is a Pauli-$X$ or Pauli-$Z$ operator, then this is ordinary bit-flip or phase-flip noise.
However, $U$ does not have to be a Pauli or even Clifford unitary.
Instead of a unitary $U$, we could also use any channel $C$.

We can also represent a coherent qubit error where we directly replace the identity channel by a unitary $U=e^{i\epsilon H}$ with $\|H\|=O(1)$ and small $\epsilon$,
\begin{equation}
\widetilde M=
\begin{tikzpicture}
\atoms{circ,lab={t=$U$,p=0:0}}{x0/p={-0.25,0}, x1/p={0.25,0}}
\draw (x0)--++(90:0.6) (x0)--++(-90:0.6) (x1)--++(90:0.6) (x1)--++(-90:0.6);
\end{tikzpicture}\;.
\end{equation}
This is a perturbation of strength $O(\epsilon)$ for small $\epsilon$.
However, since $\widetilde M$ is extremal, the incoherent perturbation strength is $1$ no matter how small $\epsilon$ is, and we do not show fault-tolerance under such perturbations.

Measurement errors can be modeled by inserting a classical identity stochastic map at each measurement outcome,
\begin{equation}
\begin{tikzpicture}
\atoms{square,xscale=4}{0/}
\draw[quantum] ([sx=-0.5]0-b)--++(-90:0.4) ([sx=0.5]0-b)--++(-90:0.4) ([sx=-0.5]0-t)--++(90:0.4) ([sx=0.5]0-t)--++(90:0.4);
\draw[classical] (0-t)--++(90:0.8);
\draw[rc,gray,dashed] (-0.2,0.4)rectangle (0.2,0.8);
\node[gray] at (0.4,0.8){$M$};
\end{tikzpicture}\;,\quad
M=
\begin{tikzpicture}
\draw[classical] (0,0)--++(90:0.8);
\end{tikzpicture}\;.
\end{equation}
Then we perturb this identity channel to
\begin{equation}
\widetilde M=
\begin{tikzpicture}
\atoms{circ,lab={t=$C$,p=0:0}}{c/p={0,0.7}}
\draw[classical] (c)--++(90:0.4) (c)--++(-90:0.4);
\end{tikzpicture}
=
\begin{pmatrix}
1-p & p\\ p & 1-p
\end{pmatrix}
\;,
\end{equation}
that is, we flip the measurement outcome with probability $p$.
Again this perturbation has incoherent strength $p$.

Perturbations of circuits cannot only describe single-qubit errors before or after syndrome measurements, but also during the syndrome readout.
That is, they can describe circuit-level errors.
As an explicit example, assume that our basic 2-qubit instrument is a projective $ZZ$ measurement.
We perform this measurement by preparing a $\ket0$ auxiliary qubit, performing $CX$ operations on it controlled by each qubit, and then performing a $Z$ measurement on the auxiliary qubit,
\begin{equation}
\begin{tikzpicture}
\atoms{square,xscale=4}{0/}
\draw[quantum] ([sx=-0.5]0-b)--++(-90:0.4) ([sx=0.5]0-b)--++(-90:0.4) ([sx=-0.5]0-t)--++(90:0.4) ([sx=0.5]0-t)--++(90:0.4);
\draw[classical] (0-t)--++(90:0.6);
\end{tikzpicture}
=
\begin{tikzpicture}
\atoms{square,xscale=1.6,yscale=1.4,lab={t=$\ket0$,p={0,0}}}{0/}
\atoms{square,xscale=4,lab={t=$CX$,p={0,0}}}{1/p={-0.5,0.6}, 2/p={0.5,1.6}}
\atoms{square,xscale=1.6,yscale=1.4,lab={t=$M_Z$,p={0,0}}}{3/p={0,2.2}}
\draw[quantum] (0-t)--([sx=0.5]1-b) ([sx=0.5]1-t)--([sx=-0.5]2-b) ([sx=-0.5]2-t)--(3-b) ([sx=-0.5]1-b)--++(-90:0.8) ([sx=-0.5]1-t)--++(90:1.8) ([sx=0.5]2-b)--++(-90:1.8) ([sx=0.5]2-t)--++(90:0.8);
\draw[classical] (3-t)--++(90:0.4);
\draw[gray,dashed,rc] (-0.3,0.9)rectangle(0.3,1.3);
\node[gray] at (0.5,1.1){$M$};
\end{tikzpicture}
\;.
\end{equation}
As shown, we insert an identity channel $M$ in between the two $CX$ unitary channels, and use this identity as the channel to perturb.
That is, we replace $M$ by some incoherent single-qubit noise as in Eq.~\eqref{eq:single_incoherent_perturbation}.

All of the perturbations described so far were just special cases of perturbing the fill instrument as shown in Eq.~\eqref{eq:whole_channel_perturbation}.
As an example for noise with a larger locality, consider 2-qubit noise.
To represent 2-qubit noise, we insert a 2-qubit identity channel acting on two nearest-neighbor qubits,
\begin{equation}
\begin{tikzpicture}
\atoms{square,xscale=4}{0/, 1/p={2,0}}
\draw[quantum] ([sx=-0.5]0-b)--++(-90:0.4) ([sx=0.5]0-b)--++(-90:0.8) ([sx=-0.5]0-t)--++(90:0.4) ([sx=0.5]0-t)--++(90:0.4) ([sx=-0.5]1-b)--++(-90:0.8) ([sx=0.5]1-b)--++(-90:0.4) ([sx=-0.5]1-t)--++(90:0.4) ([sx=0.5]1-t)--++(90:0.4);
\draw[classical] (0-t)--++(90:0.4) (1-t)--++(90:0.4);
\draw[rc,gray,dashed] (0.3,-0.4)rectangle (1.7,-0.8);
\node[gray] at (1.9,-0.6){$M$};
\end{tikzpicture}\;,\quad
M=
\begin{tikzpicture}
\draw[quantum] (0,0)--++(90:0.8) (0.5,0)--++(90:0.8);
\end{tikzpicture}\;.
\end{equation}
Note that in our toy example circuit in Eq.~\eqref{eq:qec_circuit}, the two qubits on which $M$ acts are outputs of a single preceding 2-qubit instrument, so this is again a special case of the perturbation in Eq.~\eqref{eq:whole_channel_perturbation}.
However, this might not be the case for other circuit geometries or, for example, for next-nearest-neighbor pairs of qubits.
We then perturb $M$ with some 2-qubit noise channel $C$,
\begin{equation}
\widetilde M=
(1-p)\cdot
\begin{tikzpicture}
\draw[quantum] (0,0)--++(90:0.8) (0.5,0)--++(90:0.8);
\end{tikzpicture}+
p\cdot
\begin{tikzpicture}
\atoms{square,xscale=2.5,lab={t=$C$,p={0,0}}}{0/}
\draw[quantum] ([sx=-0.25]0-b)--++(-90:0.4) ([sx=0.25]0-b)--++(-90:0.4) ([sx=-0.25]0-t)--++(90:0.4) ([sx=0.25]0-t)--++(90:0.4);
\end{tikzpicture}\;.
\end{equation}
Analogously, we can also introduce 3 or more-qubit errors.
To do this for the toy example circuit shown in Eq.~\eqref{eq:qec_circuit}, we would have to break the translation symmetry of the circuit and choose a larger unit cell.

Another possibility is to consider errors that occur simultaneously at two time-like separated places in the circuit.
For example, consider a spin-flip error that occurs both at one qubit, and at another neighboring qubit one time step later.
To this end, we rewrite our circuit as follows,
\begin{equation}
\begin{tikzpicture}
\atoms{square,xscale=4}{0/, 1/p={2,0}}
\atoms{z2}{z/p={0.9,-0.6}}
\atoms{square,dcross,small}{x/p={1.5,0.5}}
\draw[quantum] ([sx=-0.5]0-b)--++(-90:0.4) ([sx=0.5]0-b)--++(-90:0.8) ([sx=-0.5]0-t)--++(90:0.4) ([sx=0.5]0-t)--++(90:0.4) ([sx=-0.5]1-b)--++(-90:0.8) ([sx=0.5]1-b)--++(-90:0.4) ([sx=-0.5]1-t)--(x) (x)--++(90:0.4) ([sx=0.5]1-t)--++(90:0.4);
\draw[classical,rc] (z)|-(x);
\draw[rc,gray,dashed] (0.3,-0.4)rectangle (1.1,-0.8);
\node[gray] at (1.3,-0.6){$M$};
\draw[classical] (0-t)--++(90:0.3) (1-t)--++(90:0.3);
\end{tikzpicture}
\;,\quad
M
=
\begin{tikzpicture}
\draw[quantum] (0,0)--++(90:0.6);
\atoms{z2}{0/p={0.4,0.3}}
\draw[classical] (0)--++(90:0.3);
\end{tikzpicture}
\;.
\end{equation}
The empty circle tensor denotes preparation of a classical bit in the $0$ state.
The box with the cross is a classically controlled single-qubit channel $C_1$,
\begin{equation}
\begin{tikzpicture}
\atoms{square,small,dcross}{0/}
\draw[quantum] (0)--++(90:0.4) (0)--++(-90:0.4);
\draw[rc,classical] (0)--++(180:0.4)--++(-90:0.4);
\end{tikzpicture}
\coloneqq
\begin{tikzpicture}
\draw[quantum] (0,0)--++(90:0.6);
\atoms{z2}{0/p={-0.4,0.3}}
\draw[classical] (0)--++(-90:0.3);
\end{tikzpicture}
+
\begin{tikzpicture}
\atoms{square,xscale=1.4,yscale=1.2,lab={t=$C_1$,p=0:0}}{c/}
\draw[quantum] (c-b)--++(-90:0.4) (c-t)--++(90:0.4);
\atoms{z2,lab={t=$1$,p=90:0.25}}{0/p={-0.6,0}}
\draw[classical] (0)--++(-90:0.3);
\end{tikzpicture}
\end{equation}
Then we replace $M$ by the perturbed channel
\begin{equation}
\widetilde M
=
(1-p)\cdot
\begin{tikzpicture}
\draw[quantum] (0,0)--++(90:0.6);
\atoms{z2}{0/p={0.4,0.3}}
\draw[classical] (0)--++(90:0.3);
\end{tikzpicture}
+p\cdot
\begin{tikzpicture}
\atoms{square,xscale=1.4,yscale=1.2,lab={t=$C_0$,p=0:0}}{c/}
\draw[quantum] (c-b)--++(-90:0.4) (c-t)--++(90:0.4);
\atoms{z2,lab={t=$1$,p=-90:0.25}}{0/p={0.6,0}}
\draw[classical] (0)--++(90:0.3);
\end{tikzpicture}
\;.
\end{equation}
That is, we apply the single-qubit noise channel $C_0$ with probability $p$.
If we do apply it, then we also apply $C_1$ to another qubit at a later time.
This perturbation has incoherent strength $p$.

Another way to introduce correlated noise is to couple the circuit to a weakly correlated classical process.
To this end, we rewrite the circuit such that we regularly prepare classical bits in state $0$, and erase them in the next time step.
Depending on the state of these bits, we perform a single-qubit error channel $C_1$ or not,
\begin{equation}
\begin{tikzpicture}
\atoms{square,xscale=4}{0/, 1/p={1,1.4}}
\atoms{square,dcross,small}{x0/p={0.5,0.7}, x1/p={0.5,2.1}, x2/p={0.5,-0.7}}
\atoms{z2,scale=0.7}{z0/p={0.8,0.35}, z1/p={0.8,1.75}}
\atoms{delta,scale=0.7}{d0/p={0.8,-0.35}, d1/p={0.8,1.05}, dm0/p={0.8,-0.7}, dm1/p={0.8,0.7}, dm2/p={0.8,2.1}}
\draw[quantum] (x2)--([sx=0.5]0-b) ([sx=0.5]0-t)--(x0) (x0)--([sx=-0.5]1-b) ([sx=-0.5]1-t)--(x1) (x1)--++(90:0.3) (x2)--++(-90:0.3) ([sx=-0.5]0-b)--++(-90:0.3) ([sx=-0.5]0-t)--++(90:0.3) ([sx=0.5]1-b)--++(-90:0.3) ([sx=0.5]1-t)--++(90:0.3);
\draw[classical] (0-t)--++(90:0.3) (1-t)--++(90:0.3);
\draw[classical] (dm0)--(d0) (dm0)--++(-90:0.3) (z0)--(dm1) (dm1)--(d1) (z1)--(dm2) (dm2)--++(90:0.3) (dm0)--(x2) (dm1)--(x0) (dm2)--(x1);
\draw[rc,gray,dashed] ($(d0)+(-0.2,-0.2)$)rectangle ($(z0)+(0.2,0.2)$) ($(d1)+(-0.2,-0.2)$)rectangle ($(z1)+(0.2,0.2)$);
\node[gray] at (1.2,0){$M$};
\node[gray] at (1.2,1){$M$};
\end{tikzpicture}\;,\quad
M=
\begin{tikzpicture}
\atoms{z2}{z/p={0,0.5}}
\atoms{delta}{d/}
\draw[classical] (d)--++(-90:0.4) (z)--++(90:0.4);
\end{tikzpicture}\;.
\end{equation}
Here, the filled circle tensors denote erasure of a classical bit (if there is one index), or copying of a classical bit (if there are three indices).
Now consider a perturbation where we prepare each classical bit in state $1$ instead of $0$ with probability $p_1$ if the last bit was $0$, and with probability $p_2$ if the last bit was $1$,
\begin{equation}
\widetilde M
=
\begin{pmatrix}
1-p_1 & p_1\\
1-p_2 & p_2
\end{pmatrix}\;.
\end{equation}
If $p_2>p_1$, the bit-flip errors are correlated, meaning that if an error occurs at a time $t$, it is more likely to again occur at time $t+1$, and at $t+n$ where the bias is exponentially decaying in $n$.
That is, we can introduce errors with a finite correlation length in time.

Correlations between errors may also spread in space.
To this end, we rewrite the circuit by adding bits that are initialized in state $0$, moved by one lattice site, used to control an action on a qubit, and then discarded,
\begin{equation}
\begin{tikzpicture}
\atoms{square,xscale=4}{0/, 1/p={2.6,0}, 2/p={1.3,1.5}}
\atoms{delta}{d0/p={1.1,-0.2},d1/p={1.5,-0.2},d2/p={2.4,1.3},d5/p={0.2,1.3},d6/p={0.4,1},d7/p={2.2,1},d8/p={1.7,-0.5},d9/p={0.9,-0.5}}
\atoms{z2}{z0/p={1.1,0.2},z1/p={1.5,0.2}}
\atoms{square,dcross,small}{x0/p={0.5,-0.5}, x1/p={2.1,-0.5}, x2/p={0.8,1}, x3/p={1.8,1}, x4/p={-0.5,-0.5}, x5/p={3.1,-0.5}}
\draw[quantum] ([sx=-0.5]0-b)--(x4) (x4)--++(-110:0.4) ([sx=0.5]0-b)--(x0) (x0)--++(-70:0.4) ([sx=-0.5]0-t)--++(110:0.4) ([sx=0.5]0-t)--(x2) (x2)--([sx=-0.5]2-b) ([sx=0.5]2-t)--++(90:0.3) ([sx=-0.5]2-t)--++(90:0.3) ([sx=-0.5]1-t)--(x3) (x3)--([sx=0.5]2-b) ([sx=0.5]1-b)--(x5) (x5)--++(-70:0.4) ([sx=-0.5]1-b)--(x1) (x1)--++(-110:0.5) ([sx=0.5]1-t)--++(70:0.4);
\draw[rc,gray,dashed] (0.9,-0.4)rectangle (1.7,0.4);
\node[gray] at (1.3,0.6){$M$};
\draw[classical] (0-t)--++(90:0.3) (1-t)--++(90:0.3) (2-t)--++(90:0.3);
\draw[classical] (z0)--(d6) (d6)--(d5) (d6)--(x2) (z1)--(d7) (d7)--(x3) (d7)--(d2) (d0)--(d9)--(x0) (d9)--++(-135:0.5) (d1)--(d8)--(x1) (d8)--++(-45:0.5) (x4)--++(180:0.3) (x5)--++(0:0.3);
\end{tikzpicture}\;.
\end{equation}
An example for a perturbation is given by
\begin{equation}
M=
(1-p_1-2p_2)\cdot
\begin{tikzpicture}
\atoms{z2}{z0/p={0,0.5}, z1/p={0.5,0.5}}
\atoms{delta}{d0/, d1/p={0.5,0}}
\draw[classical] (d0)--++(-90:0.4) (d1)--++(-90:0.4) (z0)--++(90:0.4) (z1)--++(90:0.4);
\end{tikzpicture}
+p_1\cdot
\begin{tikzpicture}
\atoms{z2odd}{z0/p={0,0.5}, z1/p={0.5,0.5}}
\atoms{delta}{d0/, d1/p={0.5,0}}
\draw[classical] (d0)--++(-90:0.4) (d1)--++(-90:0.4) (z0)--++(90:0.4) (z1)--++(90:0.4);
\end{tikzpicture}
+p_2\cdot
\begin{tikzpicture}
\draw[classical] (0,0)--(0.5,0.7) (0.5,0)--(0,0.7);
\end{tikzpicture}
+p_2\cdot
\begin{tikzpicture}
\draw[classical] (0,0)--++(0,0.7) (0.5,0)--++(0,0.7);
\end{tikzpicture}
\;.
\end{equation}
Roughly, this corresponds to bit-flip errors which are generated with probability $p_1$, and then persist at later times performing a random walk with decay rate $\sim p_2$.
Instead of introducing correlations using auxiliary classical bits, we could also introduce ``quantum'' correlations using auxiliary qubits.

\end{document}